\documentclass[usenatbib]{mnras}

\usepackage{graphicx}
\usepackage{natbib}
\usepackage{rotating}
\usepackage[toc,page]{appendix}
\usepackage{aas_macros}
\usepackage{multirow}
\usepackage{placeins}
\usepackage{epstopdf}
\usepackage{epsfig}
\usepackage{amssymb}
\usepackage{amsmath}
\usepackage{caption}
\usepackage{float}
\usepackage[maxfloats=100]{morefloats}
\def\onecolumn{%
  \global\columnwidth\textwidth
  \global\hsize\columnwidth
  \global\linewidth\columnwidth
  }

\newcommand{\Atm}{\ensuremath{A_{TM}}}
\newcommand{\Btm}{\ensuremath{B_{TM}}}

\newcommand{\gamtm}{\ensuremath{\gamma_{TM}}}

\newcommand{\Alm}{\ensuremath{A_{LM}}}
\newcommand{\Blm}{\ensuremath{B_{LM}}}
\newcommand{\gamlm}{\ensuremath{\gamma_{LM}}}
\newcommand{\intlm}{\ensuremath{\delta_{LM}}}

\newcommand{\fbcs}{\ensuremath{F_{BCS}}}
\newcommand{\lbcs}{\ensuremath{L_{BCS}}}
\newcommand{\lcxo}{\ensuremath{L_{CXO}}}

\newcommand{\fcxo}{\ensuremath{F_{CXO}}}
\newcommand{\lbol}{\ensuremath{L_{bol}}}
\newcommand{\llim}{\ensuremath{L_{lim}}}
\newcommand{\lbcsh}{\ensuremath{\hat{L}_{BCS}}}
\newcommand{\lcxoh}{\ensuremath{\hat{L}_{CXO}}}
\newcommand{\Mh}{\ensuremath{\hat{M}}}

\newcommand{\rf}{\ensuremath{r_{500}}}

\title[The LM Relation of Luminous Clusters]{Chandra Measurements of a Complete Sample of X-ray Luminous
  Galaxy Clusters: The Luminosity--Mass Relation}
\author[P. A. Giles et al.]{
\parbox[h]{\textwidth}{P. A. Giles$^1$\thanks{E-mail:
P.Giles@bristol.ac.uk}, B. J. Maughan$^1$, H. Dahle$^2$,
  M. Bonamente$^{3,4}$, D. Landry$^3$, C. Jones$^5$, M. Joy$^4$,
  S. S. Murray$^5$, N. van der Pyl$^1$
}
\vspace*{12pt} \\
\parbox[h]{\textwidth}{
$^1$ HH Wills Physics Laboratory, Tyndall Avenue, Bristol, BS8 1TL,
  UK \\
$^2$ Institute of Theoretical Astrophysics, University of Oslo,
P.O. Box 1029, Blindern, N-0315 Oslo, Norway \\
$^3$ Physics Department, University of Alabama in Huntsville,
  Huntsville, AL, U.S.A \\
$^4$ NASA National Space Science and Technology Center, Huntsville,
  AL, U.S.A \\
$^5$ Harvard-Smithsonian Center for Astrophysics, 60 Garden Street,
Cambridge, MA 02138, USA
}}
\begin{document}

\date{Accepted 2016 October 9.}

\pagerange{\pageref{firstpage}--\pageref{lastpage}} \pubyear{2016}

\maketitle

\label{firstpage}

\begin{abstract}

We present the results of work involving a statistically complete
sample of 34 galaxy clusters, in the redshift range 0.15$\le z \le$0.3 
observed with {\em Chandra}.  We investigate the luminosity-mass
($LM$) relation for the cluster sample, with the masses obtained via a
full hydrostatic mass analysis.  We utilise a method to fully
account for selection biases when modeling the $LM$ relation, and find
that the $LM$ relation is significantly different than the relation
modelled when not account for selection effects.  We find that
  the luminosity of our clusters is 2.2$\pm$0.4 times higher (when
  accounting for selection effects) than the average for a given mass,
  its mass is 30\% lower than the population average for a given
  luminosity. Equivalently, using the $LM$ relation measured from this
  sample without correcting for selection biases would lead to the
  underestimation by 40\% of the average mass of a cluster with a
  given luminosity.  Comparing the hydrostatic masses to mass
  estimates determined from the $Y_{X}$ parameter, we find that they
  are entirely consisent, irrespective of the dynamical state of the
  cluster.           

\end{abstract}


\section{Introduction}

Clusters of galaxies are the largest gravitationally-collapsed
structures in the Universe.  Studying properties such as the number 
density of clusters and details of their growth from the highest
density perturbations in the early Universe, offers insight into the 
underlying cosmology
\citep[e.g.][]{2008MNRAS.387.1179M,2009ApJ...692.1060V,2015arXiv150201597P}.
The study of galaxy clusters has been transformed with the launch of
powerful X-ray telescopes such as {\em Chandra} and {\em XMM}, which have
allowed the study of
the X-ray emitting intracluster medium (ICM) with unprecedented detail 
and accuracy.  Cluster properties have been used widely in the
determination of cosmological parameters. Cosmological studies
utilising clusters include investigating the cluster temperature
function
\citep[e.g.][]{1991ApJ...372..410H,1997ApJ...489L...1H,1998MNRAS.298.1145E,2002A&A...383..773I},
scaling relations such as the luminosity-mass
\citep[e.g][]{2002ApJ...567..716R,2006ApJ...648..956S} and the
temperature-mass \citep[e.g.][]{2006ApJ...640..691V} relations, using
the gas mass fraction, f$_{\rm gas}$
\citep{2008MNRAS.383..879A,2014MNRAS.440.2077M}, the cluster 
mass function \citep[e.g.][]{2009ApJ...692.1060V} and the cluster
luminosity function
\citep[e.g.][]{2014A&A...570A..31B,2016A&A...592A...2P}, to place
constraints on various cosmological parameters.  Since one of the most 
important ingredients of these cosmological studies is the cluster
mass, large efforts have been undertaken to accurately determine this 
quantity.  One such method involves the construction of radial
temperature and gas density profiles of the ICM, and under the
assumption of hydrostatic equilibrium, the cluster mass can be
determined.      

Observations using X-rays have become a well established method of
estimating cluster masses,  However, constructing temperature profiles 
for individual clusters for use in a hydrostatic mass analysis
generally require long telescope exposure times, and not is not
feasible for large samples of clusters. Therefore, deriving well
calibrated scaling relations between simple cluster observables and
mass is of crucial importance for using clusters as cosmological
probes \citep[e.g.][]{2010MNRAS.406.1773M}.  The X-ray luminosity
($L$) is one of the easiest cluster properties to obtain, and has had
a rich history in its scaling with mass ($M$).  Under the assumption
of self-similarity \citep{1986MNRAS.222..323K}, the $LM$ relation
(throughout this work we use the notation $LM$ when generally
discussing the luminosity-mass relation) is expected to follow a
relationship of $L \propto M^{4/3}$.  However, observational studies
of the $LM$ relation have found a slope steeper than the self-similar
expectation
\citep[e.g.][]{2008MNRAS.387L..28R,2009A&A...498..361P,2014ApJ...794...48C}.
The most widely accepted theory for the steep slope of the $LM$
relation, is due to heating from sources such as supernovae (Sne) and
active galactic nuclei (AGN) feedback
\citep[e.g][]{2010MNRAS.408.2213S,2011MNRAS.412.1965M,2014MNRAS.441.1270L}.
This causes gas to be expelled from the inner region,
hence suppressing the luminosity.  This effect should be larger in
lower mass systems due to the shallower potential well, which
therefore causes the observed steepening of the $LM$ relation.  

A complication of measuring the $LM$ relation is that the cluster
samples used are traditionally X-ray selected, with X-ray flux limited
samples suffering from two forms of selection bias, Malmquist bias, where
higher luminosity clusters are detectable out to higher redshifts and
so occupy a larger survey volume, and Eddington bias, where in the
presence of intrinsic or statistical scatter in luminosity for a given
mass, objects above a flux limit will have above-average luminosities for
their mass. Due to the steep slope of the cluster mass function, the
Eddington bias is amplified, resulting in a net movement of lower mass
objects into a flux limited sample.  The consequence of biases on the
observed $LM$ relation is to bias the normalisation high and the slope
low \citep[see][]{2011ARA&A..49..409A}.  Therefore, taking these biases
into account is paramount when modeling cluster scaling
relations, in order to uncover the true nature of any
non-gravitational heating which drives departures from self-similar
behavior with mass or redshift.  Although scaling relation
studies have had a rich history, at the present time only a small
number of published relations attempt to account for selection biases
\citep[e.g][]{2006ApJ...648..956S,2007MNRAS.382.1289P,2009A&A...498..361P,2009ApJ...692.1033V,2012A&A...546A...6A,2015A&A...573A..75B,2015A&A...573A.118L,2015MNRAS.450.3675S},
while \cite{2010MNRAS.406.1759M} (hereafter M10a) provides the most
robust handling of selection effects to date.

Deriving cluster masses through X-ray observations with the assumption
that the ICM is in hydrostatic equilibrium, is not always valid, as
some clusters have complex temperature structures due to processes
such as merger events.  Relaxed systems have traditionally been
  used for the determination of X-ray masses, as departures from
  hydrostatic equilibrium are minimized for these systems
\citep[e.g.][]{2007ApJ...668....1N}.  Therefore methods have
been developed to infer the cluster dynamical state 
\citep[e.g][]{2006MNRAS.373..881P,2015MNRAS.449..199M}.  Furthermore,
clusters that appear to host a cool core (CC) are frequently used for
mass derivations as they are believed to be dynamically relaxed.
However, the presence of a CC alone cannot be used to accurately
determined the dynamical state.

Many methods are used to infer the presence of a CC 
\citep[see][for a comprehensive study]{2010A&A...513A..37H}, including
measuring the central temperature drop
\citep[e.g.][]{2006ApJ...639...64O}, the central cooling time 
\citep[e.g.][]{2005MNRAS.359.1481B,2010A&A...521A..64S,2011A&A...532A.133M},
the core entropy \citep[e.g.][]{2011MNRAS.418.1089C} and the cuspiness
of the gas density profile \citep{2007hvcg.conf...48V}.  Frequently,
cheaply obtainable cluster properties such as luminosity and
temperature are used as a mass proxy, using well calibrated scaling
relations, to calculate the masses of large cluster samples.  This
first requires constructing scaling relations for dynamically relaxed
clusters, and then inferring the cluster masses from these relations.
It has been found however that using CC clusters sometimes results in
larger scatter of the $LM$ scaling relation compared to non-Cool Core
(NCC) clusters \citep[e.g.][]{2006ApJ...639...64O}.
\cite{2012MNRAS.421.1583M} found that by defining a cluster sub-sample
using clusters appearing both dynamically relaxed and hosting a cool
core, the luminosity-temperature ($LT$) relation appears self-similar,
compared to unrelaxed and non-cool core clusters. While this method
may be the preferred choice for defining sub-samples of clusters for
mass derivations, this limits the cluster sample size used for mass
calculations, thus the derived scaling relation may not be
representative of the whole cluster population.

This paper aims to measure the masses for a complete sample of 34
clusters to measure the X-ray luminosity--mass scaling relation,
utilising hydrostatic mass estimates and fully accounting for
selection effects.  The outline of the paper is as follows.  In
Sect.~\ref{sec:data} we discuss the sample selection and data
analysis.  Sect.~\ref{sec:analysis} details the cluster analysis and
determines the dynamical state of individual clusters.  Notes on
individual clusters are given in 
Sect.~\ref{sec:notes}.  Our results are presented in
Sect.~\ref{sec:xscaling}.  The discussion and conclusions are
presented in Sect.~\ref{sec:disc} and Sect.~\ref{sec:conc}
respectively.  Throughout this paper we assume a {\em WMAP}9 cosmology
of H$_{0}$=69.7 km s$^{-1}$ Mpc$^{-1}$, $\Omega_{\rm M}$=0.282,
$\Omega_{\Lambda}$=0.718 and $\sigma_8$=0.817
\citep{2013ApJS..208...19H}. 

\section{Sample and Data Preparation}
\label{sec:data}

\begin{figure}
\begin{center}
\includegraphics[width=8.8cm, clip=true]{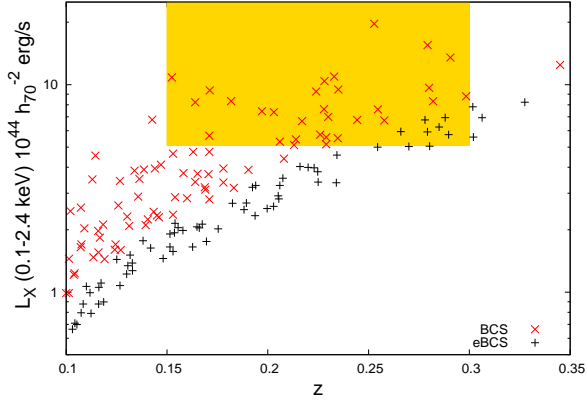} 
\end{center}
\vspace{-1.0cm}
\caption[]{\small{Plot of the luminosity-redshift distribution of the
    BCS and eBCS clusters.  The yellow shaded region highlights the
    region enclosed by the luminosity and redshift cuts imposed to
    defined our cluster sample (see Sect.~\ref{sec:data}).}\label{fig:lz}}
\end{figure}

The sample of clusters used in our analysis was defined by the
conditions given in \cite{2006ApJ...653..954D}.  The sample of
clusters represents a complete sample of X-ray luminous clusters
taken from the RASS-based, X-ray flux limited ROSAT Brightest Cluster
Sample (BCS) of \cite{1998MNRAS.301..881E} and its low-flux extension
(eBCS, \cite{2000MNRAS.318..333E}).  \cite {2006ApJ...653..954D}
imposed a lower cutoff in X-ray luminosity of  
L$_{\rm X,0.1-2.4keV}$ = 6 $\times$ 10$^{44}$ erg s$^{-1}$ (the limit
based upon a cosmology assuming $\Omega_{\rm M}$=0.3,
$\Omega_{\Lambda}$=0.7 and h=0.7), corresponding to
a sample of 36 clusters within the redshift range 0.15 $\leq$ z $\leq$
0.30. Figure~\ref{fig:lz} plots the luminosity-redshift distribution
of the BCS and eBCS, with the yellow shaded region highlighting the
region enclosed by the luminosity and redshift cuts defining our
sample selection.  Note that two of the clusters were dropped from the
cluster sample, as detailed below, leading to a final sample of 34
clusters.  

The cluster A689 satisfies these selection criteria, but was noted in
the original detection as having a large portion of its flux
coming from embedded point sources, and was therefore excluded from
the cluster sample. The source of the embedded point source was found
to be a central BL-Lac object \citep{2012MNRAS.419..503G}, with the
re-analysis determining a cluster X-ray luminosity $\sim$10 times
lower than quoted in the BCS, well below the sample cutoff X-ray
luminosity. Furthermore, we found that the redshift given for the
cluster Zw5768 in \cite{1998MNRAS.301..881E} was incorrect. With
  the correct (lower) redshift, the cluster drops below the luminosity
  limit of our sample and was rejected (see Sect.~\ref{sec:notes}).
For the X-ray analysis we obtained {\em Chandra} observations to
complete the sample and downloaded archived observations of the
remaining clusters from the {\em Chandra} data archive.

\begin{table*}
\caption[]{\small{Cluster Sample and {\em Chandra}
    Observations.}\label{tab:clustsamp}}
\begin{tabular}{lccccccc}
\hline\hline
 & R.A. & Dec. & & Lx$_{\rm BCS,0.1-2.4 keV}$ & \multicolumn{3}{c}{{\em Chandra} Observations} \\
\cline{6-8}
Cluster & (J2000.0) & (J2000.0) & z & (10$^{44}$ ergs s$^{-1}$) & ObsID & Aim Point & Exposure$^{\dagger}$
(ks) \\
\hline
A2204 & 16 32 47.04 & +05 34 31.26 & 0.152 & 12.51$\pm$1.34 & 6104 7940 & I & 86.8 \\
RXJ1720.1+2638 & 17 20 10.08 & +26 37 29.28 & 0.164 & 9.58$\pm$1.08 & 1453 3224 4361 &
I & 45.2 \\
A586 & 07 32 20.40 & +31 37 56.28 & 0.171 & 6.64$\pm$1.30 & 11723 & I & 9.91 \\
A1914 & 14 23 00.96 & +37 49 33.96 & 0.171 & 10.99$\pm$1.11 & 3593 & I & 18.9 \\
A665 & 08 30 57.36 & +65 50 33.36 & 0.182 & 9.84$\pm$1.54 & 12286 13201 & I & 26.4 \\
A115 & 00 55 50.69 & +26 24 37.80 & 0.197 & 8.90$\pm$2.13 & 13458 13459 15578 & I & 282 \\
A520 & 04 54 10.00 & +02 55 18.16 & 0.203 & 8.85$\pm$1.99 & 9424 9425 9426 9430 & I & 447 \\
A963 & 10 17 30.36 & +39 02 53.88 & 0.206 & 6.39$\pm$1.18 & 903 & S & 35.8 \\
A1423 & 11 57 19.28 & +33 36 41.08 & 0.213 & 6.19$\pm$1.34 & 538 11724 & I & 35.6 \\
A773 & 09 17 53.04 & +51 43 39.36 & 0.217 & 8.99$\pm$1.35 & 533 3588 5006 & I & 40.1 \\
A1763 & 13 35 18.24 & +40 59 59.28 & 0.223 & 9.32$\pm$1.33 & 3591 & I & 19.6 \\
A2261 & 17 22 27.12 & +32 07 56.64 & 0.224 & 11.32$\pm$1.55 & 5007 & I & 24.1 \\
A1682 & 13 06 50.40 & +43 33 26.28 & 0.226 & 7.02$\pm$1.37 & 11725 & I & 19.9 \\
A2111 & 15 39 41.52 & +34 25 50.16 & 0.229 & 6.83$\pm$1.66 & 544 11726 & I & 31.2 \\
Zw5247 & 12 34 18.96 & +09 46 12.86 & 0.229 & 6.32$\pm$1.58 & 11727 & I & 19.7 \\
A267 & 01 52 42.14 & +01 00 41.30 & 0.230 & 8.57$\pm$1.80 & 1448 & I & 7.29 \\
A2219 & 16 40 20.40 & +46 42 29.52 & 0.230 & 12.74$\pm$1.37 & 14355 14356 14431 & I & 118 \\
A2390 & 21 53 36.72 & +17 41 44.52 & 0.233 & 13.43$\pm$3.14 & 4193 & S & 22.0 \\
Zw2089 & 09 00 36.96 & +20 53 40.20 & 0.235 & 6.79$\pm$1.76 & 10463 & S & 40.1 \\
RXJ2129.6+0005 & +21 29 40.10 & 00 05 20.91 & 0.235 & 11.67$\pm$2.92 & 552 9370 & I & 39.6 \\
A1835 & 14 01 10.92 & +02 52 42.47 & 0.253 & 24.49$\pm$3.35 & 6880 6881 7370 & I & 193.2 \\
A68 & 00 37 60.09 & +09 09 33.05 & 0.255 & 9.48$\pm$2.61 & 3250 & I & 9.99 \\
MS1455.0+2232 & 14 57 15.12 & +22 20 35.52 & 0.258 & 8.41$\pm$2.10 & 4192 & I & 91.37 \\
A2631 & 23 37 38.16 & +00 16 90.11 & 0.278 & 8.57$\pm$1.80 & 3248 11728 & I & 25.9 \\
A1758N & 13 32 38.88 & +50 33 38.88 & 0.279 & 7.51$\pm$1.61 & 13997 15538 15540 & I & 147 \\
A1576 & 12 36 58.32 & +63 11 19.68 & 0.279 & 7.20$\pm$1.80 & 7938 15127 & I & 43.1 \\
A697 & 08 42 57.60 & +36 21 55.80 & 0.282 & 10.57$\pm$3.28 & 4217 & I & 17.4 \\
RXJ0439.0+0715 & 04 39 00.67 & +07 16 30.76 & 0.285 & 8.37$\pm$2.55 & 1449 3583 & I & 20.5 \\
RXJ0437.1+0043 & 04 37 90.46 & +00 43 54.15 & 0.285 & 7.96$\pm$2.34 & 7900 11729 & I & 42.5 \\
A611 & 08 00 56.64 & +36 03 23.40 & 0.288 & 8.86$\pm$2.53 & 3194 & S & 15.4 \\
Zw7215 & 15 01 23.04 & +42 20 54.96 & 0.290 & 7.34$\pm$1.91 & 7899 & I & 13.0 \\
Zw3146 & 10 23 39.60 & +04 11 12.88 & 0.291 & 17.27$\pm$2.94 & 909 9371 & I & 78.7 \\
A781 & 09 20 26.16 & +30 30 20.52 & 0.298 & 11.29$\pm$2.82 & 534 15128 & I & 45.0 \\
A2552 & 23 11 33.12 & +03 38 60.93 & 0.302 & 10.08$\pm$2.88 & 11730 & I & 22.6 \\
\hline
\end{tabular}
\begin{flushleft}
\small{{\bf Notes}. Column 1: Cluster name; Column 2:
   R.A.; Column 3: DEC; Column 4: Cluster Redshift; Column 5:
   Luminosity from \cite{1998MNRAS.301..881E,2000MNRAS.318..333E},
   converted to a $\Lambda$CDM cosmology; Column 6: {\em Chandra}
   ObsID; Column 7: {\em Chandra} aimpoint; Column 8: Cleaned exposure
 time.}
\end{flushleft}
\hspace{0.01cm}
\end{table*}

All 34 galaxy clusters in this sample were analysed with the
\textsc{CIAO}\footnote{See http://cxc.harvard.edu/ciao/} 4.6 software package
and \textsc{CALDB}\footnote{See http://cxc.harvard.edu/caldb/} version 4.5.9.
We applied standard processing techniques to
the level 1 photon lists to generate a level 2 photon lists.  We
inspected background lightcurves of the observations following the
recommendations of \cite{2003ApJ...583...70M}, to search for possible
background fluctuations.  The lightcurves were cleaned by 3$\sigma$
clipping and periods with count rates $>$20$\%$ deviation from the
mean rate were rejected.  The final cleaned exposure times are listed
in Table~\ref{tab:clustsamp}.

In order to take into account the background of each observation,
appropriate blank-sky backgrounds were obtained (which are processed
identically to the cluster observations) and reprojected onto the sky to
match the cluster observation.  For background data sets taken after
01 December 2001, the background observations were telemered in VFAINT
mode. Therefore, the additional VFAINT cleaning procedure was applied
to the source and background data sets\footnote{See
  http://cxc.harvard.edu/ciao/why/aciscleanvf.html}.
 
We followed a method outlined in \cite{2005ApJ...628..655V} in
order to improve the accuracy of the background by applying small
adjustments to the baseline model. We first corrected for the rate of
charged particle events, which has a secular and short-term variation
by as much as 30\%. We renormalise the background in the 9.5--12 keV
band, where the {\em Chandra} effective area is nearly zero and the
observed flux is due entirely to the particle background events. The
renormalisation factor was derived by taking the ratio of the observed
count rate in the source and background observations respectively. In
addition to the particle background, the blank-sky and source
observations contain differing contributions from the soft X-ray
background, containing a mixture of the Galactic and geocoronal
backgrounds, significant at energies $\leq$1 keV. To take into account
any difference in this background component between the blank-sky and
source observations, spectra were extracted in regions of the field of
view free from cluster emission.  The blank-sky spectrum was then
  subtracted from that of the local background, and the residuals
modeled in the 0.4-1keV band using an APEC thermal plasma model
\citep{2001ApJ...556L..91S}, with the abundance set to solar and
assuming zero redshift. This component is usually adequately described
with a temperature 0.18 keV, however in cases when this produced a
poor fit to the residuals the temperature was allowed to be free and
then fixed at the value which produced the best fit (see
Sect.~\ref{sec:notes} for cases when this was applied). This component
was then included in the spectral modeling of the cluster (see
Sect.~\ref{sec:props}).
         
\section{Data Analysis}
\label{sec:analysis}

In this section we detail the data analysis performed on our sample of
clusters. The analysis follows closely the analysis presented in
\cite{2012MNRAS.421.1583M}, which was closely based in turn on
  \cite{2005ApJ...628..655V}. Any deviations from this standard
  analysis are described in the following sections.

\subsection{Cluster Spectral Properties}
\label{sec:props}

Cluster spectra were extracted and fits performed in the 0.6 - 9.0 keV
band with an absorbed APEC plasma model \citep[using ATOMDB version
2.0.1, and relative abundances fixed to the solar ratios
of][]{1989GeCoA..53..197A}. The absorbing column was fixed at the
Galactic value \citep{2005A&A...440..775K} and the abundance
  allowed to vary. The fits were performed in {\tt XSPEC}
  \citep{1996ASPC..101...17A} using the $C$-statistic (the use of the
  $C$-statistic is discussed further in Sect~\ref{sec:landrycomp}),
  with the spectra grouped to contain at least one count per bin.
When determining the uncertainties on the temperature, the
  uncertainty due to the modelling of the soft background component
  was estimated as the variation in the temperature of the cluster
  component when the normalisation of the soft background component
  (see Sect~\ref{sec:data}) was set to $\pm 1\sigma$ of the fitted
  value. This error term was then added in quadrature to the original
  statistical error on the temperature to produce the final
  temperature error bar. Since many of the clusters in the sample
contained multiple {\em Chandra} observations, the individual
observations were analysed separately as outlined below. The data were
then combined for certain stages of the analysis. Source and
background spectra were extracted as below for individual observations
and fit simultaneously with the temperature, abundance and
  normalisations of the APEC components tied together and the redshift
  and absorbing column fixed.

The cluster properties were derived within $r_{500}$ (including the
cluster core), the radius at which the density of the cluster becomes
500 times the critical density of the Universe at the cluster
redshift.  Estmimates of the cluster $r_{500}$ were estimated from the
cluster mass, based on a hydrostatic mass ($M_{H}$) analysis (see
Sections~\ref{sec:tempmodel} and \ref{sec:xmass}).  We denote
$L_{CXO}$ and $L_{bol}$ as the unabsorbed 0.1--2.4 keV (rest
frame) and bolometric luminosities respectively.   

\subsection{Gas Density Modeling}
\label{sec:gasdens}

We make use of the observed projected emissivity profile to accurately
measure and model the gas density profile.  We converted each annular
bin in the background subtracted, exposure-corrected surface
brightness profile (measured in the 0.7--2.0 keV band, constructed
such that each bin contained at least 50 cluster counts)
into an integrated emission measure for each annulus.  The conversion
factor was determined by extracting an ARF and RMF in each annular bin
and using these, we simulate a spectrum assuming an absorbed APEC model.
The absorption was set at the Galactic value
\citep{2005A&A...440..775K} and the metal abundance set to 0.3 solar.
As the data in each annular bin were not sufficient to measure a
temperature, the temperature of the model in each bin was obtained by
utilizing the average temperature profile found by
\cite{2006ApJ...640..691V}, depending on the radius of the bin, the
determined $r_{500}$ and the global temperature for each cluster.  The
normalisation of the spectral model was set to 1 and re-arranged to
determine the emission integral for each bin given the derived count
rate.

The gas density profile was then fit with a modified version of the
standard 1D $\beta$-model proposed by \cite{2006ApJ...640..691V},
hereafter V06; 
\begin{align}
n_pn_e & = n^2_0\frac{(r/r_c)^{-\alpha}}{(1 +
  r^2/r^2_c)^{3\beta-\alpha/2}}\frac{1}{(1 +
  r^{\gamma}/r^{\gamma}_s)^{\epsilon/\gamma}}
\label{equ:gasprof}
\end{align}
We employ the same constraints as employed by V06 i.e. $\gamma$ is
fixed at 3 and $\epsilon$ $<$ 5 to exclude nonphysical sharp density
breaks.  We simplify the model slightly by excluding the second
$\beta$-model component outlined in V06, so that the model could be
used to fit to higher and lower quality data in our sample.

This model was then projected along the line of sight and fit to the
observed projected emission measure profile. The parameters in
equation (\ref{equ:gasprof}) are strongly correlated and therefore the
individual parameters degenerate.  For this reason the uncertainty
  on the derived density profile was estimated by generating synthetic
  emissivity profiles, where each data point in the original profile
  was replaced by a value sampled from a Gaussian centered on the value
  of the best-fitting model with a standard deviation equal to the
  measurement error for that point. 1,000 such synthetic datasets were
  generated and fit as before to give 1,000 output density profiles.
  These were used in all subsequent analyses to propagate the
  uncertainties on the gas density profile.  The individual parameters
  for each cluster can be found in the appendix
  (Table~\ref{tab:gasparams}). 

\subsection{Temperature Profile Modeling}
\label{sec:tempmodel}

To determine the total hydrostatic mass of a cluster, we use the method
outlined in \cite{2006ApJ...640..691V}.  This requires the use of a
projected temperature profile.  The temperature profile is constructed
such that it describes the temperature decline in the central regions
of most clusters, and a description for the profile in the outer
regions of a cluster.  The profile in the central regions of a cluster
can be described as:
\begin{align}
T_{cool}(r) & = \frac{(x + T_{min}/T_0)}{(x + 1)}, x = \left(\frac{r}{r_{cool}}\right)^{a_{cool}}
\label{equ:tempdecl}
\end{align}
Outside the cooling region, the temperature profile can be represented
by:
\begin{align}
T(r) & =\frac{(r/r_t)^{-a}}{[1 + (r/r_t)^b]^{c/b}}
\label{equ:tempoutside}
\end{align}
The final three-dimensional temperature profile is then given by:
\begin{align}
T_{3D}(r) & = T_0 T_{cool}(r)T(r)
\label{equ:vik3d}
\end{align}

For our clusters, the temperature profiles were constructed by creating
concentric annuli centered on the cluster such that each annuli was a
specific fraction of the determined $r_{500}$.  To determine the
$r_{500}$ for the temperature profile binning, we constructed
temperature profiles with each bin containing a minimum of 700 cluster
counts.  A mass analysis was performed (following
Sect.~\ref{sec:xmass}) using these initial temperature profiles, and an
initial $r_{500}$ calculated.  For clusters with greater than 10
temperature bins in this initial temperature profile, the profiles
were rescaled to simply contain 10 bins, with the bins recaled to
specific fractions of the initial $r_{500}$.  The fractions of
$r_{500}$ were calculated based on having a minimum signal-to-noise
(S/N) of 20, and were calculated based on the lowest S/N cluster with
greater than 10 temperature bins.  For clusters with fewer than 10
temperature bins in the initial profile, the initial number of bins
were simply rescaled to fractions of the initial $r_{500}$.   With
temperature profiles constructed such that the bins are defined in
fractions of $r_{500}$, a second mass analysis is performed to
determined the final $r_{500}$ and hydrostatic mass.  This
method ensures that the mass estimates for the clusters are derived in
a consistent way.   Furthermore, the errors on the temperature for
each radial bin are converted to account for the fact that the
likelihood curve for a measured temperature is approximately Gaussian
in log space.  We use the method of \cite{2012A&A...546A...6A} to
convert the generally asymmetric errors reported by {\tt XSPEC} into a
log-normal likelihood.

The temperature profile model (eq~\ref{equ:vik3d}) was fit to the data
by projecting it along the line of sight \citep[using a method
  outlined in][]{2006ApJ...640..710V} and computing the $\chi^2$ in the
log of the temperature  (Table~\ref{tab:tempparams} lists the
individual fit parameters for our clusters and the figures of the
temperature profiles with the corresponding fit are presented).  This
model has great functional freedom with nine free parameters and can
describe many smooth temperature distributions.  To take into account
the uncertainties on the temperature profile, we follow the same
Monte-Carlo method as that employed in Sect.~\ref{sec:gasdens},
generating and refitting synthetic temperature profiles based on the
initial model.  For cases when the number of bins in the temperature
profile is less than the nine free parameters, one or more of the
parameters are frozen at values given by the average temperature
profile given in V06. The constraints imposed when fitting to a
temperature profile with low temperature bins are as follows
\begin{description}
\item[(1)] No cool core: $x$=1 and $T_{0}$=$T_{\rm min}$ in 
eq~\ref{equ:tempdecl}.
\item[(2)] $b$=$c$ and $a$=0.
\item[(3)] No cool core, $b$=c and $a$=0.
\end{description}
The constrains (1), (2) and (3) were employed when a cluster's
temperature profile had 7-9, 6 and 5 bins respectively.

\subsection{X-ray Hydrostatic Mass Derivation}
\label{sec:xmass}  

To derive the total hydrostatic mass of the cluster, within a radius
r, we use the three-dimensional models of the temperature profile,
$T(r)$, and gas density profile, obtained by a fit to the emission
measure profile converted to a gas density, $\rho_{g}(r)$, and the
hydrostatic equilibrium equation  
\citep{1988xrec.book.....S},
\begin{align}
M(r) & = -\frac{k T(r) r}{\mu m_{\rm p} G}
\left(\frac{d\hspace{2pt}log\hspace{2pt}T(r)}{d\hspace{2pt}log\hspace{2pt}r}
  +
\frac{d\hspace{2pt}log\hspace{2pt}\rho_{\rm g}(r)}{d\hspace{2pt}log\hspace{2pt}r}\right)
\label{equ:hydromass}
\end{align} 
where k is the Boltzmann constant, $\mu$ corresponds to the mean
molecular weight in unit of $m_p$ (where $\mu$=0.5954), where $m_p$ is
the mass of a proton.  The gas density profile and
temperature profiles are constructed using the method outlined in
sections \ref{sec:gasdens} and \ref{sec:tempmodel} respectively.

\subsection{Determining the dynamical state of a cluster}
\label{sec:struct}

Here we wish to determine which clusters in our sample both appear
dynamically relaxed and host a CC (RCC).

We first determine which clusters in our sample are dynamically
relaxed. The dynamical state of the cluster was measured using the
centroid shift ($\langle$w$\rangle$), following the method of
\cite{2006MNRAS.373..881P}. The centroid shift was defined as the
standard deviation of the distance between the X-ray peak and the
centroid. The centroid was measured within a series of circular
apertures centered on the X-ray peak, with the apertures
  decreasing in size from from $r_{500}$ to $0.05\,r_{500}$, in steps
  of $0.05\,r_{500}$.  The errors on $\langle$w$\rangle$ were
  derived by producing 100 Monte-Carlo randomisations of the input
  source and background images with pixels randomised under a Poisson
  distribution centered on the observed counts in each pixel. These
  were then analysed in the same way as the real images to give a
  distribution of $\langle$w$\rangle$, from which we used the standard
  deviation as an estimate of the error on $\langle$w$\rangle$. The
values of $\langle$w$\rangle$ are given in
table~\ref{tab:structprops}. We make a cut at
$\langle$w$\rangle$=0.009, above which clusters are classed as
dynamically unrelaxed, and below which clusters are classed as
dynamically relaxed. This value was chosen when visually inspecting
images of each cluster ranked in order of $\langle$w$\rangle$, and
seeing a clear change in the structure of the clusters above this
value.  This value is close to the value determined by
  \cite{2013A&A...549A..19W}, who found that $\langle$w$\rangle$=0.01
  was the value of choice to split between relaxed and unrelaxed
  clusters for a sample of 121 simulated clusters. A value of
  $\langle$w$\rangle$=0.01 was also used in \cite{2009A&A...498..361P}
  to split between relaxed and unrelaxed clusters for the REXCESS
  sample of clusters.

We next determine which clusters in our sample contain a CC.  In a
comprehensive study, \cite{2010A&A...513A..37H} tested 16 CC probes,
and concluded that for high quality data a direct measurement of the
central cooling time (CCT) is the preferred probe.  Many of the
clusters within our sample have high quality data and a reliable
measurement of the CCT can be obtained.  However, for lower
quality data the cuspiness of the gas density profile is the preferred
choice. We utilise both of these probes to derive our cool-core
sub-sample of clusters.

To derive the CCT of our clusters, we use the equation given in
\cite{1988xrec.book.....S}.
\begin{align}
t_{\rm cool} & = 8.5\times10^{10} {\rm yr} \left(\frac{n_{\rm
    p}}{10^{-3}}\right)^{-1}(0.079 kT_{\rm CCT})^{1/2}
\label{equ:tcool}
\end{align}
where $n_{\rm p}$=$\sqrt{\rm (1.17n_{p}n_{e})}$, and $n_{p}n_{e}$
is measured using the best fitting gas density models
given in Sect.~\ref{sec:gasdens}.  kT$_{\rm CCT}$ is measured by
extracting a spectrum within [0-0.048]$r_{500}$ \citep[the radius
  defined in][]{2010A&A...513A..37H} and fit with an absorbed APEC
model and the addition of the background model from
Section~\ref{sec:data}.  As in Sect.~\ref{sec:tempmodel}, the
  temperature errors were transformed via the method of
  \cite{2012A&A...546A...6A}.  The errors on $t_{\rm cool}$ were derived
from log-normal randomisations centered on the kT$_{\rm CCT}$
and within the transformed error bars determined from the spectral fit.
Cuspiness is defined as the logarithmic slope of the gas density
profile at a radius of 0.04$r_{500}$, and is modeled using the best
fitting gas density models (Sect.~\ref{sec:gasdens}).  We note
  that while the errors on both $t_{cool}$ and cuspiness reflect the
  statistical quality of the data, they may be underestimated due to
  the assumption of a parametric form of the gas density profile.

In order to determine which clusters in our sample contained a
cool core, we used the cuts defined in \cite{2010A&A...513A..37H} for
the cuspiness and t$_{\rm cool}$ parameters.  Clusters in our sample
are determined to have a cool core if they have a cuspiness value
greater than 0.7, and t$_{\rm cool}$ less than 7.7Gyr.  The values of
cuspiness and t$_{\rm cool}$ are given in Table~\ref{tab:structprops}. 

\begin{figure*}
\begin{center}
\setlength{\unitlength}{1in}
\begin{picture}(7.0,2.5)
\put(-0.6,0.01){\scalebox{1.25}{\includegraphics[width=7cm, clip=true, origin=c]{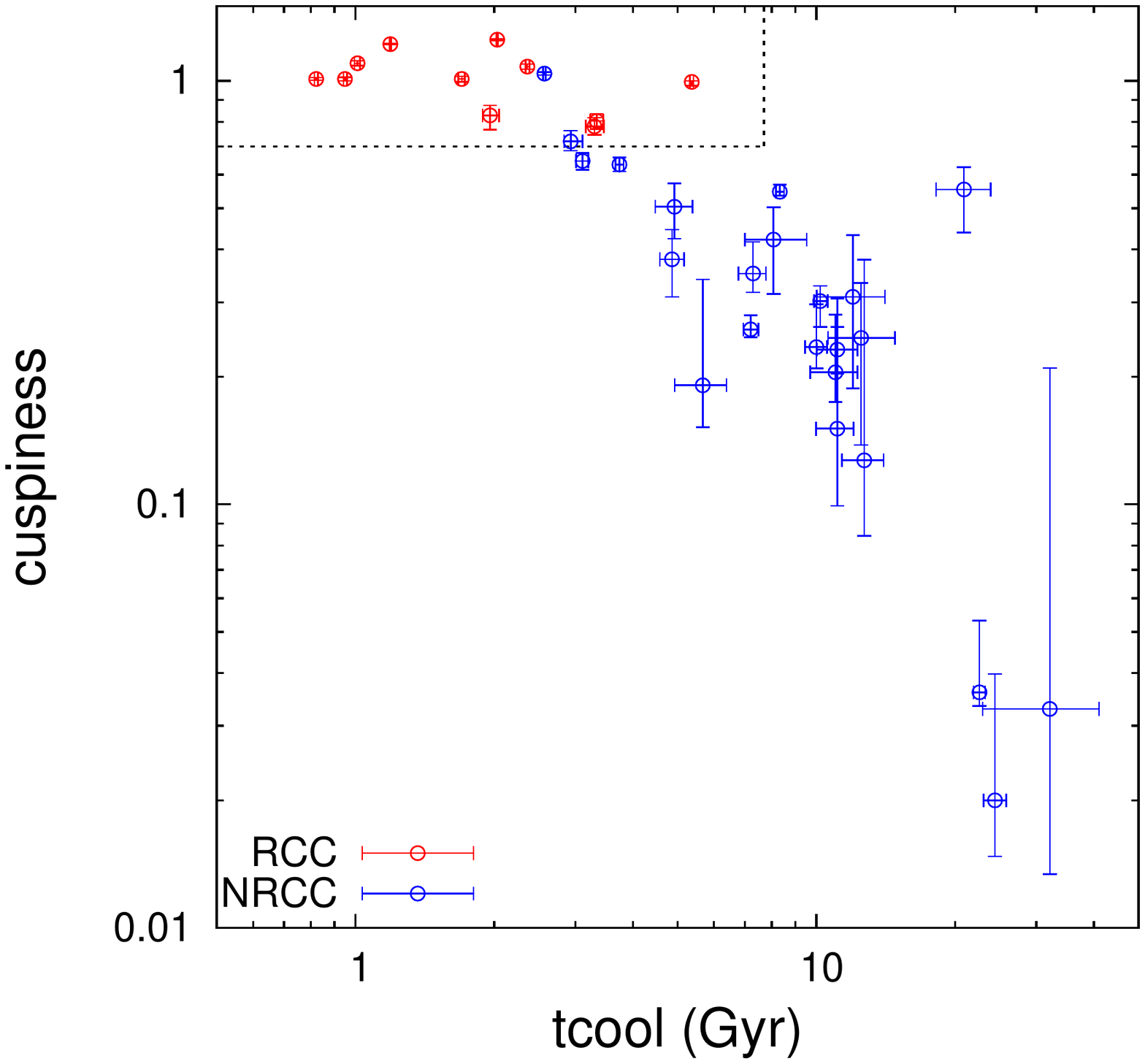}}}
\put(1.7,0.01){\scalebox{1.25}{\includegraphics[width=7cm, clip=true, origin=c]{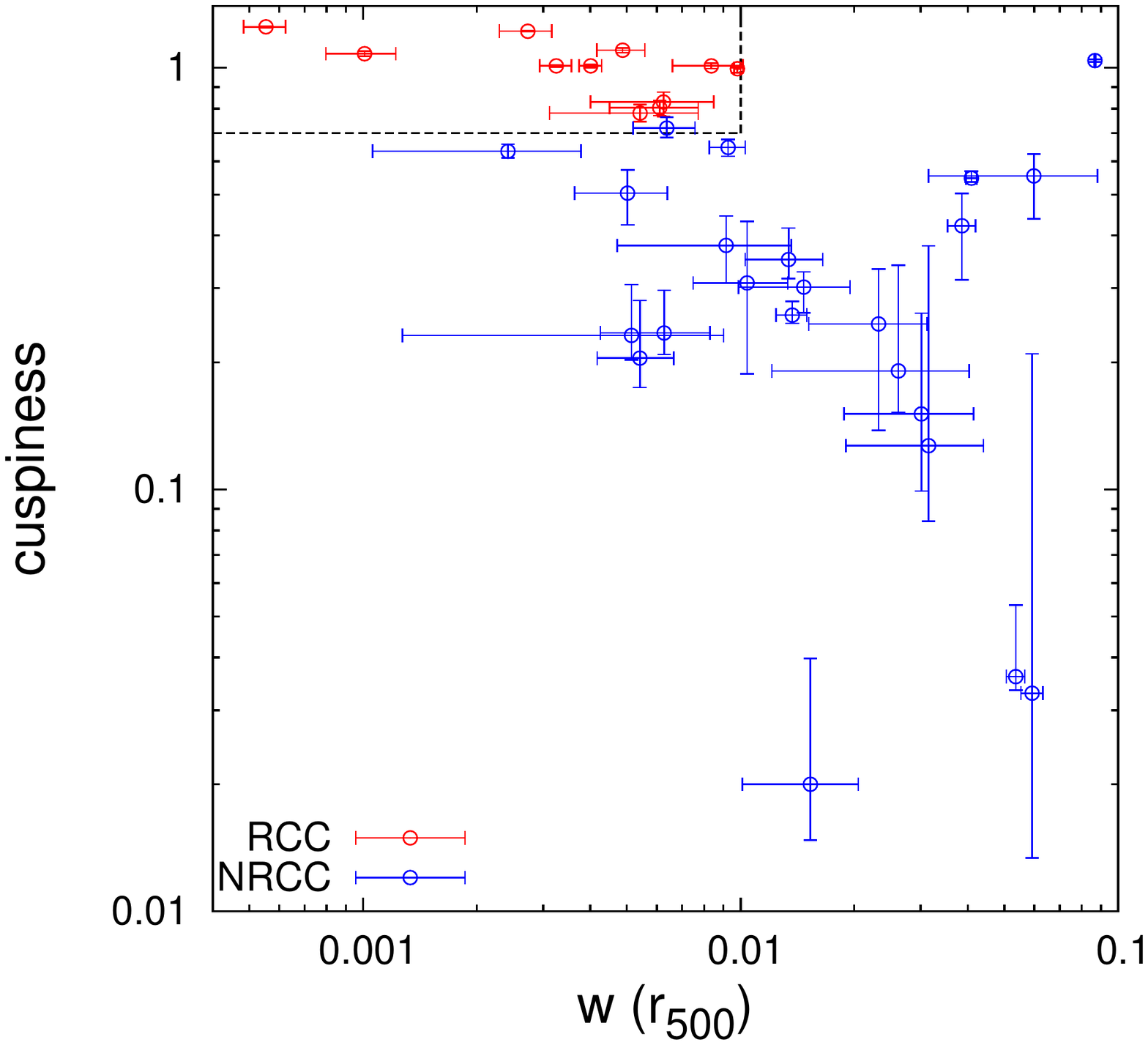}}}
\put(4.0,0.01){\scalebox{1.25}{\includegraphics[width=7cm, clip=true, origin=c]{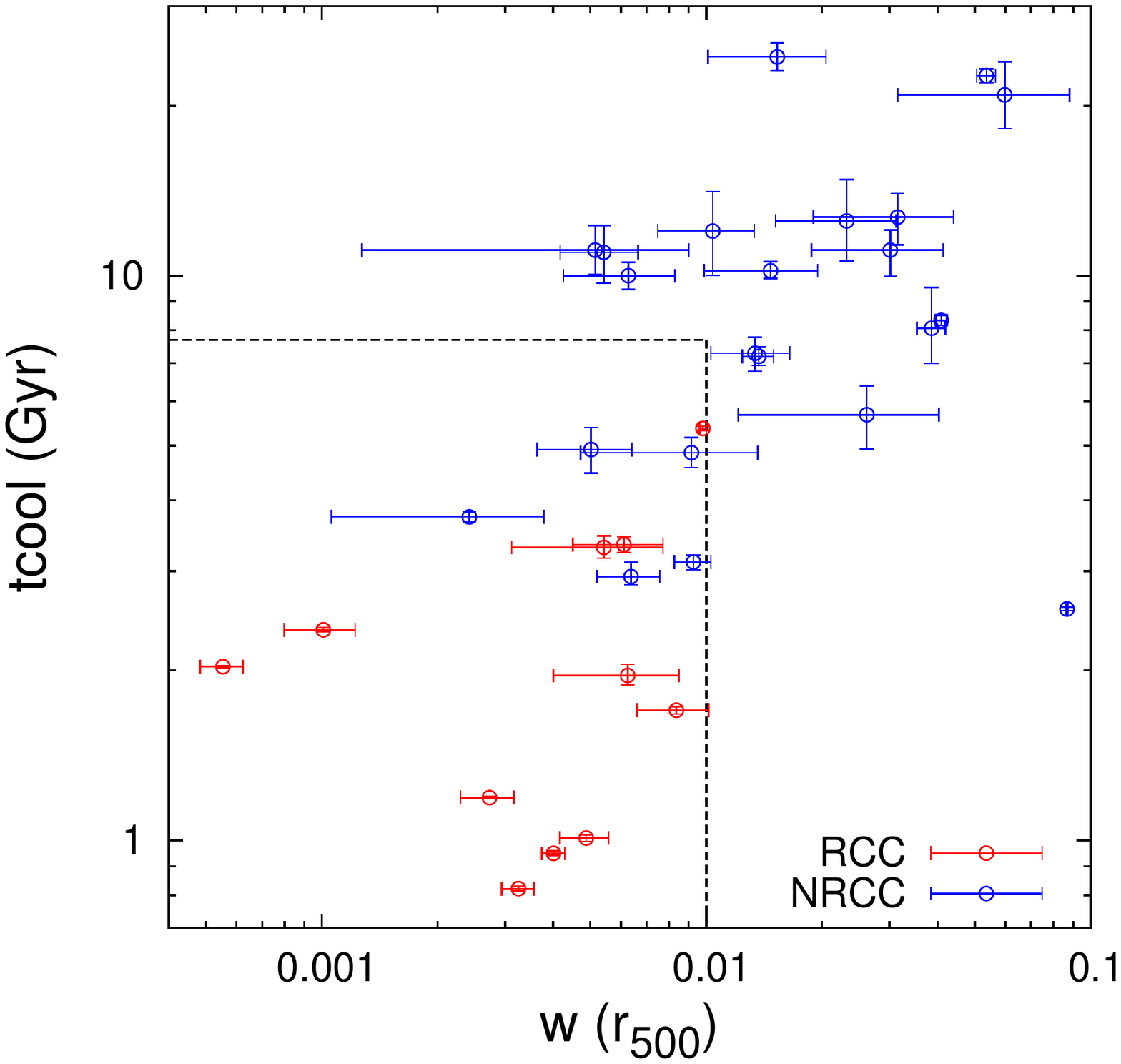}}}
\end{picture}
\end{center}
\vspace{-0.5cm}
\caption[]{\small{Left: Plot of t$_{\rm cool}$ against
    $\langle$w$\rangle$ with the dashed horizontal line the cut at
    t$_{\rm cool}$=6.5 and the dashed vertical line the cut at
    $\langle$w$\rangle$=0.009$r_{500}$, Middle: Plot of cuspiness
    against $\langle$w$\rangle$ with the dashed horizontal line the
    cut at cuspiness=0.65 and the dashed vertical line the cut at
    $\langle$w$\rangle$=0.009$r_{500}$, Right: Plot of cuspiness
    against t$_{\rm cool}$ with the dashed horizontal line the cut at
    cuspiness=0.65 and the dashed vertical line the cut at t$_{\rm
      cool}$=6.5.  The red hollow circles represent the RCC clusters,
    blue hollow squares represent the NRCC clusters.  The cluster at
    $\langle$w$\rangle$=0.087 is the double cool-core cluster
    A115.}\label{fig:coolcore}}
\end{figure*}

We plot t$_{\rm cool}$ against cuspiness in Figure~\ref{fig:coolcore},
left plot, and cuspiness and t$_{\rm cool}$ against
$\langle$w$\rangle$ in the middle and right plots respectively.  In
each plot, the RCC sample and NRCC sample are given by the red and
blue open circles respectively, and the cuts in cuspiness, t$_{\rm
  cool}$ and $\langle$w$\rangle$ are shown by the black dashed lines.
We note that the cluster A611 had residual flaring in the background
lightcurve and therefore the properties of the cluster could only be
derived out to a radius of $\approx$150$^{\prime\prime}$ (see
Sect.~\ref{sec:notes}).  A reliable hydrostatic mass estimate for
this cluster could not be determined, and it was therefore dropped
from the RCC sample. Using the cuts described above, we find 10/34
clusters classed as RCC and 24/34 clusters classed as NRCC.    

\begin{table}
\begin{center}
\caption[]{\small{Dynamical properties of the cluster
    sample.  $^{\dagger}$ Denotes clusters in the RCC sample.
    (see Sect.~\ref{sec:struct}).}\label{tab:structprops}}
\begin{tabular}{lccc}
\hline\hline
Cluster & Cuspiness & t$_{\rm cool}$ & $\langle$w$\rangle$ \\
 & & (Gyr) & 10$^{-3}r_{500}$ \\
\hline
  A2204$^{\dagger}$          & 1.25$^{+0.01}_{-0.01}$ & 2.03$^{+0.01}_{-0.01}$ & 0.55$\pm$0.07 \\        
  RXJ1720.1+2638$^{\dagger}$ & 1.08$^{+0.01}_{-0.01}$ & 2.36$^{+0.03}_{-0.03}$ & 1.01$\pm$0.21 \\ 
  A586           & 0.38$^{+0.07}_{-0.07}$ & 4.86$^{+0.44}_{-0.37}$ & 9.16$\pm$4.45 \\        
  A1914          & 0.26$^{+0.02}_{-0.01}$ & 7.20$^{+0.41}_{-0.43}$ & 13.7$\pm$1.28 \\        
  A665           & 0.55$^{+0.02}_{-0.01}$ & 8.33$^{+0.22}_{-0.23}$ & 40.9$\pm$1.33 \\        
  A115           & 1.04$^{+0.01}_{-0.01}$ & 2.57$^{+0.03}_{-0.02}$ & 86.9$\pm$0.33 \\        
  A520           & 0.04$^{+0.02}_{-0.01}$ & 22.6$^{+0.90}_{-1.00}$ & 53.6$\pm$3.04 \\        
  A963           & 0.63$^{+0.02}_{-0.02}$ & 3.74$^{+0.11}_{-0.12}$ & 2.42$\pm$1.36 \\        
  A1423$^{\dagger}$          & 0.80$^{+0.03}_{-0.03}$ & 3.34$^{+0.15}_{-0.14}$ & 6.11$\pm$1.61 \\        
  A773           & 0.24$^{+0.06}_{-0.03}$ & 10.0$^{+0.82}_{-0.80}$ & 6.27$\pm$2.02 \\        
  A1763          & 0.23$^{+0.07}_{-0.03}$ & 11.1$^{+1.59}_{-2.60}$ & 5.14$\pm$3.87 \\        
  A2261          & 0.65$^{+0.03}_{-0.03}$ & 3.11$^{+0.13}_{-0.13}$ & 9.27$\pm$1.02 \\        
  A1682          & 0.42$^{+0.08}_{-0.11}$ & 8.07$^{+1.83}_{-1.54}$ & 38.6$\pm$3.31 \\        
  A2111          & 0.15$^{+0.11}_{-0.05}$ & 11.1$^{+1.27}_{-1.38}$ & 30.1$\pm$11.3 \\        
  Zw5247        & 0.55$^{+0.07}_{-0.12}$ & 20.9$^{+4.89}_{-3.87}$ & 59.8$\pm$28.3 \\        
  A267           & 0.19$^{+0.15}_{-0.04}$ & 5.67$^{+1.04}_{-0.88}$ & 26.2$\pm$14.1 \\        
  A2219          & 0.30$^{+0.03}_{-0.04}$ & 10.2$^{+0.49}_{-0.50}$ & 14.7$\pm$4.81 \\        
  A2390          & 0.99$^{+0.01}_{-0.02}$ & 5.37$^{+0.06}_{-0.07}$ & 9.81$\pm$0.20 \\         
  Zw2089$^{\dagger}$        & 1.10$^{+0.02}_{-0.01}$ & 1.01$^{+0.02}_{-0.02}$ & 4.87$\pm$0.71 \\               
  RXJ2129.6+0005$^{\dagger}$ & 1.01$^{+0.01}_{-0.01}$ & 1.70$^{+0.04}_{-0.04}$ & 8.37$\pm$1.78 \\        
  A1835$^{\dagger}$          & 1.22$^{+0.01}_{-0.01}$ & 1.19$^{+0.01}_{-0.01}$ & 2.73$\pm$0.43 \\        
  A68            & 0.31$^{+0.12}_{-0.12}$ & 12.0$^{+3.37}_{-2.95}$ & 10.4$\pm$2.92 \\        
  MS1455.0+2232$^{\dagger}$  & 1.01$^{+0.01}_{-0.01}$ & 0.95$^{+0.01}_{-0.01}$ & 4.01$\pm$0.28 \\        
  A2631          & 0.25$^{+0.09}_{-0.11}$ & 12.5$^{+2.99}_{-2.69}$ & 23.2$\pm$8.04 \\        
  A1758N          & 0.02$^{+0.02}_{-0.01}$ & 24.4$^{+1.95}_{-2.20}$ & 15.3$\pm$5.20 \\ 
  A1576          & 0.35$^{+0.07}_{-0.03}$ & 7.29$^{+0.70}_{-0.72}$ & 13.4$\pm$3.11 \\        
  A697           & 0.20$^{+0.08}_{-0.03}$ & 11.0$^{+1.89}_{-1.75}$ & 5.41$\pm$1.24 \\        
  RXJ0439.0+0715$^{\dagger}$ & 0.78$^{+0.04}_{-0.04}$ & 3.30$^{+0.23}_{-0.21}$ & 5.42$\pm$2.30 \\         
  RXJ0437.1+0043$^{\dagger}$ & 0.83$^{+0.05}_{-0.06}$ & 1.96$^{+0.11}_{-0.09}$ & 6.25$\pm$2.24 \\        
  A611          & 0.72$^{+0.04}_{-0.03}$ & 2.93$^{+0.17}_{-0.19}$ & 6.38$\pm$1.19 \\        
  Zw3146$^{\dagger}$        & 1.01$^{+0.01}_{-0.01}$ & 0.82$^{+0.01}_{-0.01}$ & 3.25$\pm$0.31 \\        
  Zw7215        & 0.13$^{+0.25}_{-0.04}$ & 12.7$^{+1.78}_{-1.66}$ & 31.5$\pm$12.5 \\        
  A781           & 0.03$^{+0.18}_{-0.02}$ & 32.1$^{+19.1}_{-12.7}$ & 59.2$\pm$3.94 \\        
  A2552          & 0.50$^{+0.07}_{-0.08}$ & 4.92$^{+0.62}_{-0.53}$ & 5.02$\pm$1.38 \\        
\hline
\end{tabular}
\begin{flushleft}
\small{{\bf Notes}. Column 1: Cluster name; Column 2:
   Cuspiness measured as the logarithmic slope of the gas density
    profile at 0.04$r_{500}$.; Column 3: Cooling time measured within
    0.048$r_{500}$; Column 4: Centroid shift.}
\end{flushleft}
\hspace{0.01cm}
\end{center}
\end{table}

\section{Notes on Individual Clusters}
\label{sec:notes}
In this section we note any peculiarities or points of interest for
observations in which we departed from the described analysis process.
\begin{description}
\item[A586]-- ObsID 530 was rejected due to long, low level flaring, leaving ObsID 11723.
\item[A665]-- We reject ObsID 531 due to large temperature required to
  fit soft background residuals (0.6 keV). We reject ObsID 3586 due to
  several periods of high background.  We reject ObsID 7700 due to
  energy filters placed on the observation.  This leaves ObsIDs 12286
  and 13201, both $\approx$50ks observations.
\item[A115]-- This cluster is undergoing a major off-axis merger, with
  two sub-clusters separated by 300$^{\prime\prime}$ (1 Mpc) in
  projection.  Two regions were manually excluded from the analysis to
  exclude emission from the southern subcluster.  Sources were
  excluded at $\alpha$[2000.0] = 00$^h$55$^m$58.89$^s$,
  $\delta$[2000.0] = +26$^{\circ}$19$^{\prime}$28.45$^{\prime\prime}$
  and $\alpha$[2000.0] = 00$^h$56$^m$03.78$^s$, $\delta$[2000.0] =
  +26$^{\circ}$22$^{\prime}$44.48$^{\prime\prime}$, with radii
  182$^{\prime\prime}$ and 63$^{\prime\prime}$ respectively.
\item[A963]-- A temperature of 0.32 keV was used when fitting an APEC
  model to the soft background residuals.
\item[A1763]-- An extended source at $\alpha$[2000.0] =
  13$^h$34$^m$55.29$^s$, $\delta$[2000.0] =
  +40$^{\circ}$57$^{\prime}$22.93$^{\prime\prime}$,
  304$^{\prime\prime}$ from the cluster core was manually excluded.
  The extended emission is likely associated with a known X-ray source
  \citep{2010ApJS..189...37E}.
\item[A2261]-- A small extended source at $\alpha$[2000.0] =
  17$^h$22$^m$12.15$^s$, $\delta$[2000.0] =
  +32$^{\circ}$06$^{\prime}$54.0$^{\prime\prime}$,
  200$^{\prime\prime}$ from the cluster core was manually excluded.
  The extended emission is associated with a known galaxy at a
  photo-z=0.304 \citep{2010ApJS..191..254H}.  
\item[A1682]-- ObsID 2344 was rejected due to high flare periods,
  leaving ObsID 11725.
\item[A2111]-- This cluster has been shown to be undergoing a head-on
  merger with a subcluster, appearing as a comet-shaped X-ray
  subcomponent and hotter than the surrounding gas
  \citep{1997MNRAS.288..702W}.  A region at $\alpha$[2000.0] =
  15$^h$39$^m$32.68$^s$, $\delta$[2000.0] =
  +34$^{\circ}$28$^{\prime}$04.79$^{\prime\prime}$,
  210$^{\prime\prime}$ from the cluster core was manually excluded to
  exclude the emission from the merger.
\item[Zw5247]-- ObsID 539 was rejected due to long, low level flaring,
  leaving ObsID 11727.  This system consists of a binary merger of two
  clusters of similar mass.  Only one redshift is given in
  \cite{1998MNRAS.301..881E}, corresponding to the position of the
  southern subcluster.  For this reason the northern sub cluster was
  manually excluded from our analysis using a box region at
  $\alpha$[2000.0] = 12$^h$34$^m$33.18$^s$, $\delta$[2000.0] =
  +09$^{\circ}$49$^{\prime}$52.84$^{\prime\prime}$ of length
  428$\times$421$^{\prime\prime}$.
\item[A267]-- ObsID 3580 was rejected due to long, low level flaring
  leaving ObsID 1448.
\item[A2390]--  We discard ObsID 500 and 501 as both observations were
  taken in FAINT mode, which results in poorer background rejection,
  leaving ObsID 4192.  A temperature of 0.25 keV was used when fitting
  an APEC model to the soft background residuals.
\item[MS1455.0+2232]-- For ObsID 4192 we used a temperature of 0.21
  keV when fitting an APEC model to the soft background residuals.
\item[Zw5768]-- The cluster redshift was given as z=0.266 in
  \cite{2000MNRAS.318..333E}. This was found to be incorrect, with
  \cite{2000ApJS..129..435B} reporting a spectroscopic redshift of
  z=0.171. To check this we searched the SDSS DR7 release
  \citep{2011ApJS..193...29A}, and found the redshift of the BCG of
  the cluster to be z=0.172.  With this updated redshift, the BCS
    flux for the cluster corresponds to a luminosity well below the
    limit used to define our sample, and so this cluster was dropped
    from our analysis.
\item[A1758N]-- ObsId 7710 was excluded due to energy filters placed
  on the observation and residual flaring in the observation.  ObsID
  2213 was excluded due to large residual flaring in the observation.
  The cluster A1758S was excluded manually from the analysis, using a
  circle region centered at $\alpha$[2000.0] = 13$^h$32$^m$32.04$^s$,
  $\delta$[2000.0] = +50$^{\circ}$24$^{\prime}$32.95$^{\prime\prime}$.      
\item[RXJ0439.0+0715]-- We excluded the first 6ks of the observation
  of ObsID 3583 due to flaring.
\item[A611] -- We excluded the first 22ks of the observation due to
  long low level flaring.  Periods of high background were
  still present in the observation and therefore the cluster
  temperature was extracted out to a radius of
  $\approx$150$^{\prime\prime}$, and this was assumed to
  be the average cluster temperature (kT = 8.41$^{+0.93}_{-0.75}$
  keV).  The cluster properties were then extracted within a radius of
  $r_{500}$ determined following the procedure outlined in
  Sect.~\ref{sec:props}, with the temperature fixed at the value above. 
\item[Zw3146]-- For ObsID 9371 we used a temperature of 0.26 keV when
  fitting an APEC model to the soft background residuals. 
\item[A2552]-- ObsID 3288 was rejected due to high flare periods,
  leaving ObsID 11730.  An extended source, identified as the galaxy
  cluster NSCS J231153+034038 at z=0.36 \citep{2004AJ....128.1017L},
  at $\alpha$[2000.0] = 23$^h$11$^m$48.3$^s$, $\delta$[2000.0] =
  +03$^{\circ}$40$^{\prime}$47.1$^{\prime\prime}$,
  276$^{\prime\prime}$ from the cluster core was manually excluded.
\end{description}       

\section{X-ray Scaling Relations}
\label{sec:xscaling}

Establishing the relationship between total mass and observable
quantities is a critical step for the derivation of cosmological
parameters using galaxy clusters.  Cluster properties such as the
X-ray luminosity, gas mass, temperature and the Y$_{\rm X}$ parameter
(the product of the gas mass and temperature), provide useful proxies
for cluster mass, via the use of well calibrated scaling relations.
In this work we focus on the scaling of the luminosity with mass.  We
investigate the form of the luminosity-mass relation, focusing on the
sample relation (not accounting for biases), and the bias-corrected
relation.  The scaling relations are split between the relaxed and
unrelaxed sub-samples defined in section~\ref{sec:struct}.   

\begin{table*}
\begin{center}
\vspace{0.5cm}
\caption[]{\small{X-ray Properties of our cluster sample. The
    properties are derived within $r_{500}$ determined from the
    hydrostatic mass analysis (see
    Sect.~\ref{sec:xmass}). $^{\dagger}$ Indicates relaxed
    clusters.}\label{tab:ktresults}}
\begin{tabular}{lcccc}
\hline\hline
 & $r_{H,500}$ & $L_{CXO}$ & $L_{bol}$ & $M_{H}$ \\
Cluster & (Mpc) & (10$^{44}$ ergs s$^{-1}$) & (10$^{44}$ ergs s$^{-1}$) & (10$^{14}$M$_{\odot}$) \\ 
\hline
A2204$^{\dagger}$          & 1.38$^{+0.03}_{-0.03}$ & 16.74$\pm$0.06 & 43.77$\pm$0.16 &  8.55$^{+0.59}_{-0.48}$ \\  
RXJ1720.1+2638$^{\dagger}$ & 1.36$^{+0.11}_{-0.07}$ & 9.69$\pm$0.10  & 23.70$\pm$0.25 &  8.20$^{+2.13}_{-1.21}$ \\  
A586           & 1.11$^{+0.13}_{-0.07}$ & 5.46$\pm$0.13  & 13.37$\pm$0.32 &  4.49$^{+1.75}_{-0.81}$ \\  
A1914          & 1.52$^{+0.14}_{-0.11}$ & 11.92$\pm$0.13 & 38.88$\pm$0.43 &  11.50$^{+3.54}_{-2.28}$ \\  
A665           & 1.70$^{+0.02}_{-0.03}$ & 8.38$\pm$0.06  & 24.08$\pm$0.18 &  16.37$^{+0.67}_{-0.93}$ \\  
A115           & 1.13$^{+0.01}_{-0.02}$ & 5.60$\pm$0.04  & 12.43$\pm$0.08 &  4.85$^{+0.17}_{-0.25}$ \\  
A520           & 1.33$^{+0.02}_{-0.02}$ & 7.04$\pm$0.03  & 19.14$\pm$0.09 &  8.00$^{+0.36}_{-0.28}$ \\  
A963           & 1.11$^{+0.04}_{-0.03}$ & 6.74$\pm$0.07  & 15.93$\pm$0.17 &  4.75$^{+0.53}_{-0.41}$ \\  
A1423$^{\dagger}$          & 1.09$^{+0.06}_{-0.04}$ & 5.25$\pm$0.08  & 12.05$\pm$0.19 &  4.42$^{+0.71}_{-0.53}$ \\  
A773           & 1.38$^{+0.12}_{-0.06}$ & 7.07$\pm$0.11  & 20.48$\pm$0.31 &  9.12$^{+2.59}_{-1.13}$ \\  
A1763          & 1.42$^{+0.15}_{-0.11}$ & 8.23$\pm$0.13  & 24.00$\pm$0.38 &  10.01$^{+3.59}_{-2.14}$ \\  
A2261          & 1.25$^{+0.08}_{-0.04}$ & 11.38$\pm$0.13 & 31.65$\pm$0.36 &  6.89$^{+1.41}_{-0.71}$ \\  
A1682          & 1.13$^{+0.09}_{-0.07}$ & 4.36$\pm$0.11  & 10.93$\pm$0.28 &  5.02$^{+1.29}_{-0.84}$ \\  
A2111          & 1.23$^{+0.12}_{-0.04}$ & 4.67$\pm$0.09  & 12.65$\pm$0.24 &  6.46$^{+2.02}_{-0.74}$ \\  
Zw 5247        & 0.94$^{+0.12}_{-0.07}$ & 2.76$\pm$0.12  & 6.17$\pm$0.26  &  2.90$^{+1.15}_{-0.59}$ \\  
A267           & 0.99$^{+0.15}_{-0.08}$ & 5.89$\pm$0.22  & 12.74$\pm$0.49 &  3.37$^{+1.76}_{-0.76}$ \\  
A2219          & 1.51$^{+0.04}_{-0.02}$ & 17.11$\pm$0.12 & 59.16$\pm$0.42 &  12.14$^{+0.94}_{-0.51}$ \\      
A2390          & 1.61$^{+0.10}_{-0.05}$ & 18.93$\pm$0.10 & 57.04$\pm$0.31 &  14.81$^{+3.00}_{-1.48}$ \\     
Zw 2089$^{\dagger}$        & 0.94$^{+0.15}_{-0.07}$ & 6.19$\pm$0.10  & 11.22$\pm$0.18 &  2.96$^{+1.67}_{-0.65}$ \\  
RXJ2129.6+0005$^{\dagger}$ & 1.22$^{+0.10}_{-0.07}$ & 9.68$\pm$0.10  & 22.78$\pm$0.25 &  6.42$^{+1.78}_{-1.04}$ \\  
A1835$^{\dagger}$          & 1.50$^{+0.05}_{-0.04}$ & 22.68$\pm$0.08 & 60.98$\pm$0.21 &  12.23$^{+1.40}_{-1.05}$ \\   
A68            & 1.15$^{+0.19}_{-0.12}$ & 6.64$\pm$0.20  & 20.97$\pm$0.62 &  9.30$^{+2.46}_{-1.42}$ \\  
MS1455.0+2232$^{\dagger}$  & 1.06$^{+0.04}_{-0.03}$ & 11.00$\pm$0.09 & 22.60$\pm$0.17 &  4.33$^{+0.57}_{-0.32}$ \\  
A2631          & 1.28$^{+0.11}_{-0.07}$ & 8.06$\pm$0.14  & 22.38$\pm$0.40 &  7.68$^{+2.15}_{-1.15}$ \\  
A1758          & 1.64$^{+0.20}_{-0.10}$ & 8.79$\pm$0.10  & 25.06$\pm$0.30 &  16.19$^{+7.40}_{-2.83}$ \\  
A1576          & 1.19$^{+0.11}_{-0.06}$ & 6.16$\pm$0.10  & 17.62$\pm$0.28 &  6.29$^{+1.98}_{-0.92}$ \\  
A697           & 1.55$^{+0.21}_{-0.13}$ & 13.40$\pm$0.26 & 45.61$\pm$0.87 &  13.85$^{+6.47}_{-3.18}$ \\     
RXJ0439.0+0715$^{\dagger}$ & 1.17$^{+0.11}_{-0.07}$ & 7.36$\pm$0.13  & 17.72$\pm$0.30 &  5.56$^{+1.71}_{-0.92}$ \\  
RXJ0437.1+0043$^{\dagger}$ & 1.17$^{+0.13}_{-0.06}$ & 7.71$\pm$0.12  & 19.11$\pm$0.29 &  5.88$^{+2.13}_{-0.93}$ \\  
A611           & 1.12$^{+0.15}_{-0.06}$ & 6.79$\pm$0.13  & 19.15$\pm$0.37 &  5.36$^{+2.45}_{-0.92}$ \\  
Zw 3146$^{\dagger}$        & 1.27$^{+0.37}_{-0.17}$ & 20.55$\pm$0.12 & 50.19$\pm$0.30 &  8.85$^{+1.74}_{-1.09}$ \\  
Zw 7215        & 1.33$^{+0.08}_{-0.06}$ & 5.00$\pm$0.19  & 13.39$\pm$0.51 &  7.69$^{+9.00}_{-2.67}$ \\  
A781           & 1.13$^{+0.07}_{-0.05}$ & 5.34$\pm$0.10  & 13.25$\pm$0.25 &  6.89$^{+1.07}_{-0.95}$ \\  
A2552          & 1.22$^{+0.08}_{-0.07}$ & 9.03$\pm$0.18  & 26.99$\pm$0.53 &  6.90$^{+1.49}_{-1.13}$ \\  
\hline
\end{tabular}
\hspace{0.01cm}
\end{center}
\end{table*}

\subsection{The sample \lcxo-$M_{H}$ relation}
\label{sec:unbiaslm}

\begin{figure}
\begin{center}
\includegraphics[width=9.0cm]{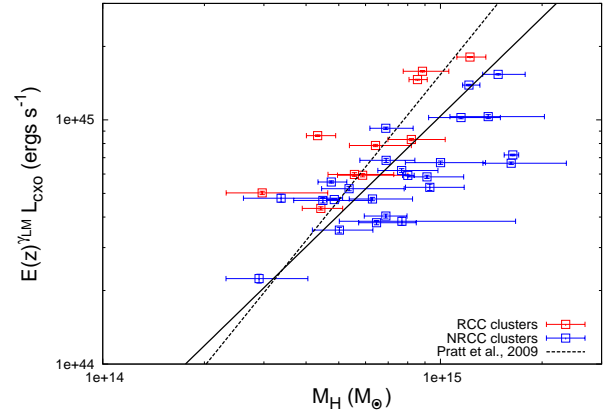}
\end{center}
\vspace{-1cm}
\caption[]{\small{The $L_{CXO}-M_{H}$ relation for our
    clusters, with the masses derived from the relaxed $M_{H}-T$
    relation.  The clusters are split between the relaxed (red open
    squares) and unrelaxed (blue open squares) samples.  The corresponding
    BCES fit (see Sect.~\ref{sec:unbiaslm}) to all the clusters (RCC
    and NRCC) is given by the black dashed line (sec
    Sect.~\ref{sec:unbiaslm}).  The $LM$ relation of
    \cite{2009A&A...498..361P} is given by the dotted-dashed cyan
    line.}\label{fig:Lx-Mx}}
\end{figure}

Here we derive the luminosity--mass ($LM$) relation
for our clusters.  Due to the relative ease of measuring the X-ray
luminosity of clusters, scaling relations involving the luminosity
have had a rich history
\citep[e.g.][]{1977MNRAS.181P..25M,1999MNRAS.305..631A,2002ApJ...567..716R,2009A&A...498..361P,2012MNRAS.421.1583M}.
Figure~\ref{fig:Lx-Mx} shows the derived $L_{CXO}-M_{H}$
relation (where $M_{H}$ is the mass derived from the hydrostatic mass
analysis) split between the relaxed (red open circles) and unrelaxed
(blue open squares) sub-samples.  The luminosities are derived in the
[0-1]$r_{500}$ range for consistency with the comparison to the
bias-corrected \lcxo-$M_{H}$ relation (see Sect.~\ref{sec:biascorrlm}).

We fit to the data a power law relation of the form 
\begin{align}
E(z)^{-\gamma_{\rm LM}}\left(\frac{L_{CXO}}{L_{0}}\right)=A_{\rm LM}\left(\frac{M_{H}}{M_{\rm 0}}\right)^{B_{LM}} 
\end{align}
assuming $L_{0}=10^{45}$ ergs s$^{-1}$, $M_{0}=10^{15}M_{\odot}$ and
$\gamma_{LM}=2$.  Note that the expected self-similar value of
  $\gamma_{LM}$ depends on the energy band in which the luminosities
  are measured. The use of $\gamma_{LM}=2$ is appropriate for
  soft-band ($0.1-2.4$ keV) luminosities.
  \citep{2015MNRAS.446.2629E}. The power law was fit to the data
using the BCES orthogonal regression in log space
\citep{1996ApJ...470..706A}. We find a normalisation and slope of
$A_{LM}=(1.82\pm0.66)\times10^{45}$ ergs s$^{-1}$ and
\Blm=1.42$\pm$0.60 for the relaxed sample, and
$A_{LM}=(0.92\pm0.13)\times10^{45}$ ergs s$^{-1}$ and
\Blm=1.16$\pm$0.27 for the unrelaxed sample.  We find that the
normalisation of the unrelaxed sample is 1.19$\pm$0.09 times lower than
that of the relaxed sample (significant at the 1.3$\sigma$ level), however
this is unsurprising due to the large increase in luminosity towards
the centers of cool core clusters.  We compare to the $LM$ relation
given in \cite{2009A&A...498..361P}, appropriate for core included
luminosities in the 0.1--2.4 keV band \citep[see Table A.2
  in][]{2009A&A...498..361P}.  The \cite{2009A&A...498..361P} $LM$
relation is given by the dashed-dotted cyan line in
Figure~\ref{fig:Lx-Mx}.  Although the \cite{2009A&A...498..361P} $LM$
relation appears somewhat steeper than our \lcxo-$M_{H}$, the
difference is only significant at the 1.3$\sigma$ level.

\subsection{Selection Function}
\label{sec:sfunc}

The sample was selected to match that in \cite{2006ApJ...653..954D},
who selected the clusters from the (e)BCS in the redshift range
$0.150<z<0.303$ based on the soft-band ($0.1-2.4$ keV) luminosity as
measured in the BCS, $\lbcs^{EdS}>10^{45}$ erg s$^{-1}$. However (as
we indicate with the EdS superscript), the BCS luminosities were
computed assuming an Einstein de-Sitter cosmology ($\Omega_M=1$,
$\Omega_\Lambda=0$, $H_0=50$ km s$^{-1}$ Mpc$^{-1}$), so in order to
work in our preferred $\Lambda$CDM cosmology it was necessary to
convert the BCS luminosities and our selection function to this
cosmology. The $\Lambda$CDM selection function $\llim(z)$ is well
approximated by 
$\lbcs>5.26\times10^{44}\times10^{0.324z}$ erg s$^{-1}$, where $\lbcs$
(i.e. without an EdS superscript) indicates the soft band BCS
luminosity in our $\Lambda$CDM cosmology.

\begin{figure}
\begin{center}
\scalebox{0.31}{\includegraphics*{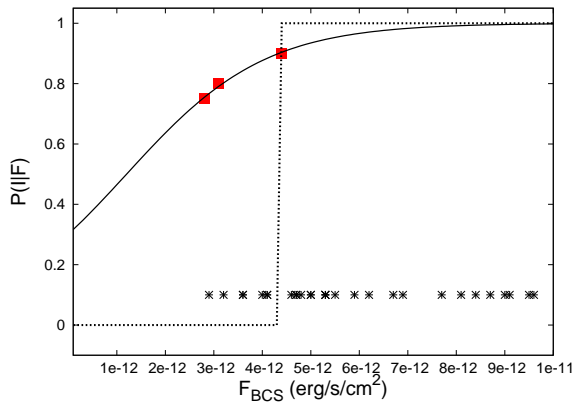}}
\vspace{-1cm}
\caption[]{\label{f.selfn} Our logistic function approximation to the
  (e)BCS completeness as a function of (e)BCS flux is plotted with the
  published completeness values for the (e)BCS survey (red
  squares). The asterisk symbols indicate the $\fbcs$ of the
  clusters in our sample, and are at arbitrary completeness.  The
  dotted line represents a step function used to test a limiting case
  of the selection function (see~\ref{sec:syseffects}).}
\end{center}
\end{figure}

The completeness of the BCS survey is a function of flux, but the full
selection function has not been published. However, completeness
estimates at specific fluxes are given in \cite{1998MNRAS.301..881E} and
\cite{2000MNRAS.318..333E}, and so we modeled the survey completeness as a
logistic function of the form
\begin{align}
P(I|f) & = \left(1+e^{-0.7(f-1.2)}\right)^{-1}
\label{eq:pif}
\end{align}
where $P(I|f)$ is the probability that a cluster with a normalised
(e)BCS $0.1-2.4$keV flux $f=F_{BCS}/(10^{-12}$ erg s$^{-1}$) is
included in the sample.  The numerical constants in this model were
determined from a simple fit by eye of the logistic function to the
published (e)BCS completeness values, as illustrated in Figure
\ref{f.selfn}. This functional approximation is within $0.01$ in
$P(I|f)$ of the published completeness values.  We show in
  Sect.~\ref{sec:syseffects} that our results are not sensitive to the
  details of the assumed model for the selection function.

\subsection{The bias-corrected \lcxo-$M_{H}$ relation}
\label{sec:biascorrlm}

The preceding fit of the observed $L_{CXO}-M_{H}$ relation
(Sect.~\ref{sec:unbiaslm}) represents an accurate description of the
correlation between luminosity and mass for our subsample. However, in 
order to compute an unbiased estimate of the population
$L_{CXO}-M_{H}$ relation, care must be taken to avoid the
effects of selection biases \citep[see][for a
discussion]{2011ARA&A..49..409A}. Our analysis is based closely on
that of M10a, which presents the most complete treatment of selection
biases in X-ray cluster surveys and their effects on scaling relations
and cosmological studies.

One way to visualise the steps required to correct for the selection
biases is to consider how one would realistically generate a
synthetic population like that being studied. In our case the steps
would be: (i) to use a model mass function $\phi=dN/dMdV$ to predict
the number of clusters as a function of mass and redshift in the
volume studied; (ii) to then generate a Poissonian realisation of that
population, and assign each cluster a luminosity \lbcs\ based on its
mass and redshift, according to a model $LM$ relation; (iii) next,
intrinsic scatter at the appropriate level (a lognormal with standard
deviation $\intlm$) would be added to the assigned $L$ values;
(iv) fluxes could then be computed for each cluster, using assumptions
about the temperature and metal abundance (and their mass dependence)
to allow k-correction of the fluxes into the observed frame
($F_{BCS}$); (v) statistical scatter would then be added to those 
luminosities and fluxes, requiring a model predicting the size of the
statistical error for a given flux (e.g. by converting the flux to
number of counts for a mean exposure time, and assuming Poisson
errors); (vi) finally, the sample selection would be applied,
rejecting all clusters fainter than $\llim(z)$, and
discarding clusters probabilistically as a function of their flux
according to $P(I|f)$. At stage (v) secondary
observations (such as our {\em Chandra} follow-up observations) can be
generated by computing a {\em Chandra} luminosity with its own statistical
uncertainty, and possibly including a cross-calibration scaling
between {\em ROSAT} and {\em Chandra}.

Considering this procedure for simulating data with properties close
to the true population gives insight into the likelihood function of
the data (the probability of the data being observed given a model and
its parameter values).  For the LM relation, the final likelihood
thus depends on the likelihood of the number of clusters detected in
the subsample (and by extension the number omitted), and the
likelihood of the detected clusters having their observed
properties. In the following sections we derive the likelihood for the
LM relation and describe our specific implementation of this in
fitting to our sample.

\subsubsection{The likelihood function for the $LM$ relation}
\label{sec:likelihood}

The number of clusters predicted by our model to be observed in the
subsample defined by our selection function is the integral of the
mass function over the volume over the survey, weighted by the
probability that a cluster of a given mass would be included in the
subsample given the $LM$ relation and the intrinsic and statistical
scatter on the luminosity.  Following M10a and using the notation that
observed quantities are denoted with a hat, this is expressed as
\begin{align}
\left<N_{det}\right> & = \int dM \int dz \, \phi(M,z)\frac{dV}{dzd\Omega}\Omega
\nonumber \\
  &  \times \int d\lbcs \, P(\lbcs|M)  \nonumber \\
  &  \times \int d\lbcsh \, P(\lbcsh|\lbcs) P(I|\lbcsh,z)
\label{equ:ndet}
\end{align}
where $\Omega$ is the survey area. In this expression, the first
probability $P(\lbcs|M)$ is the probability that a cluster of mass $M$
has some intrinsically scattered luminosity \lbcs\ (so is a function of
our $L_{CXO}-M_{H}$ relation parameters \Alm, \Blm, \intlm). The
second probability $P(\lbcsh|\lbcs)$ is the probability that a cluster
of luminosity \lbcs\ would be observed to have a luminosity \lbcsh,
and so depends on a model of the measurement error on a cluster of
arbitrary luminosity. The measurement error is expected to be
dominated by counting statistics, but a direct conversion from
luminosity to flux to counts would require an exposure time, which is
not uniform across the survey. Instead we derived an empirical
function to predict the measurement error on a cluster of given flux
by fitting a power law to the measurement errors of the (e)BCS fluxes
as a function of flux. The best fitting relation had the form
\begin{align}
\frac{\sigma_F}{F_0} & = 0.49 \left(\frac{\fbcs}{F_0}\right)^{0.53}
\end{align}
with the normalisation factor $F_0=10^{-12}$ erg s$^{-1}$ cm$^{-2}$.
The observed scaling is thus very close to the square-root scaling
expected for Poisson errors. 

The final probability in equation \ref{equ:ndet}, $P(I|\lbcsh,z)$ is
the probability that a cluster with an observed BCS luminosity
$\lbcsh$ at a redshift $z$ would be included in the subsample. This is
a combination of the step function associated with $\llim(z)$, and
$P(I|f)$ describing the BCS completeness. Note that in principal, the
probability of inclusion should depend on $T$ in addition to $\lbcsh$
and $z$, since the k-correction for the flux is temperature
dependent. However, since the BCS luminosities were estimated from
{\em ROSAT} fluxes without temperature measurements, a reference $LT$
relation was used to provide the
temperature for the k-correction. Since our inclusion probability must
match as closely as possible the BCS completeness function, we use the
same method to k-correct $\lbcs$ when estimating $f$ for the selection
function (equation \ref{eq:pif}), which removes the $T$ dependence.

The likelihood of a cluster in our sample having the observed
properties ($\lbcsh,\lcxoh,\Mh$) is given by
\begin{align}
P(\lbcsh,\lcxoh,\Mh|I,z) & = \int dM\, \int d\lbcs\,
\frac{\phi(M,z)}{\left<N\right>} \nonumber \\
 & \times P(\lbcs|M) P(\lbcsh|\lbcs) \nonumber \\
 & \times P(\lcxoh|\lbcs) P(\Mh|M).
\label{eq:pobs}
\end{align}
The quantity $\left<N\right>$ is the total number of clusters
predicted by the model, and is given by the integral of the mass
function $\phi$ over the mass range of interest, and normalises the
mass function to a probability distribution for an arbitrary cluster
to have a mass $M$ at redshift $z$. We note that $\left<N\right>$ is
not a parameter of our model, but is a useful parameter to monitor.
$P(\lbcs|M)$ is as defined above,  and the remaining terms are the
probability of each of the observables, using the measured uncertainty
for that observable. Here we have treated each of the observables as
independent, although in principal a covariance will exist between
$\lcxoh$ and $\Mh$ as the luminosity is determined within an aperture
derived from the observed mass. In practice, this effect will be weak
as the luminosity is centrally concentrated and is insensitive to the
precise choice of aperture. The joint probability of the full set of
observed cluster properties is the product of $P(\lbcsh,\lcxoh,\Mh)$
over all $N_{det}$ observed clusters in the sample. Note that we
neglect any observational uncertainty on $z$.

M10a showed that the final likelihood for the sample of clusters and
their observed properties is the product of a Poisson likelihood of
$N$ total (detected plus undetected) clusters given the model
prediction $\left<N\right>$, a binomial coefficient accounting for the
number of ways of drawing $N_{det}$ detected clusters from the total
$N$, the joint probability of the set of observed cluster properties
(the product of equation \ref{eq:pobs} over the $N_{det}$ clusters)
and the probability of not detecting the remaining $N-N_{det}$
clusters. Neglecting terms not dependent on the model parameters, the
likelihood simplifies to 
\begin{align}\label{eq:lik}
P(\lbcsh,\lcxoh,\Mh,z) \propto e^{-\left<N_{det}\right>}\prod_{i=1}^{N_{det}}\left<\tilde{n}_{det,i}\right>
\end{align}
where
\begin{align}
\left<\tilde{n}_{det,i}\right>=P(\lbcsh,\lcxoh,\Mh)\left<N\right>
\end{align}
for the $i$th cluster.

\subsubsection{Implementation and nuisance parameters}
\label{sec:nuis}

With the likelihood in equation \ref{eq:lik}, and priors on the
model parameters, we can compute the posterior probability
distribution for each parameter using standard Markov Chain Monte
Carlo (MCMC) techniques. Our final set of model parameters consists of
those parameters describing the $L_{CXO}-M_{H}$ relation
(\Alm,\Blm,\intlm), each of which were assigned uniform priors, along
with a nuisance parameter described below (naturally marginalised over
in the MCMC procedure).

  A cross-calibration factor $X_{cal}$ describing the uncertainty
  in the calibration between {\em Chandra} and {\em ROSAT} fluxes was
  introduced as a nuisance parameter in the model. $X_{cal}$ is
  defined as the ratio of the {\em Chandra} flux measured within \rf\
  to \fbcs. This parameter thus encompasses several factors: cross
  calibration between {\em ROSAT} and {\em Chandra} fluxes; a mean
  aperture correction from \rf\ to the 1.43Mpc radius to which the
  (e)BCS fluxes were extrapolated; and exclusion of point sources in
  {\em Chandra} data that may have been unresolved or only partially
  excluded in the {\em ROSAT} data. We assigned a weak prior to
  $X_{cal}$, using a log-normal distribution with mean 0 and
  standard deviation of 1 in natural log space i.e. a 100\%
  uncertainty, although the results are insensitive to this
  choice, $X_{cal}$ is well-constrained by the data.

The analysis also requires a mass function to describe the cluster
number density (above some threshold mass), and for this we used the
mass function of \cite{2008ApJ...688..709T}. In our analysis we have
treated all cosmological parameters as constant, 
and we take the same approach with the mass function, using a
tabulated mass function produced by {\em HMFcalc}
\citep{2013arXiv1306.6721M} for our {\em WMAP}9 cosmology with a
virial mass $M_{500c}$ (the mass $r_{500}$ with respect to the
critical density of the Universe). We have fixed all cosmological
parameters since we are focusing on the scaling relations, and do not
expect useful cosmological constraints from our sample. Fixing the
cosmological and mass function parameters means that our scaling
relations are not marginalised over the uncertainties in those
parameters, and should be regarded as estimates for a fixed
cosmology \citep[in contrast, the combined cosmology and scaling
relation study of][does include cosmological and mass function
parameters in the analysis]{2010MNRAS.406.1773M}. 

Our final model thus consists of three parameters describing the $LM$
relation (\Alm,\Blm,\intlm), and the nuisance parameter $X_{cal}$
describing the conversion between \fbcs\ and \fcxo.

Our fits were performed using the {\em R} statistical computing
environment\footnote{http://www.r-project.org/}, and the posterior
probability distribution was analysed using the Bayesian inference
package {\em Laplace's
  Demon}\footnote{http://www.bayesian-inference.com/software} within 
R, which contains many MCMC algorithms.  The fits were performed with
a lower bount flux cut of 10$^{12}$ erg s$^{-1}$ cm$^{-2}$ (i.e. the
lower bound of the integration), corresponding to the count rate cut
of 0.07 cnts s$^{-1}$ employed in \cite{1998MNRAS.301..881E}.  The
best fitting model parameters are described by the mean and standard
deviation of the posterior probability distribution, as estimated from
the MCMC chain after excluding the start of the chain before the
parameter values became stationary.

\subsubsection{The \lcxo-$M_{H}$\ relation}
\label{sec:lcxo-mf-relation}

\begin{figure*}
\begin{center}
\scalebox{0.7}{\includegraphics*{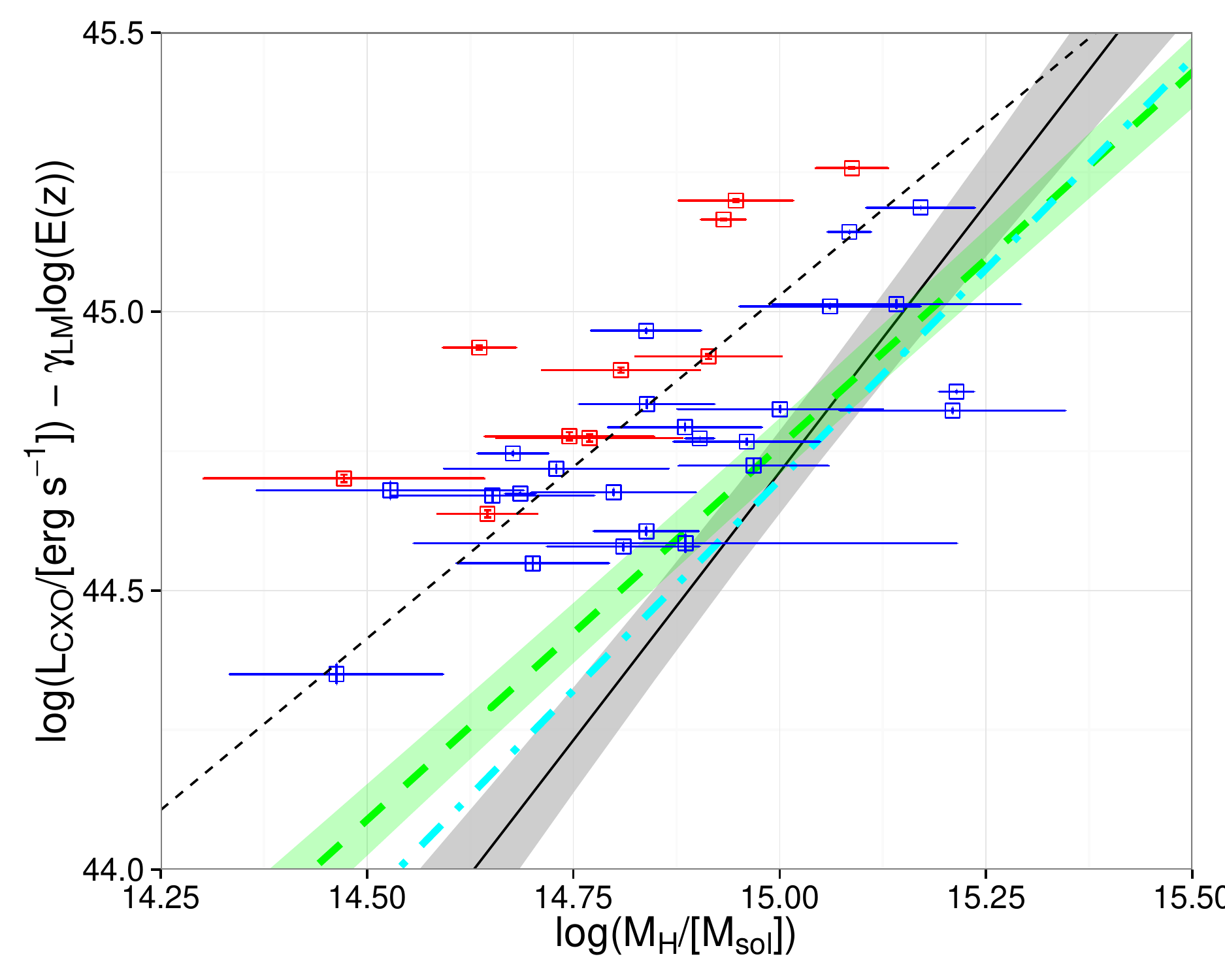}}
\caption{\small{$L_{CXO}-M_{H}$ relation with best fitting
  model. The hollow squares show the \lcxo\ luminosities split between
  the relaxed (red) and unrelaxed (blue) clusters, calibrated to
  ROSAT reference using the calibration factor $X_{cal}$. Our best
  fitting model is shown as the solid
  black line with the grey shading indicating the $1\sigma$
  uncertainty. The bold dashed green line and shaded region indicate the best
  fitting LM relation for the ``all data'' sample of M10b with its
  $1\sigma$ uncertainty, scaled by a factor of 1.10 to scale the {\em
    ROSAT} PSPC luminosties used in M10b onto our {\em Chandra}
  reference (see Sect.~\ref{sec:lcxo-mf-relation}).  The bold cyan
  dashed-dotted indicates the best fitting $LM$ relation for the BCS only
  sample of M10b (see Sect.~\ref{sec:mantzcomp}).}\label{f.lm500}}
\end{center}
\end{figure*}

\begin{figure*}
\begin{center}
\scalebox{0.9}{\includegraphics*{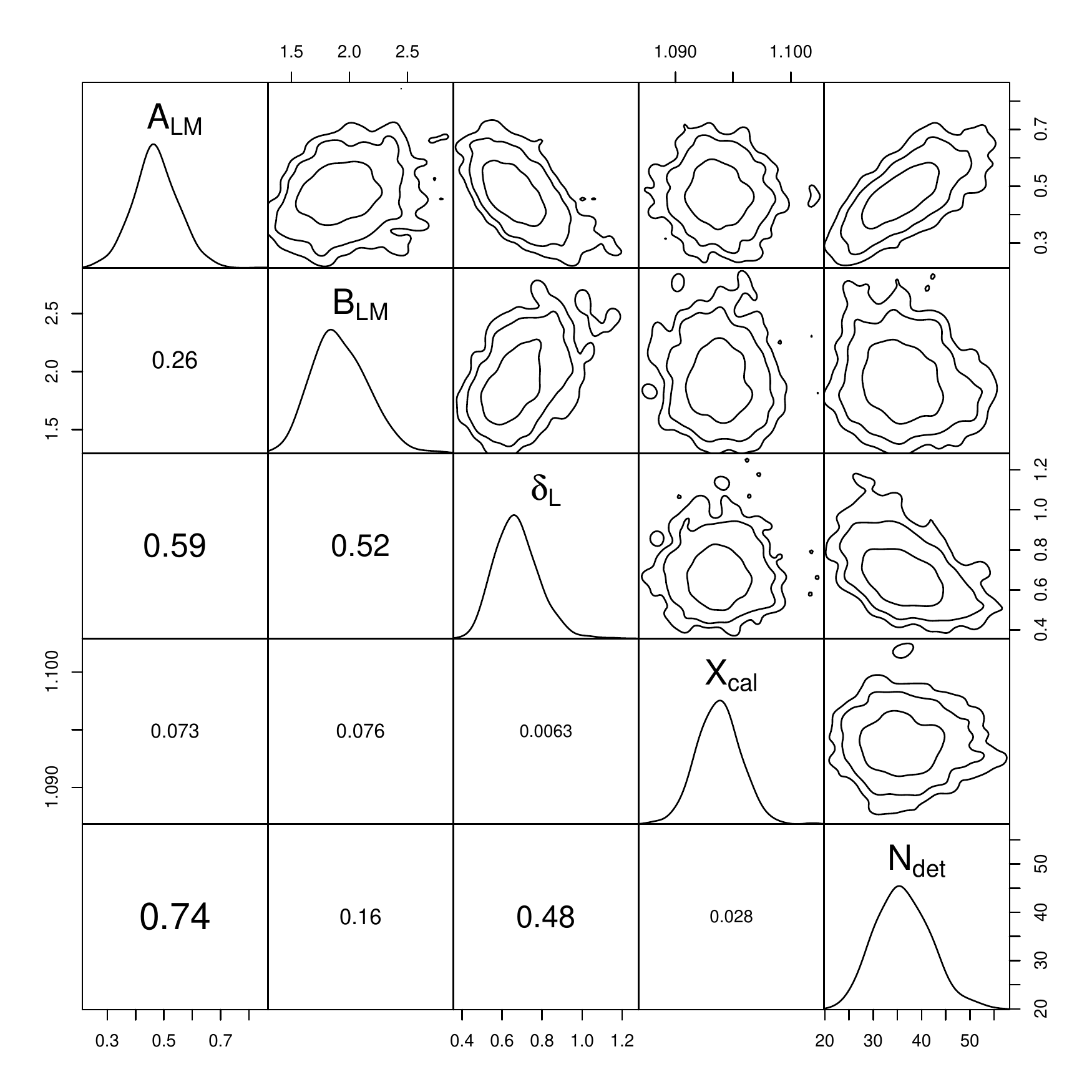}}
\caption[]{\label{f.mat} Correlation matrix of the $L_{CXO}-M_{H}$
  relation model. The posterior densities are shown along the
  diagonal, with $1\sigma$, $2\sigma$, and $3\sigma$ confidence
  contours for the pairs of parameters shown on the upper triangle
  panels. The lower triangle panels show the Pearson's correlation
  coefficient for the corresponding pair of parameters (with a text
  size proportional to the correlation strength).}
\end{center}
\end{figure*}

The $L_{CXO}-M_{H}$ relation is plotted in Figure~\ref{f.lm500}, and the
best fit parameters are given in Table~\ref{t.lmhydro}, which include the
uncertainty due to that on $X_{cal}$ (which has a best fit value of
$X_{cal}$=1.094$\pm$0.002).  The posterior probability distributions of the model
parameters and the correlations between parameters are shown in
Figure~\ref{f.mat}.  Figure~\ref{f.mat} illustrates that the $X_{cal}$
parameter is not degenerate with the parameters of interest.

\begin{table*}
\begin{center}
\caption{\label{t.lmhydro} Best fitting parameters for the $LM$ relations
  modeled here.  For the fits performed using the M10a method, the $LM$
  relations were modelled in the soft band, with
   the Chandra luminosities calibrated to the ROSAT luminosities by
   the $X_{cal}$ factor (denoted as \lbcs-$M_{H}$). In the table we
   also give the relation calibrated to $Chandra$ soft band
   luminosities (simply scaling by $X_{cal}$) and Chandra bolometric
   luminosities, by applying a bolometric correction (see
   Sect.~\ref{sec:l_bol-mf-relation}).  These relations are denoted by
   \lcxo-$M_{H}$ and \lbol-$M_{H}$ respectively.  \intlm\ is the intrinsic
   scatter measured in natural log space so represents a fractional value.}
\begin{tabular}{l|ccccc}
  \hline
  \hline
  \multicolumn{6}{c}{r$_{\rm H,500}$} \\
  Relation & Method & \Alm\ & \Blm\ & \gamlm & \intlm \\
  \hline
  \lcxo-$M_{H}$ & BCES & $1.15\pm0.15$ & $1.34\pm0.29$ & 2 & $0.38\pm0.07$  \\
  \lbcs-$M_{H}$ & M10a & $0.47\pm0.08$ & $1.92\pm0.24$ & 2 & $0.68\pm0.11$ \\
  \lcxo-$M_{H}$ & M10a & $0.52\pm0.09$ & $1.92\pm0.24$ & 2 & $0.68\pm0.11$ \\
  \lbol-$M_{H}$ & M10a & $1.45\pm0.24$ & $2.22\pm0.24$ & 7/3 & $0.68\pm0.11$ \\
\hline
\end{tabular}
\end{center}
\end{table*}

We compare to the BCES fit outlined in Section~\ref{sec:unbiaslm},
given by the black dashed line in Figure~\ref{f.lm500}.  The
difference between the two fitting methods is visibly striking, with
the normalisation of the fit when accounting for selection effects
2.2$\pm$0.4 times lower when not taking into account selection effects
(significant at the 3.7$\sigma$ level).  This comparison clearly shows
the size of the biases on cluster samples selected to have very
luminous clusters, such as the LoCuSS \citep{2008A&A...482..451Z}
and CCCP \citep{2013ApJ...767..116M} cluster samples, and is an
extreme illustration of the importance of modeling selection biases.
We also compare to the $LM$ relation given in
\citep[][hereafter M10b]{2010MNRAS.406.1773M}, which uses the method
outlined in M10a to account for selection effects.  This comparison is
discussed further in Sect.~\ref{sec:mantzcomp}.     

\subsubsection{The $L_{bol}-M_{H}$ relation}
\label{sec:l_bol-mf-relation}

So far we have been considering the scaling of soft-band luminosity
with mass, but it is often useful to refer to the bolometric
luminosity $L_{bol}$. It is not possible to fit our model directly to
bolometric luminosities, since the selection function is defined in
terms of the soft-band luminosity. Instead we can convert the
$L_{CXO}-M_{H}$ relation to an $L_{bol}-M_{H}$ relation by using a
bolometric correction.

Using {\tt XSPEC} simulations, we find this correction can be approximated
by
\begin{align}
\frac{\lbol}{\lcxo} & = A_{bol} \left(\frac{T}{T_0}\right)^{B_{bol}}
\end{align}
with $A_{bol}=2.08$ and $B_{bol}=0.54$ for $T_0=5$keV, giving bolometric
luminosities accurate to $\lesssim$3$\%$ across the range
$3-15$keV. Combined with a temperature-mass (TM) relation of the form
\begin{align}
\frac{T}{T_0} & = A_{TM} E(z)^{\gamtm} \left(\frac{M}{M_0}\right)^{B_{TM}}
\end{align}
then the bolometric $LM$ relation becomes
\begin{align}
\frac{\lbol}{L_0} & = E(z)^{\gamlm+B_k\gamtm} \Alm A_{bol} \Atm^{B_{bol}}
\left(\frac{M}{M_0}\right)^{\Blm+B_{bol}\Btm} \nonumber \\
 & = E(z)^{\gamma_{LM,bol}}A_{LM,bol}\left(\frac{M}{M_0}\right)^{B_{LM,bol}}
\end{align}
Note that the self-similar evolution of the $TM$ relation alters the
evolution of the bolometric $L_{bol}-M_{H}$ relation from that of the soft-band
$L_{CXO}-M_{H}$ relation.

\begin{figure}
\begin{center}
\scalebox{0.43}{\includegraphics*{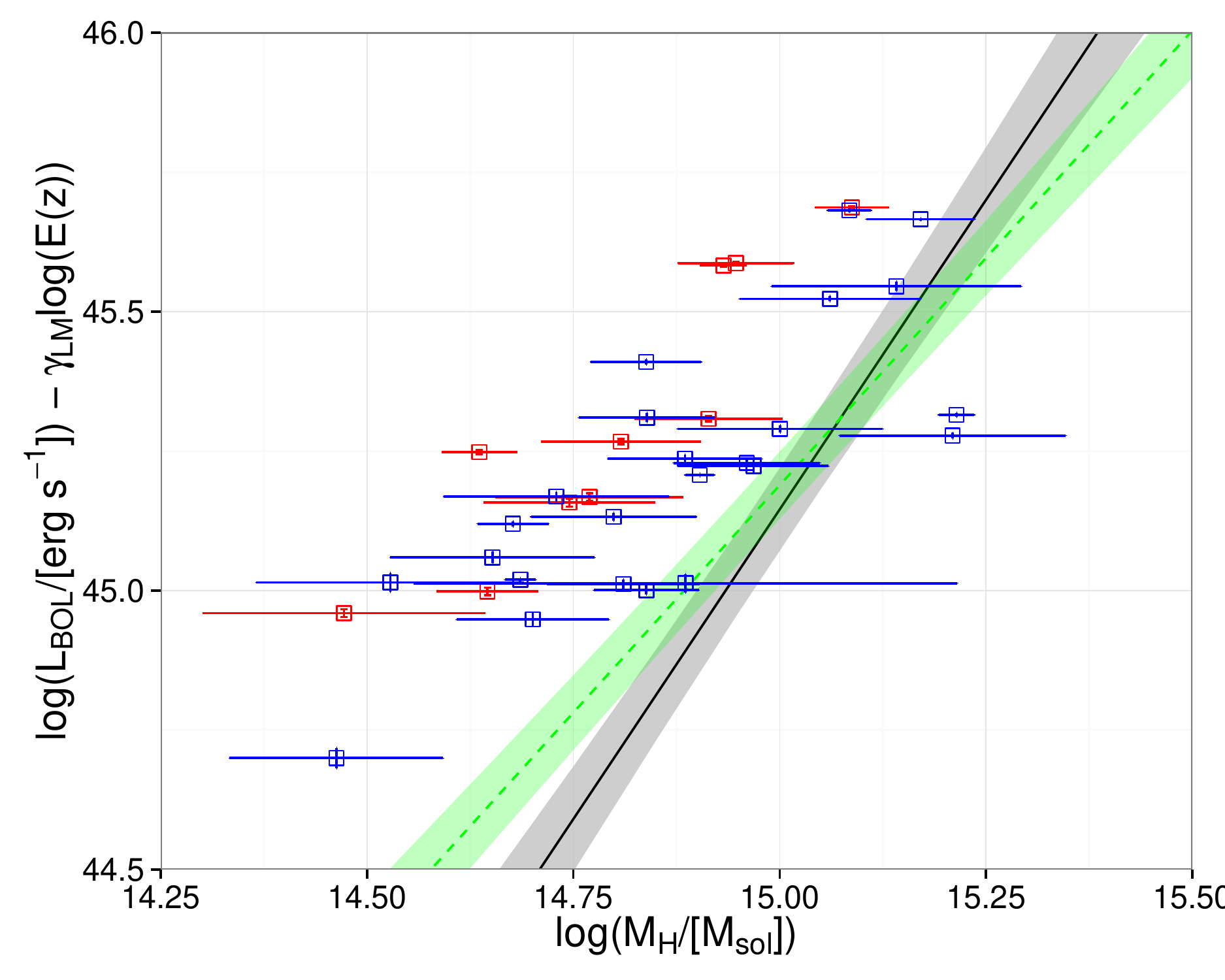}}
\caption[]{\label{f.lmbol} $L_{bol}-M_{H}$ relation with best fitting
  model. The hollow squares show the \lbol\ luminosities, split
  between the relaxed (red) and unrelaxed (blue) clusters. Our best
  fitting model is shown as the solid black line with grey shading
  indicating the $1\sigma$ uncertainty (transformed from the $L_{CXO}-M_{H}$
  using the corrections outlined in
  Sect.~\ref{sec:l_bol-mf-relation}). The dashed green line and shaded
  region indicate the best fitting bolometric $LM$ relation for the
  ``all data'' sample of M10b with its $1\sigma$ uncertainty.}
\end{center}
\end{figure}

To derive the $L_{bol}-M_{H}$ relation, we used the $TM$ relation presented
in section~\ref{sec:syseffects}.  The $L_{bol}-M_{H}$ relation is shown in
Figure~\ref{f.lmbol}, with the best-fitting model coefficients given
in Table~\ref{t.lmhydro}. The uncertainties on the model parameters include
correlated uncertainties on $\Alm$, $\Blm$, and $X_{cal}$ from the
posterior chains of the MCMC analysis, and the uncertainty on the
slope and normalisation of the $TM$ relation, treating those as
independent. This is justified since there is not a strong covariance
between the $TM$ and $LM$ relations
\citep{2010MNRAS.406.1773M,2014MNRAS.437.1171M}.

\section{Discussion}
\label{sec:disc}

\subsection{Comparison with Mantz et al. (2010b)}
\label{sec:mantzcomp}

Figure~\ref{f.lm500} shows the best fitting relation of the ``all
data'' sample of M10b. This relation was derived from a sample of
238 clusters at $z<0.5$ and with {\em ROSAT} luminosities for all clusters,
with {\em Chandra} follow-up observations providing luminosities and
masses (estimated from the gas mass) for a subset of 66 objects.
This $LM$ relation was derived as part of a cosmological analysis
(which includes non-cluster cosmological data to constrain their
derived cosmological parameters), rather than having cosmological
parameters fixed as in our analysis. In spite of these differences,
the M10b $LM$ relation is the most suitable comparison for our work as
it is the only other example of a $LM$ relation with full corrections
for selection biases. 

The luminosities in M10b were calibrated to a {\em ROSAT} PSPC
reference, so in order to compare with our $L_{CXO}-M_{H}$ relation,
we derived a calibration of the M10b luminosities onto our {\em
  Chandra} reference by comparing luminosities for 24 objects in
common between the samples. We found a low scatter correlation between
the values, with the M10b luminosities higher by a factor of 1.10 on
average. This difference is in the opposite sense to that found in
M10b, where {\em Chandra} luminosities (using CALDB 4.1.2) were found
to be $14\%$ higher than PSPC luminosities. This difference is due to
the evolving {\em Chandra} calibration (we used CALDB 4.6.2), and
systematic differences in the analyses. We do not pursue these
calibration differences further, but simply scale the M10b $LM$
relation normalisation by a factor $1.10$ and note that this
difference is typical of cross-calibration uncertainties in X-ray
telescope effective areas \citep{2010A&A...523A..22N}.  With this
  scaling in place we find a reasonably good agreement between the
  relations, with the slope of our relation being steeper at the
  2.4$\sigma$ level.

One might expect a better agreement between the relations since
a) the fit method used in this work is based upon that used in M10a,
and b) our sample contains 24 clusters in common with the M10b sample.
 We investigate this difference by first comparing the masses of
  the 24 clusters in common, noting that M10b use the gas mass as a
  proxy for cluster mass.  We find that the masses of the clusters in
  common are entirely consistent (when fitting a power-law relation
  with the slope fixed at unity).  The scatter between the mass
  estimates is 22$\pm$5\%, which is consistent with the larger scatter
  we find in the $L_{CXO}-M_{H}$ relation compared with that in the
  M10b $L-M_{Mgas}$ relation.  Secondly, the samples that were used is
  that M10b used clusters drawn from three different parent 
  surveys, while our sample is derived from the (e)BCS. To eliminate
  this difference, we compared our $L_{CXO}-M_{H}$ relation to a
  version of the M10b $LM$ relation derived using only clusters from
  the (e)BCS survey (A. Mantz, private communication).

The bold cyan dashed-dotted line Figure~\ref{f.lm500} shows the
  relation based upon using (e)BCS-only clusters in M10b, using the
  M10a analysis.  When these consistent cluster
subsets are used the agreement is improved and the slopes and
normalisation are both within $\approx$1.5$\sigma$.

\subsection{Mass Comparisons}
\label{sec:landrycomp}

The sample of clusters presented in this work were also studied in
\cite{2013MNRAS.433.2790L}, hereafter L13.  We compare here the
masses derived in this work to those presented in L13.  We note that
although this work and L13 use the \cite{2006ApJ...640..691V} method
to derive cluster masses, the implementation was performed separately.
There have also been three updates from the L13 paper, to this
work. As stated, we use {\em WMAP}9 cosmology throughout, whereas L13
use {\em WMAP}7 \citep[$\Omega_{\rm M}$=0.27, $\Omega_{\Lambda}$=0.73 and
H$_{0}$=70.2,][]{2011ApJS..192...18K}.  The next change is the
versions of CIAO and CALDB used in the separate analysis, we have used
CIAO 4.6 and CALDB v4.5.9, whereas L13 use CIAO 4.2 and CALDB v4.3.1.
Finally, in this work we used the $C$-statistic in spectral fits,
while L13 used the $\chi^2$ statistic with binned spectra.

\begin{figure}
\begin{center}
\includegraphics[width=9.0cm]{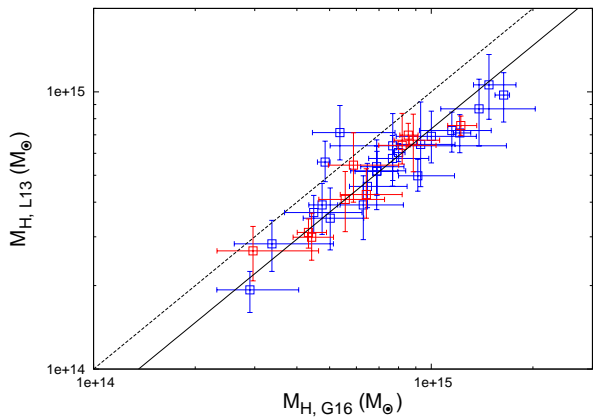}
\end{center}
\vspace{-1cm}
\caption[]{\small{Comparison of the mass estimate determined in this
    work and those presented in \cite{2013MNRAS.433.2790L}.  The
    dashed line represents a 1:1 relation, with the black line a fit
    to the data assuming a power-law relation with the slope fixed at
    unity.}\label{fig:landrycomp}}
\end{figure}

Figure~\ref{fig:landrycomp} compares the masses given in L10 to those
derived in this work.  The black dashed line represents a 1:1
relationship. The black solid line is a fit to the data (using a power
-law with the slope fixed at unity), where we find that our masses are
on average 29$\pm$3.0\% higher than those in L13.  We should note that
the masses are not compared within the same radii, but within their
respective $r_{500}$.  An analysis of our cluster sample with the same
cosmology, CIAO and CALDB versions, and $\chi^2$ statistic yields a
1:1 mass comparison. We investigated the impact of these differences
in analysis methods. Changing the cosmology from WMAP7 to WMAP9 changed
  $r_{500}$ and $M_{500}$ by $\approx$1\%. The choice of statistic in the
  spectra fitting had a larger impact. If we used the $\chi^2$
  statistic, with spectra grouped to at least 30 counts per bin, the
  inferred temperatures were systematically lower, and the derived
  $M_{500}$ were lower by $10\%$ on average. The remaining 20\%
  difference in mass compared to L13 is thus due to the different
  CIAO/CALDB versions used. There have been several major updates to
  the CALDB between the two versions used in these studies, and we do
  not attempt to investigate which are responsible for the observed
  shift in masses.

We further compare our derived masses to those presented in
\cite{2014MNRAS.443.2342M}. \cite{2014MNRAS.443.2342M} studied a
sample of 50 clusters from the LoCuSS cluster sample, calculating
masses based on {\em Chandra} observations and utilising temperature
profiles to calculate hydrostatic masses. Using 21 clusters in common
between the LoCuSS sample and the clusters in this work, we find our
masses are 11$\pm$5\% higher, consistent with our use of the
  $C$-statistic \citep[][used the $\chi^2$
  statistic]{2014MNRAS.443.2342M}.

\subsection{Systematic Effects}
\label{sec:syseffects}

\subsubsection{Reliability of hydrostatic masses}

  Throughout this work we have used masses determined assuming
  hydrostatic equilibrium of the ICM. However, non-thermal pressure
  sources associated with bulk and turbulent motions of the cluster
  gas lead to what is known as the hydrostatic mass bias.
  Hydrodynamical simulations have shown that these processes can lead
  to under-estimates of the hydrostatic cluster mass by $\sim$10-30\%
  \citep[e.g][]{2004MNRAS.355.1091K,2008ApJ...681..167J,2009ApJ...705.1129L,2014MNRAS.442..521S}.
  Observationally however, the level of hydrostatic bias is less
  clear. Several recent studies have attempted to measure the amount
  of bias by comparing hydrostatic mass estimates to estimates based
  upon other techniques (e.g. weak-lensing, caustics) that are
  independent of the equilibrium state of the ICM. Some have found
  evidence for a level of bias similar to that of simulations
  \citep[e.g][]{2014MNRAS.443.1973V,2015MNRAS.449..685H}, while others
  found results consistent with no hydrostatic bias
  \citep[e.g.][]{2015arXiv151107872M,2016MNRAS.457.1522A,2016MNRAS.456L..74S}.
  Intriguingly, \cite{2016MNRAS.456L..74S} showed that results of
  \cite{2014MNRAS.443.1973V} and \cite{2015MNRAS.449..685H}
  converged on a low ($5-10\%$) bias when only $z<0.3$ clusters were
  used (the same redshift range as our sample). 

  In addition to the question of an overall hydrostatic bias, we must
  consider whether the hydrostatic masses will be more biased or
  scattered relative to the true mass for unrelaxed clusters.
  Hydrodynamical simulations tend to agree that disturbed clusters
  show a larger bias than relaxed ones, although disagreements exist
  over the size of the biases
  \citep[e.g.][]{2009ApJ...705.1129L,2014ApJ...792...25N,2016MNRAS.455.2936S}.
  However, for our sample we find observational evidence that the
  hydrostatic masses of the unrelaxed clusters are not differently
  biased than those of the relaxed clusters. 

  Firstly, in \cite{2015arXiv151107872M}, we compared the hydrostatic
  and caustic mass profiles of 16 clusters from the sample studied in
  this work. The caustic masses are taken from
  \cite{2013ApJ...767...15R}, and are not effected by the dynamical
  state of the cluster. The comparison implies that the hydrostatic
  masses cannot be biased low by more than $10\%$ (at the $3\sigma$
  level), and shows no evidence for a dependence on the dynamical
  state of the clusters (albeit based on a relatively small subset of
  clusters).

\begin{figure}
\begin{center}
\includegraphics[width=9.0cm]{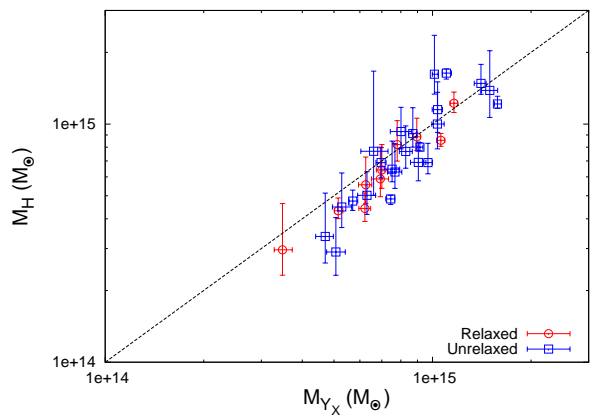}
\end{center}
\vspace{-1cm}
\caption[]{\small{Comparison of the masses calculated via our
    hydrostatic mass analysis and those calculated via the $YM$
    relation of \cite{2009ApJ...692.1033V}, split between the relaxed
    (red open squares) and unrelaxed (blue open squares) samples.  The
  black dashed line represents a 1:1 relationship.}\label{fig:YM}} 
\end{figure}

  A second piece of observational evidence that our hydrostatic masses 
  are reliable comes from the comparison to the masses calculated via
  the $Y_{X}$-Mass ($Y_{X}-M$) relation of \cite{2009ApJ...692.1033V}.
  $Y_X$ is the product of the gas mass and core-excised temperature
  measured.  Simulations have shown the $Y_{X}$ parameter to be a
  low-scatter proxy for cluster mass, regardless of its dynamical
  state \citep{2006ApJ...650..128K}, however, observational evidence
  is so far lacking.  Nonetheless, if hydrostatic masses
  for unrelaxed clusters were significantly effected by biases
  compared to relaxed clusters, a comparison of hydrostatic masses to
  $Y_{X}$ based masses should highlight this.  We iterated on the
  $Y_{X}-M$ relation until $\rf$ converged, and the masses ($M_{Y_{X}}$)
  were calculated from the $Y_{X}$ determined within this radius.
  Figure~\ref{fig:YM} shows the resulting comparison of the masses
  determined from our hydrostatic analysis, used throughout this work,
  and $M_{Y_{X}}$ as described above, split between the relaxed (red
  open squares) and unrelaxed (blue open squares) clusters.  The black
  dashed line represents a 1:1 relationship.  Although the clusters
  appear to differ slightly from the 1:1 correlation, the data do not
  exclude a 1:1 relationship. 

  The resulting mass comparison for the relaxed and unrelaxed
  subsamples are in excellent agreement, and both show a similarly low
  scatter, supporting the idea that our hydrostatic masses are not
  more biased or scattered for the unrelaxed clusters compared with
  their relaxed counterparts. A possible reason for this is that the
  dynamical activity of the clusters is more important in the inner
  parts of the cluster, while the ICM around $\rf$ is close to
  equilibrium.

  Overall, these observational studies support the use of hydrostatic
  masses as a calibrator of the $LM$ relation, but clearly the
  question of hydrostatic bias remains open and is an important
  possible source of systematic uncertainty in our results.

\subsubsection{Dependence on cosmological parameters}
\label{sec:depend-cosm-param}
Throughout this paper we have assumed a {\em WMAP}9 cosmology.
Recently however, data from {\em Planck} has found support for a
different cosmology \citep[][]{2015arXiv150201589P}. Since the
  cosmology is held fixed in our analysis, this is a source of
  systematic uncertainty. As stated in Sect.~\ref{sec:landrycomp}, the
effect of changing the cosmology is negligible on the measured cluster
properties when changing from {\em WMAP}7 to {\em WMAP}9. Although the
change is larger for the {\em Planck} cosmology, the derived
properties will change only at the 1\% level. Therefore, the $LM$
relation of the observed clusters will remain largely unchanged. The
cosmology will impact more strongly the bias-corrected fitting of the
$L_{CXO}-M_{H}$ relation. The {\em Planck} cosmology currently predicts a
higher value of $\sigma_{8}$ (the fluctuation amplitude at 8 $h^{-1}$
Mpc) compared to {\em WMAP}9.  An increase in value of
  $\sigma_{8}$ would lead to an increased number density of clusters
  in the Universe. In this situation, in order for the model to
  correctly predict the number of clusters observed in our sample, the
  underlying cluster population, inferred from the mass function,
  would have to be less luminous on average for their given mass. This
  would lead to a lower inferred normalisation of the $LM$ relation.
  Incidently, if the underlying population were on average less
  luminous, our cluster sample would be more extremely biased, falling
further into the tail of the luminosity intrinsic scatter.

\subsubsection{Uncertainty on the selection function}
\label{sec:uncert-select-funct}

When modelling the selection function (see Sect.~\ref{sec:sfunc}),
  we use three completeness estimates at specific fluxes and model
  with a logistic function.  However, this selection function is
  likely a simplified form of the true (e)BCS selection function
  (which is unavailable).  We tested our use of the logistic function to
model the selection function by considering some limiting cases of the
behavior of P(I$|$F).  For this test we considered step functions at two
different flux cuts.  The first step function uses a flux cut
calculated from the luminosity of the least luminous cluster in our
sample.   The second flux cut is taken from the 90\% completeness
level ($F_{X,90\%}$=4.4$\times$10$^{-12}$ erg s$^{-1}$ cm$^{-2}$) as
given in \cite{1998MNRAS.301..881E}.  We note that $F_{X,90\%}$ is
larger than the eBCS flux limit and thus we remove 5 clusters
from our sample when considering this second step function. The $L_{CXO}-M_{H}$
relation modelled using these step functions, and the logistic
function used throughout this anaylsis, are all entirely consistent.

\section{Summary and Conclusions}
\label{sec:conc}

Using a statistically complete sample of 34 high luminosity galaxy
clusters, we have derived the X-ray hydrostatic masses of the cluster
sample, and investigated the form of the luminosity-mass scaling
relation.  The form of the relation is fit using two methods, one
using a simple regression fit to the data, and another accounting for
selection effects.  Our main conclusions are as follows.
\begin{enumerate}
\item Using the central cooling time, the cuspiness of the gas density
  profile, and the centroid shift, we separate the cluster sample into
  relaxed cool core (relaxed) and non-cool core (unrelaxed) clusters.
  We find 10/34 relaxed clusters and 24/34 unrelaxed clusters.
\item We derive hydrostatic mass estimates for the cluster sample,
  irrespective of the dynamical state of the cluster, utilising gas
  density and temperature profiles.
\item Taking fully into account selection effects, we fit for the
  soft-band luminosity-mass relation, finding a slope of $B_{\rm
    LM}$=1.92$\pm$0.24 and scatter $\delta_{\rm LM}$=0.68$\pm$0.11.
  Comparing this relation to one that does not account for selection
  effects, we find that accounting for selection effects lowers the
  normalisation of the $L_{CXO}-M_{H}$ relation by a factor of
  2.2$\pm$0.4.  
\item Throughout the analysis we use the $C$-statistic when fitting
  cluster spectra.  Although the $C$-statistic has been shown to more
  accurately recover the cluster temperature, the $\chi^2$ statistic
  has been more commonly employed.  Comparing the hydrostatic masses
  determined using both statistics, we find the $C$-statistic masses are
  10$\pm$2.3\% higher that those found using the $\chi^2$ statistic.  
\item Testing the use of step functions to model the selection
  function, we find that the fitted $L_{CXO}-M_{H}$ relation is
  consistent with the relation when using our logistic function to
  model the selection function.
\end{enumerate}

We have studied a highly biased cluster sample, where the selection
has a profound effect on the derived scaling relations.  We have shown
the importance of taking into account the selection effects when
fitting for the observed luminosity-mass scaling relation.  This is
crucial for the understanding of scaling laws of cluster samples when
used for the purposes of cosmology.  Current and upcoming cluster surveys
(e.g. {\em XXL}, {\em Planck}, clusters detected with {\em e-ROSITA})
will all require a method of determining the cluster mass for
cosmological studies.  This will most likely come in the form of a
mass-observable scaling relation, for which the selection effects will
need to be fully accounted for.  Furthermore, although high luminosity
clusters are observationally cheaper to follow-up in order to derive
X-ray hydrostatic masses, and hence the construction of scaling
relations, they lead to highly biased cluster samples, as shown
throughout this work.      

\section*{Acknowledgments}

PG acknowledges support from the UK Science and Technology Facilities
Council.  We thank A. Mantz for useful discussions and for providing
additional fits used in the comparison to his work.  

\bibliographystyle{mnras}

\begin{thebibliography}{100}
\makeatletter
\relax
\def\mn@urlcharsother{\let\do\@makeother \do\$\do\&\do\#\do\^\do\_\do\%\do\~}
\def\mn@doi{\begingroup\mn@urlcharsother \@ifnextchar [ {\mn@doi@}
  {\mn@doi@[]}}
\def\mn@doi@[#1]#2{\def\@tempa{#1}\ifx\@tempa\@empty \href
  {http://dx.doi.org/#2} {doi:#2}\else \href {http://dx.doi.org/#2} {#1}\fi
  \endgroup}
\def\mn@eprint#1#2{\mn@eprint@#1:#2::\@nil}
\def\mn@eprint@arXiv#1{\href {http://arxiv.org/abs/#1} {{\tt arXiv:#1}}}
\def\mn@eprint@dblp#1{\href {http://dblp.uni-trier.de/rec/bibtex/#1.xml}
  {dblp:#1}}
\def\mn@eprint@#1:#2:#3:#4\@nil{\def\@tempa {#1}\def\@tempb {#2}\def\@tempc
  {#3}\ifx \@tempc \@empty \let \@tempc \@tempb \let \@tempb \@tempa \fi \ifx
  \@tempb \@empty \def\@tempb {arXiv}\fi \@ifundefined
  {mn@eprint@\@tempb}{\@tempb:\@tempc}{\expandafter \expandafter \csname
  mn@eprint@\@tempb\endcsname \expandafter{\@tempc}}}

\bibitem[\protect\citeauthoryear{{Aihara} et~al.,}{{Aihara}
  et~al.}{2011}]{2011ApJS..193...29A}
{Aihara} H.,  et~al., 2011, \mn@doi [\apjs] {10.1088/0067-0049/193/2/29}, \href
  {http://adsabs.harvard.edu/abs/2011ApJS..193...29A} {193, 29}

\bibitem[\protect\citeauthoryear{{Akritas} \& {Bershady}}{{Akritas} \&
  {Bershady}}{1996}]{1996ApJ...470..706A}
{Akritas} M.~G.,  {Bershady} M.~A.,  1996, \mn@doi [\apj] {10.1086/177901},
  \href {http://adsabs.harvard.edu/abs/1996ApJ...470..706A} {470, 706}

\bibitem[\protect\citeauthoryear{{Allen}, {Rapetti}, {Schmidt}, {Ebeling},
  {Morris}  \& {Fabian}}{{Allen} et~al.}{2008}]{2008MNRAS.383..879A}
{Allen} S.~W.,  {Rapetti} D.~A.,  {Schmidt} R.~W.,  {Ebeling} H.,  {Morris}
  R.~G.,   {Fabian} A.~C.,  2008, \mn@doi [\mnras]
  {10.1111/j.1365-2966.2007.12610.x}, \href
  {http://adsabs.harvard.edu/abs/2008MNRAS.383..879A} {383, 879}

\bibitem[\protect\citeauthoryear{{Allen}, {Evrard}  \& {Mantz}}{{Allen}
  et~al.}{2011}]{2011ARA&A..49..409A}
{Allen} S.~W.,  {Evrard} A.~E.,   {Mantz} A.~B.,  2011, \mn@doi [\araa]
  {10.1146/annurev-astro-081710-102514}, \href
  {http://ukads.nottingham.ac.uk/abs/2011ARA%26A..49..409A} {49, 409}

\bibitem[\protect\citeauthoryear{{Anders} \& {Grevesse}}{{Anders} \&
  {Grevesse}}{1989}]{1989GeCoA..53..197A}
{Anders} E.,  {Grevesse} N.,  1989, \mn@doi [\gca]
  {10.1016/0016-7037(89)90286-X}, \href
  {http://adsabs.harvard.edu/abs/1989GeCoA..53..197A} {53, 197}

\bibitem[\protect\citeauthoryear{{Andreon}}{{Andreon}}{2012}]{2012A&A...546A..%
.6A}
{Andreon} S.,  2012, \mn@doi [\aap] {10.1051/0004-6361/201219194}, \href
  {http://adsabs.harvard.edu/abs/2012A%26A...546A...6A} {546, A6}

\bibitem[\protect\citeauthoryear{{Applegate} et~al.,}{{Applegate}
  et~al.}{2016}]{2016MNRAS.457.1522A}
{Applegate} D.~E.,  et~al., 2016, \mn@doi [\mnras] {10.1093/mnras/stw005},
  \href {http://adsabs.harvard.edu/abs/2016MNRAS.457.1522A} {457, 1522}

\bibitem[\protect\citeauthoryear{{Arnaud}}{{Arnaud}}{1996}]{1996ASPC..101...17%
A}
{Arnaud} K.~A.,  1996, in {Jacoby} G.~H.,  {Barnes} J.,  eds,  Astronomical
  Society of the Pacific Conference Series Vol. 101, Astronomical Data Analysis
  Software and Systems V. p.~17

\bibitem[\protect\citeauthoryear{{Arnaud} \& {Evrard}}{{Arnaud} \&
  {Evrard}}{1999}]{1999MNRAS.305..631A}
{Arnaud} M.,  {Evrard} A.~E.,  1999, \mn@doi [\mnras]
  {10.1046/j.1365-8711.1999.02442.x}, \href
  {http://adsabs.harvard.edu/abs/1999MNRAS.305..631A} {305, 631}

\bibitem[\protect\citeauthoryear{{Bauer}, {Fabian}, {Sanders}, {Allen}  \&
  {Johnstone}}{{Bauer} et~al.}{2005}]{2005MNRAS.359.1481B}
{Bauer} F.~E.,  {Fabian} A.~C.,  {Sanders} J.~S.,  {Allen} S.~W.,   {Johnstone}
  R.~M.,  2005, \mn@doi [\mnras] {10.1111/j.1365-2966.2005.08999.x}, \href
  {http://adsabs.harvard.edu/abs/2005MNRAS.359.1481B} {359, 1481}

\bibitem[\protect\citeauthoryear{{Bharadwaj}, {Reiprich}, {Lovisari}  \&
  {Eckmiller}}{{Bharadwaj} et~al.}{2015}]{2015A&A...573A..75B}
{Bharadwaj} V.,  {Reiprich} T.~H.,  {Lovisari} L.,   {Eckmiller} H.~J.,  2015,
  \mn@doi [\aap] {10.1051/0004-6361/201424586}, \href
  {http://adsabs.harvard.edu/abs/2015A%26A...573A..75B} {573, A75}

\bibitem[\protect\citeauthoryear{{B{\"o}hringer} et~al.,}{{B{\"o}hringer}
  et~al.}{2000}]{2000ApJS..129..435B}
{B{\"o}hringer} H.,  et~al., 2000, \mn@doi [\apjs] {10.1086/313427}, \href
  {http://adsabs.harvard.edu/abs/2000ApJS..129..435B} {129, 435}

\bibitem[\protect\citeauthoryear{{B{\"o}hringer}, {Chon}  \&
  {Collins}}{{B{\"o}hringer} et~al.}{2014}]{2014A&A...570A..31B}
{B{\"o}hringer} H.,  {Chon} G.,   {Collins} C.~A.,  2014, \mn@doi [\aap]
  {10.1051/0004-6361/201323155}, \href
  {http://adsabs.harvard.edu/abs/2014A%26A...570A..31B} {570, A31}

\bibitem[\protect\citeauthoryear{{Comis}, {de Petris}, {Conte}, {Lamagna}  \&
  {de Gregori}}{{Comis} et~al.}{2011}]{2011MNRAS.418.1089C}
{Comis} B.,  {de Petris} M.,  {Conte} A.,  {Lamagna} L.,   {de Gregori} S.,
  2011, \mn@doi [\mnras] {10.1111/j.1365-2966.2011.19562.x}, \href
  {http://adsabs.harvard.edu/abs/2011MNRAS.418.1089C} {418, 1089}

\bibitem[\protect\citeauthoryear{{Connor}, {Donahue}, {Sun}, {Hoekstra},
  {Mahdavi}, {Conselice}  \& {McNamara}}{{Connor}
  et~al.}{2014}]{2014ApJ...794...48C}
{Connor} T.,  {Donahue} M.,  {Sun} M.,  {Hoekstra} H.,  {Mahdavi} A.,
  {Conselice} C.~J.,   {McNamara} B.,  2014, \mn@doi [\apj]
  {10.1088/0004-637X/794/1/48}, \href
  {http://adsabs.harvard.edu/abs/2014ApJ...794...48C} {794, 48}

\bibitem[\protect\citeauthoryear{{Dahle}}{{Dahle}}{2006}]{2006ApJ...653..954D}
{Dahle} H.,  2006, \mn@doi [\apj] {10.1086/508654}, \href
  {http://adsabs.harvard.edu/abs/2006ApJ...653..954D} {653, 954}

\bibitem[\protect\citeauthoryear{{Ebeling}, {Edge}, {Bohringer}, {Allen},
  {Crawford}, {Fabian}, {Voges}  \& {Huchra}}{{Ebeling}
  et~al.}{1998}]{1998MNRAS.301..881E}
{Ebeling} H.,  {Edge} A.~C.,  {Bohringer} H.,  {Allen} S.~W.,  {Crawford}
  C.~S.,  {Fabian} A.~C.,  {Voges} W.,   {Huchra} J.~P.,  1998, \mn@doi
  [\mnras] {10.1046/j.1365-8711.1998.01949.x}, \href
  {http://adsabs.harvard.edu/abs/1998MNRAS.301..881E} {301, 881}

\bibitem[\protect\citeauthoryear{{Ebeling}, {Edge}, {Allen}, {Crawford},
  {Fabian}  \& {Huchra}}{{Ebeling} et~al.}{2000}]{2000MNRAS.318..333E}
{Ebeling} H.,  {Edge} A.~C.,  {Allen} S.~W.,  {Crawford} C.~S.,  {Fabian}
  A.~C.,   {Huchra} J.~P.,  2000, \mn@doi [\mnras]
  {10.1046/j.1365-8711.2000.t01-1-03549.x}, \href
  {http://adsabs.harvard.edu/abs/2000MNRAS.318..333E} {318, 333}

\bibitem[\protect\citeauthoryear{{Eke}, {Cole}, {Frenk}  \& {Patrick
  Henry}}{{Eke} et~al.}{1998}]{1998MNRAS.298.1145E}
{Eke} V.~R.,  {Cole} S.,  {Frenk} C.~S.,   {Patrick Henry} J.,  1998, \mn@doi
  [\mnras] {10.1046/j.1365-8711.1998.01713.x}, \href
  {http://adsabs.harvard.edu/abs/1998MNRAS.298.1145E} {298, 1145}

\bibitem[\protect\citeauthoryear{{Ettori}}{{Ettori}}{2015}]{2015MNRAS.446.2629%
E}
{Ettori} S.,  2015, \mn@doi [\mnras] {10.1093/mnras/stu2292}, \href
  {http://adsabs.harvard.edu/abs/2015MNRAS.446.2629E} {446, 2629}

\bibitem[\protect\citeauthoryear{{Evans} et~al.,}{{Evans}
  et~al.}{2010}]{2010ApJS..189...37E}
{Evans} I.~N.,  et~al., 2010, \mn@doi [\apjs] {10.1088/0067-0049/189/1/37},
  \href {http://adsabs.harvard.edu/abs/2010ApJS..189...37E} {189, 37}

\bibitem[\protect\citeauthoryear{{Giles}, {Maughan}, {Birkinshaw}, {Worrall}
  \& {Lancaster}}{{Giles} et~al.}{2012}]{2012MNRAS.419..503G}
{Giles} P.~A.,  {Maughan} B.~J.,  {Birkinshaw} M.,  {Worrall} D.~M.,
  {Lancaster} K.,  2012, \mn@doi [\mnras] {10.1111/j.1365-2966.2011.19715.x},
  \href {http://adsabs.harvard.edu/abs/2012MNRAS.419..503G} {419, 503}

\bibitem[\protect\citeauthoryear{{Hao} et~al.,}{{Hao}
  et~al.}{2010}]{2010ApJS..191..254H}
{Hao} J.,  et~al., 2010, \mn@doi [\apjs] {10.1088/0067-0049/191/2/254}, \href
  {http://adsabs.harvard.edu/abs/2010ApJS..191..254H} {191, 254}

\bibitem[\protect\citeauthoryear{{Henry}}{{Henry}}{1997}]{1997ApJ...489L...1H}
{Henry} J.~P.,  1997, \mn@doi [\apjl] {10.1086/310949}, \href
  {http://adsabs.harvard.edu/abs/1997ApJ...489L...1H} {489, L1+}

\bibitem[\protect\citeauthoryear{{Henry} \& {Arnaud}}{{Henry} \&
  {Arnaud}}{1991}]{1991ApJ...372..410H}
{Henry} J.~P.,  {Arnaud} K.~A.,  1991, \mn@doi [\apj] {10.1086/169987}, \href
  {http://adsabs.harvard.edu/abs/1991ApJ...372..410H} {372, 410}

\bibitem[\protect\citeauthoryear{{Hinshaw} et~al.,}{{Hinshaw}
  et~al.}{2013}]{2013ApJS..208...19H}
{Hinshaw} G.,  et~al., 2013, \mn@doi [\apjs] {10.1088/0067-0049/208/2/19},
  \href {http://adsabs.harvard.edu/abs/2013ApJS..208...19H} {208, 19}

\bibitem[\protect\citeauthoryear{{Hoekstra}, {Herbonnet}, {Muzzin}, {Babul},
  {Mahdavi}, {Viola}  \& {Cacciato}}{{Hoekstra}
  et~al.}{2015}]{2015MNRAS.449..685H}
{Hoekstra} H.,  {Herbonnet} R.,  {Muzzin} A.,  {Babul} A.,  {Mahdavi} A.,
  {Viola} M.,   {Cacciato} M.,  2015, \mn@doi [\mnras] {10.1093/mnras/stv275},
  \href {http://adsabs.harvard.edu/abs/2015MNRAS.449..685H} {449, 685}

\bibitem[\protect\citeauthoryear{{Hudson}, {Mittal}, {Reiprich}, {Nulsen},
  {Andernach}  \& {Sarazin}}{{Hudson} et~al.}{2010}]{2010A&A...513A..37H}
{Hudson} D.~S.,  {Mittal} R.,  {Reiprich} T.~H.,  {Nulsen} P.~E.~J.,
  {Andernach} H.,   {Sarazin} C.~L.,  2010, \mn@doi [\aap]
  {10.1051/0004-6361/200912377}, \href
  {http://adsabs.harvard.edu/abs/2010A%26A...513A..37H} {513, A37}

\bibitem[\protect\citeauthoryear{{Ikebe}, {Reiprich}, {B{\"o}hringer}, {Tanaka}
   \& {Kitayama}}{{Ikebe} et~al.}{2002}]{2002A&A...383..773I}
{Ikebe} Y.,  {Reiprich} T.~H.,  {B{\"o}hringer} H.,  {Tanaka} Y.,   {Kitayama}
  T.,  2002, \mn@doi [\aap] {10.1051/0004-6361:20011769}, \href
  {http://adsabs.harvard.edu/abs/2002A%26A...383..773I} {383, 773}

\bibitem[\protect\citeauthoryear{{Jeltema}, {Hallman}, {Burns}  \&
  {Motl}}{{Jeltema} et~al.}{2008}]{2008ApJ...681..167J}
{Jeltema} T.~E.,  {Hallman} E.~J.,  {Burns} J.~O.,   {Motl} P.~M.,  2008,
  \mn@doi [\apj] {10.1086/587502}, \href
  {http://adsabs.harvard.edu/abs/2008ApJ...681..167J} {681, 167}

\bibitem[\protect\citeauthoryear{{Kaiser}}{{Kaiser}}{1986}]{1986MNRAS.222..323%
K}
{Kaiser} N.,  1986, \mnras, \href
  {http://adsabs.harvard.edu/abs/1986MNRAS.222..323K} {222, 323}

\bibitem[\protect\citeauthoryear{{Kalberla}, {Burton}, {Hartmann}, {Arnal},
  {Bajaja}, {Morras}  \& {P{\"o}ppel}}{{Kalberla}
  et~al.}{2005}]{2005A&A...440..775K}
{Kalberla} P.~M.~W.,  {Burton} W.~B.,  {Hartmann} D.,  {Arnal} E.~M.,  {Bajaja}
  E.,  {Morras} R.,   {P{\"o}ppel} W.~G.~L.,  2005, \mn@doi [\aap]
  {10.1051/0004-6361:20041864}, \href
  {http://adsabs.harvard.edu/abs/2005A%26A...440..775K} {440, 775}

\bibitem[\protect\citeauthoryear{{Kay}, {Thomas}, {Jenkins}  \& {Pearce}}{{Kay}
  et~al.}{2004}]{2004MNRAS.355.1091K}
{Kay} S.~T.,  {Thomas} P.~A.,  {Jenkins} A.,   {Pearce} F.~R.,  2004, \mn@doi
  [\mnras] {10.1111/j.1365-2966.2004.08383.x}, \href
  {http://adsabs.harvard.edu/abs/2004MNRAS.355.1091K} {355, 1091}

\bibitem[\protect\citeauthoryear{{Komatsu} et~al.,}{{Komatsu}
  et~al.}{2011}]{2011ApJS..192...18K}
{Komatsu} E.,  et~al., 2011, \mn@doi [\apjs] {10.1088/0067-0049/192/2/18},
  \href {http://adsabs.harvard.edu/abs/2011ApJS..192...18K} {192, 18}

\bibitem[\protect\citeauthoryear{{Kravtsov}, {Vikhlinin}  \&
  {Nagai}}{{Kravtsov} et~al.}{2006}]{2006ApJ...650..128K}
{Kravtsov} A.~V.,  {Vikhlinin} A.,   {Nagai} D.,  2006, \mn@doi [\apj]
  {10.1086/506319}, \href {http://adsabs.harvard.edu/abs/2006ApJ...650..128K}
  {650, 128}

\bibitem[\protect\citeauthoryear{{Landry}, {Bonamente}, {Giles}, {Maughan},
  {Joy}  \& {Murray}}{{Landry} et~al.}{2013}]{2013MNRAS.433.2790L}
{Landry} D.,  {Bonamente} M.,  {Giles} P.,  {Maughan} B.,  {Joy} M.,   {Murray}
  S.,  2013, \mn@doi [\mnras] {10.1093/mnras/stt901}, \href
  {http://adsabs.harvard.edu/abs/2013MNRAS.433.2790L} {433, 2790}

\bibitem[\protect\citeauthoryear{{Lau}, {Kravtsov}  \& {Nagai}}{{Lau}
  et~al.}{2009}]{2009ApJ...705.1129L}
{Lau} E.~T.,  {Kravtsov} A.~V.,   {Nagai} D.,  2009, \mn@doi [\apj]
  {10.1088/0004-637X/705/2/1129}, \href
  {http://adsabs.harvard.edu/abs/2009ApJ...705.1129L} {705, 1129}

\bibitem[\protect\citeauthoryear{{Le Brun}, {McCarthy}, {Schaye}  \&
  {Ponman}}{{Le Brun} et~al.}{2014}]{2014MNRAS.441.1270L}
{Le Brun} A.~M.~C.,  {McCarthy} I.~G.,  {Schaye} J.,   {Ponman} T.~J.,  2014,
  \mn@doi [\mnras] {10.1093/mnras/stu608}, \href
  {http://adsabs.harvard.edu/abs/2014MNRAS.441.1270L} {441, 1270}

\bibitem[\protect\citeauthoryear{{Lopes}, {de Carvalho}, {Gal}, {Djorgovski},
  {Odewahn}, {Mahabal}  \& {Brunner}}{{Lopes}
  et~al.}{2004}]{2004AJ....128.1017L}
{Lopes} P.~A.~A.,  {de Carvalho} R.~R.,  {Gal} R.~R.,  {Djorgovski} S.~G.,
  {Odewahn} S.~C.,  {Mahabal} A.~A.,   {Brunner} R.~J.,  2004, \mn@doi [\aj]
  {10.1086/423038}, \href {http://adsabs.harvard.edu/abs/2004AJ....128.1017L}
  {128, 1017}

\bibitem[\protect\citeauthoryear{{Lovisari}, {Reiprich}  \&
  {Schellenberger}}{{Lovisari} et~al.}{2015}]{2015A&A...573A.118L}
{Lovisari} L.,  {Reiprich} T.~H.,   {Schellenberger} G.,  2015, \mn@doi [\aap]
  {10.1051/0004-6361/201423954}, \href
  {http://adsabs.harvard.edu/abs/2015A%26A...573A.118L} {573, A118}

\bibitem[\protect\citeauthoryear{{Mahdavi}, {Hoekstra}, {Babul}, {Bildfell},
  {Jeltema}  \& {Henry}}{{Mahdavi} et~al.}{2013}]{2013ApJ...767..116M}
{Mahdavi} A.,  {Hoekstra} H.,  {Babul} A.,  {Bildfell} C.,  {Jeltema} T.,
  {Henry} J.~P.,  2013, \mn@doi [\apj] {10.1088/0004-637X/767/2/116}, \href
  {http://adsabs.harvard.edu/abs/2013ApJ...767..116M} {767, 116}

\bibitem[\protect\citeauthoryear{{Mantz}, {Allen}, {Ebeling}  \&
  {Rapetti}}{{Mantz} et~al.}{2008}]{2008MNRAS.387.1179M}
{Mantz} A.,  {Allen} S.~W.,  {Ebeling} H.,   {Rapetti} D.,  2008, \mn@doi
  [\mnras] {10.1111/j.1365-2966.2008.13311.x}, \href
  {http://adsabs.harvard.edu/abs/2008MNRAS.387.1179M} {387, 1179}

\bibitem[\protect\citeauthoryear{{Mantz}, {Allen}, {Rapetti}  \&
  {Ebeling}}{{Mantz} et~al.}{2010a}]{2010MNRAS.406.1759M}
{Mantz} A.,  {Allen} S.~W.,  {Rapetti} D.,   {Ebeling} H.,  2010a, \mn@doi
  [\mnras] {10.1111/j.1365-2966.2010.16992.x}, \href
  {http://ukads.nottingham.ac.uk/abs/2010MNRAS.406.1759M} {406, 1759}

\bibitem[\protect\citeauthoryear{{Mantz}, {Allen}, {Ebeling}, {Rapetti}  \&
  {Drlica-Wagner}}{{Mantz} et~al.}{2010b}]{2010MNRAS.406.1773M}
{Mantz} A.,  {Allen} S.~W.,  {Ebeling} H.,  {Rapetti} D.,   {Drlica-Wagner} A.,
   2010b, \mn@doi [\mnras] {10.1111/j.1365-2966.2010.16993.x}, \href
  {http://ukads.nottingham.ac.uk/abs/2010MNRAS.406.1773M} {406, 1773}

\bibitem[\protect\citeauthoryear{{Mantz}, {Allen}, {Morris}, {Rapetti},
  {Applegate}, {Kelly}, {von der Linden}  \& {Schmidt}}{{Mantz}
  et~al.}{2014}]{2014MNRAS.440.2077M}
{Mantz} A.~B.,  {Allen} S.~W.,  {Morris} R.~G.,  {Rapetti} D.~A.,  {Applegate}
  D.~E.,  {Kelly} P.~L.,  {von der Linden} A.,   {Schmidt} R.~W.,  2014,
  \mn@doi [\mnras] {10.1093/mnras/stu368}, \href
  {http://adsabs.harvard.edu/abs/2014MNRAS.440.2077M} {440, 2077}

\bibitem[\protect\citeauthoryear{{Mantz}, {Allen}, {Morris}, {Schmidt}, {von
  der Linden}  \& {Urban}}{{Mantz} et~al.}{2015}]{2015MNRAS.449..199M}
{Mantz} A.~B.,  {Allen} S.~W.,  {Morris} R.~G.,  {Schmidt} R.~W.,  {von der
  Linden} A.,   {Urban} O.,  2015, \mn@doi [\mnras] {10.1093/mnras/stv219},
  \href {http://adsabs.harvard.edu/abs/2015MNRAS.449..199M} {449, 199}

\bibitem[\protect\citeauthoryear{{Markevitch} et~al.,}{{Markevitch}
  et~al.}{2003}]{2003ApJ...583...70M}
{Markevitch} M.,  et~al., 2003, \mn@doi [\apj] {10.1086/345347}, \href
  {http://adsabs.harvard.edu/abs/2003ApJ...583...70M} {583, 70}

\bibitem[\protect\citeauthoryear{{Martino}, {Mazzotta}, {Bourdin}, {Smith},
  {Bartalucci}, {Marrone}, {Finoguenov}  \& {Okabe}}{{Martino}
  et~al.}{2014}]{2014MNRAS.443.2342M}
{Martino} R.,  {Mazzotta} P.,  {Bourdin} H.,  {Smith} G.~P.,  {Bartalucci} I.,
  {Marrone} D.~P.,  {Finoguenov} A.,   {Okabe} N.,  2014, \mn@doi [\mnras]
  {10.1093/mnras/stu1267}, \href
  {http://adsabs.harvard.edu/abs/2014MNRAS.443.2342M} {443, 2342}

\bibitem[\protect\citeauthoryear{{Maughan}}{{Maughan}}{2014}]{2014MNRAS.437.11%
71M}
{Maughan} B.~J.,  2014, \mn@doi [\mnras] {10.1093/mnras/stt1931}, \href
  {http://ukads.nottingham.ac.uk/abs/2014MNRAS.437.1171M} {437, 1171}

\bibitem[\protect\citeauthoryear{{Maughan}, {Giles}, {Randall}, {Jones}  \&
  {Forman}}{{Maughan} et~al.}{2012}]{2012MNRAS.421.1583M}
{Maughan} B.~J.,  {Giles} P.~A.,  {Randall} S.~W.,  {Jones} C.,   {Forman}
  W.~R.,  2012, \mn@doi [\mnras] {10.1111/j.1365-2966.2012.20419.x}, \href
  {http://adsabs.harvard.edu/abs/2012MNRAS.421.1583M} {421, 1583}

\bibitem[\protect\citeauthoryear{{Maughan}, {Giles}, {Rines}, {Diaferio},
  {Geller}, {Van Der Pyl}  \& {Bonamente}}{{Maughan}
  et~al.}{2015}]{2015arXiv151107872M}
{Maughan} B.~J.,  {Giles} P.~A.,  {Rines} K.~J.,  {Diaferio} A.,  {Geller}
  M.~J.,  {Van Der Pyl} N.,   {Bonamente} M.,  2015, preprint, \href
  {http://adsabs.harvard.edu/abs/2015arXiv151107872M} {} (\mn@eprint {arXiv}
  {1511.07872})

\bibitem[\protect\citeauthoryear{{McCarthy}, {Schaye}, {Bower}, {Ponman},
  {Booth}, {Dalla Vecchia}  \& {Springel}}{{McCarthy}
  et~al.}{2011}]{2011MNRAS.412.1965M}
{McCarthy} I.~G.,  {Schaye} J.,  {Bower} R.~G.,  {Ponman} T.~J.,  {Booth}
  C.~M.,  {Dalla Vecchia} C.,   {Springel} V.,  2011, \mn@doi [\mnras]
  {10.1111/j.1365-2966.2010.18033.x}, \href
  {http://adsabs.harvard.edu/abs/2011MNRAS.412.1965M} {412, 1965}

\bibitem[\protect\citeauthoryear{{Mitchell}, {Ives}  \& {Culhane}}{{Mitchell}
  et~al.}{1977}]{1977MNRAS.181P..25M}
{Mitchell} R.~J.,  {Ives} J.~C.,   {Culhane} J.~L.,  1977, \mnras, \href
  {http://adsabs.harvard.edu/abs/1977MNRAS.181P..25M} {181, 25P}

\bibitem[\protect\citeauthoryear{{Mittal}, {Hicks}, {Reiprich}  \&
  {Jaritz}}{{Mittal} et~al.}{2011}]{2011A&A...532A.133M}
{Mittal} R.,  {Hicks} A.,  {Reiprich} T.~H.,   {Jaritz} V.,  2011, \mn@doi
  [\aap] {10.1051/0004-6361/200913714}, \href
  {http://adsabs.harvard.edu/abs/2011A%26A...532A.133M} {532, A133}

\bibitem[\protect\citeauthoryear{{Murray}, {Power}  \& {Robotham}}{{Murray}
  et~al.}{2013}]{2013arXiv1306.6721M}
{Murray} S.,  {Power} C.,   {Robotham} A.,  2013, preprint, \href
  {http://ukads.nottingham.ac.uk/abs/2013arXiv1306.6721M} {} (\mn@eprint
  {arXiv} {1306.6721})

\bibitem[\protect\citeauthoryear{{Nagai}, {Kravtsov}  \& {Vikhlinin}}{{Nagai}
  et~al.}{2007}]{2007ApJ...668....1N}
{Nagai} D.,  {Kravtsov} A.~V.,   {Vikhlinin} A.,  2007, \mn@doi [\apj]
  {10.1086/521328}, \href {http://adsabs.harvard.edu/abs/2007ApJ...668....1N}
  {668, 1}

\bibitem[\protect\citeauthoryear{{Nelson}, {Lau}  \& {Nagai}}{{Nelson}
  et~al.}{2014}]{2014ApJ...792...25N}
{Nelson} K.,  {Lau} E.~T.,   {Nagai} D.,  2014, \mn@doi [\apj]
  {10.1088/0004-637X/792/1/25}, \href
  {http://adsabs.harvard.edu/abs/2014ApJ...792...25N} {792, 25}

\bibitem[\protect\citeauthoryear{{Nevalainen}, {David}  \&
  {Guainazzi}}{{Nevalainen} et~al.}{2010}]{2010A&A...523A..22N}
{Nevalainen} J.,  {David} L.,   {Guainazzi} M.,  2010, \mn@doi [\aap]
  {10.1051/0004-6361/201015176}, \href
  {http://ukads.nottingham.ac.uk/abs/2010A%26A...523A..22N} {523, A22}

\bibitem[\protect\citeauthoryear{{O'Hara}, {Mohr}, {Bialek}  \&
  {Evrard}}{{O'Hara} et~al.}{2006}]{2006ApJ...639...64O}
{O'Hara} T.~B.,  {Mohr} J.~J.,  {Bialek} J.~J.,   {Evrard} A.~E.,  2006,
  \mn@doi [\apj] {10.1086/499327}, \href
  {http://adsabs.harvard.edu/abs/2006ApJ...639...64O} {639, 64}

\bibitem[\protect\citeauthoryear{{Pacaud} et~al.,}{{Pacaud}
  et~al.}{2007}]{2007MNRAS.382.1289P}
{Pacaud} F.,  et~al., 2007, \mn@doi [\mnras]
  {10.1111/j.1365-2966.2007.12468.x}, \href
  {http://adsabs.harvard.edu/abs/2007MNRAS.382.1289P} {382, 1289}

\bibitem[\protect\citeauthoryear{{Pacaud} et~al.,}{{Pacaud}
  et~al.}{2016}]{2016A&A...592A...2P}
{Pacaud} F.,  et~al., 2016, \mn@doi [\aap] {10.1051/0004-6361/201526891}, \href
  {http://adsabs.harvard.edu/abs/2016A%26A...592A...2P} {592, A2}

\bibitem[\protect\citeauthoryear{{Planck Collaboration} et~al.,}{{Planck
  Collaboration} et~al.}{2015b}]{2015arXiv150201589P}
{Planck Collaboration} et~al., 2015b, preprint, \href
  {http://adsabs.harvard.edu/abs/2015arXiv150201589P} {} (\mn@eprint {arXiv}
  {1502.01589})

\bibitem[\protect\citeauthoryear{{Planck Collaboration} et~al.,}{{Planck
  Collaboration} et~al.}{2015a}]{2015arXiv150201597P}
{Planck Collaboration} et~al., 2015a, preprint, \href
  {http://adsabs.harvard.edu/abs/2015arXiv150201597P} {} (\mn@eprint {arXiv}
  {1502.01597})

\bibitem[\protect\citeauthoryear{{Poole}, {Fardal}, {Babul}, {McCarthy},
  {Quinn}  \& {Wadsley}}{{Poole} et~al.}{2006}]{2006MNRAS.373..881P}
{Poole} G.~B.,  {Fardal} M.~A.,  {Babul} A.,  {McCarthy} I.~G.,  {Quinn} T.,
  {Wadsley} J.,  2006, \mn@doi [\mnras] {10.1111/j.1365-2966.2006.10916.x},
  \href {http://adsabs.harvard.edu/abs/2006MNRAS.373..881P} {373, 881}

\bibitem[\protect\citeauthoryear{{Pratt}, {Croston}, {Arnaud}  \&
  {B{\"o}hringer}}{{Pratt} et~al.}{2009}]{2009A&A...498..361P}
{Pratt} G.~W.,  {Croston} J.~H.,  {Arnaud} M.,   {B{\"o}hringer} H.,  2009,
  \mn@doi [\aap] {10.1051/0004-6361/200810994}, \href
  {http://adsabs.harvard.edu/abs/2009A%26A...498..361P} {498, 361}

\bibitem[\protect\citeauthoryear{{Reiprich} \& {B{\"o}hringer}}{{Reiprich} \&
  {B{\"o}hringer}}{2002}]{2002ApJ...567..716R}
{Reiprich} T.~H.,  {B{\"o}hringer} H.,  2002, \mn@doi [\apj] {10.1086/338753},
  \href {http://adsabs.harvard.edu/abs/2002ApJ...567..716R} {567, 716}

\bibitem[\protect\citeauthoryear{{Rines}, {Geller}, {Diaferio}  \&
  {Kurtz}}{{Rines} et~al.}{2013}]{2013ApJ...767...15R}
{Rines} K.,  {Geller} M.~J.,  {Diaferio} A.,   {Kurtz} M.~J.,  2013, \mn@doi
  [\apj] {10.1088/0004-637X/767/1/15}, \href
  {http://adsabs.harvard.edu/abs/2013ApJ...767...15R} {767, 15}

\bibitem[\protect\citeauthoryear{{Rykoff} et~al.,}{{Rykoff}
  et~al.}{2008}]{2008MNRAS.387L..28R}
{Rykoff} E.~S.,  et~al., 2008, \mn@doi [\mnras]
  {10.1111/j.1745-3933.2008.00476.x}, \href
  {http://adsabs.harvard.edu/abs/2008MNRAS.387L..28R} {387, L28}

\bibitem[\protect\citeauthoryear{{Santos}, {Tozzi}, {Rosati}  \&
  {B{\"o}hringer}}{{Santos} et~al.}{2010}]{2010A&A...521A..64S}
{Santos} J.~S.,  {Tozzi} P.,  {Rosati} P.,   {B{\"o}hringer} H.,  2010, \mn@doi
  [\aap] {10.1051/0004-6361/201015208}, \href
  {http://adsabs.harvard.edu/abs/2010A%26A...521A..64S} {521, A64}

\bibitem[\protect\citeauthoryear{{Sarazin}}{{Sarazin}}{1988}]{1988xrec.book...%
..S}
{Sarazin} C.~L.,  1988, {X-ray emission from clusters of galaxies}

\bibitem[\protect\citeauthoryear{{Sereno} \& {Ettori}}{{Sereno} \&
  {Ettori}}{2015}]{2015MNRAS.450.3675S}
{Sereno} M.,  {Ettori} S.,  2015, \mn@doi [\mnras] {10.1093/mnras/stv814},
  \href {http://adsabs.harvard.edu/abs/2015MNRAS.450.3675S} {450, 3675}

\bibitem[\protect\citeauthoryear{{Shi} \& {Komatsu}}{{Shi} \&
  {Komatsu}}{2014}]{2014MNRAS.442..521S}
{Shi} X.,  {Komatsu} E.,  2014, \mn@doi [\mnras] {10.1093/mnras/stu858}, \href
  {http://adsabs.harvard.edu/abs/2014MNRAS.442..521S} {442, 521}

\bibitem[\protect\citeauthoryear{{Shi}, {Komatsu}, {Nagai}  \& {Lau}}{{Shi}
  et~al.}{2016}]{2016MNRAS.455.2936S}
{Shi} X.,  {Komatsu} E.,  {Nagai} D.,   {Lau} E.~T.,  2016, \mn@doi [\mnras]
  {10.1093/mnras/stv2504}, \href
  {http://adsabs.harvard.edu/abs/2016MNRAS.455.2936S} {455, 2936}

\bibitem[\protect\citeauthoryear{{Short}, {Thomas}, {Young}, {Pearce},
  {Jenkins}  \& {Muanwong}}{{Short} et~al.}{2010}]{2010MNRAS.408.2213S}
{Short} C.~J.,  {Thomas} P.~A.,  {Young} O.~E.,  {Pearce} F.~R.,  {Jenkins} A.,
    {Muanwong} O.,  2010, \mn@doi [\mnras] {10.1111/j.1365-2966.2010.17267.x},
  \href {http://adsabs.harvard.edu/abs/2010MNRAS.408.2213S} {408, 2213}

\bibitem[\protect\citeauthoryear{{Smith}, {Brickhouse}, {Liedahl}  \&
  {Raymond}}{{Smith} et~al.}{2001}]{2001ApJ...556L..91S}
{Smith} R.~K.,  {Brickhouse} N.~S.,  {Liedahl} D.~A.,   {Raymond} J.~C.,  2001,
  \mn@doi [\apjl] {10.1086/322992}, \href
  {http://adsabs.harvard.edu/abs/2001ApJ...556L..91S} {556, L91}

\bibitem[\protect\citeauthoryear{{Smith} et~al.,}{{Smith}
  et~al.}{2016}]{2016MNRAS.456L..74S}
{Smith} G.~P.,  et~al., 2016, \mn@doi [\mnras] {10.1093/mnrasl/slv175}, \href
  {http://adsabs.harvard.edu/abs/2016MNRAS.456L..74S} {456, L74}

\bibitem[\protect\citeauthoryear{{Stanek}, {Evrard}, {B{\"o}hringer},
  {Schuecker}  \& {Nord}}{{Stanek} et~al.}{2006}]{2006ApJ...648..956S}
{Stanek} R.,  {Evrard} A.~E.,  {B{\"o}hringer} H.,  {Schuecker} P.,   {Nord}
  B.,  2006, \mn@doi [\apj] {10.1086/506248}, \href
  {http://adsabs.harvard.edu/abs/2006ApJ...648..956S} {648, 956}

\bibitem[\protect\citeauthoryear{{Tinker}, {Kravtsov}, {Klypin}, {Abazajian},
  {Warren}, {Yepes}, {Gottl{\"o}ber}  \& {Holz}}{{Tinker}
  et~al.}{2008}]{2008ApJ...688..709T}
{Tinker} J.,  {Kravtsov} A.~V.,  {Klypin} A.,  {Abazajian} K.,  {Warren} M.,
  {Yepes} G.,  {Gottl{\"o}ber} S.,   {Holz} D.~E.,  2008, \mn@doi [\apj]
  {10.1086/591439}, \href {http://adsabs.harvard.edu/abs/2008ApJ...688..709T}
  {688, 709}

\bibitem[\protect\citeauthoryear{{Vikhlinin}}{{Vikhlinin}}{2006}]{2006ApJ...64%
0..710V}
{Vikhlinin} A.,  2006, \mn@doi [\apj] {10.1086/500121}, \href
  {http://adsabs.harvard.edu/abs/2006ApJ...640..710V} {640, 710}

\bibitem[\protect\citeauthoryear{{Vikhlinin}, {Markevitch}, {Murray}, {Jones},
  {Forman}  \& {Van Speybroeck}}{{Vikhlinin}
  et~al.}{2005}]{2005ApJ...628..655V}
{Vikhlinin} A.,  {Markevitch} M.,  {Murray} S.~S.,  {Jones} C.,  {Forman} W.,
  {Van Speybroeck} L.,  2005, \mn@doi [\apj] {10.1086/431142}, \href
  {http://adsabs.harvard.edu/abs/2005ApJ...628..655V} {628, 655}

\bibitem[\protect\citeauthoryear{{Vikhlinin}, {Kravtsov}, {Forman}, {Jones},
  {Markevitch}, {Murray}  \& {Van Speybroeck}}{{Vikhlinin}
  et~al.}{2006}]{2006ApJ...640..691V}
{Vikhlinin} A.,  {Kravtsov} A.,  {Forman} W.,  {Jones} C.,  {Markevitch} M.,
  {Murray} S.~S.,   {Van Speybroeck} L.,  2006, \mn@doi [\apj]
  {10.1086/500288}, \href {http://adsabs.harvard.edu/abs/2006ApJ...640..691V}
  {640, 691}

\bibitem[\protect\citeauthoryear{{Vikhlinin}, {Burenin}, {Forman}, {Jones},
  {Hornstrup}, {Murray}  \& {Quintana}}{{Vikhlinin}
  et~al.}{2007}]{2007hvcg.conf...48V}
{Vikhlinin} A.,  {Burenin} R.,  {Forman} W.~R.,  {Jones} C.,  {Hornstrup} A.,
  {Murray} S.~S.,   {Quintana} H.,  2007, in {H.~B{\"o}hringer, G.~W.~Pratt,
  A.~Finoguenov, \& P.~Schuecker } ed., Heating versus Cooling in Galaxies and
  Clusters of Galaxies. p.~48 (\mn@eprint {} {arXiv:astro-ph/0611438}),
  \mn@doi{10.1007/978-3-540-73484-0_9}

\bibitem[\protect\citeauthoryear{{Vikhlinin} et~al.,}{{Vikhlinin}
  et~al.}{2009a}]{2009ApJ...692.1033V}
{Vikhlinin} A.,  et~al., 2009a, \mn@doi [\apj] {10.1088/0004-637X/692/2/1033},
  \href {http://adsabs.harvard.edu/abs/2009ApJ...692.1033V} {692, 1033}

\bibitem[\protect\citeauthoryear{{Vikhlinin} et~al.,}{{Vikhlinin}
  et~al.}{2009b}]{2009ApJ...692.1060V}
{Vikhlinin} A.,  et~al., 2009b, \mn@doi [\apj] {10.1088/0004-637X/692/2/1060},
  \href {http://adsabs.harvard.edu/abs/2009ApJ...692.1060V} {692, 1060}

\bibitem[\protect\citeauthoryear{{Wang}, {Ulmer}  \& {Lavery}}{{Wang}
  et~al.}{1997}]{1997MNRAS.288..702W}
{Wang} Q.~D.,  {Ulmer} M.~P.,   {Lavery} R.~J.,  1997, \mnras, \href
  {http://adsabs.harvard.edu/abs/1997MNRAS.288..702W} {288, 702}

\bibitem[\protect\citeauthoryear{{Wei{\ss}mann}, {B{\"o}hringer}, {{\v S}uhada}
   \& {Ameglio}}{{Wei{\ss}mann} et~al.}{2013}]{2013A&A...549A..19W}
{Wei{\ss}mann} A.,  {B{\"o}hringer} H.,  {{\v S}uhada} R.,   {Ameglio} S.,
  2013, \mn@doi [\aap] {10.1051/0004-6361/201219333}, \href
  {http://adsabs.harvard.edu/abs/2013A%26A...549A..19W} {549, A19}

\bibitem[\protect\citeauthoryear{{Zhang}, {Finoguenov}, {B{\"o}hringer},
  {Kneib}, {Smith}, {Kneissl}, {Okabe}  \& {Dahle}}{{Zhang}
  et~al.}{2008}]{2008A&A...482..451Z}
{Zhang} Y.-Y.,  {Finoguenov} A.,  {B{\"o}hringer} H.,  {Kneib} J.-P.,  {Smith}
  G.~P.,  {Kneissl} R.,  {Okabe} N.,   {Dahle} H.,  2008, \mn@doi [\aap]
  {10.1051/0004-6361:20079103}, \href
  {http://adsabs.harvard.edu/abs/2008A%26A...482..451Z} {482, 451}

\bibitem[\protect\citeauthoryear{{von der Linden} et~al.,}{{von der Linden}
  et~al.}{2014}]{2014MNRAS.443.1973V}
{von der Linden} A.,  et~al., 2014, \mn@doi [\mnras] {10.1093/mnras/stu1423},
  \href {http://adsabs.harvard.edu/abs/2014MNRAS.443.1973V} {443, 1973}

\makeatother
\end{thebibliography}

\appendix

\section{Parameters of the Gas Density and Temperature Profiles}
\label{sec:profileparams}

Here we give the parameters of the gas density and temperature
profiles of the cluster sample.  Tables~\ref{tab:gasparams} and
\ref{tab:tempparams} lists the individual parameters of the fits to
the gas density and temperature profiles respectively.

\begin{table*}
\begin{center}
\caption{\small{Table listing the individual parameters of the fit to the gas density profile for each of the clusters in our sample.}\label{tab:gasparams}}
\begin{tabular}{lcccccc}
\hline\hline
 & n$_{\rm 0}$ & r$_{\rm c}$ & r$_{\rm s}$ & & & \\
Cluster & 10$^{-3}$ cm$^{-3}$ & (kpc) & (kpc) & $\alpha$ & $\beta$ & $\epsilon$ \\
\hline
A2204$^{\dagger}$          & 24.579 & 67.778  & 1363.850 & 2.028  & 0.532 & 3.095 \\
RXJ1720.1+2638$^{\dagger}$ & 36.927 & 47.171  & 695.129  & 0.781  & 0.542 & 1.088 \\
A586           & 16.085 & 51.421  & 161.327  & 0.001  & 0.323 & 1.837 \\
A1914          & 14.004 & 154.980 & 2703.460 & 0.000  & 0.718 & 5.000 \\
A665           & 11.249 & 67.700  & 1367.420 & 0.000  & 0.397 & 5.000 \\
A115           & 29.314 & 29.186  & 3967.680 & 0.668  & 0.445 & 5.000 \\
A520           & 3.408  & 391.718 & 339.217  & 0.000  & 0.685 & 0.760  \\
A963           & 10.907 & 96.634  & 997.588  & 0.898  & 0.518 & 3.384 \\
A1423$^{\dagger}$          & 20.548 & 39.845  & 1632.000 & 0.330  & 0.446 & 3.257 \\
A773           & 8.315  & 138.667 & 846.138  & 0.000  & 0.552 & 1.365 \\
A1763          & 7.557  & 133.460 & 1351.600 & 0.000  & 0.500 & 2.582 \\
A2261          & 13.171 & 114.317 & 2382.670 & 0.922  & 0.587 & 1.629 \\
A1682          & 2.593  & 278.998 & 3046.010 & 0.757  & 0.610 & 0.000  \\
A2111          & 4.753  & 190.063 & 1061.220 & 0.161  & 0.602 & 0.000  \\
Z5247          & 0.954  & 396.838 & 1905.930 & 1.051  & 0.423 & 5.000 \\
A267           & 9.892  & 134.225 & 3333.110 & 0.000  & 0.655 & 0.106  \\
A2219          & 5.995  & 216.035 & 331.133  & 0.507  & 0.300 & 2.250 \\
A2390          & 68.229 & 18.435  & 579.989  & 0.000  & 0.368 & 2.878 \\
Z2089$^{\dagger}$          & 18.299 & 71.553  & 142.615  & 1.751  & 0.628 & 0.000  \\
RXJ2129.6+0005$^{\dagger}$ & 9.959  & 116.326 & 1760.560 & 1.800  & 0.566 & 5.000 \\
A1835$^{\dagger}$          & 100.000 & 27.147 & 469.409  & 0.368  & 0.483 & 1.531 \\
A68            & 5.240  & 259.711 & 665.144  & 0.416  & 0.783 & 0.000  \\
MS1455.0+2232$^{\dagger}$  & 13.548 & 106.649 & 100.000  & 1.865  & 0.411 & 1.412 \\
A2631          & 3.634  & 370.272 & 2472.830 & 0.453  & 0.819 & 0.000  \\
A1758          & 2.903  & 798.833 & 399.279  & 0.000  & 0.300 & 3.855 \\
A1576          & 10.987 & 89.185  & 1191.130 & 0.000  & 0.501 & 1.913 \\
A697           & 8.734  & 170.861 & 923.611  & 0.000  & 0.548 & 1.609 \\
RXJ0439.0+0715$^{\dagger}$ & 1.483  & 799.982 & 200.646  & 1.552  & 0.701 & 1.641 \\
RXJ0437.1+0043$^{\dagger}$ & 44.707 & 20.610  & 198.695  & 0.000  & 0.353 & 1.558 \\
A611           & 2.274  & 735.655 & 130.495  & 1.353  & 0.433 & 1.886 \\
Z7215          & 5.550  & 190.280 & 1148.840 & 0.000  & 0.647 & 0.000  \\
Z3146$^{\dagger}$          & 11.306 & 163.032 & 3984.170 & 1.866  & 0.705 & 0.000  \\
A781           & 2.684  & 466.725 & 234.173  & 0.000  & 0.403 & 1.674 \\
A2552          & 8.841  & 88.866  & 244.596  & 0.837  & 0.300 & 1.778 \\
\hline
\end{tabular}
\hspace{0.01cm}
\end{center}
\end{table*}

\begin{table*}
\begin{center}
\caption{\small{Table listing the individual parameters of the fit to the temperature profile for each of the clusters in our sample.  $\dagger$ indicates relaxed clusters.}\label{tab:tempparams}}
\begin{tabular}{lcccccccc}
\hline\hline
 & T$_0$ & r$_{\rm cool}$ & T$_{\rm min}$ & r$_{\rm t}$ & & & & \\
Cluster & (keV) & (kpc) & (keV) & (kpc) & a$_{\rm cool}$  & a & b & c \\
\hline
A2204$^{\dagger}$          & 12.366 & 20.585  &  2.857  & 984.922  & 4.754  & -0.110  & 5.000  & 3.827 \\
RXJ1720.1+2638$^{\dagger}$ & 18.601 & 10.014  &  0.128  & 345.408  & 0.000  & -0.292  & 4.124  & 0.478 \\
A586           & 9.316  & 50.020  &  3.474  & 235.606  & 0.193  &  0.176  & 1.790  & 0.000 \\
A1914          & 13.170 & 15.105  &  2.296  & 251.379  & 0.491  &  0.228  & 1.082  & 0.026 \\
A665           & 10.608 & 45.354  &  2.095  & 225.281  & 0.990  &  0.049  & 1.208  & 0.007 \\
A115           & 16.326 & 499.984 &  1.793  & 500.000  & 0.069  & -0.372  & 4.830  & 0.711 \\
A520           & 13.118 & 10.000  &  4.416  & 489.086  & 0.385  & -0.183  & 2.121  & 0.843 \\
A963           & 18.020 & 74.065  &  0.187  & 356.107  & 0.004  & -0.151  & 4.999  & 1.000 \\
A1423$^{\dagger}$          & 12.929 & 15.525  &  1.672  & 317.779  & 3.000  & -0.359  & 0.881  & 0.802 \\
A773           & 14.577 & 89.707  &  3.520  & 188.148  & 0.000  &  0.057  & 1.685  & 0.000 \\
A1763          & 9.235  & 10.009  &  5.975  & 474.620  & 0.538  &  0.018  & 0.475  & 0.000 \\
A2261          & 12.661 & 51.410  &  6.000  & 499.973  & 0.587  & -0.081  & 2.294  & 0.708 \\
A1682          & 22.910 & 125.064 &  4.997  & 174.871  & 3.000  &  0.000  & 0.609  & 0.609 \\
A2111          & 11.559 & 35.705  &  0.239  & 292.874  & 1.834  &  0.004  & 0.000  & 0.000 \\
Z5247          & 14.303 & 1.000   &  14.303 & 100.003  & 0.000  & -0.292  & 0.663  & 0.663 \\
A267           & 10.243 & 1.000   &  10.243 & 499.933  & 0.000  & -0.180  & 1.000  & 1.000 \\
A2219          & 21.601 & 36.384  &  6.000  & 499.996  & 0.277  & -0.106  & 1.253  & 0.556 \\
A2390          & 23.223 & 36.957  &  1.705  & 860.371  & 0.225  & -0.206  & 9.744  & 1.500 \\
Z2089$^{\dagger}$          & 8.146  & 151.381 &  2.217  & 153.531  & 1.472  &  0.068  & 5.000  & 0.304 \\
RXJ2129.6+0005$^{\dagger}$ & 9.984  & 53.821  &  2.133  & 428.921  & 0.755  & -0.038  & 5.000  & 0.614 \\
A1835$^{\dagger}$          & 13.290 & 24.435  &  3.050  & 1403.660 & 4.994  & -0.171  & 3.340  & 3.655 \\
A68            & 10.915 & 1.000   &  10.915 & 499.955  & 0.000  & -0.047  & 3.596  & 0.829 \\
MS1455.0+2232$^{\dagger}$  & 9.797  & 11.262  &  5.913  & 495.920  & 0.000  & -0.211  & 1.335  & 0.573 \\
A2631          & 15.474 & 10.001  &  4.642  & 248.360  & 0.741  & -0.056  & 0.726  & 0.476 \\
A1758          & 8.606  & 77.399  &  5.039  & 215.501  & 0.896  & -0.153  & 0.601  & 0.014 \\
A1576          & 12.311 & 36.027  &  2.215  & 218.731  & 1.114  &  0.079  & 1.131  & 0.281 \\
A697           & 17.424 & 36.662  &  2.560  & 234.165  & 0.001  & -0.047  & 1.415  & 0.000 \\
RXJ0439.0+0715$^{\dagger}$ & 11.448 & 34.978  &  4.274  & 499.998  & 0.000  & -0.059  & 5.000  & 0.853 \\
RXJ0437.1+0043$^{\dagger}$ & 14.097 & 55.522  &  5.437  & 499.921  & 0.049  & -0.234  & 1.027  & 0.438 \\
A611           & 8.105  & 26.170  &  0.100  & 470.611  & 3.000  &  0.056  & 5.000  & 0.484 \\
Z7215          & 14.172 & 1.000   &  14.172 & 158.818  & 0.000  & -0.139  & 0.000  & 0.000  \\
Z3146$^{\dagger}$          & 21.039 & 227.673 &  3.502  & 499.998  & 0.332  & -0.257  & 1.542  & 0.840 \\
A781           & 10.321 & 1.000   &  10.321 & 500.000  & 0.000  & -0.287  & 2.219  & 0.999 \\
A2552          & 15.305 & 17.238  &  1.818  & 234.280  & 0.009  & -0.006  & 0.948  & 0.000 \\
\hline
\end{tabular}
\hspace{0.01cm}
\end{center}
\end{table*} 



\section{Fit statistics}
\label{sec:fitstats}

\begin{figure*}
\begin{center}
\setlength{\unitlength}{1in}
\begin{picture}(7.0,2.5)
\put(0.0,0.01){\scalebox{0.33}{\includegraphics[clip=true, origin=c]{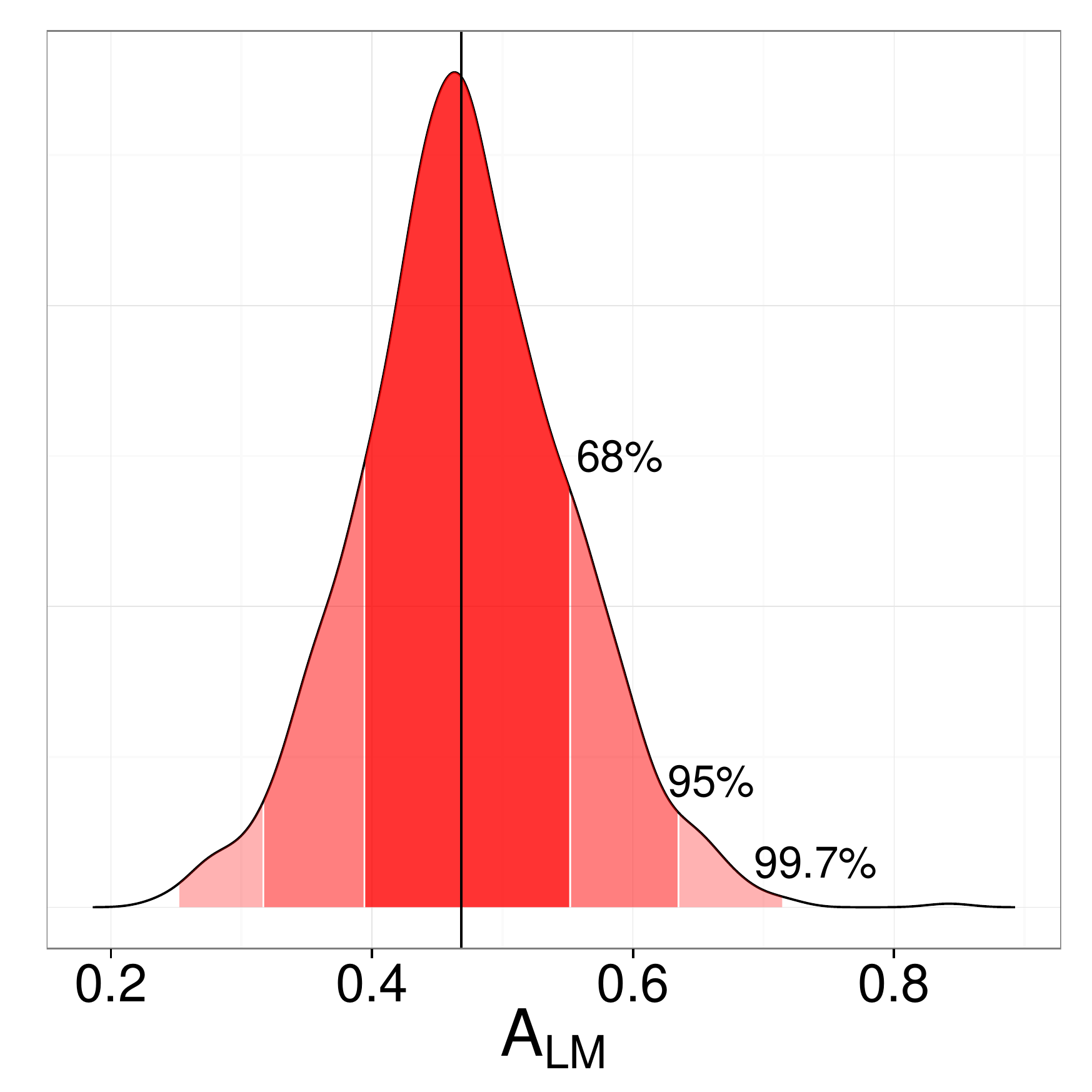}}}
\put(2.3,0.01){\scalebox{0.33}{\includegraphics[clip=true, origin=c]{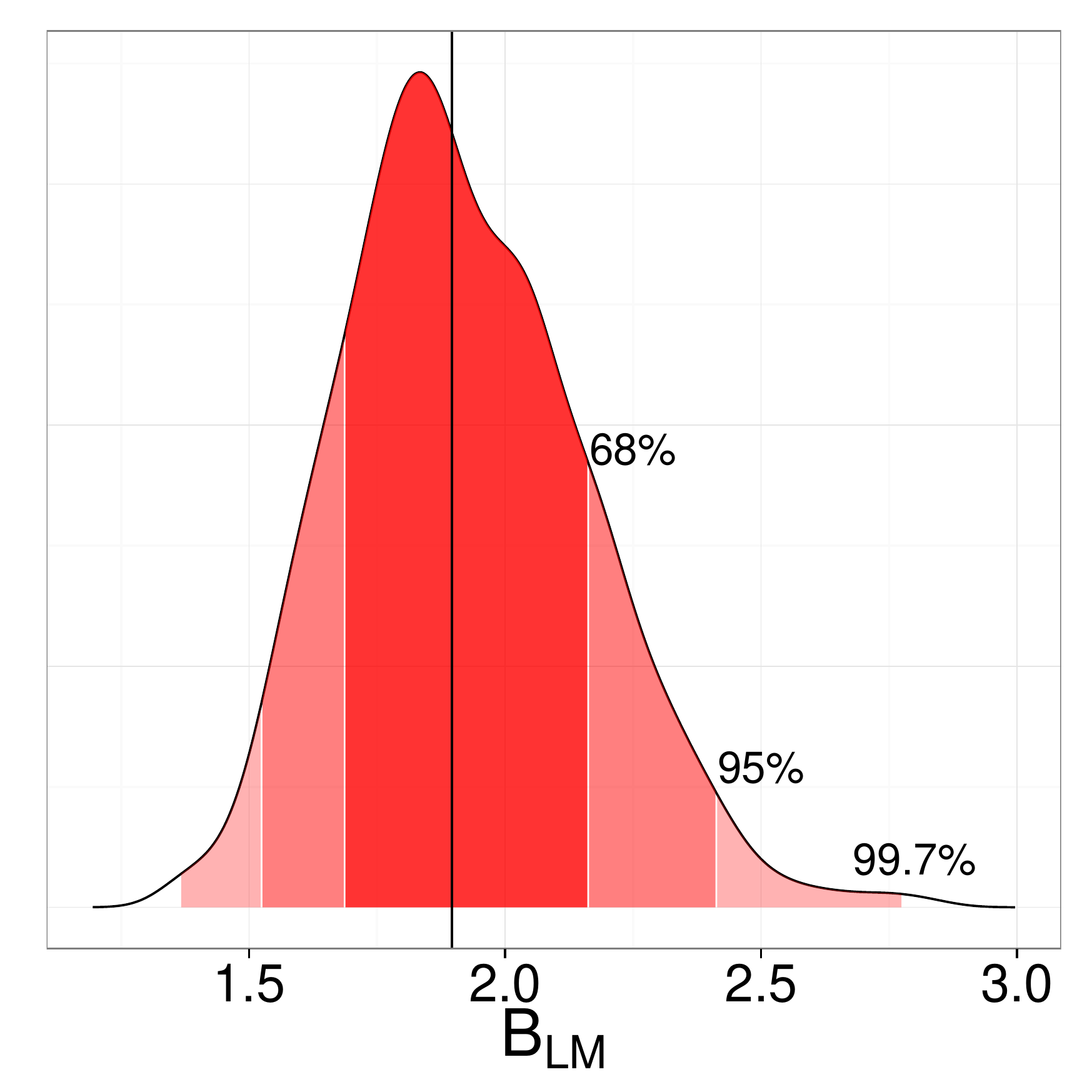}}}
\put(4.6,0.01){\scalebox{0.33}{\includegraphics[clip=true, origin=c]{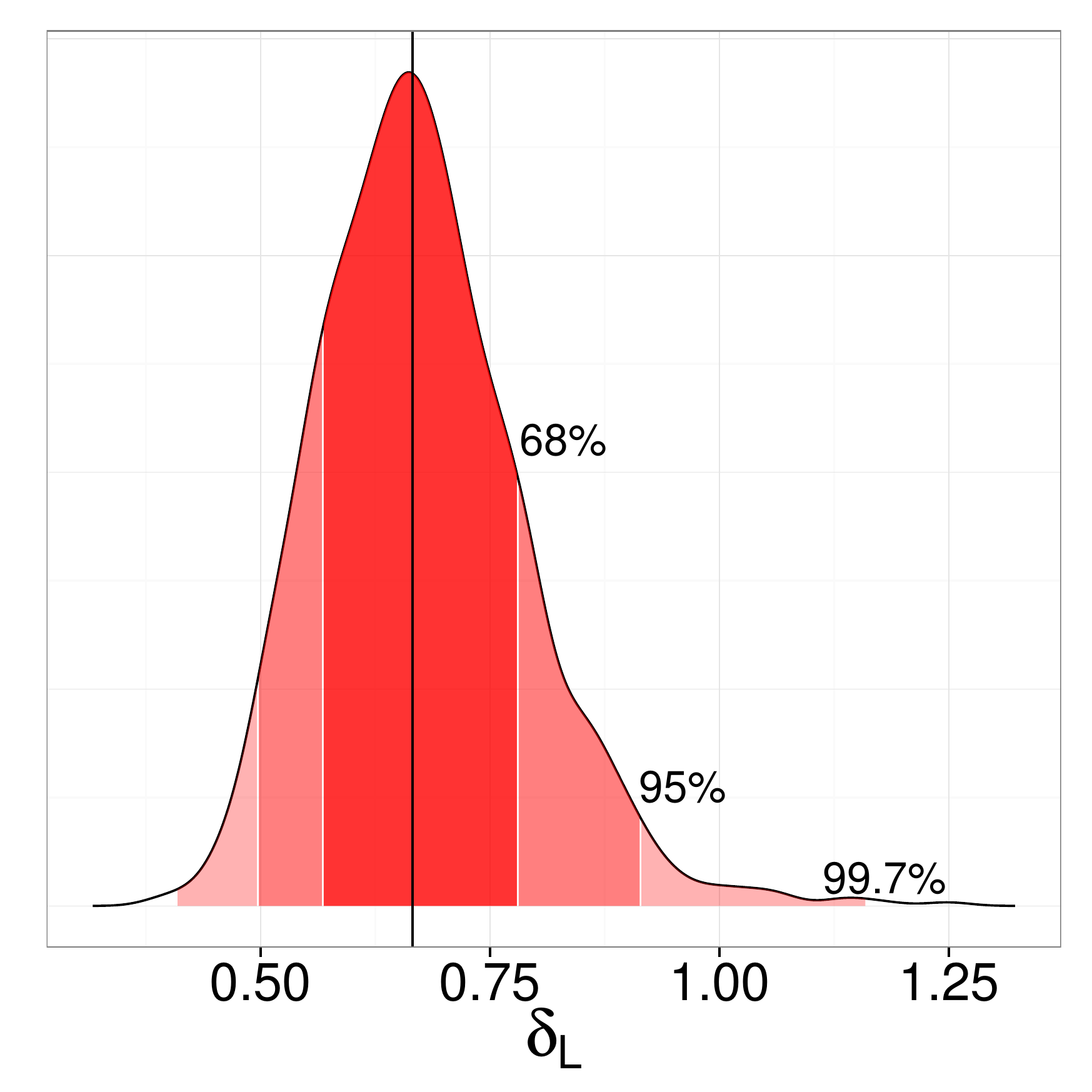}}}
\end{picture}
\end{center}
\vspace{-0.5cm}
\caption[]{\small{Plots showing the posterior densities of the fit
    parameters of the $L_{CXO}-M_{H}$ relation, with the left, middle
    and right plots showing the normalisation ($A_{LM}$), slope
    ($B_{LM}$) and scatter ($\delta_{LM}$) respectively.  The shaded
    regions in each plot represent the regions enclosing 68\%, 95\%
    and 99.7\% of the distribution, centered on the median, with the
    vertical line representing the median.}\label{fig:percentiles}}
\end{figure*}

Here we provide further information on the parameters of the
$L_{CXO}-M_{H}$ relation.  Figure~\ref{fig:percentiles} plots the
posterior densities for parameters of the $L_{CXO}-M_{H}$ relation,
with shaded regions highlighting regions enclosing 68\%, 95\% and
99.7\% of the distribution.  The vertical line in each plot represents
the median. Table~\ref{tab:percentiles} gives the medians and values
enclosing 68\%, 95\% and 99.7\% of the distribution for paramters of
the $L_{CXO}-M_{H}$ relation.

LM

\begin{table*}
\begin{center}
\caption{\small{Table listing the median and values enclosing 68\%,
    95\% and 99.7\% of the distribution for parameters of the
    $L_{CXO}-M_{H}$ relation.}\label{tab:percentiles}}
\begin{tabular}{lcccc}
\hline\hline
Parameter & Median & 68\% & 95\% & 99.7\%  \\
\hline
$A_{LM}$ & 0.47 & 0.39-0.55 & 0.32-0.63 & 0.25-0.71 \\
$B_{LM}$ & 1.90 & 1.69-2.16 & 1.53-2.41 & 1.37-2.78\\
$\delta_{LM}$ & 0.67 & 0.57-0.78 & 0.50-0.91 & 0.41-1.16 \\  
\hline
\end{tabular}
\hspace{0.01cm}
\end{center}
\end{table*} 

\section{Images, Gas Density and Temperature profiles}
\label{sec:profiles}

\begin{figure*}
\begin{center}
\setlength{\unitlength}{1in}
\begin{picture}(6.9,2.0)
\put(0.01,-0.8){\scalebox{0.34}{\includegraphics[clip=true]{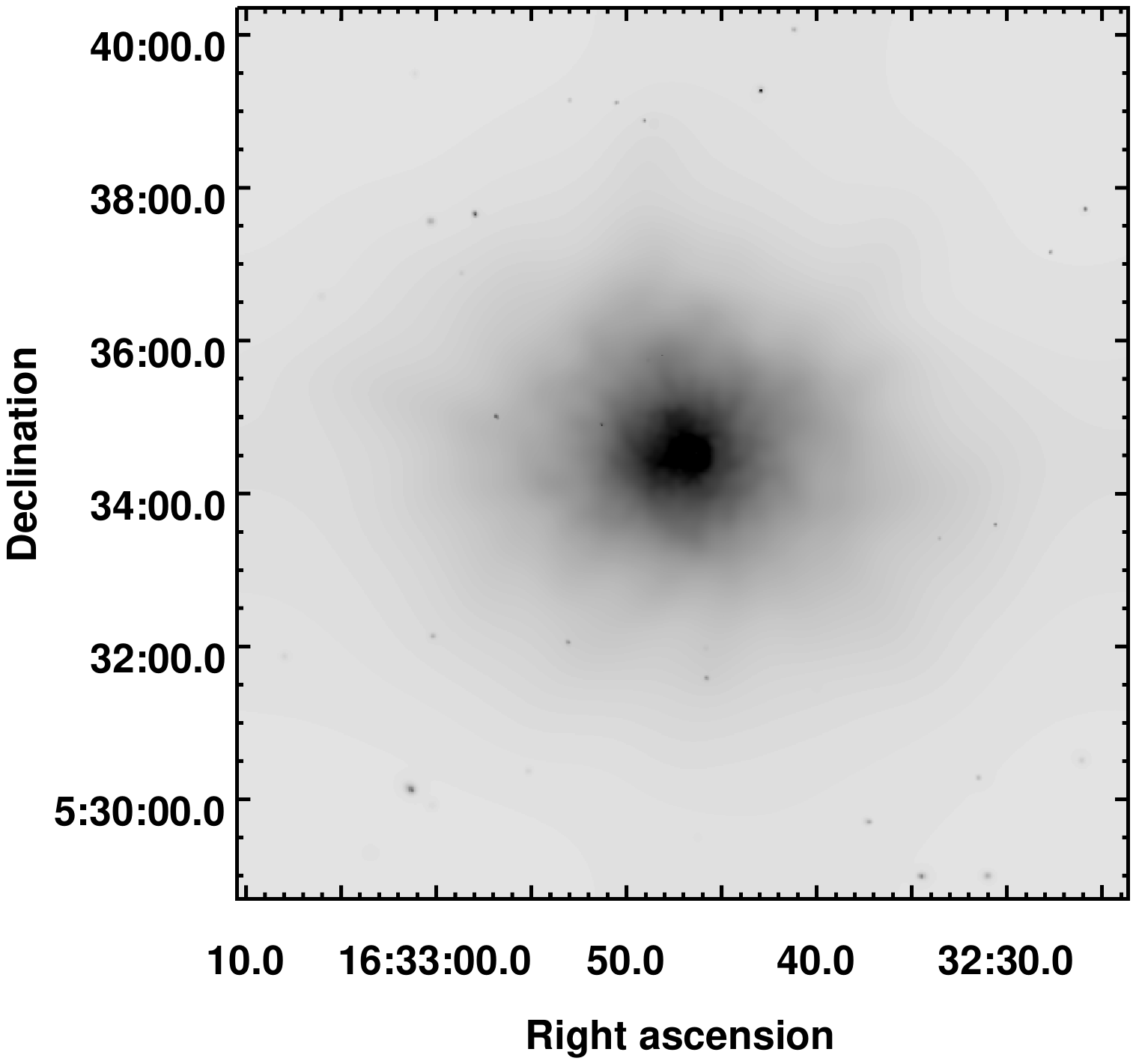}}}  
\put(2.1,-0.21){\scalebox{0.30}{\includegraphics{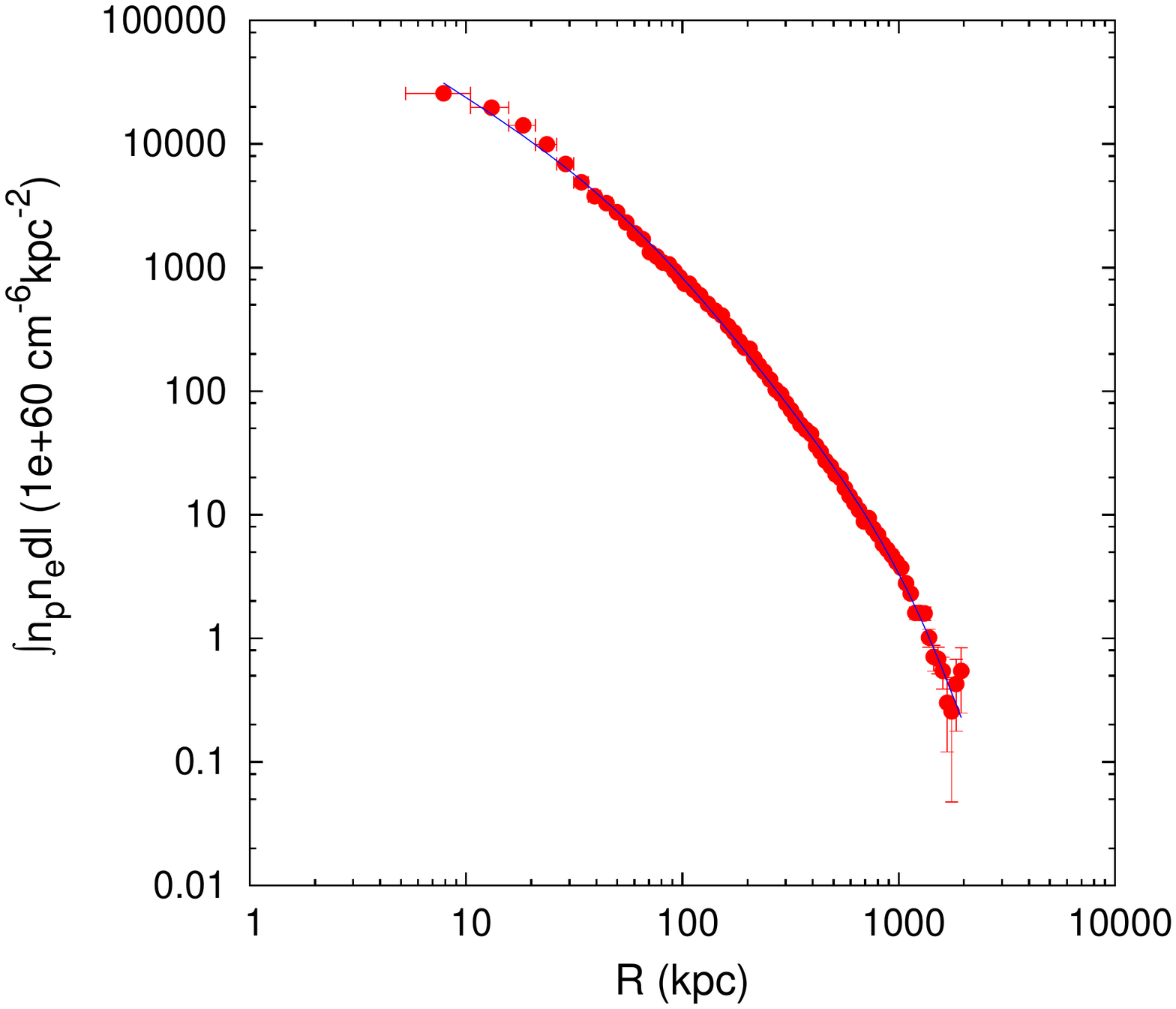}}}
\put(4.2,-0.21){\scalebox{0.30}{\includegraphics{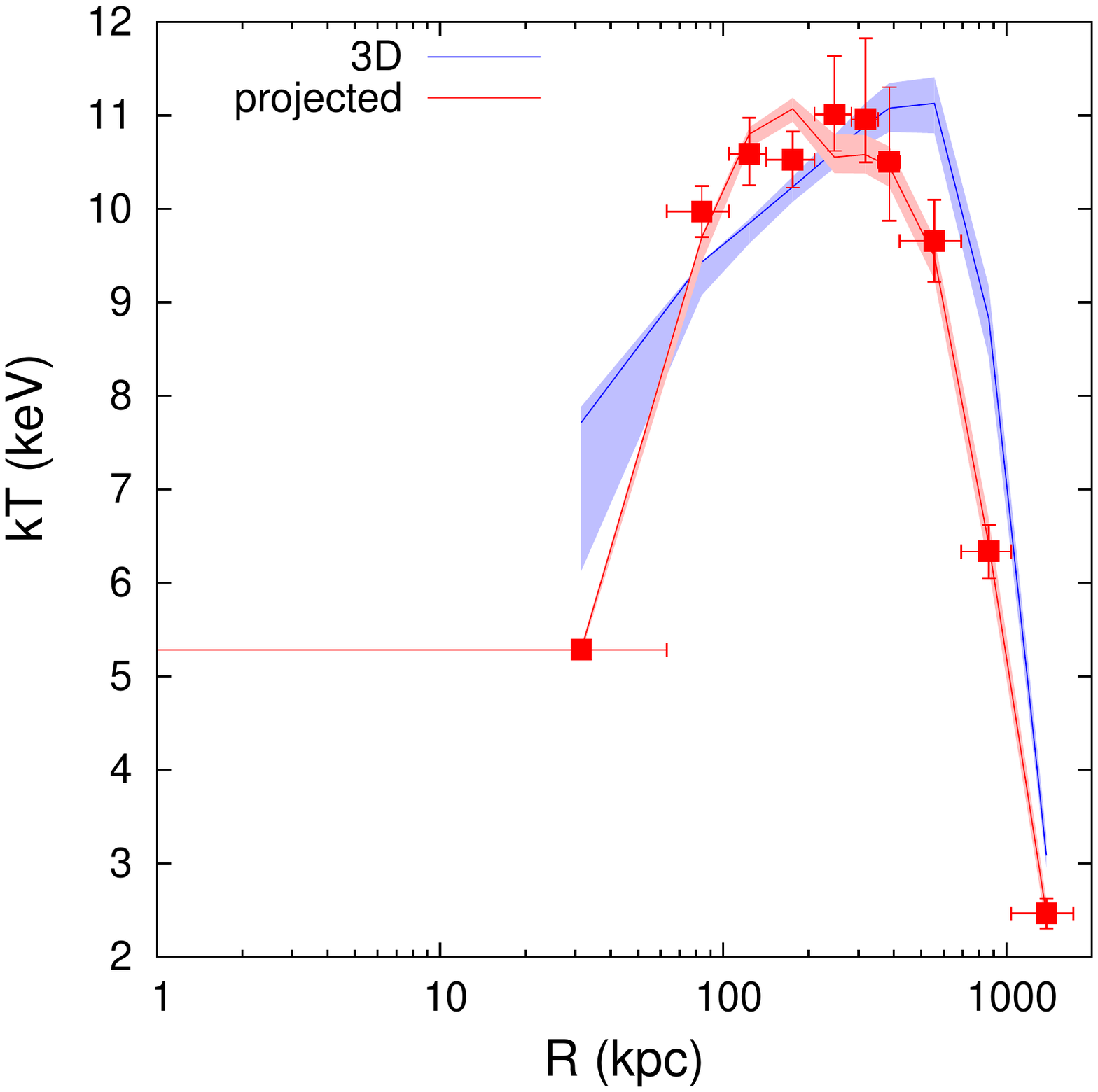}}}
\end{picture}
\end{center}
\caption{\small{Left: An adaptively smoothed image of the cluster
    A2204 (3$\times$3 Mpc on a side); Center: Observed projected
    emissivity profile for A2204, with the best fitting gas density
    profile shown by the solid blue line; Right: Temperature profile
    for A2204.  The solid red line shows the best fitting three
    dimentional model to the temperature profile, and the solid blue
    line represents the corresponding projected profile.  The red and
    blue shaded regions show the corresponding uncertainties on the
    three dimentional model and projected profile respectively,and
    obtained from Monte Carlo simulations.}\label{fig:a2204}}
\end{figure*}

\begin{figure*}
\begin{center}
\setlength{\unitlength}{1in}
\begin{picture}(6.9,2.0)
\put(0.01,-0.8){\scalebox{0.34}{\includegraphics[clip=true]{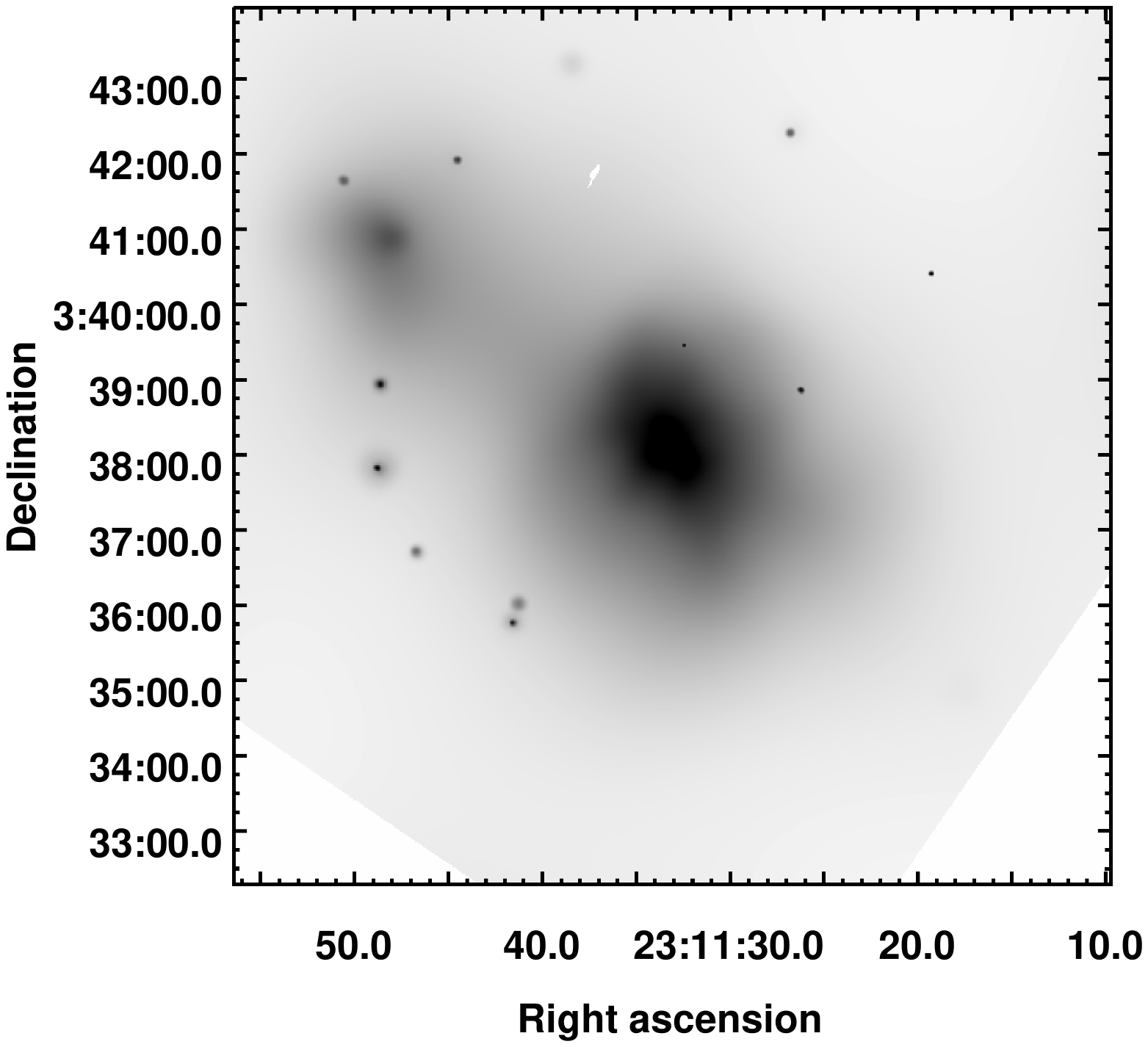}}}  
\put(2.1,-0.21){\scalebox{0.30}{\includegraphics{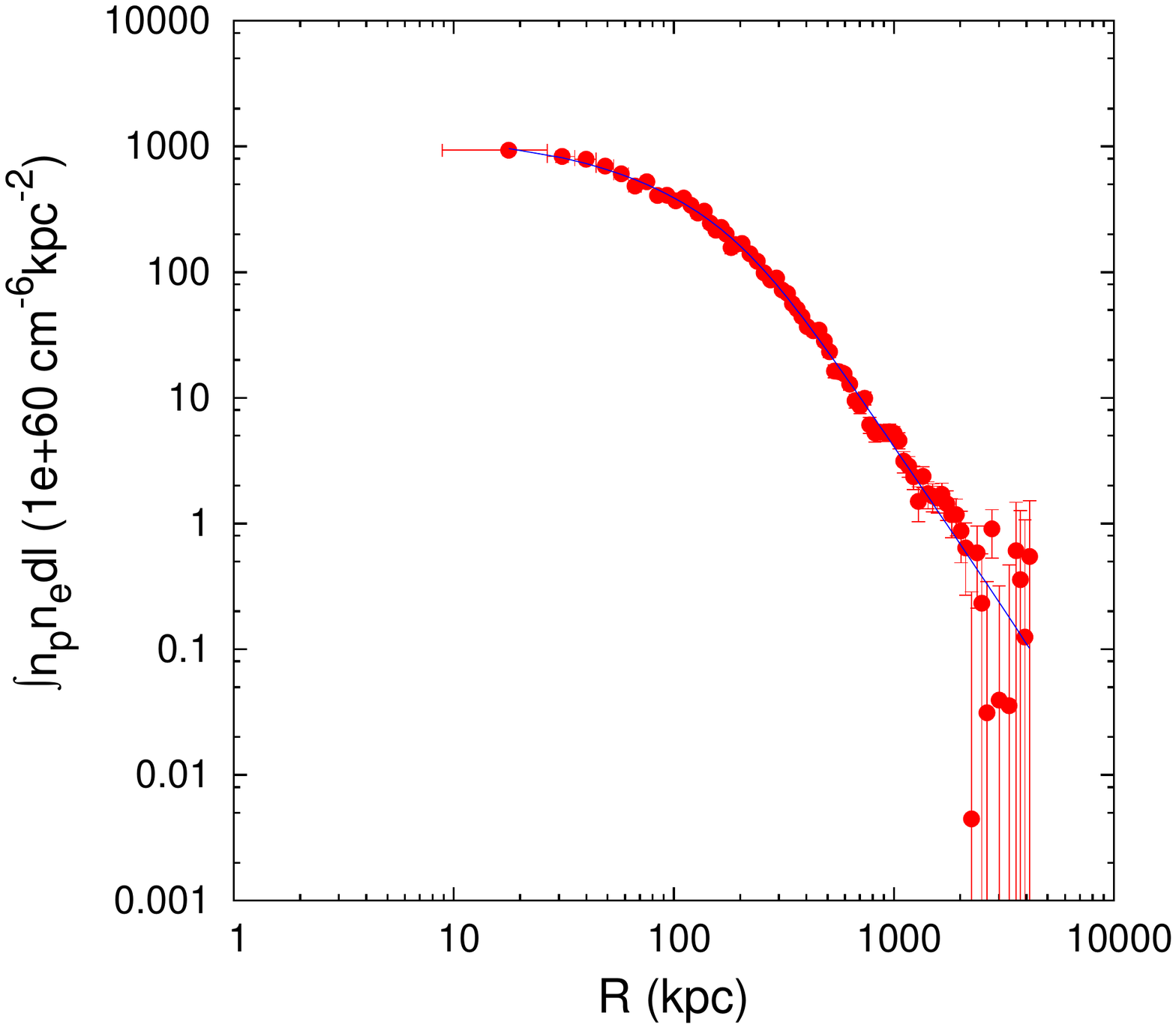}}}
\put(4.2,-0.21){\scalebox{0.30}{\includegraphics{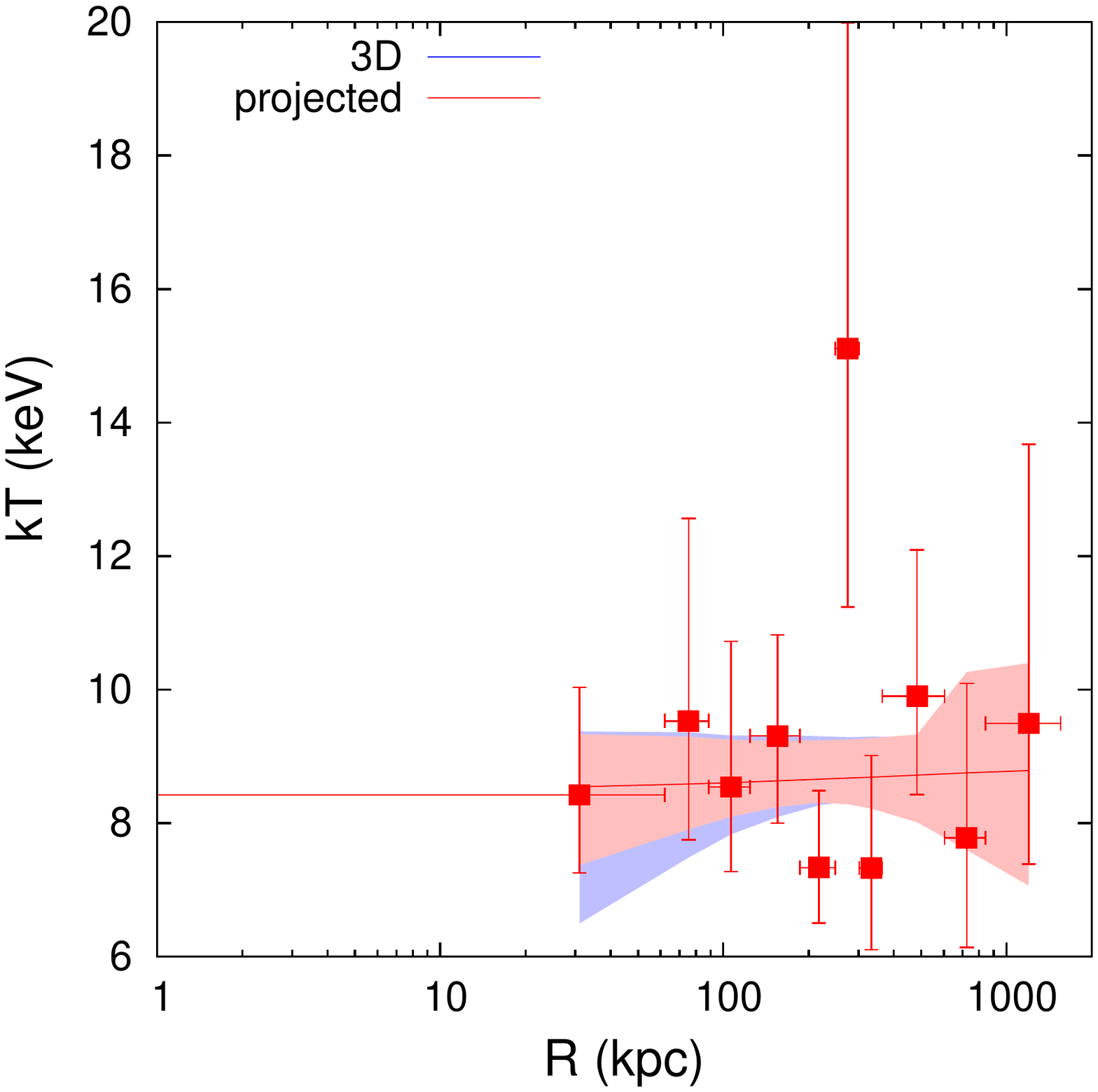}}}
\end{picture}
\end{center}
\caption{\small{Same as Figure~\ref{fig:a2204} but for A2552.}\label{fig:a2552}}
\end{figure*}

Here we show images, gas density and temperature profiles for our
cluster sample.  Figure~\ref{fig:a2204} shows an
adaptively smoothed image of the cluster measuring 3$\times$3 Mpc on a
side (left), the emmisivity profile with the best fitting gas density
profile (middle) and the temperature profile (right) with the best
fitting three dimentional model (red) and the corresponding projected
profile (blue) for the cluster A2204.
Figures~\ref{fig:a586}-\ref{fig:a2552} show the rest of the cluster
sample.  

\begin{figure*}
\begin{center}
\setlength{\unitlength}{1in}
\begin{picture}(6.9,2.0)
\put(0.01,-0.8){\scalebox{0.34}{\includegraphics[clip=true]{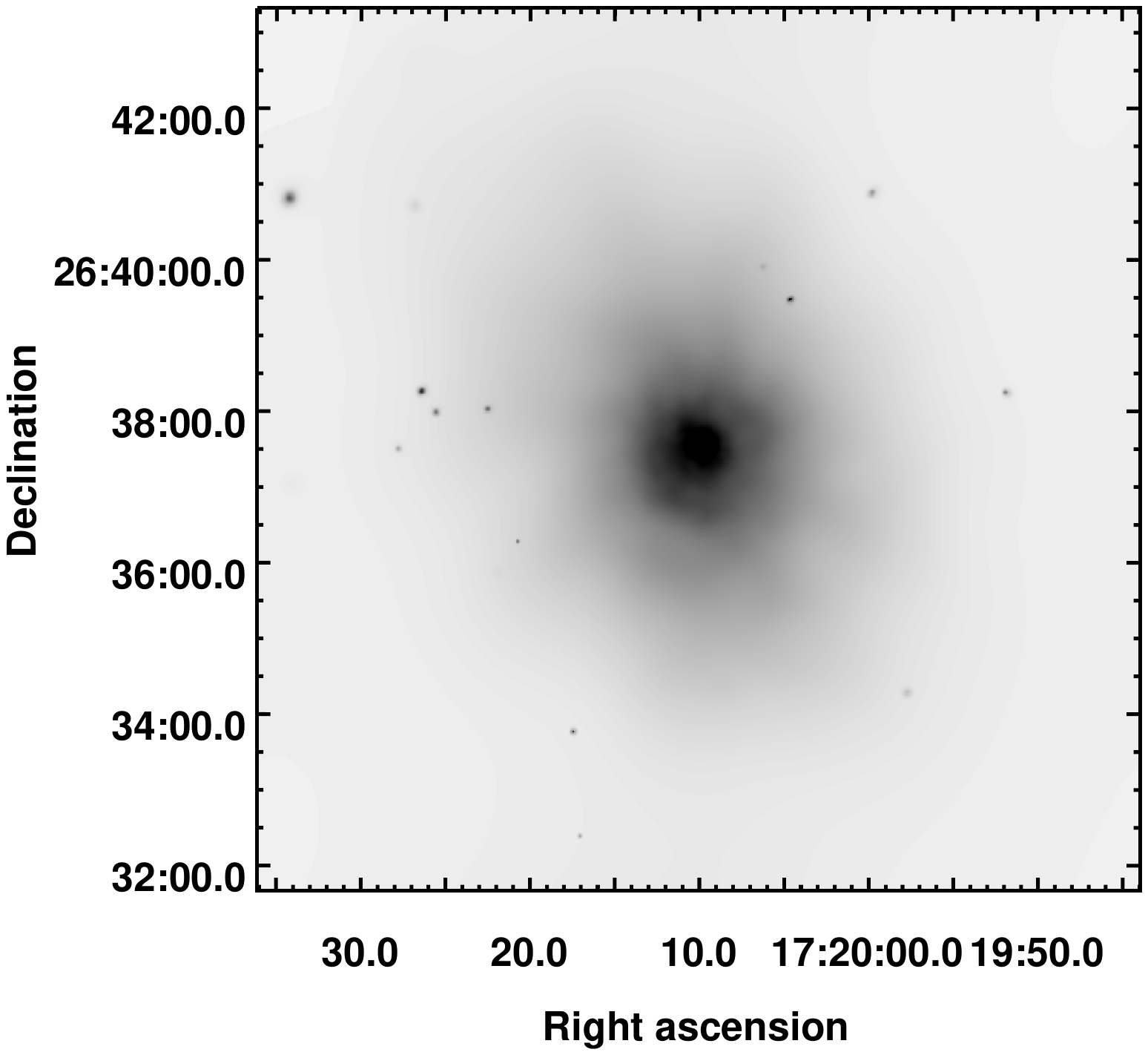}}}  
\put(2.1,-0.21){\scalebox{0.30}{\includegraphics{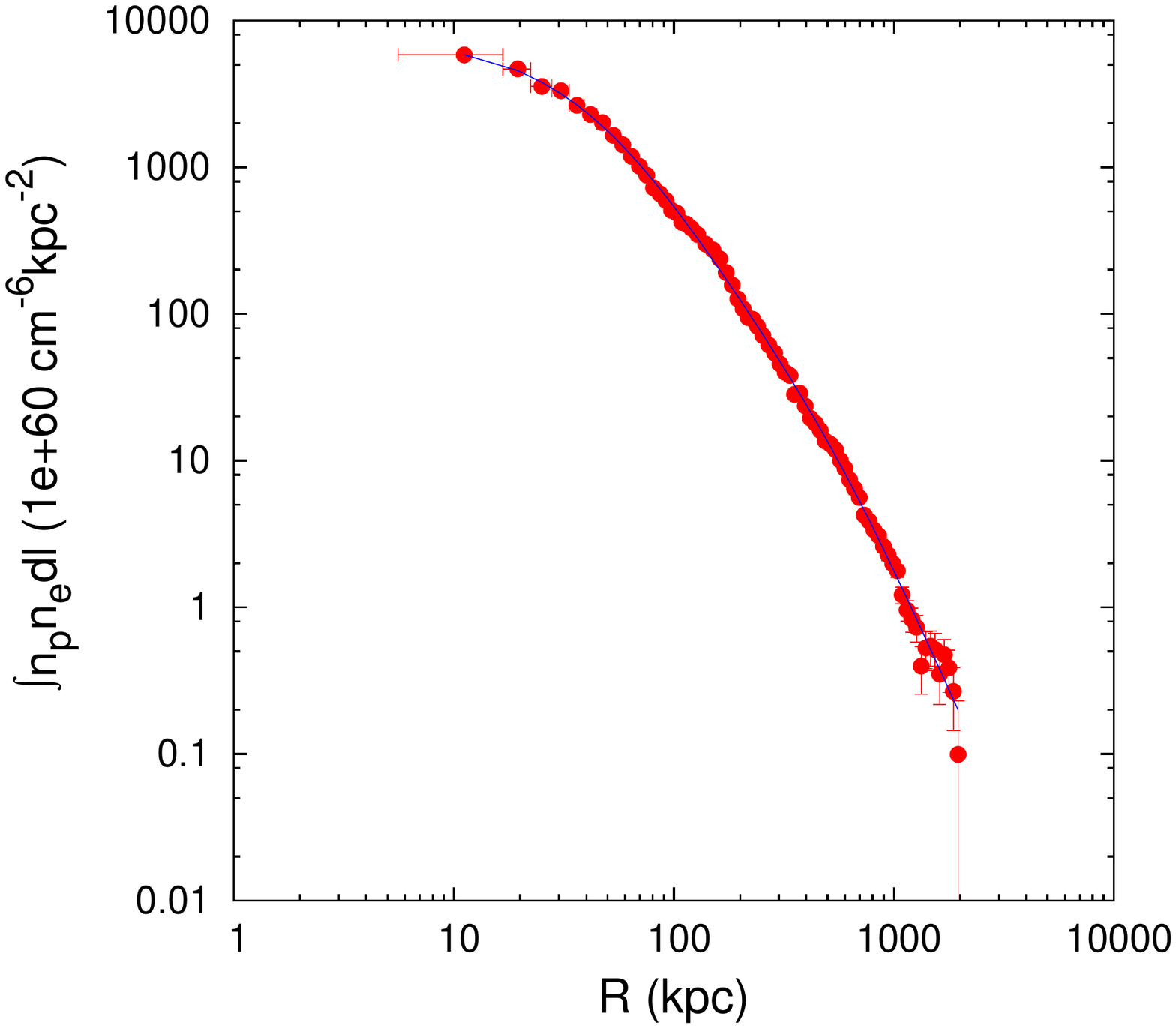}}}
\put(4.2,-0.21){\scalebox{0.30}{\includegraphics{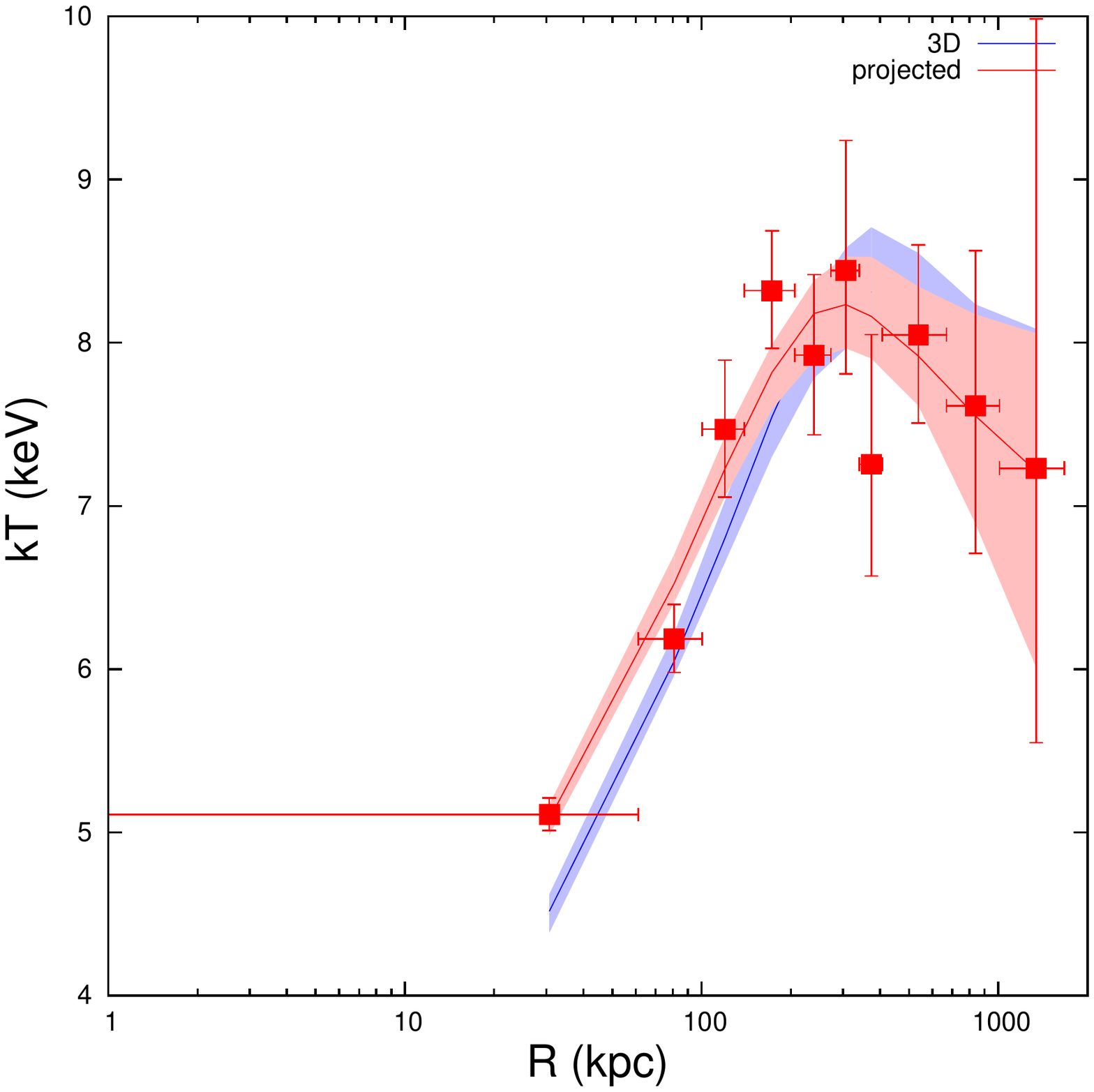}}}
\end{picture}
\end{center}
\caption{\small{Same as Figure~\ref{fig:a2204} but for RXJ1720.1+2638.} \label{fig:a586}}
\end{figure*}

\begin{figure*}
\begin{center}
\setlength{\unitlength}{1in}
\begin{picture}(6.9,2.0)
\put(0.01,-0.8){\scalebox{0.34}{\includegraphics[clip=true]{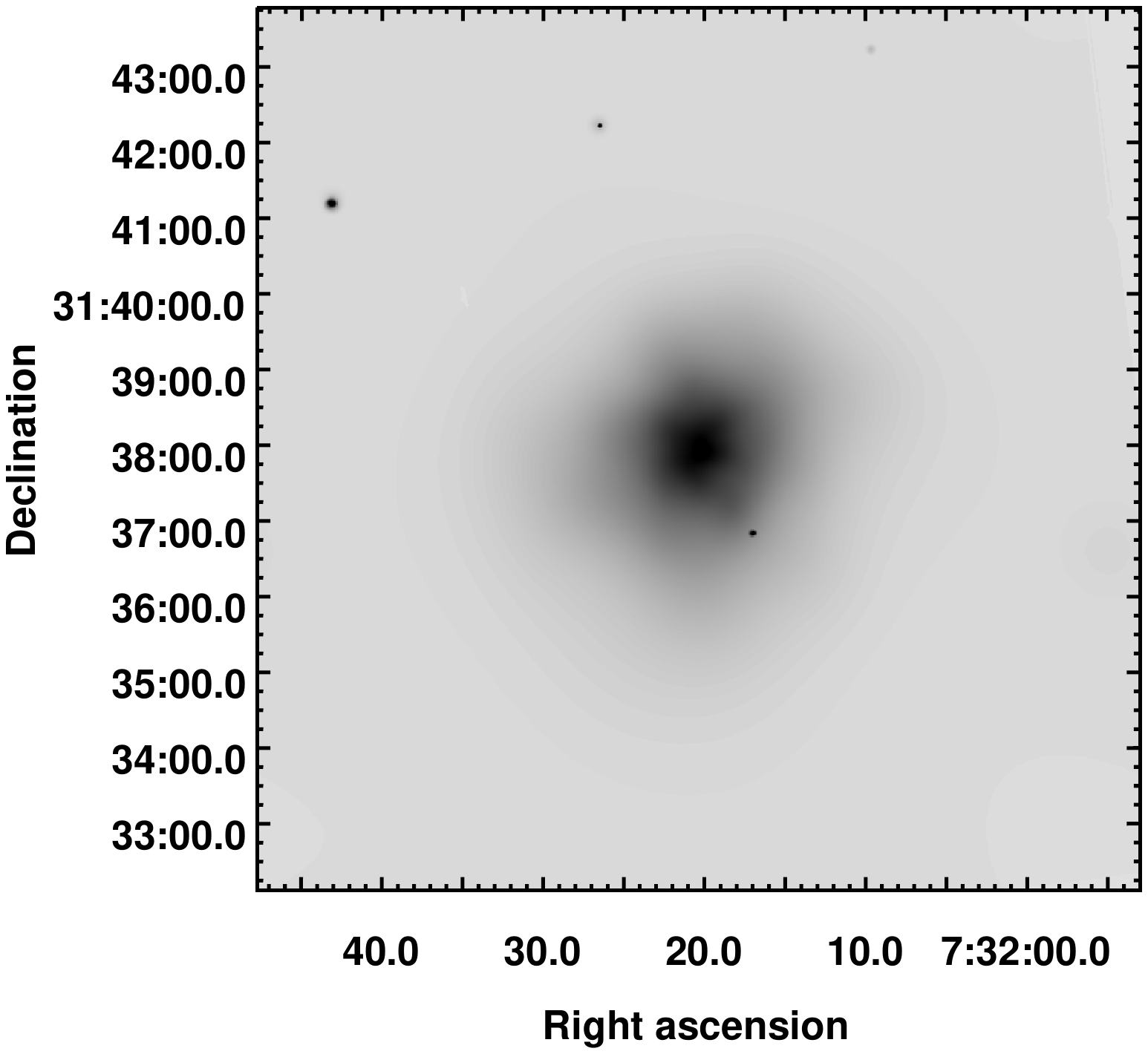}}}  
\put(2.1,-0.21){\scalebox{0.30}{\includegraphics{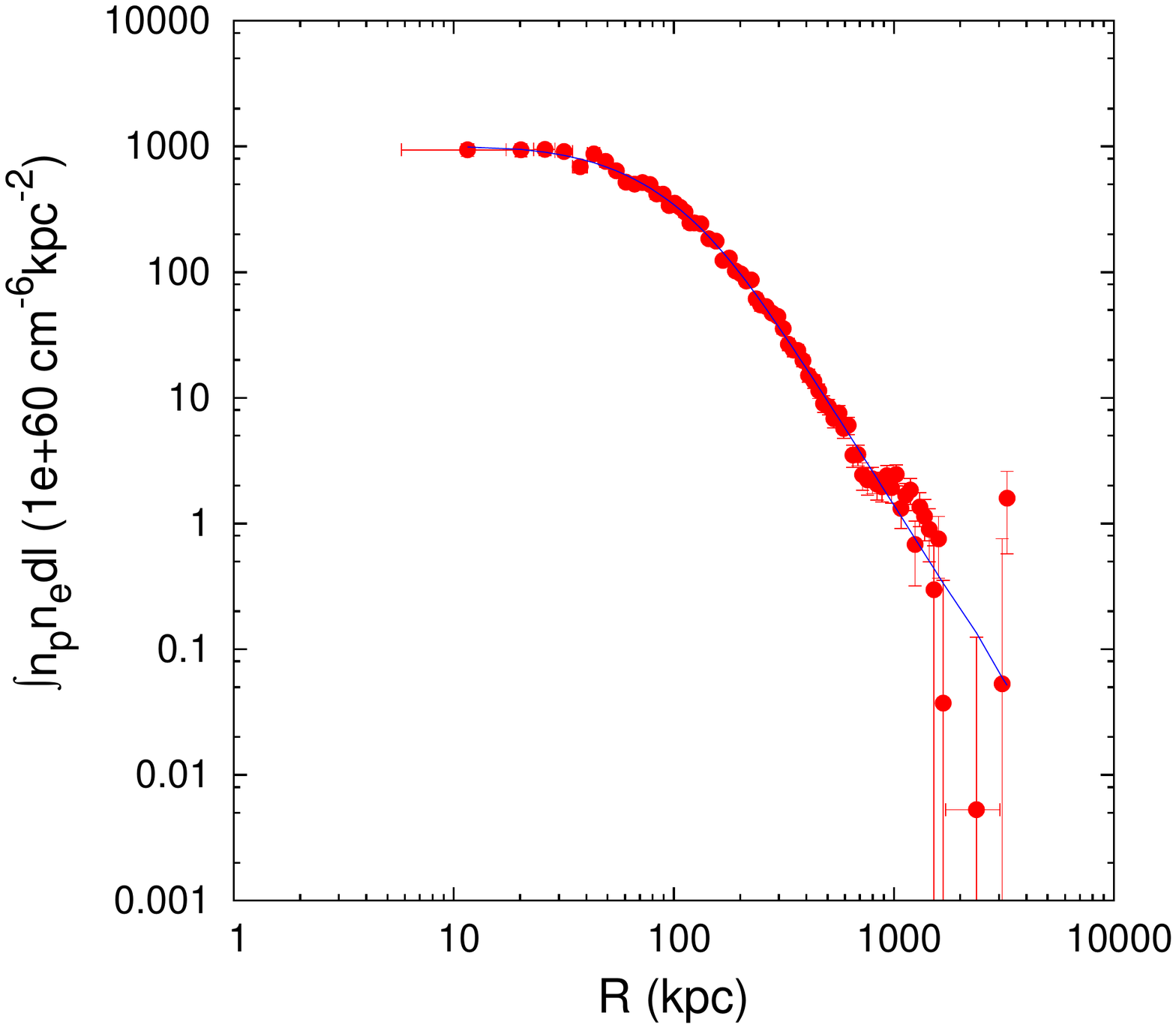}}}
\put(4.2,-0.21){\scalebox{0.30}{\includegraphics{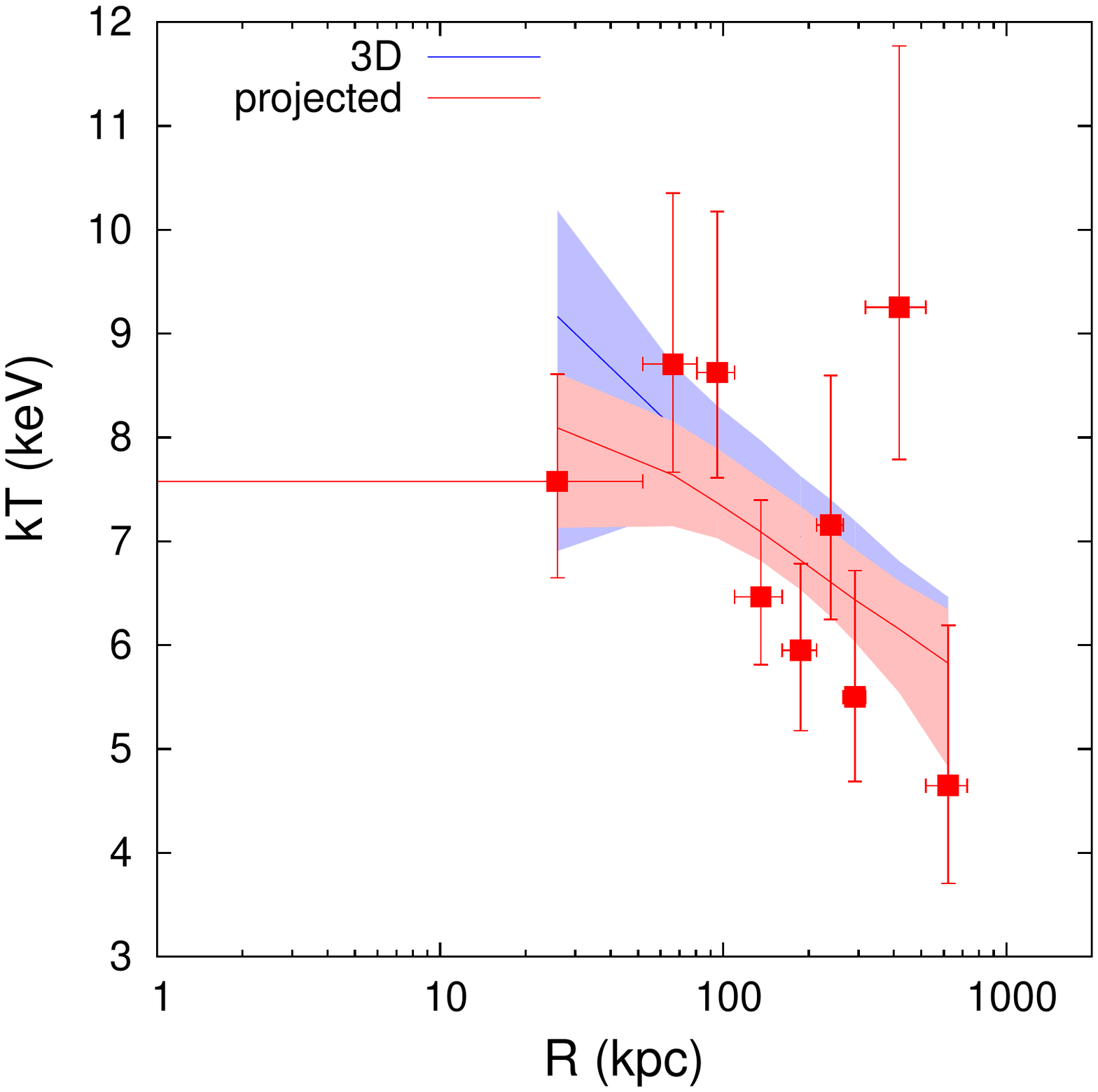}}}
\end{picture}
\end{center}
\caption{\small{Same as Figure~\ref{fig:a2204} but for A586.} \label{fig:a586}}
\end{figure*}

\begin{figure*}
\begin{center}
\setlength{\unitlength}{1in}
\begin{picture}(6.9,2.0)
\put(0.01,-0.8){\scalebox{0.34}{\includegraphics[clip=true]{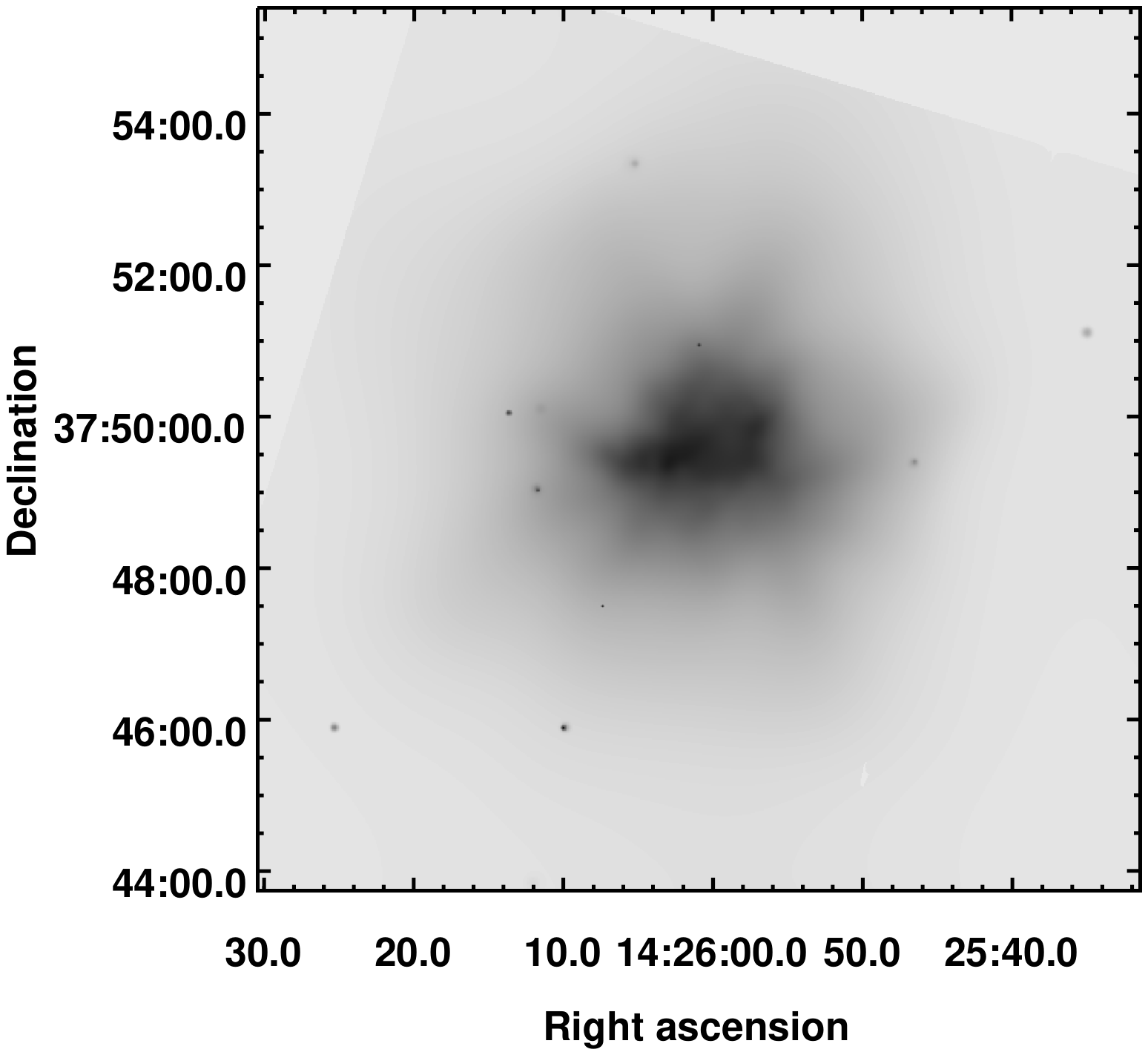}}}  
\put(2.1,-0.21){\scalebox{0.30}{\includegraphics{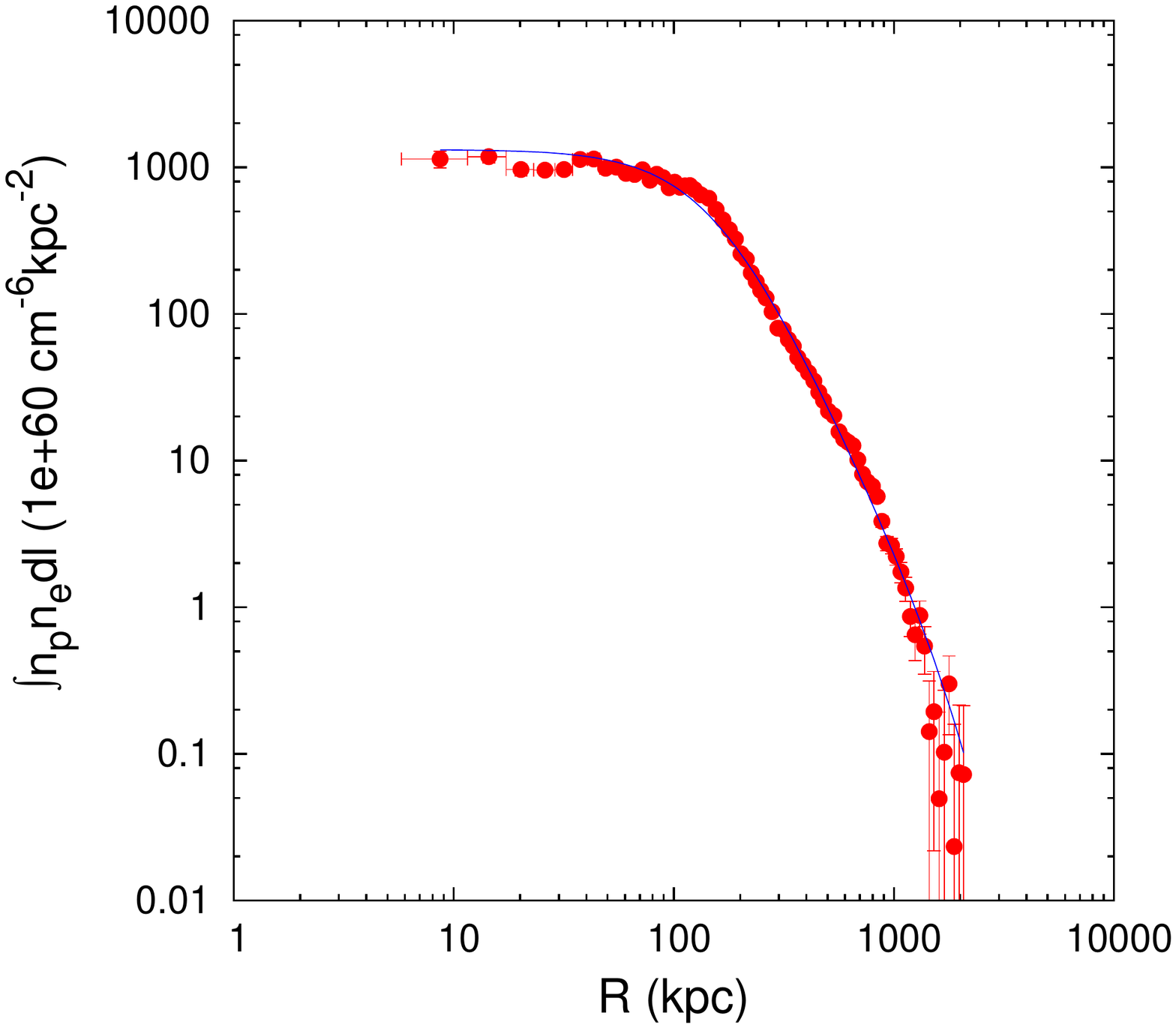}}}
\put(4.2,-0.21){\scalebox{0.30}{\includegraphics{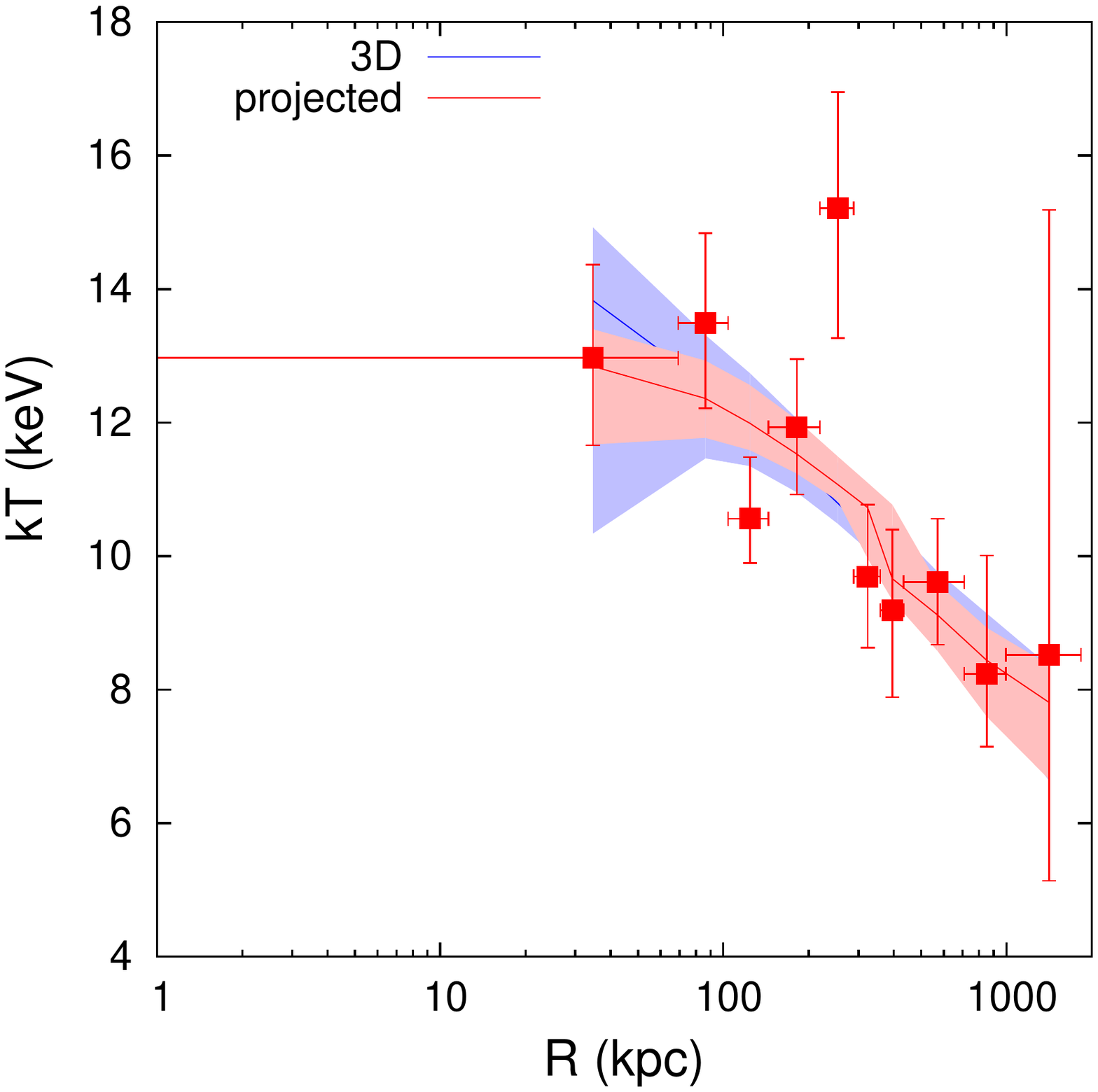}}}
\end{picture}
\end{center}
\caption{\small{Same as Figure~\ref{fig:a2204} but for A1914.}\label{fig:a1914}}
\end{figure*}

\begin{figure*}
\begin{center}
\setlength{\unitlength}{1in}
\begin{picture}(6.9,2.0)
\put(0.01,-0.8){\scalebox{0.34}{\includegraphics[clip=true]{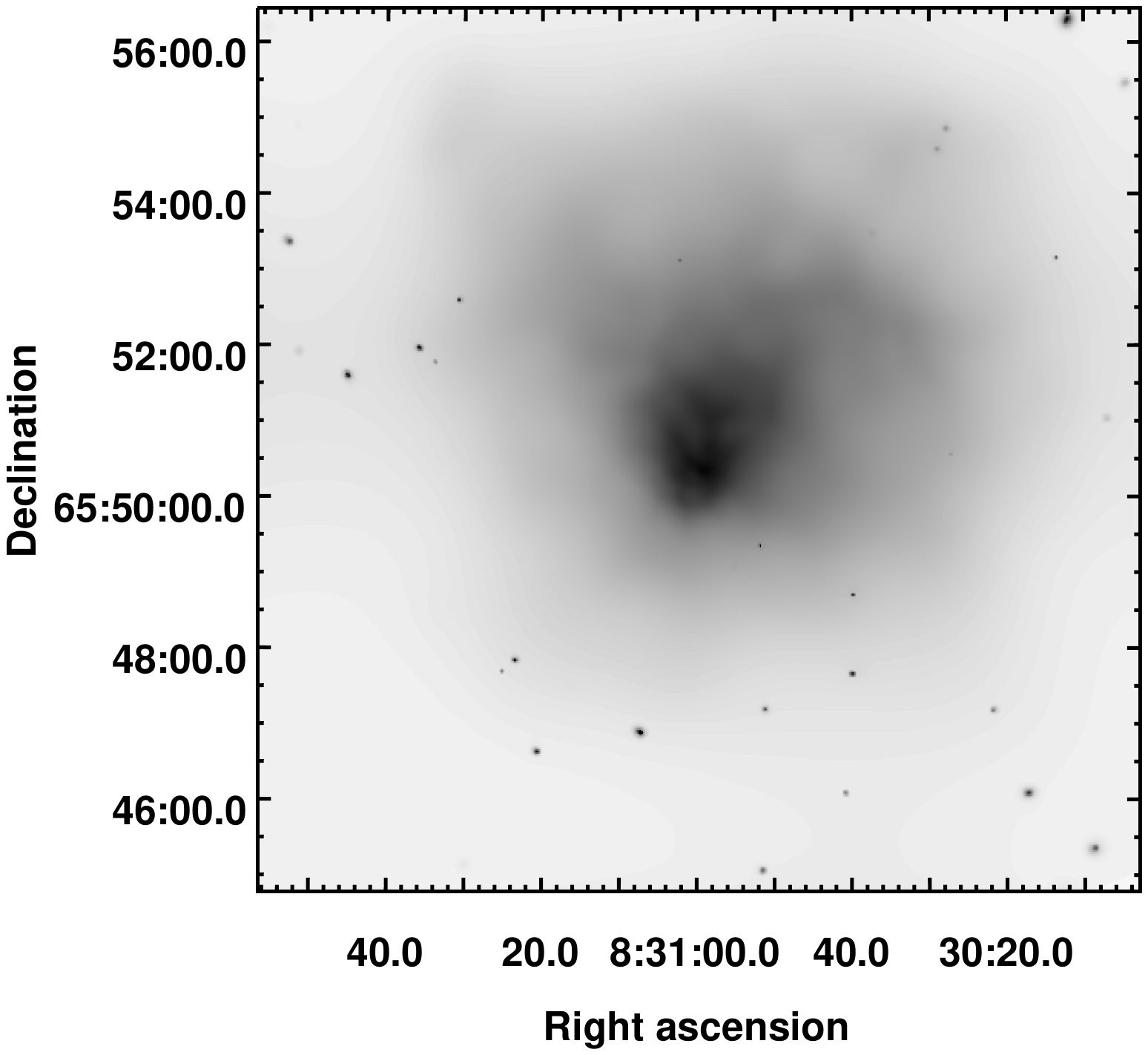}}}  
\put(2.1,-0.21){\scalebox{0.30}{\includegraphics{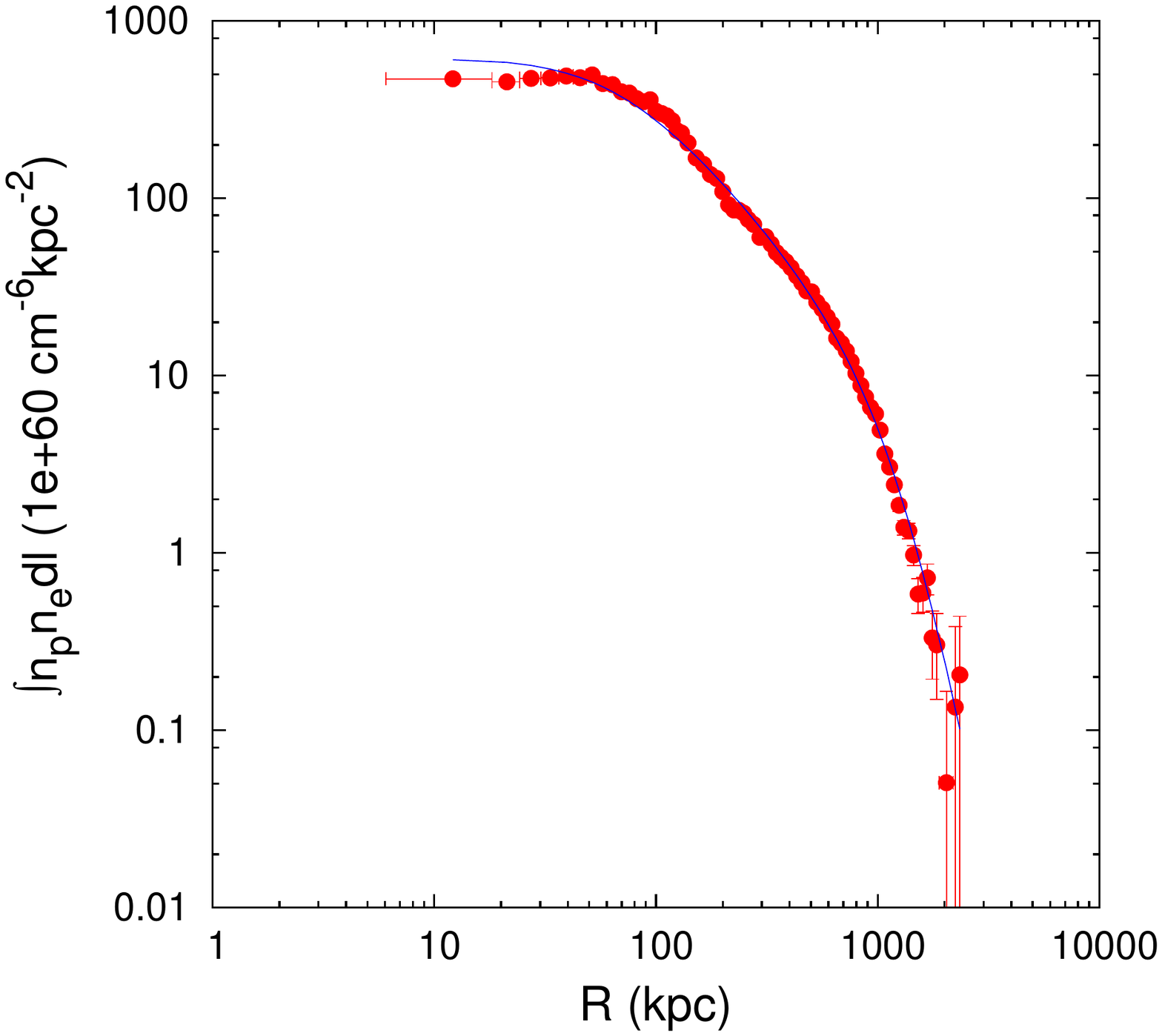}}}
\put(4.2,-0.21){\scalebox{0.30}{\includegraphics{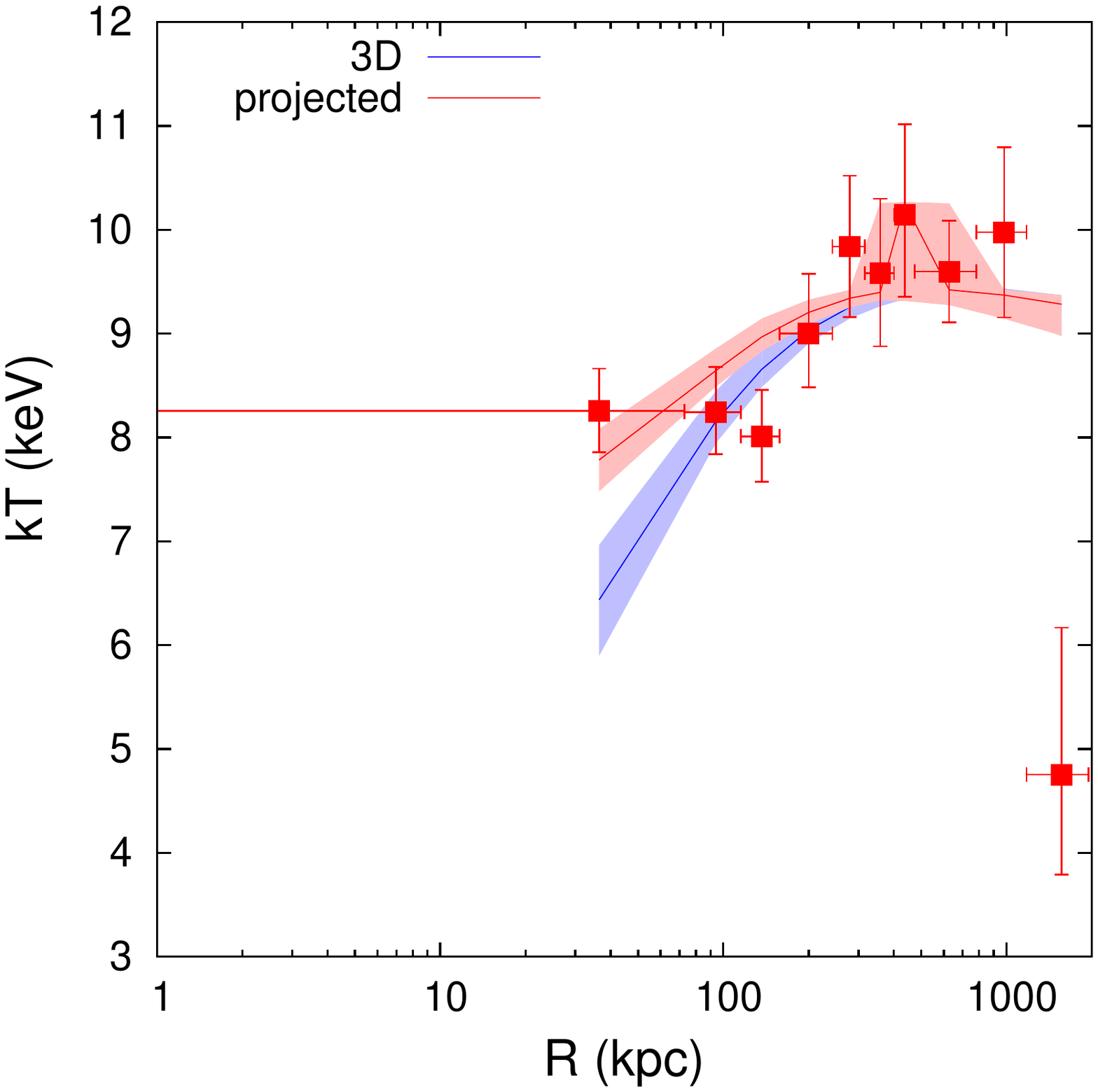}}}
\end{picture}
\end{center}
\caption{\small{Same as Figure~\ref{fig:a2204} but for A665.}\label{fig:a665}}
\end{figure*}

\begin{figure*}
\begin{center}
\setlength{\unitlength}{1in}
\begin{picture}(6.9,2.0)
\put(0.01,-0.8){\scalebox{0.34}{\includegraphics[clip=true]{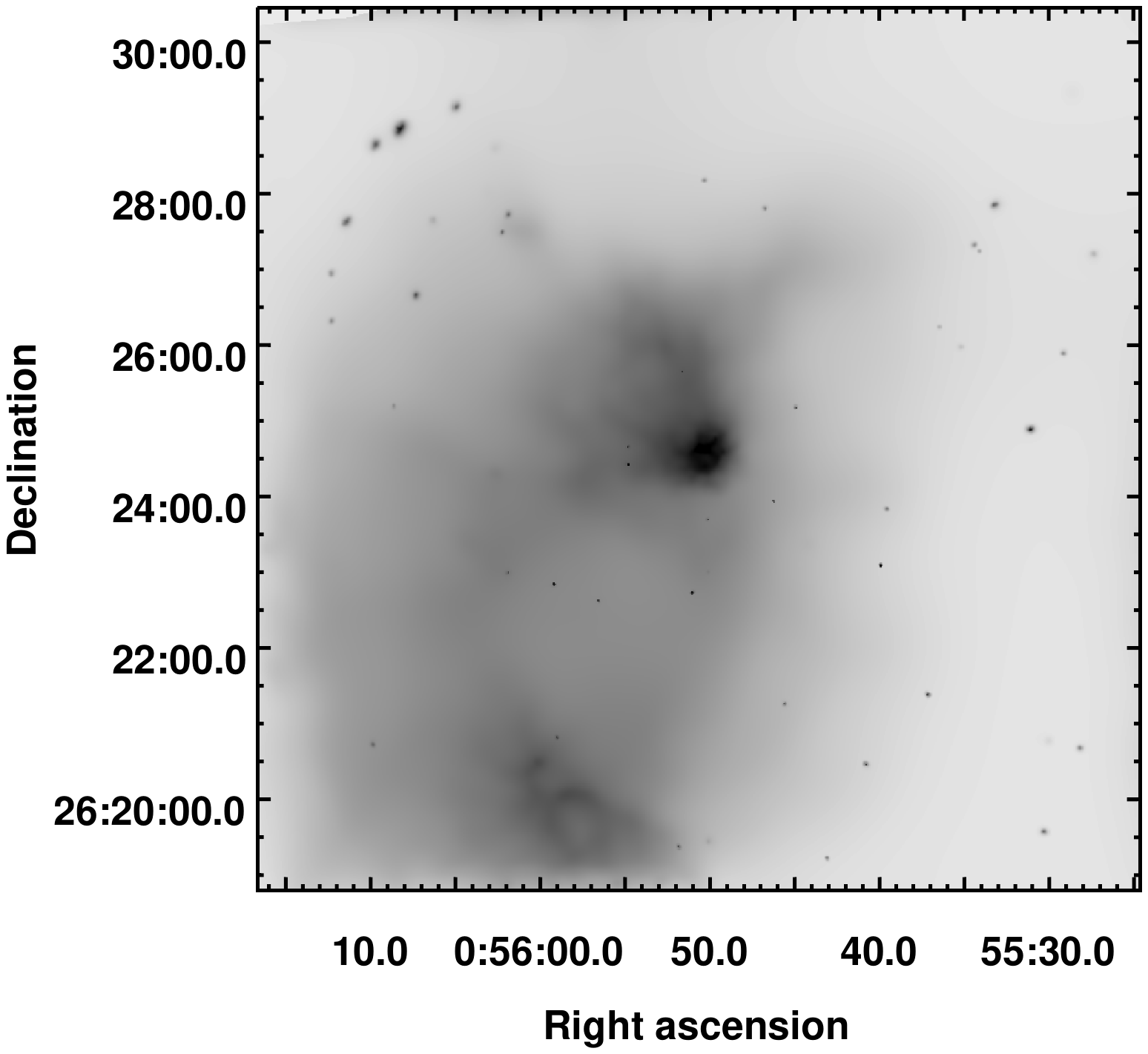}}}  
\put(2.1,-0.21){\scalebox{0.30}{\includegraphics{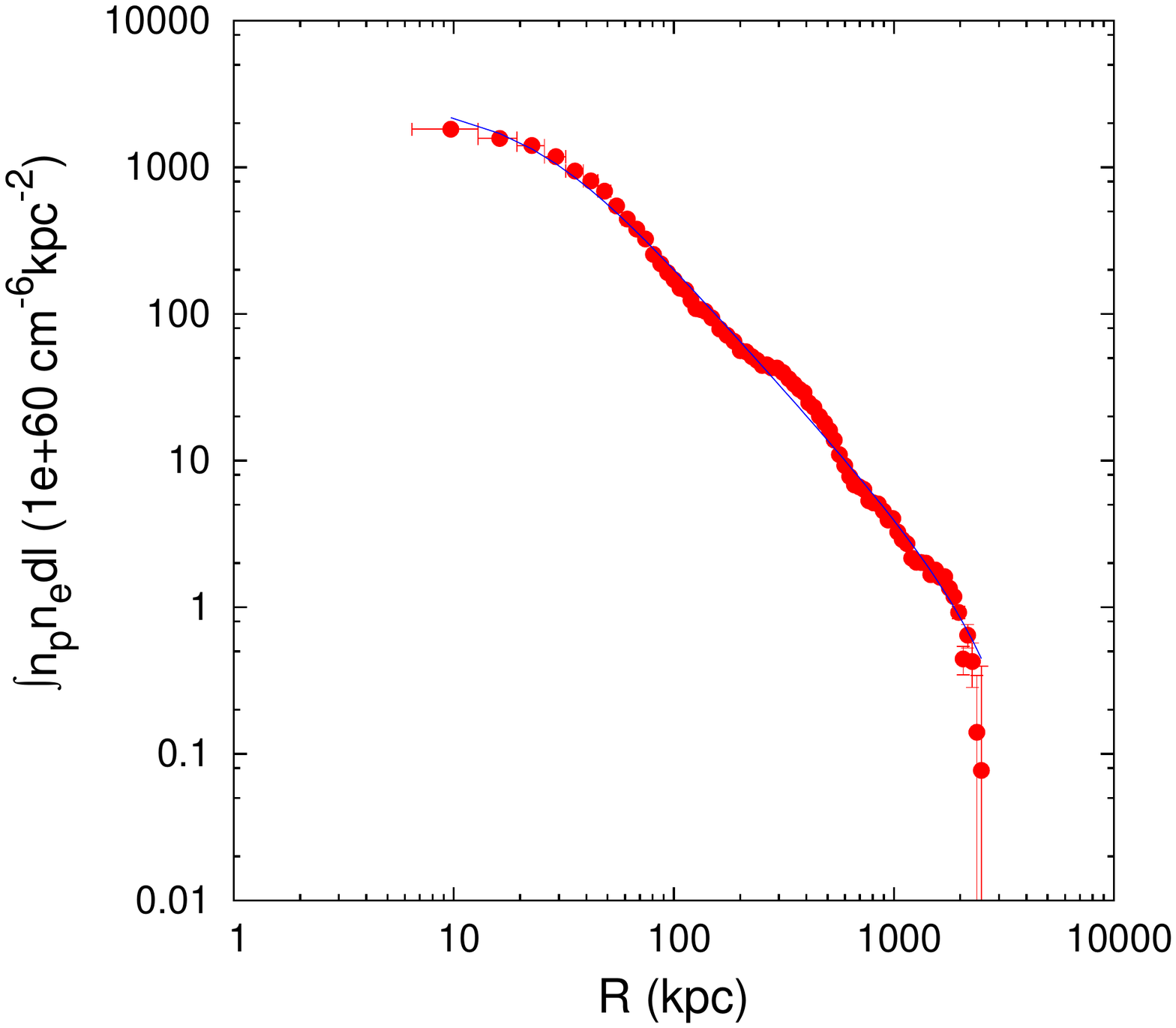}}}
\put(4.2,-0.21){\scalebox{0.30}{\includegraphics{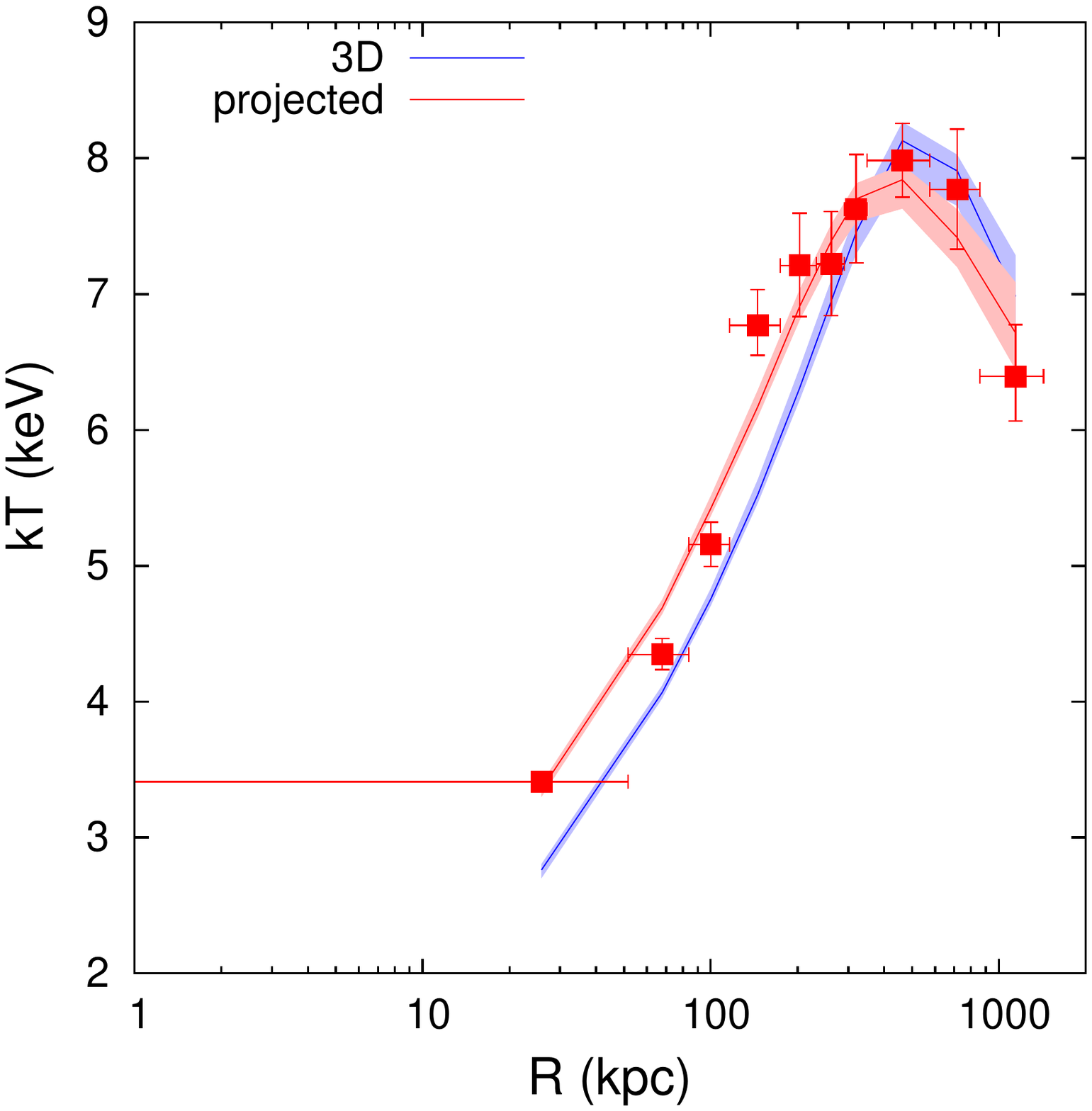}}}
\end{picture}
\end{center}
\caption{\small{Same as Figure~\ref{fig:a2204} but for A115.}\label{fig:a115}}
\end{figure*}

\begin{figure*}
\begin{center}
\setlength{\unitlength}{1in}
\begin{picture}(6.9,2.0)
\put(0.01,-0.8){\scalebox{0.34}{\includegraphics[clip=true]{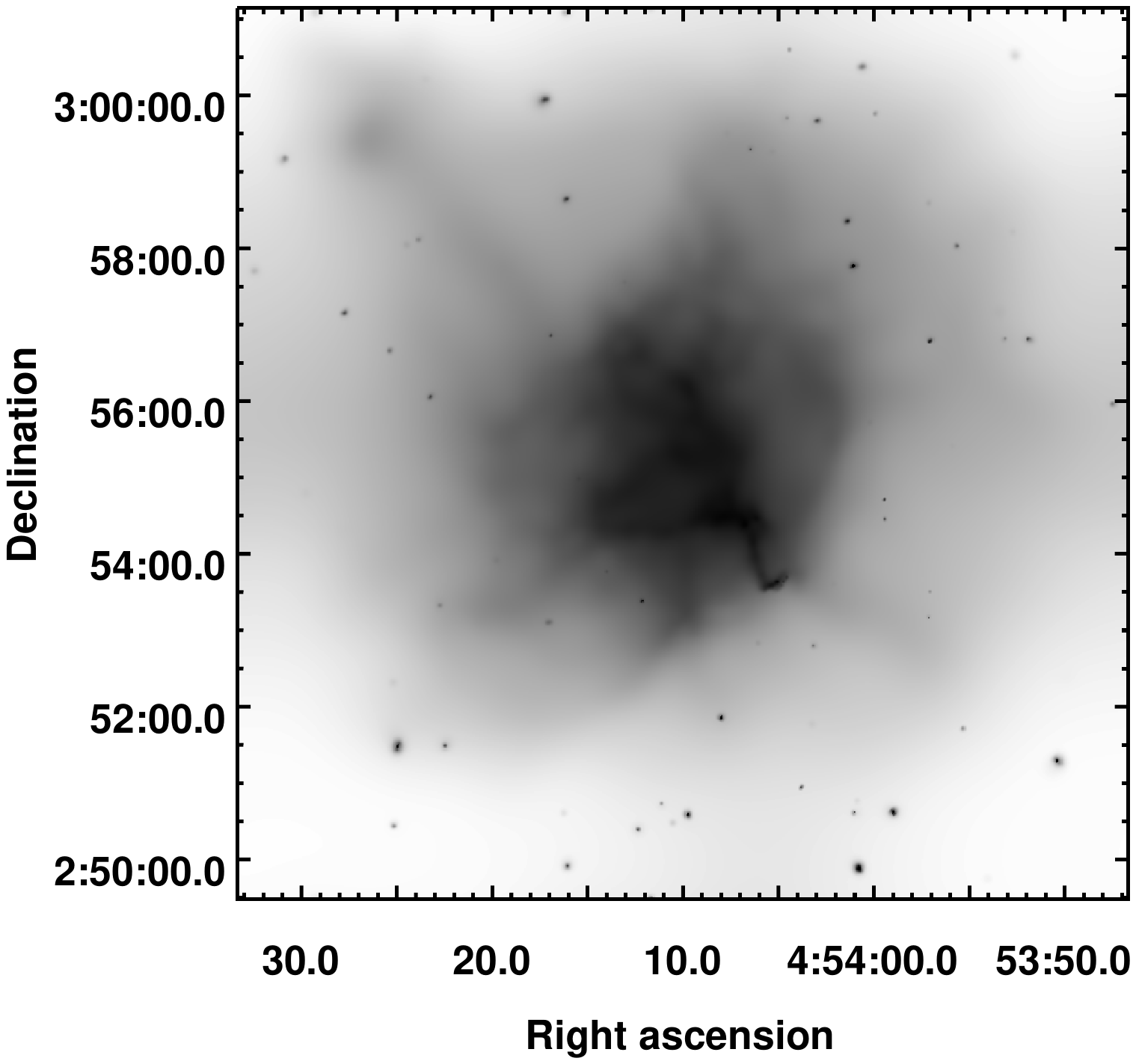}}}  
\put(2.1,-0.21){\scalebox{0.30}{\includegraphics{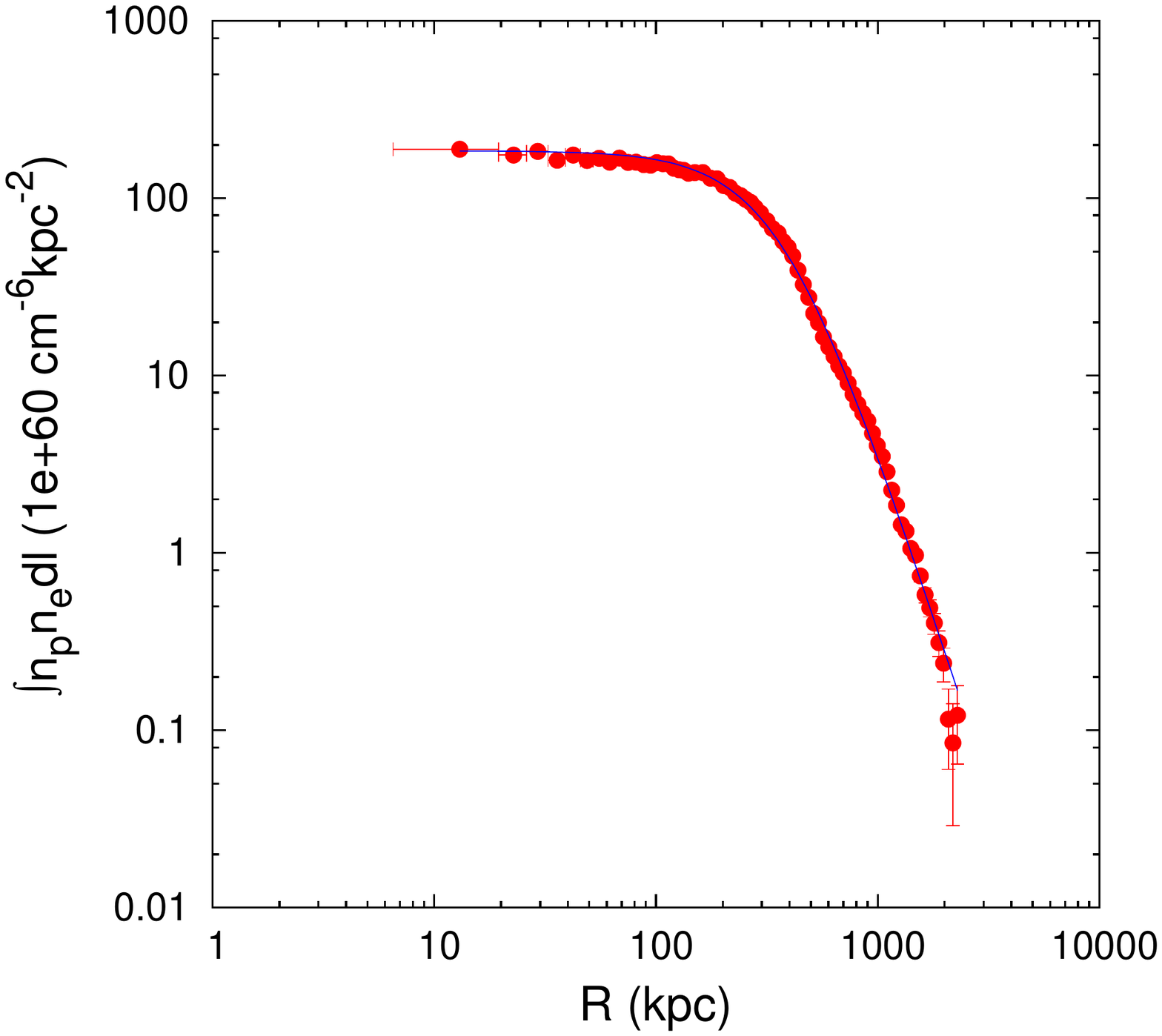}}}
\put(4.2,-0.21){\scalebox{0.30}{\includegraphics{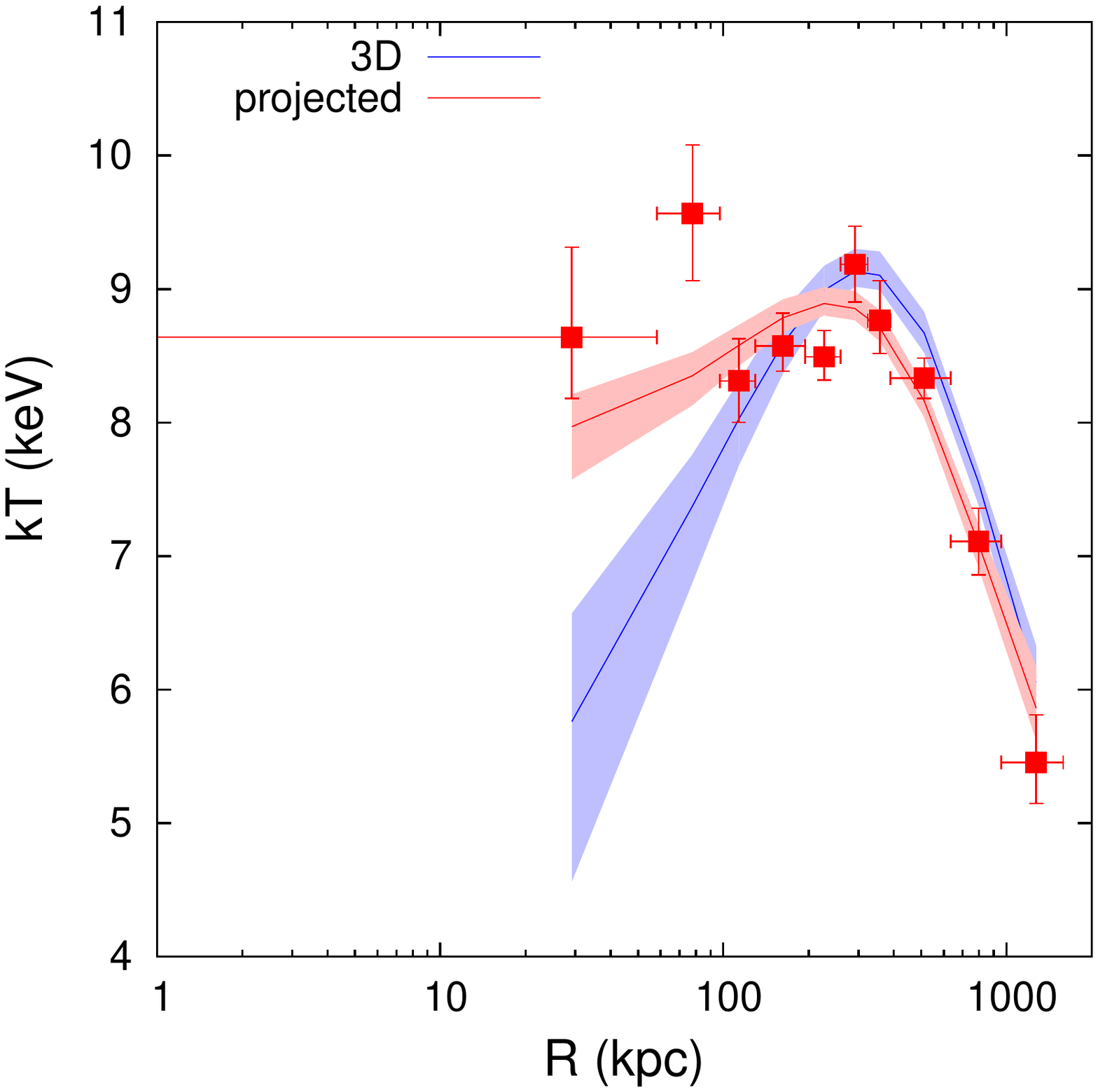}}}
\end{picture}
\end{center}
\caption{\small{Same as Figure~\ref{fig:a2204} but for A520.\label{fig:a520}}}
\end{figure*}

\begin{figure*}
\begin{center}
\setlength{\unitlength}{1in}
\begin{picture}(6.9,2.0)
\put(0.01,-0.8){\scalebox{0.34}{\includegraphics[clip=true]{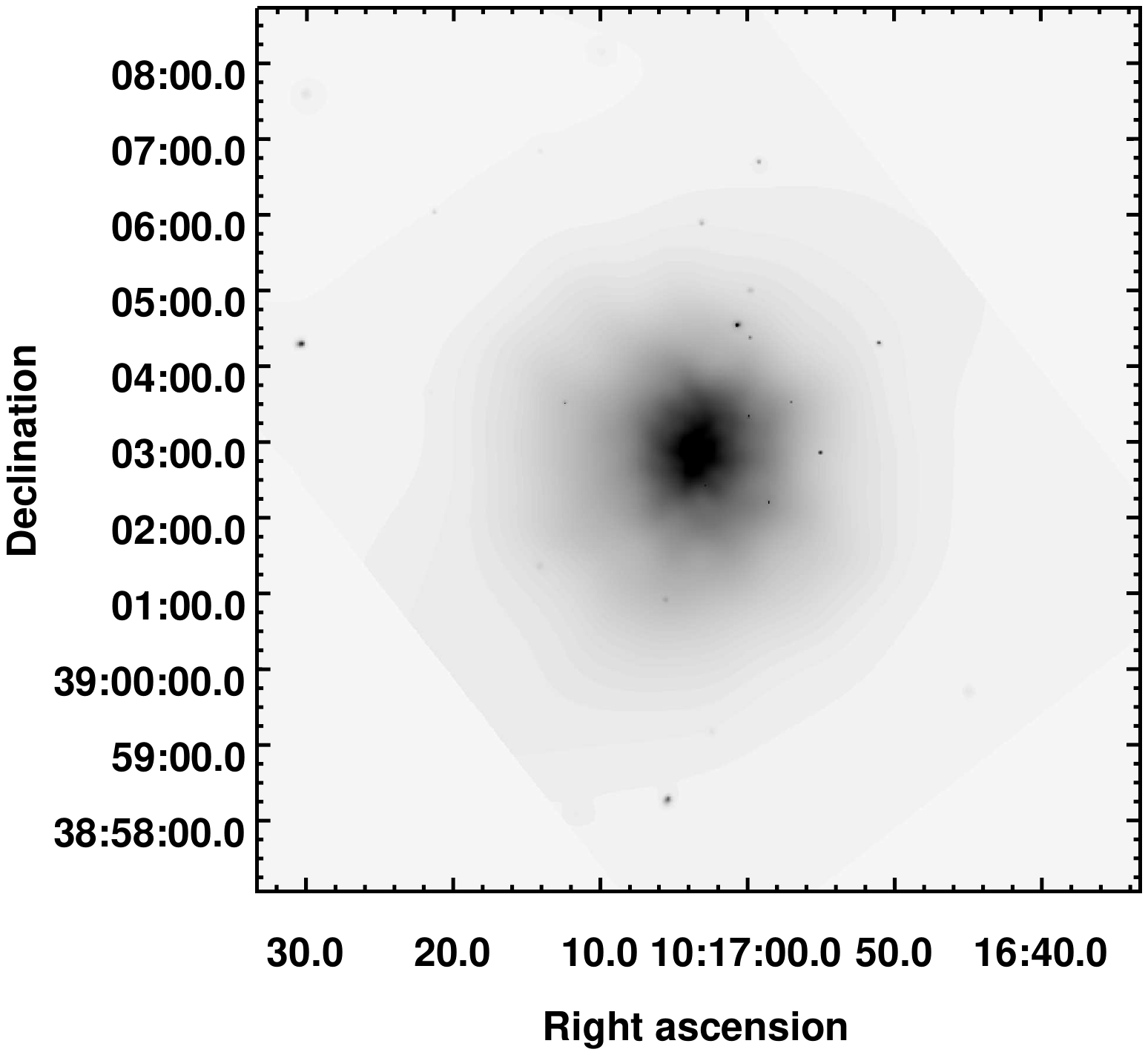}}}  
\put(2.1,-0.21){\scalebox{0.30}{\includegraphics{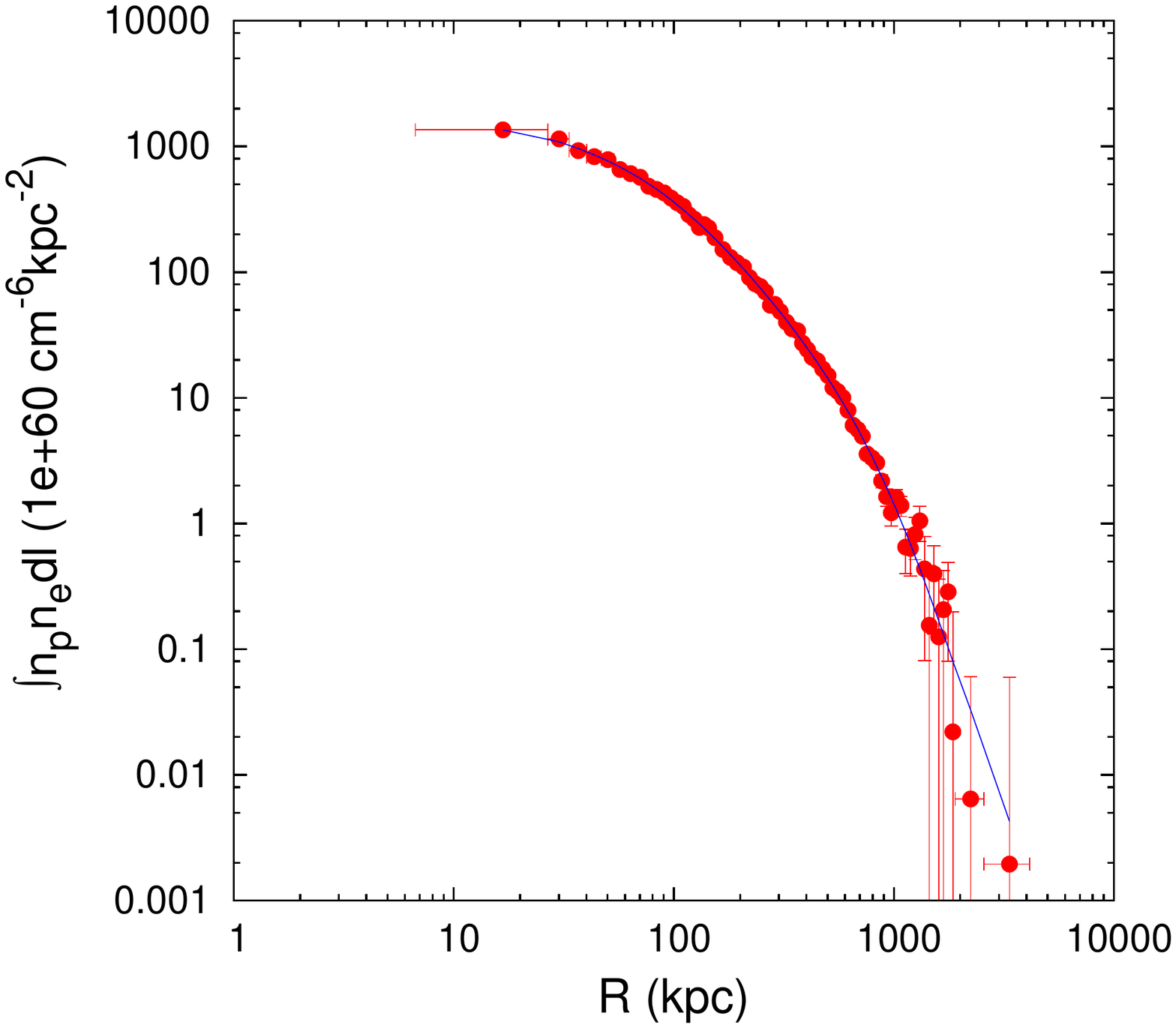}}}
\put(4.2,-0.21){\scalebox{0.30}{\includegraphics{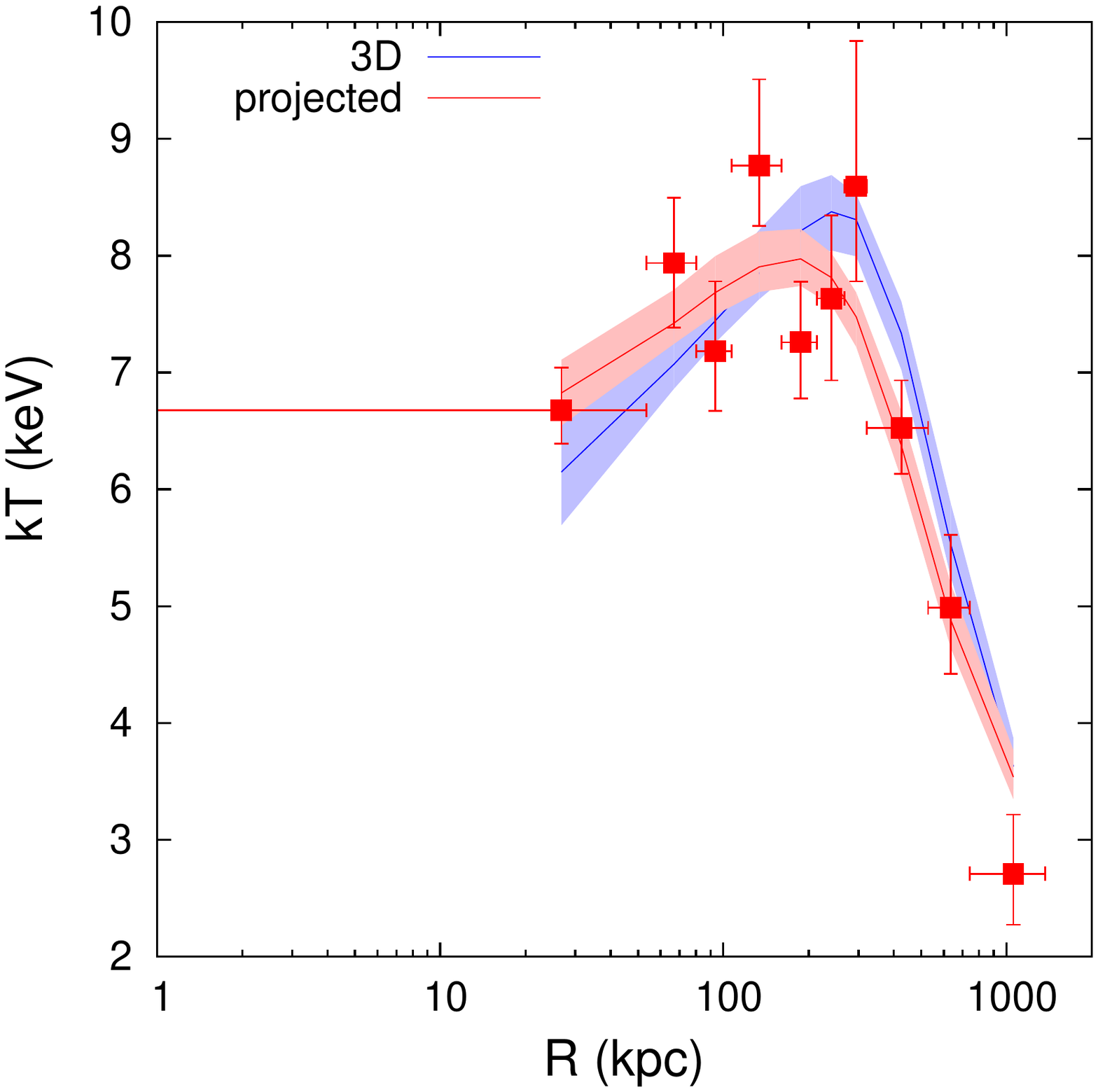}}}
\end{picture}
\end{center}
\caption{\small{Same as Figure~\ref{fig:a2204} but for A963.}\label{fig:a963}}
\end{figure*}

\begin{figure*}
\begin{center}
\setlength{\unitlength}{1in}
\begin{picture}(6.9,2.0)
\put(0.01,-0.8){\scalebox{0.34}{\includegraphics[clip=true]{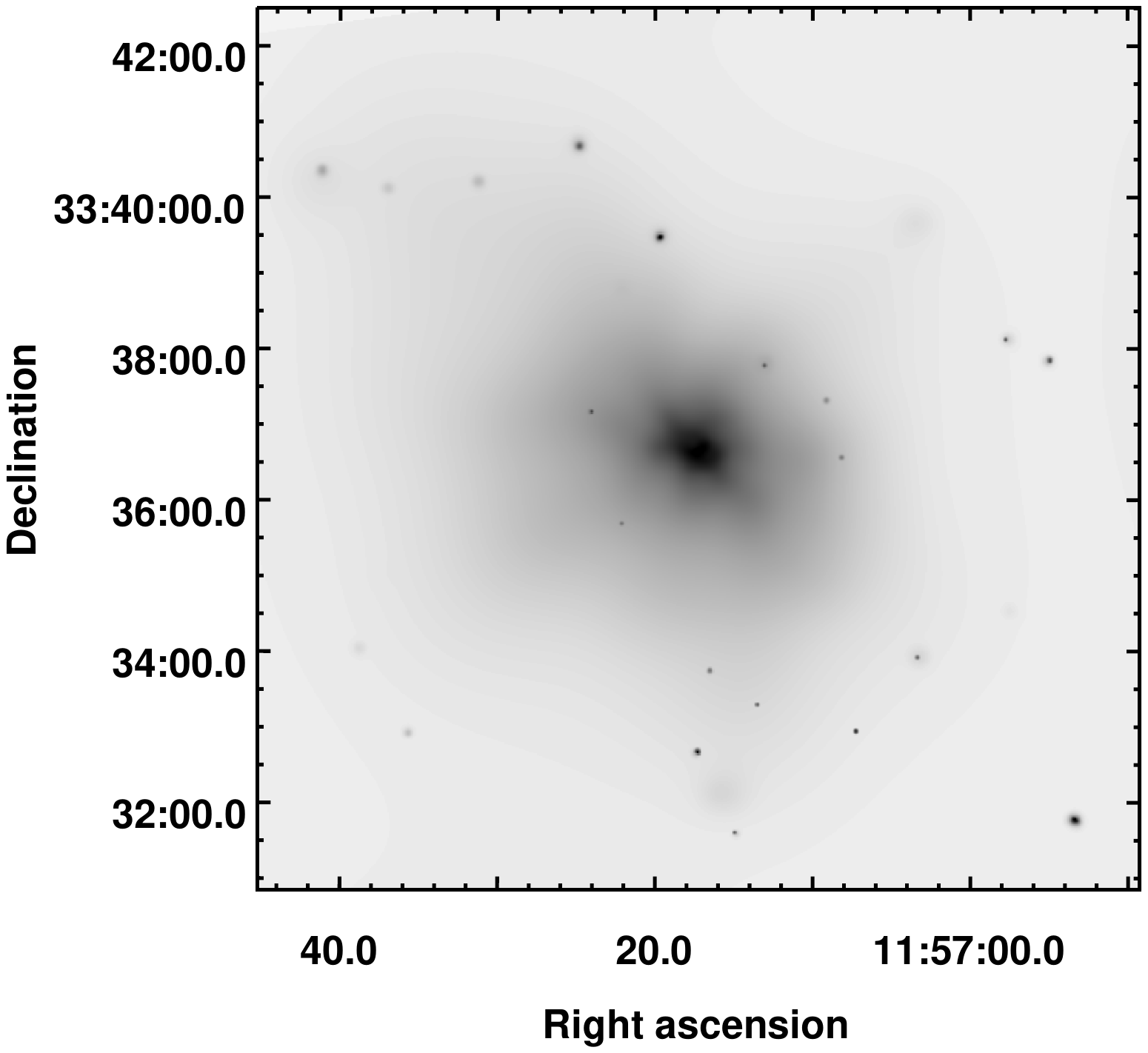}}}  
\put(2.1,-0.21){\scalebox{0.30}{\includegraphics{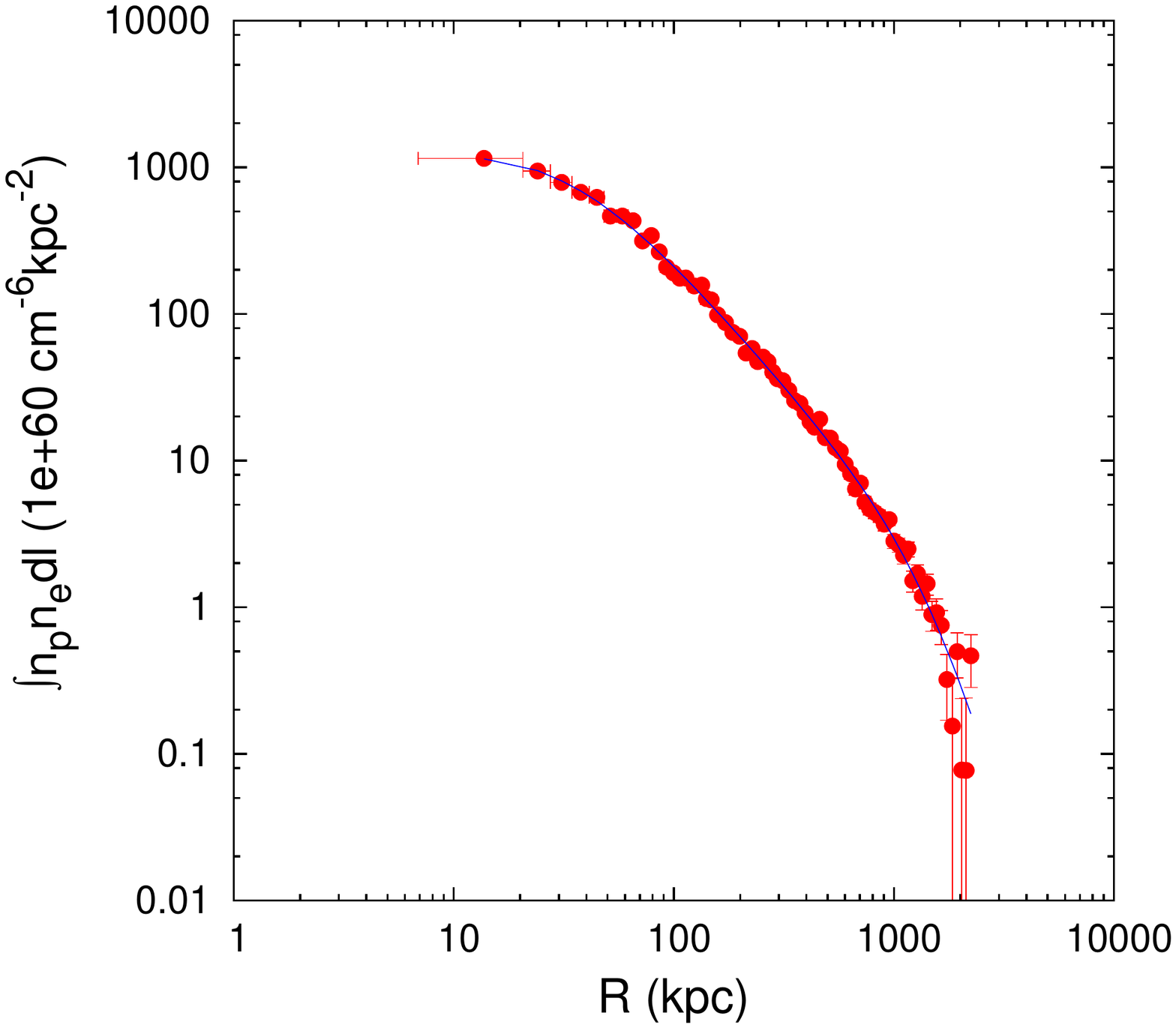}}}
\put(4.2,-0.21){\scalebox{0.30}{\includegraphics{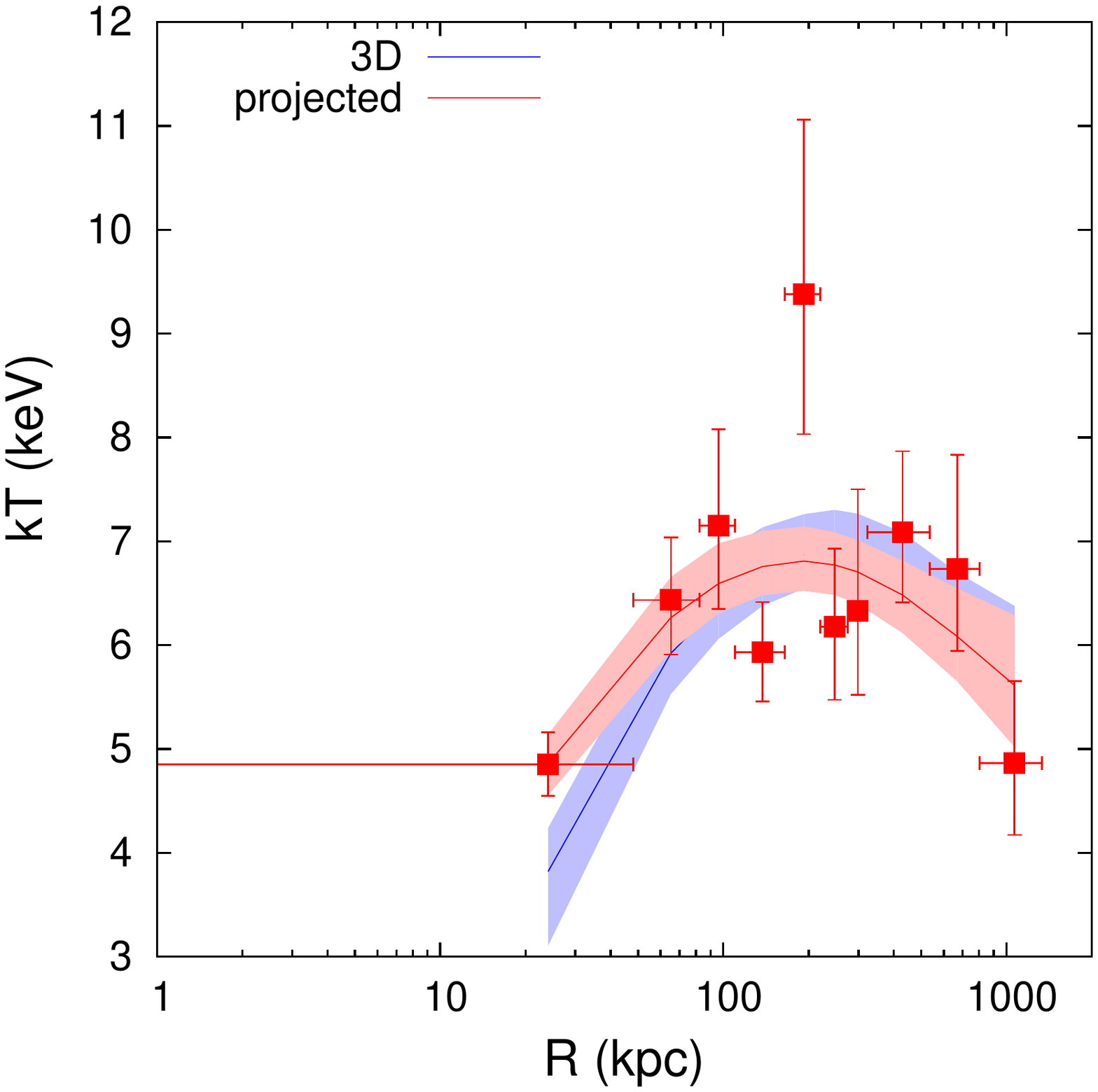}}}
\end{picture}
\end{center}
\caption{\small{Same as Figure~\ref{fig:a2204} but for A1423.}\label{fig:a1423}}
\end{figure*}

\begin{figure*}
\begin{center}
\setlength{\unitlength}{1in}
\begin{picture}(6.9,2.0)
\put(0.01,-0.8){\scalebox{0.34}{\includegraphics[clip=true]{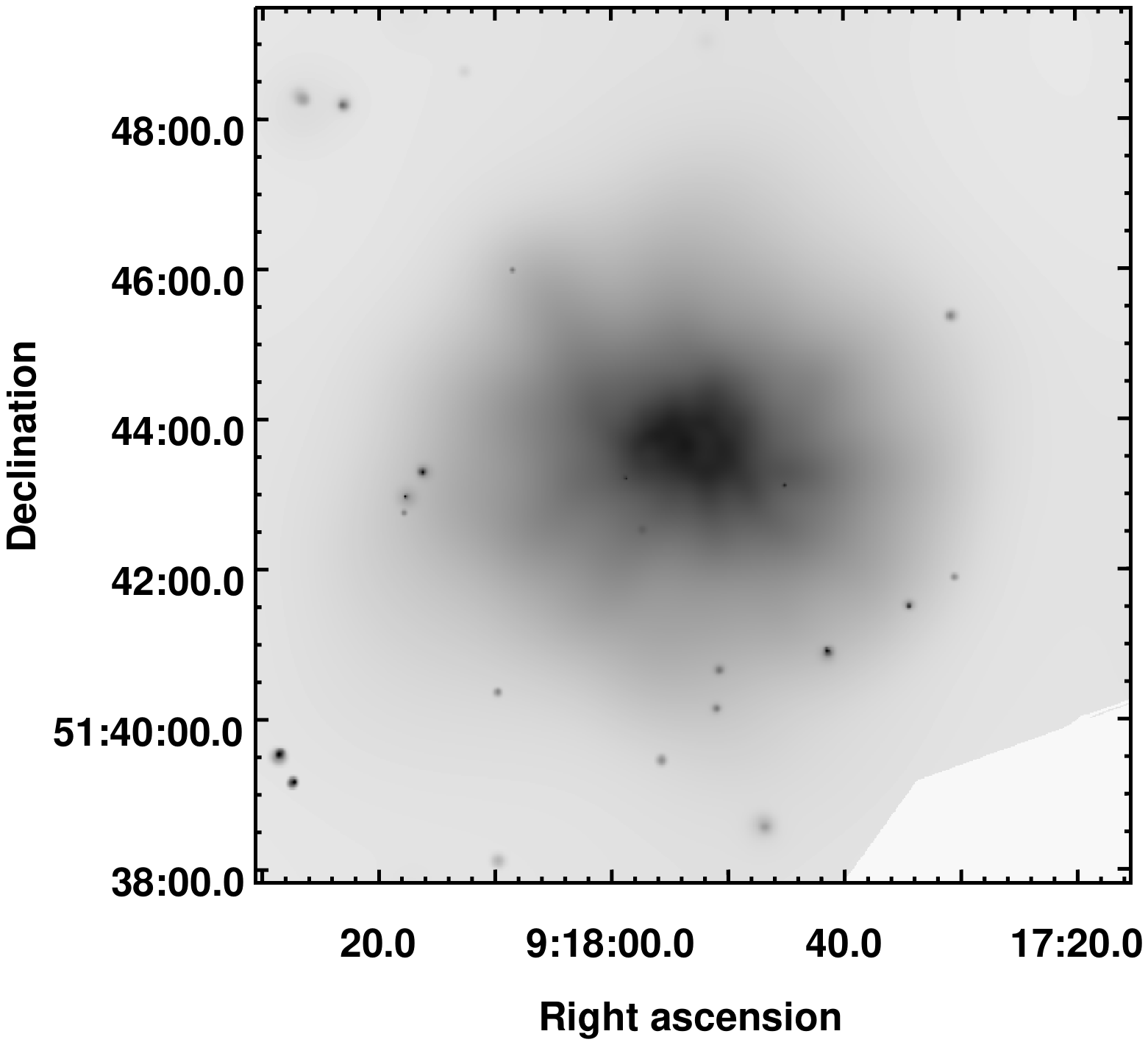}}}  
\put(2.1,-0.21){\scalebox{0.30}{\includegraphics{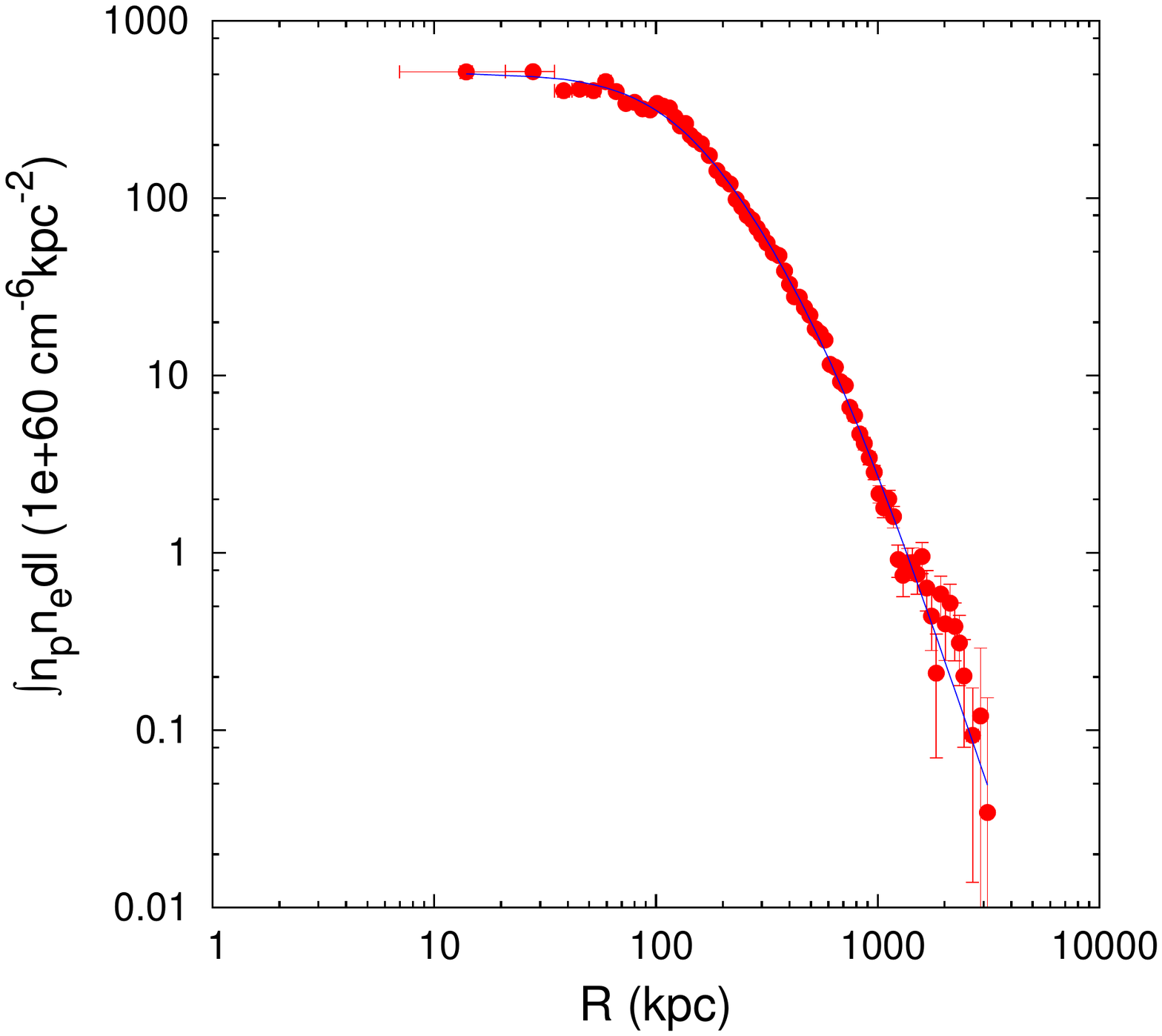}}}
\put(4.2,-0.21){\scalebox{0.30}{\includegraphics{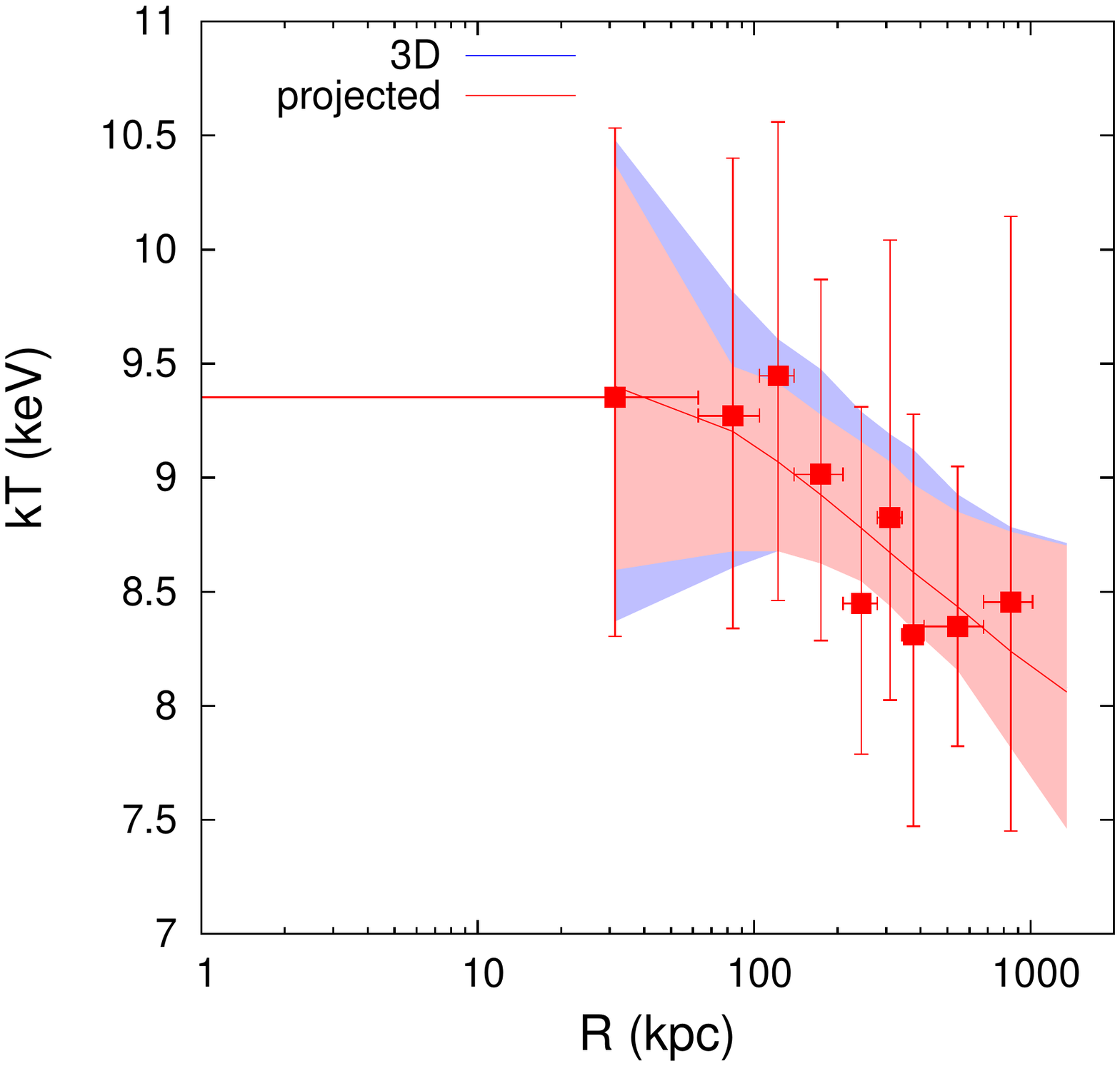}}}
\end{picture}
\end{center}
\caption{\small{Same as Figure~\ref{fig:a2204} but for A773.}\label{fig:a773}}
\end{figure*}

\begin{figure*}
\begin{center}
\setlength{\unitlength}{1in}
\begin{picture}(6.9,2.0)
\put(0.01,-0.8){\scalebox{0.34}{\includegraphics[clip=true]{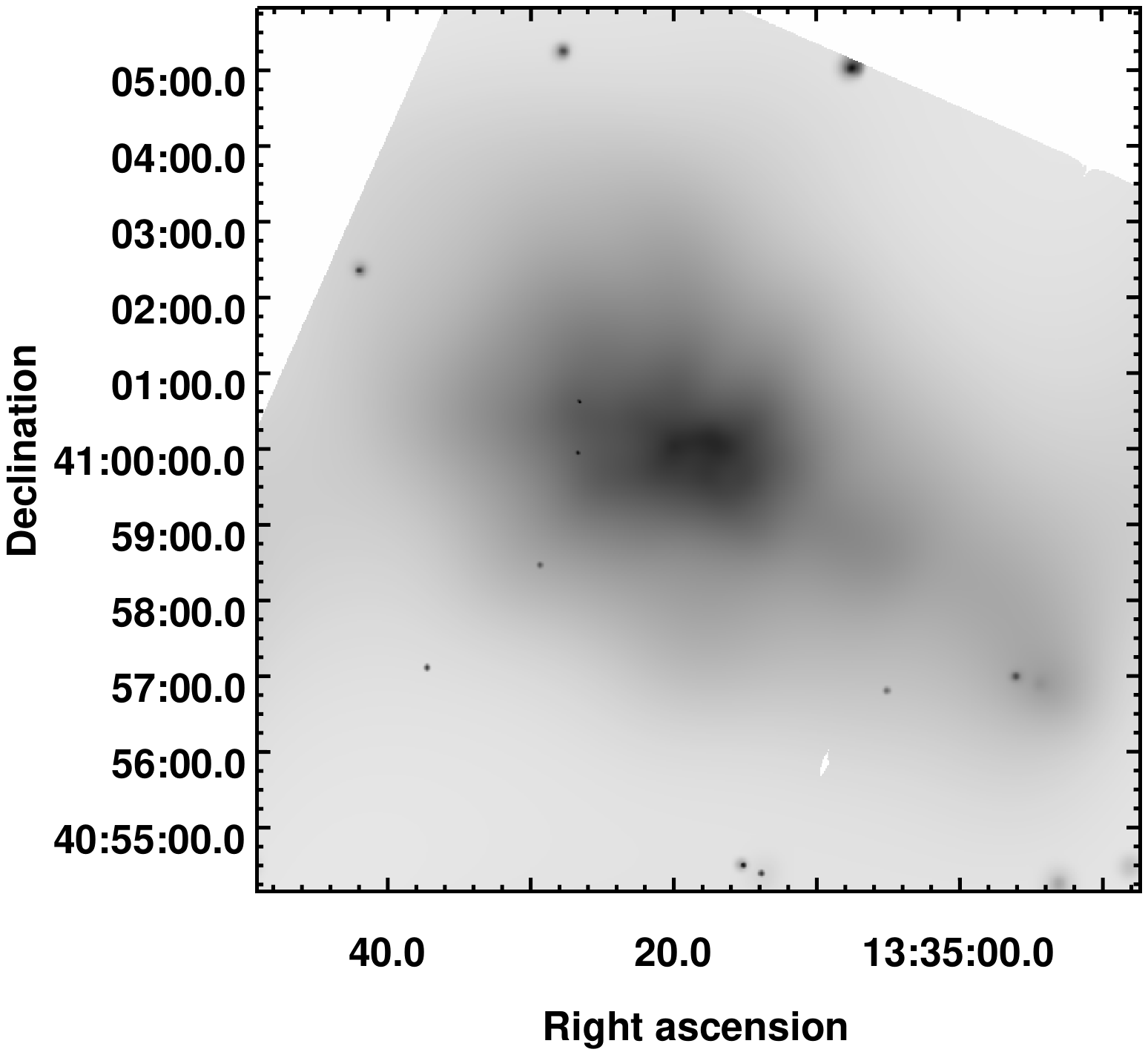}}}  
\put(2.1,-0.21){\scalebox{0.30}{\includegraphics{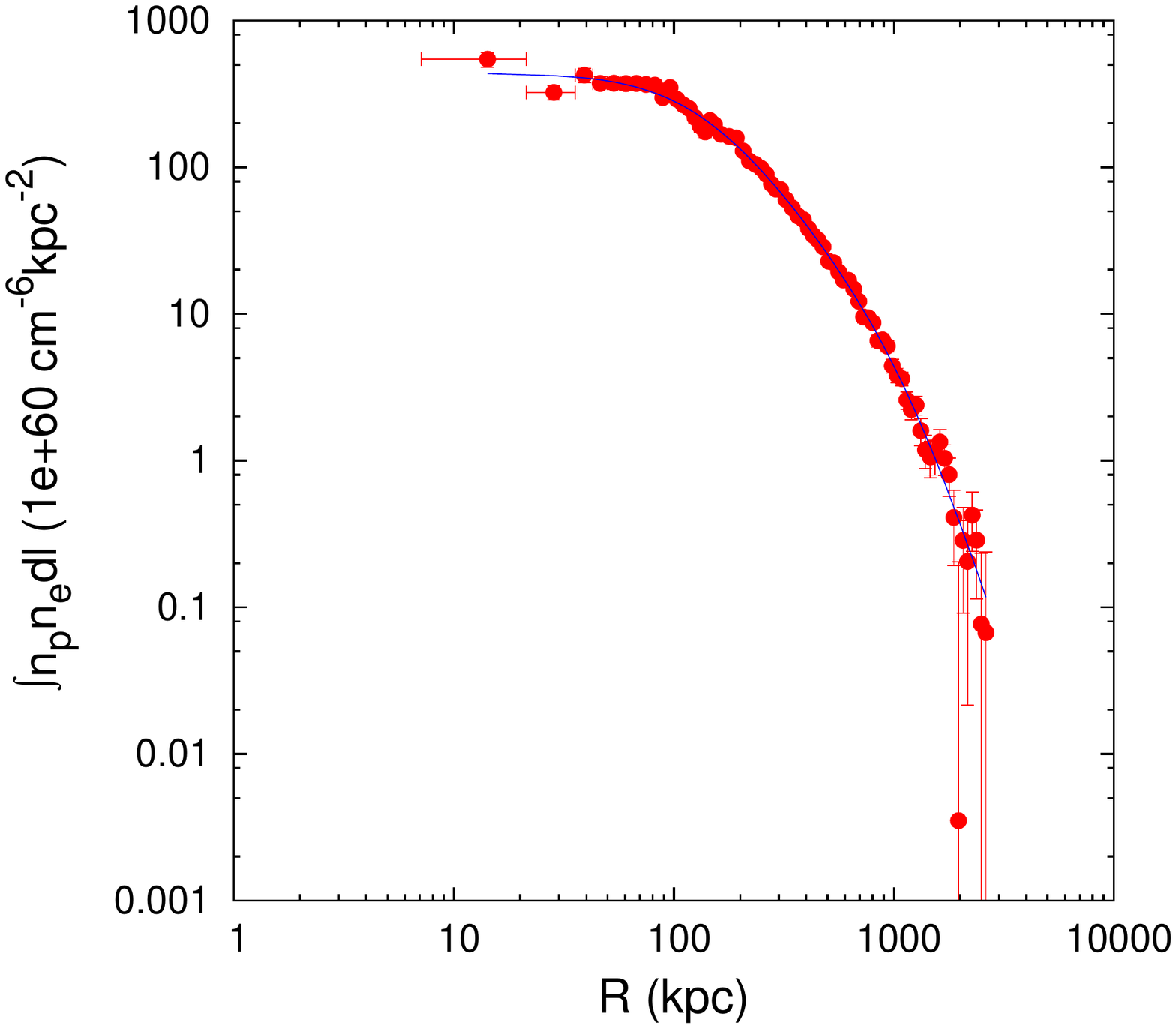}}}
\put(4.2,-0.21){\scalebox{0.30}{\includegraphics{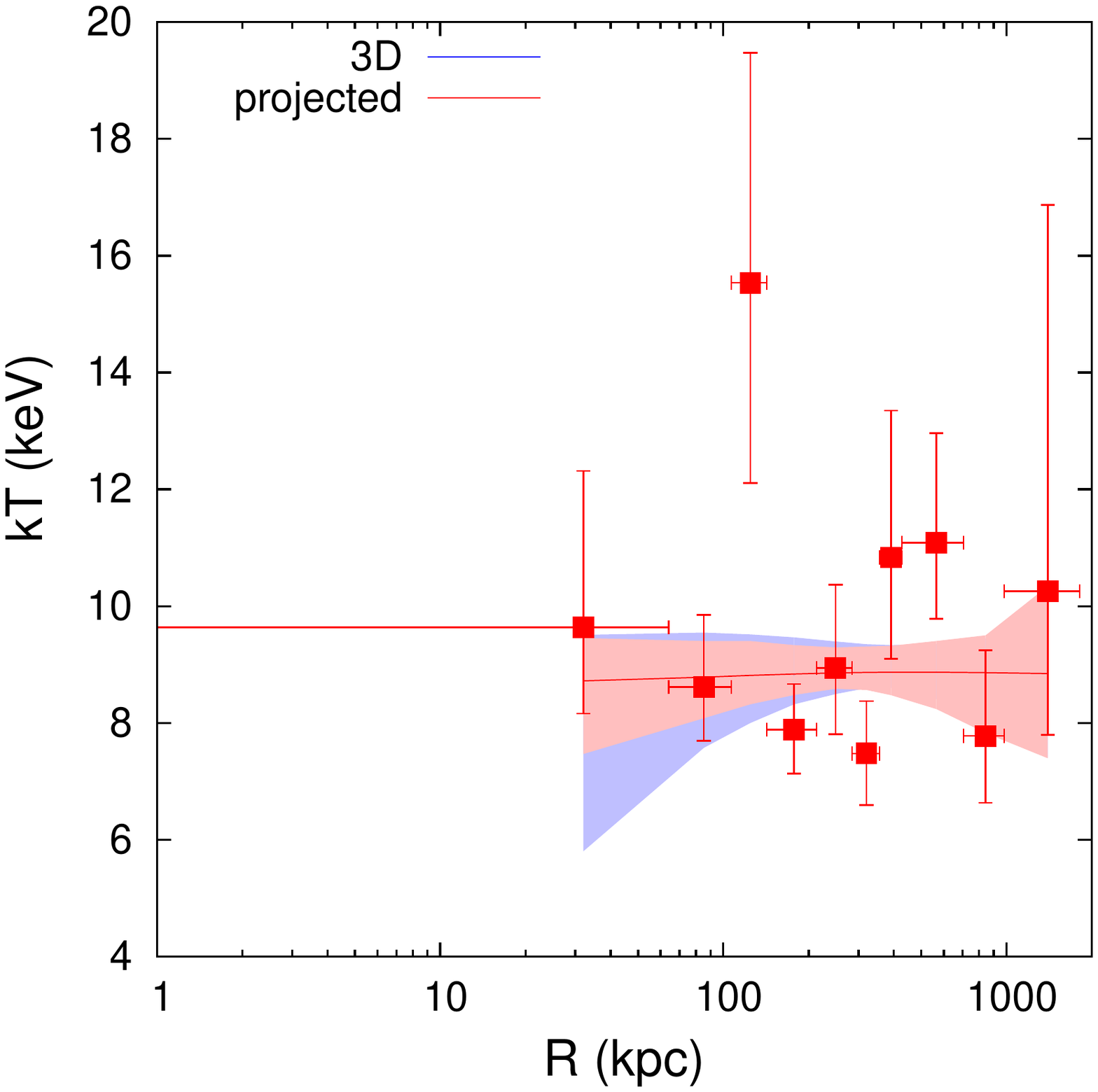}}}
\end{picture}
\end{center}
\caption{\small{Same as Figure~\ref{fig:a2204} but for A1763.}\label{fig:a1763}}
\end{figure*}

\begin{figure*}
\begin{center}
\setlength{\unitlength}{1in}
\begin{picture}(6.9,2.0)
\put(0.01,-0.8){\scalebox{0.34}{\includegraphics[clip=true]{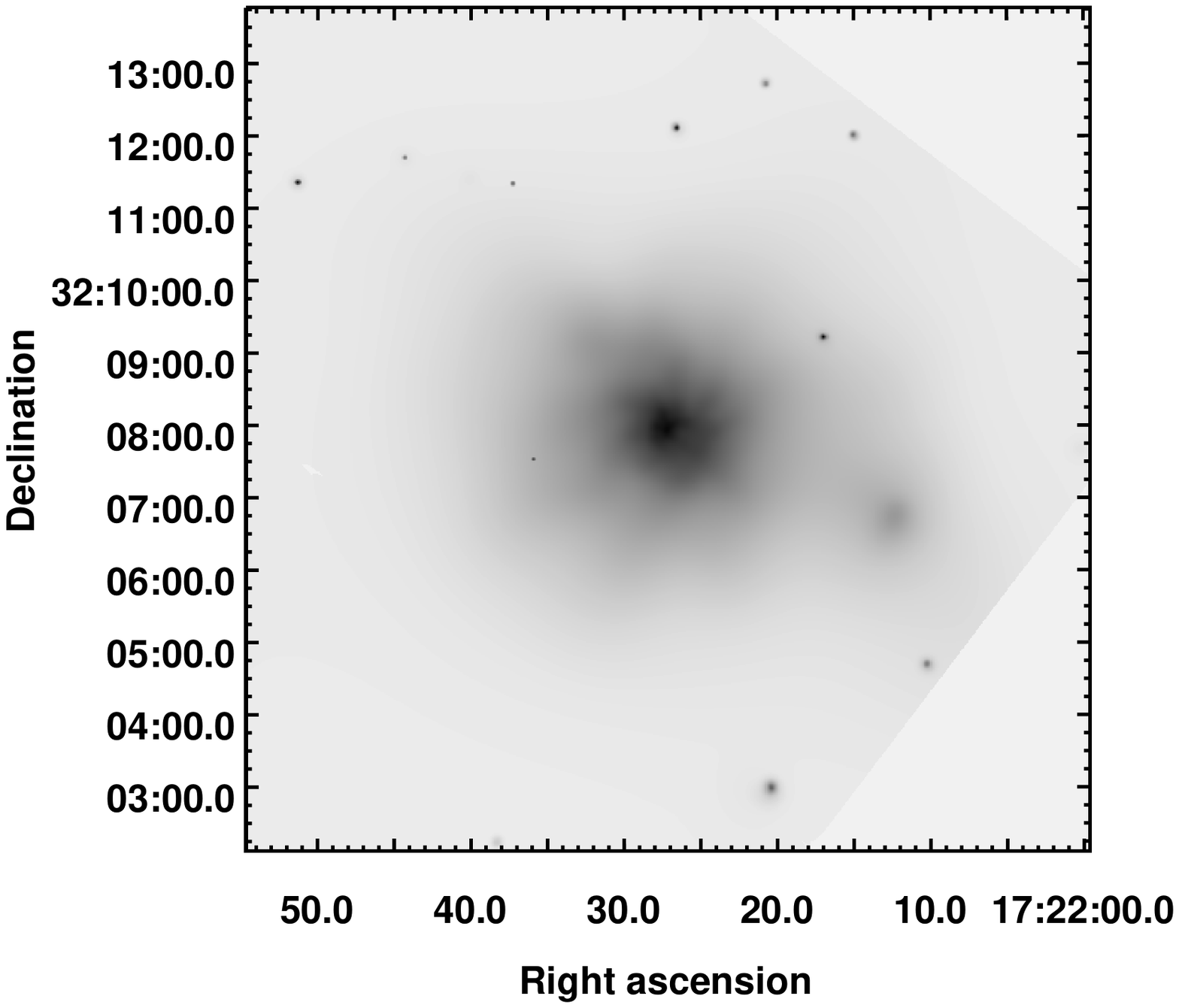}}}  
\put(2.1,-0.21){\scalebox{0.30}{\includegraphics{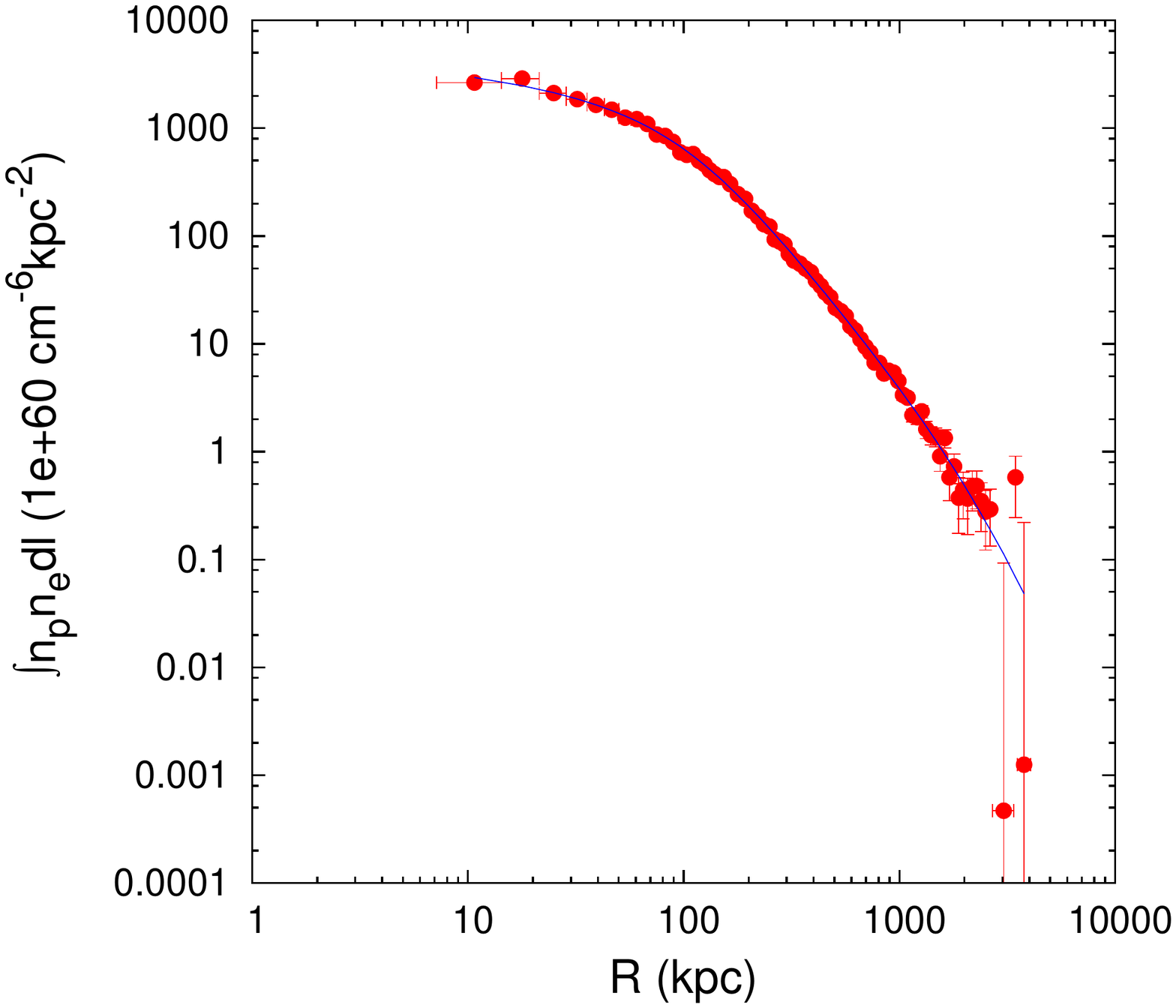}}}
\put(4.2,-0.21){\scalebox{0.30}{\includegraphics{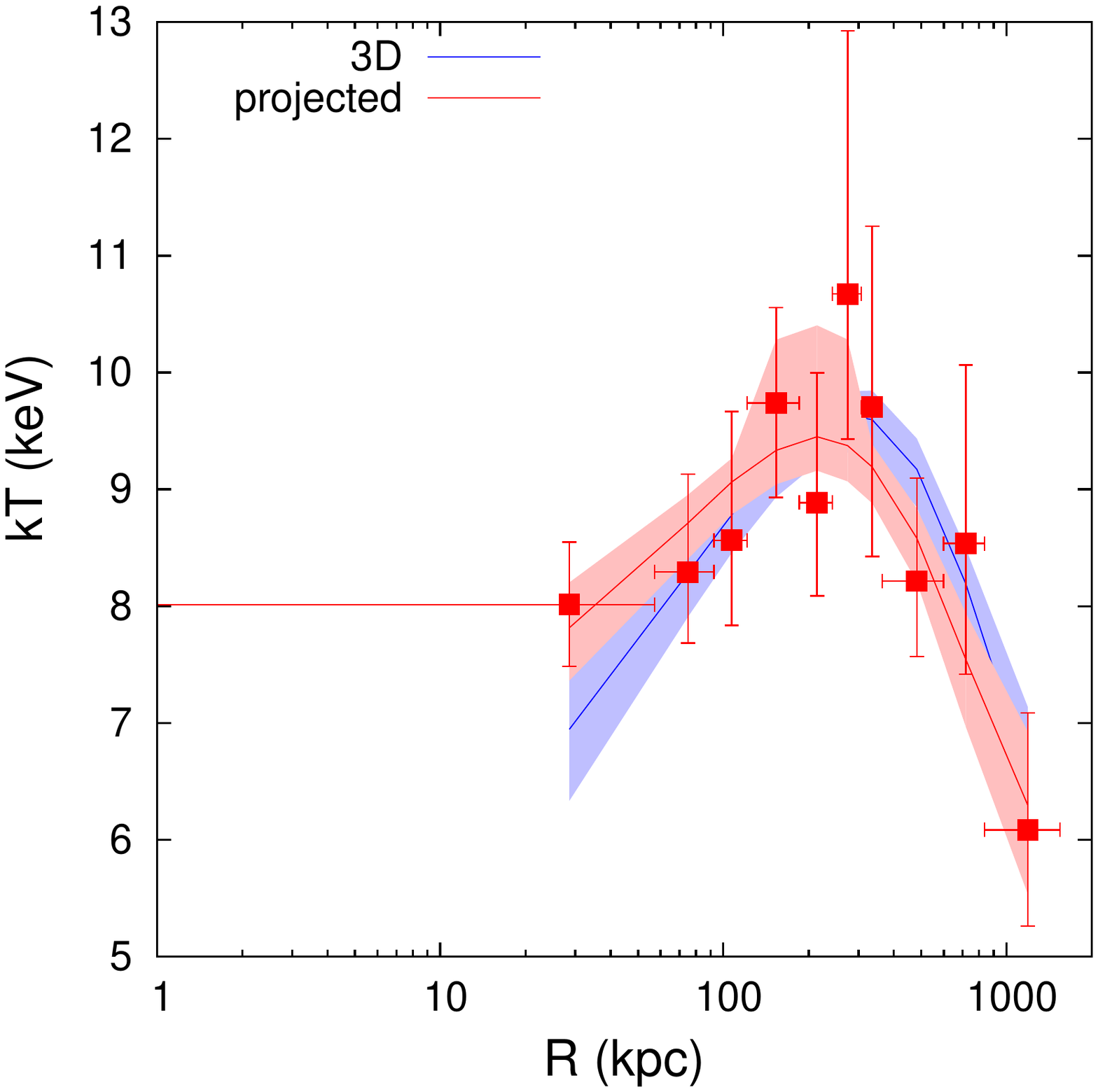}}}
\end{picture}
\end{center}
\caption{\small{Same as Figure~\ref{fig:a2204} but for A2261.}\label{fig:a2261}}
\end{figure*}

\begin{figure*}
\begin{center}
\setlength{\unitlength}{1in}
\begin{picture}(6.9,2.0)
\put(0.01,-0.8){\scalebox{0.34}{\includegraphics[clip=true]{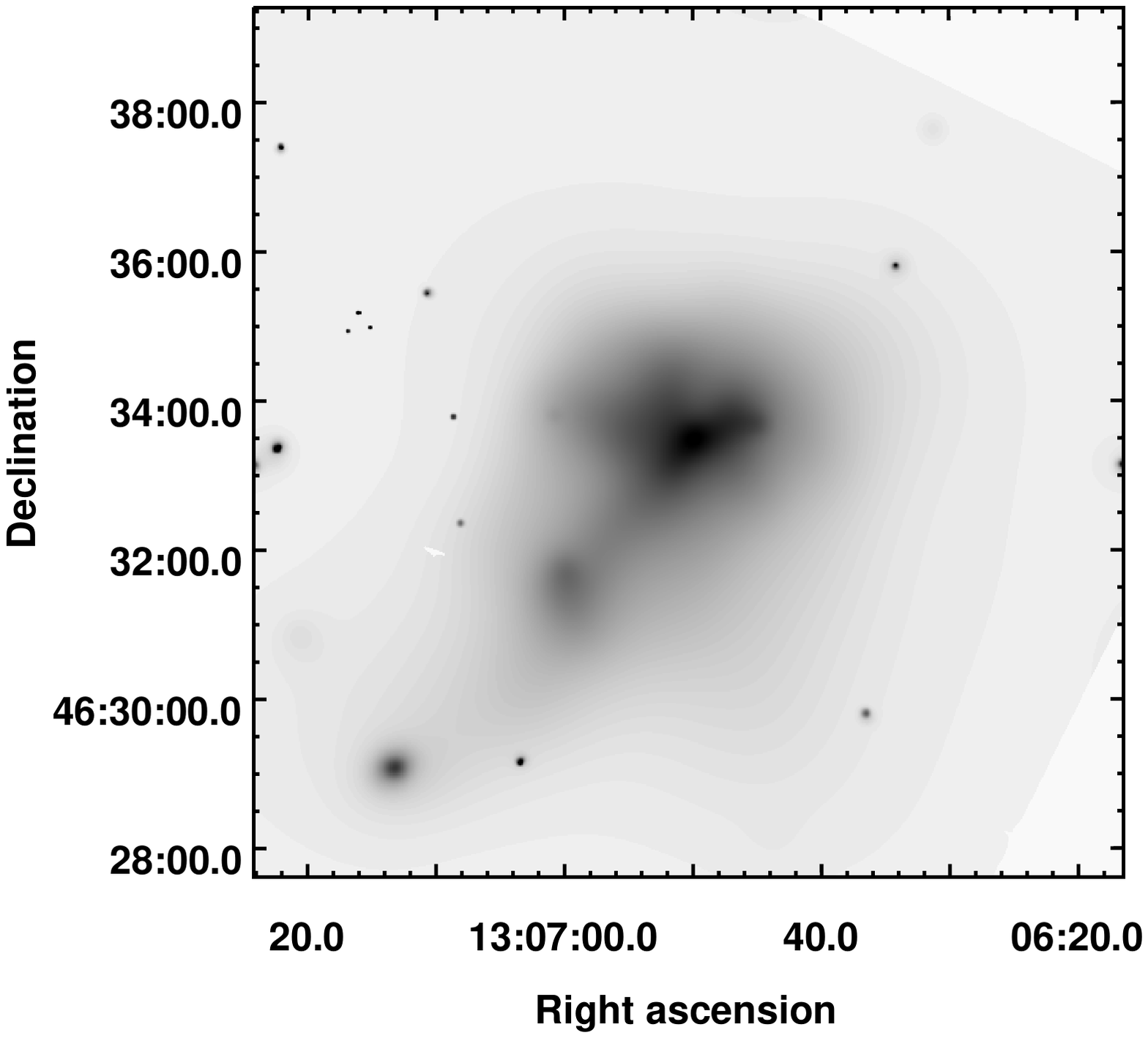}}}  
\put(2.1,-0.21){\scalebox{0.30}{\includegraphics{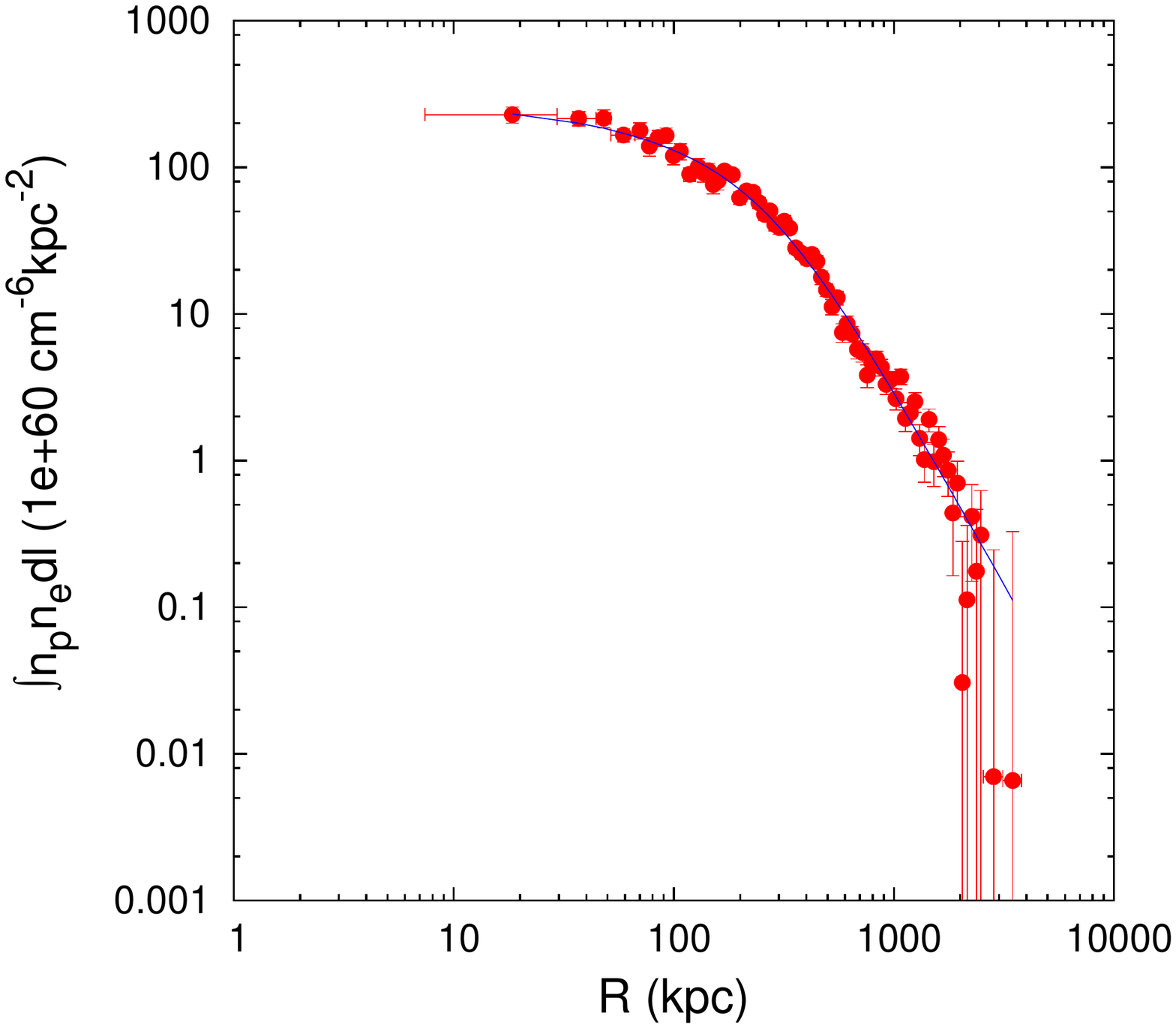}}}
\put(4.2,-0.21){\scalebox{0.30}{\includegraphics{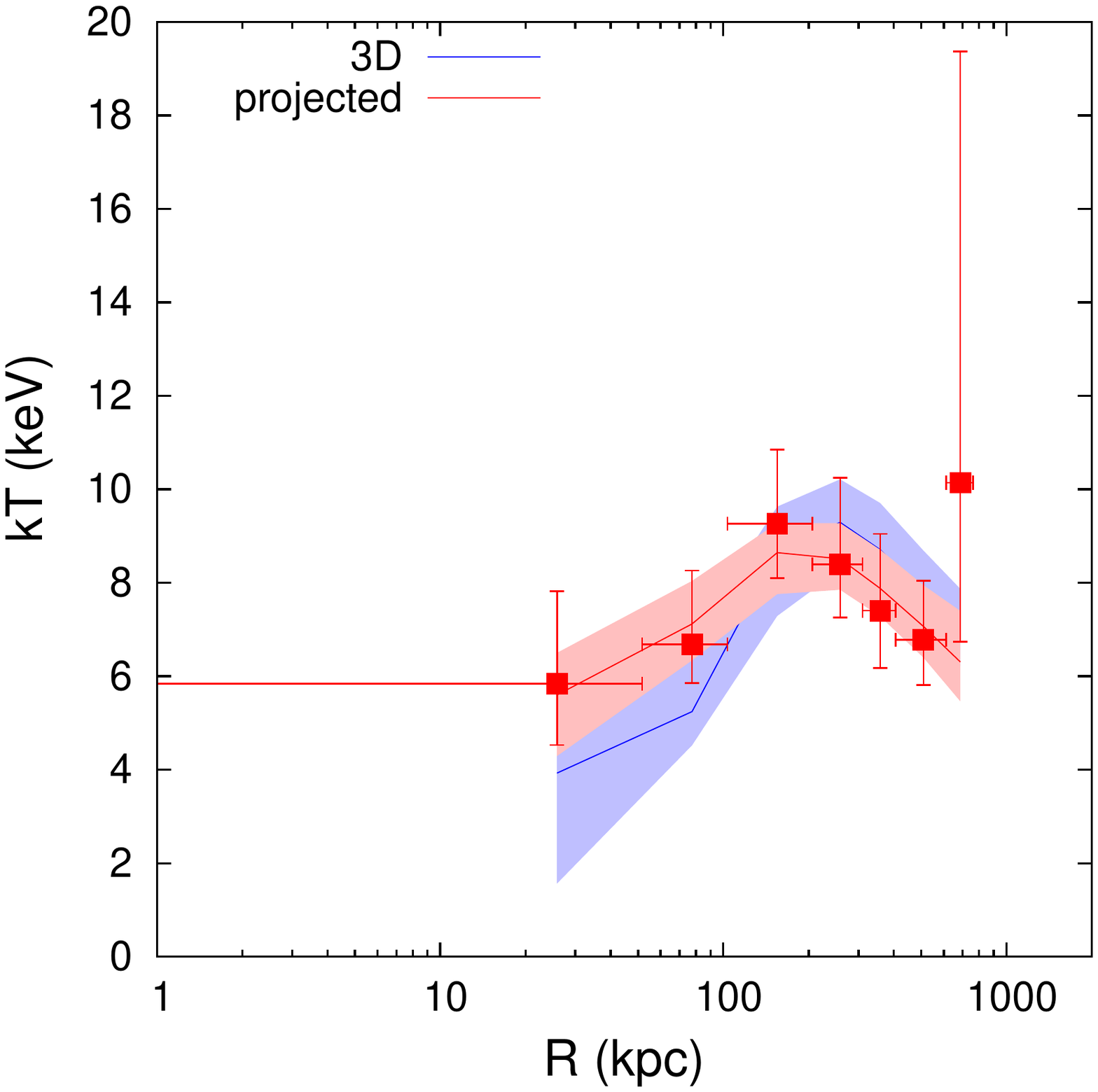}}}
\end{picture}
\end{center}
\caption{\small{Same as Figure~\ref{fig:a2204} but for A1682.}\label{fig:a1682}}
\end{figure*}

\begin{figure*}
\begin{center}
\setlength{\unitlength}{1in}
\begin{picture}(6.9,2.0)
\put(0.01,-0.8){\scalebox{0.34}{\includegraphics[clip=true]{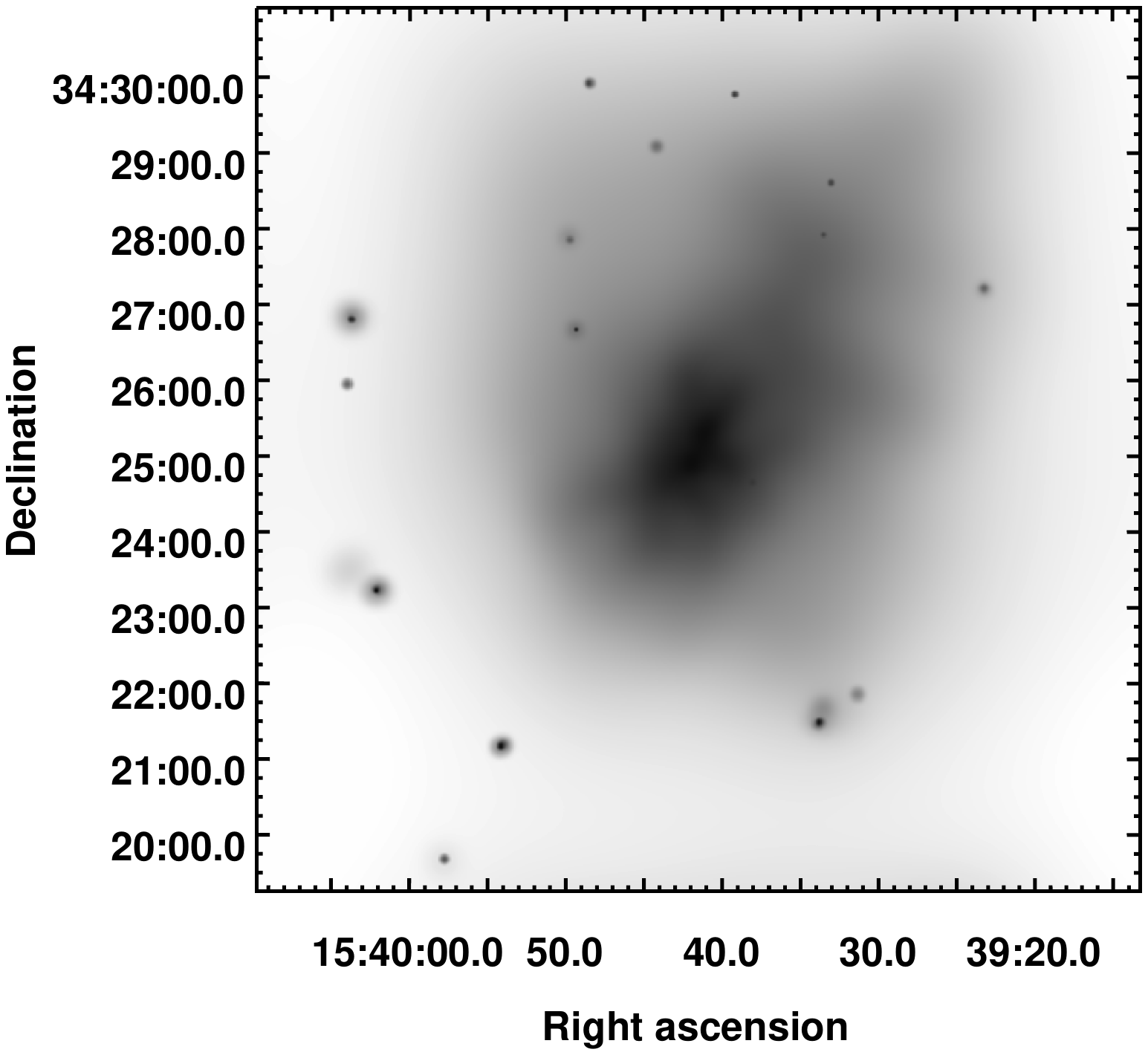}}}  
\put(2.1,-0.21){\scalebox{0.30}{\includegraphics{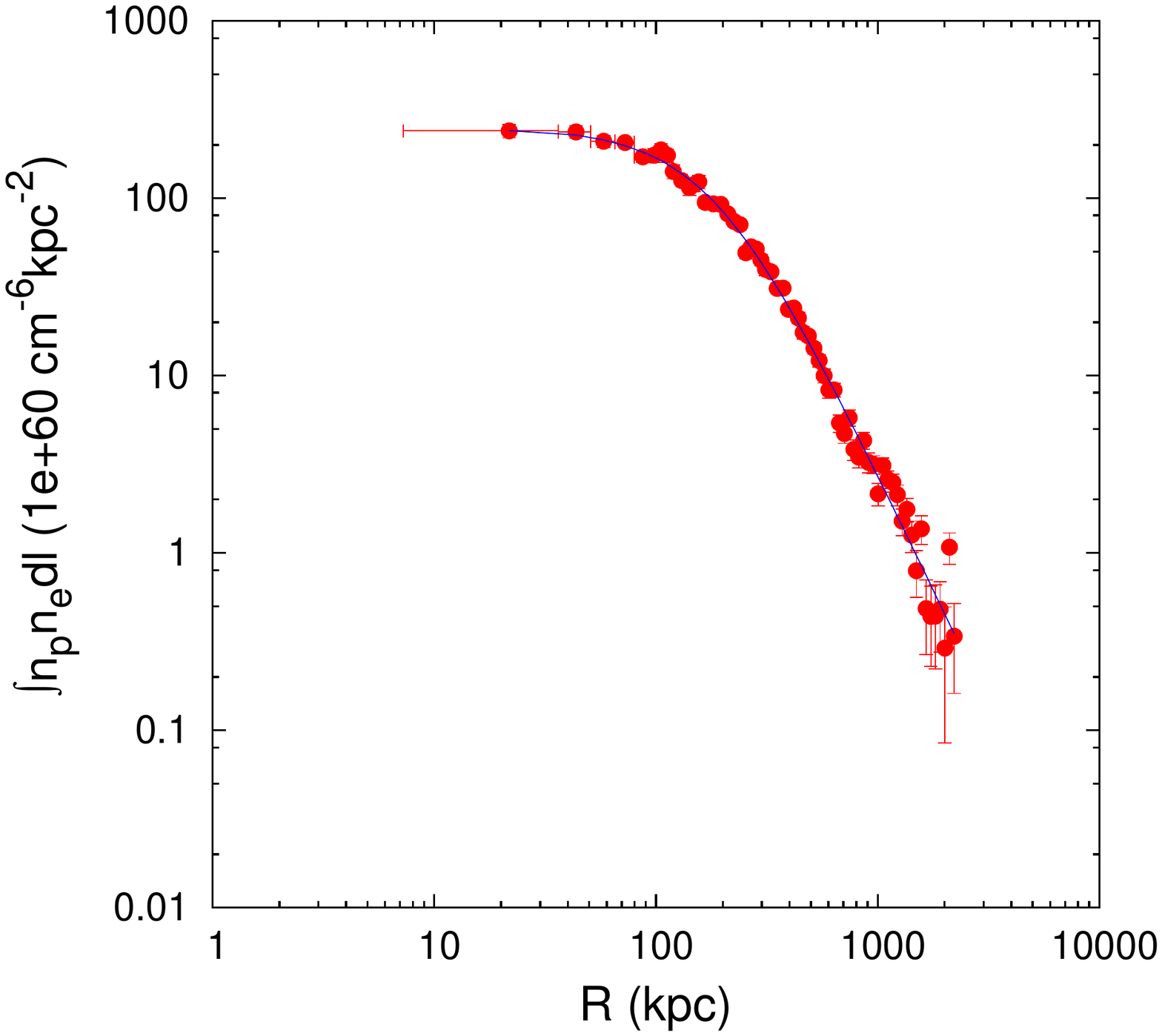}}}
\put(4.2,-0.21){\scalebox{0.30}{\includegraphics{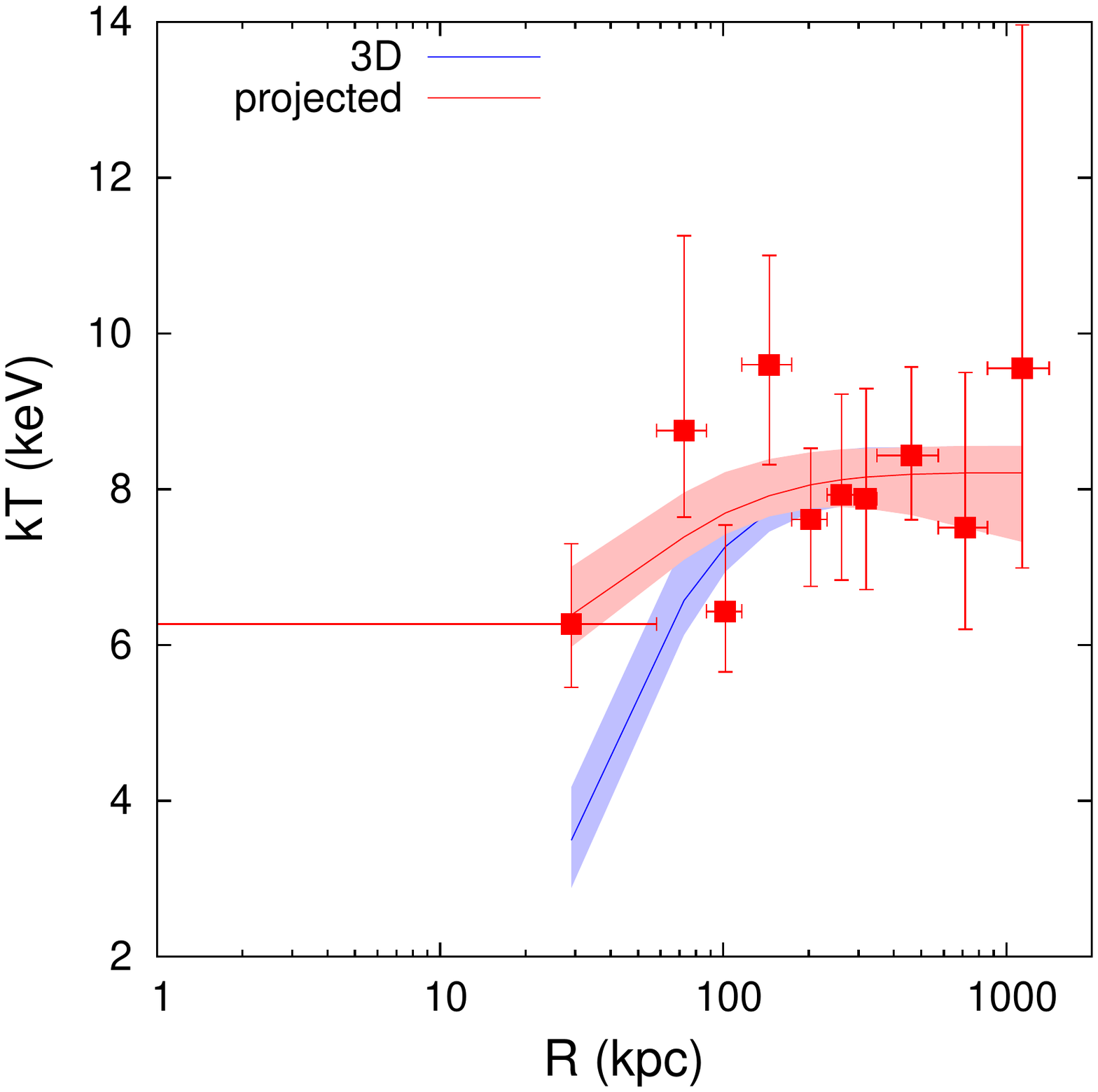}}}
\end{picture}
\end{center}
\caption{\small{Same as Figure~\ref{fig:a2204} but for A2111.}\label{fig:a2111}}
\end{figure*}

\begin{figure*}
\begin{center}
\setlength{\unitlength}{1in}
\begin{picture}(6.9,2.0)
\put(0.01,-0.8){\scalebox{0.34}{\includegraphics[clip=true]{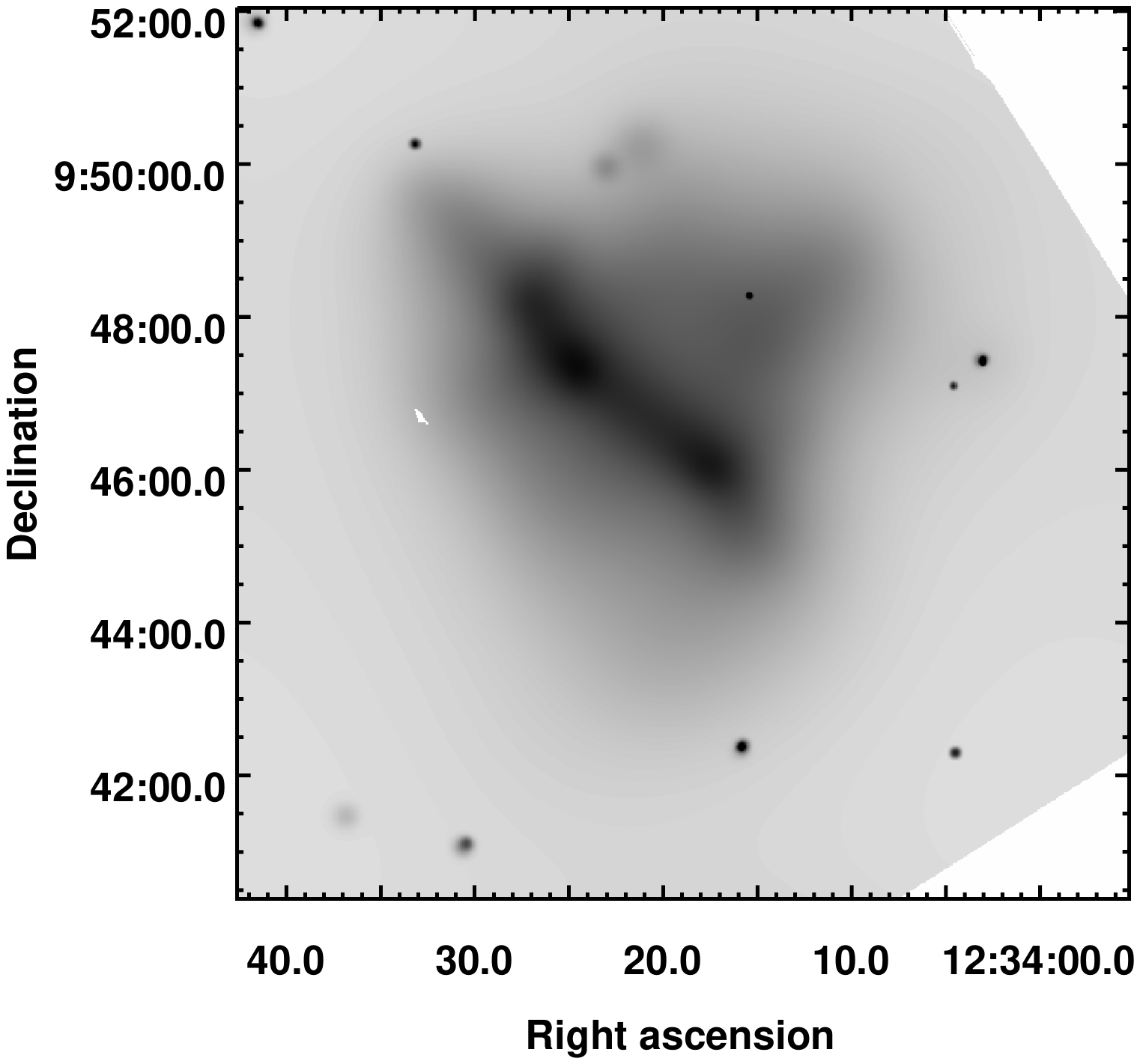}}}  
\put(2.1,-0.21){\scalebox{0.30}{\includegraphics{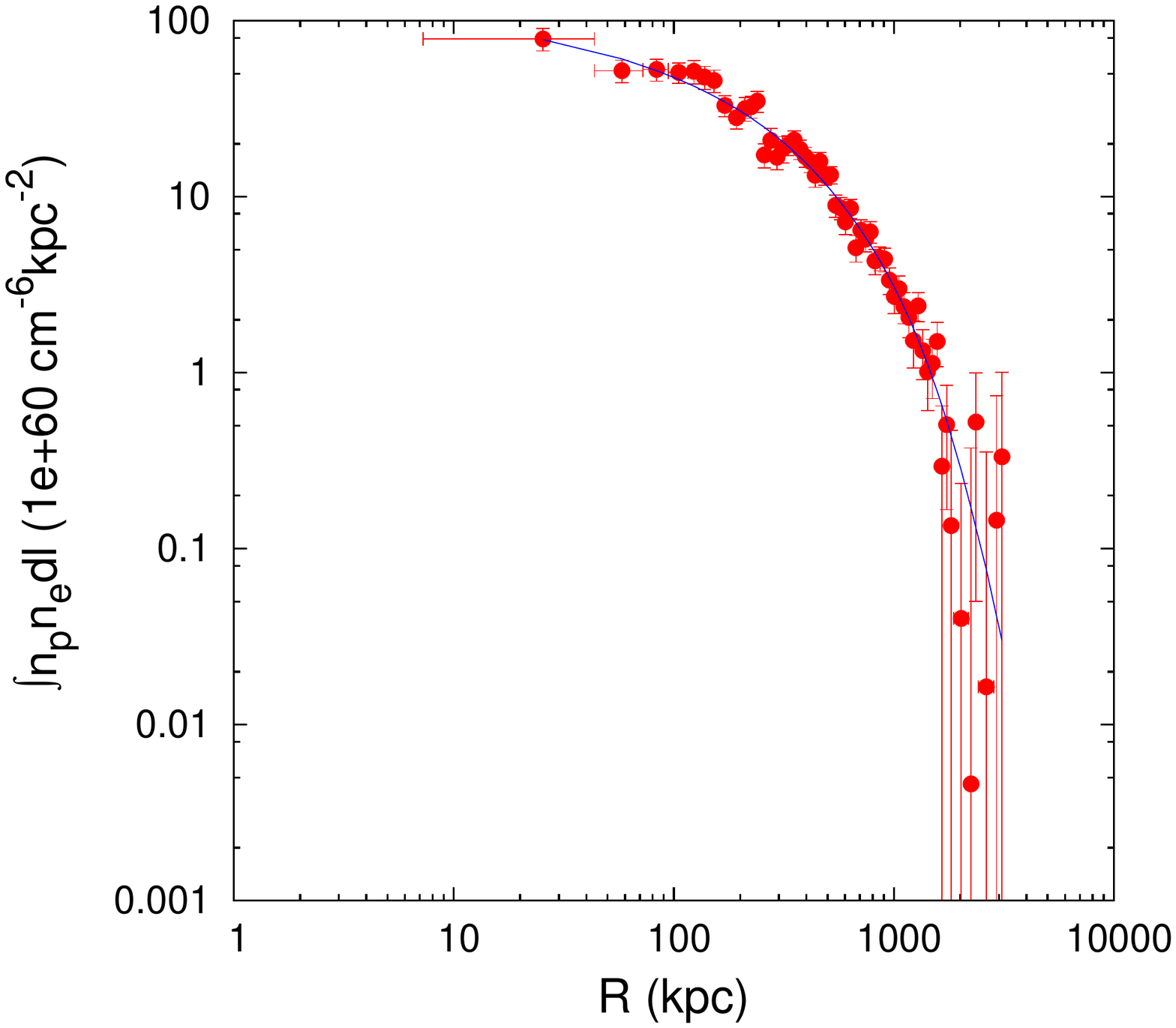}}}
\put(4.2,-0.21){\scalebox{0.30}{\includegraphics{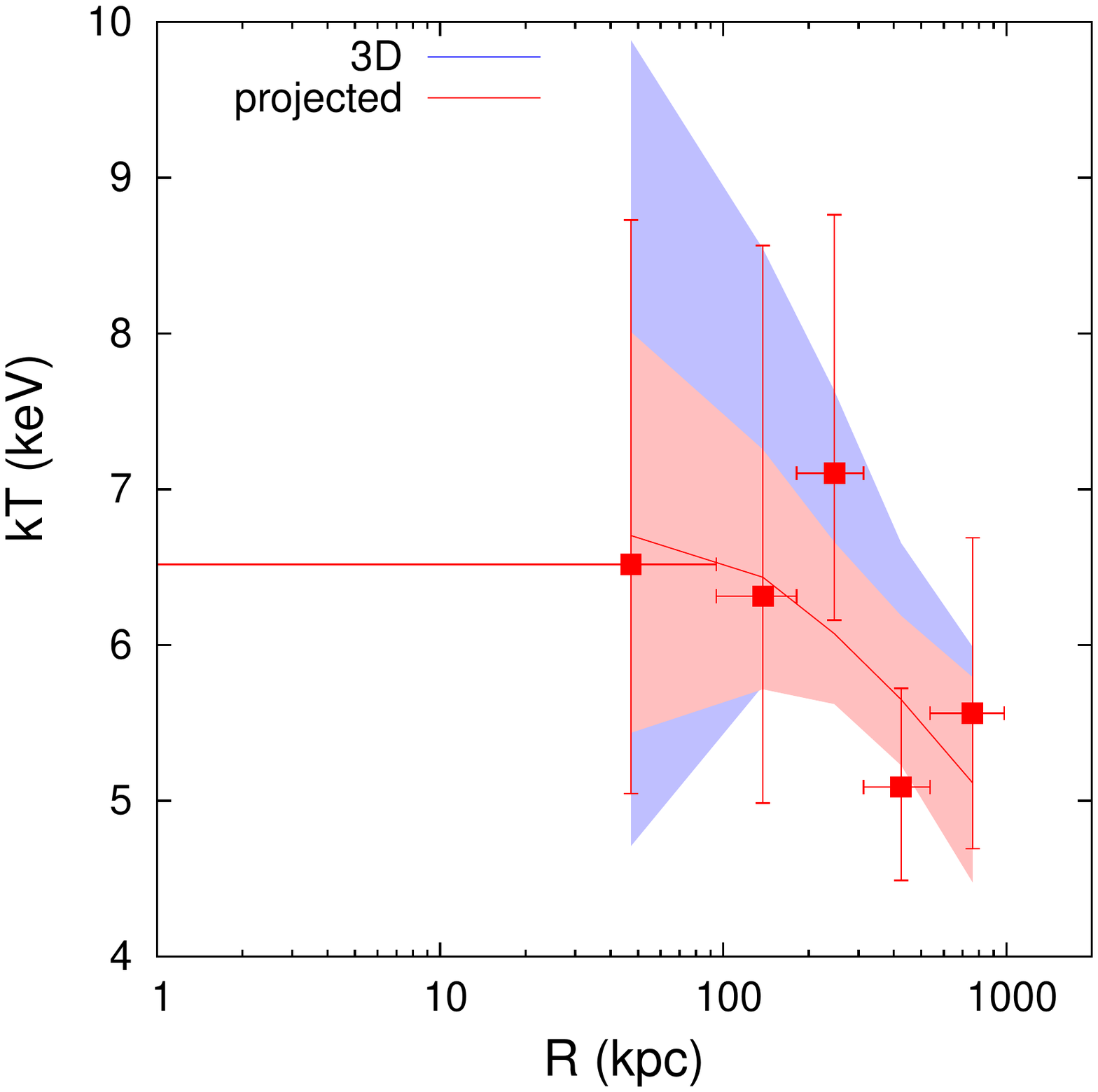}}}
\end{picture}
\end{center}
\caption{\small{Same as Figure~\ref{fig:a2204} but for Z5247.}\label{fig:z5247}}
\end{figure*}

\begin{figure*}
\begin{center}
\setlength{\unitlength}{1in}
\begin{picture}(6.9,2.0)
\put(0.01,-0.8){\scalebox{0.34}{\includegraphics[clip=true]{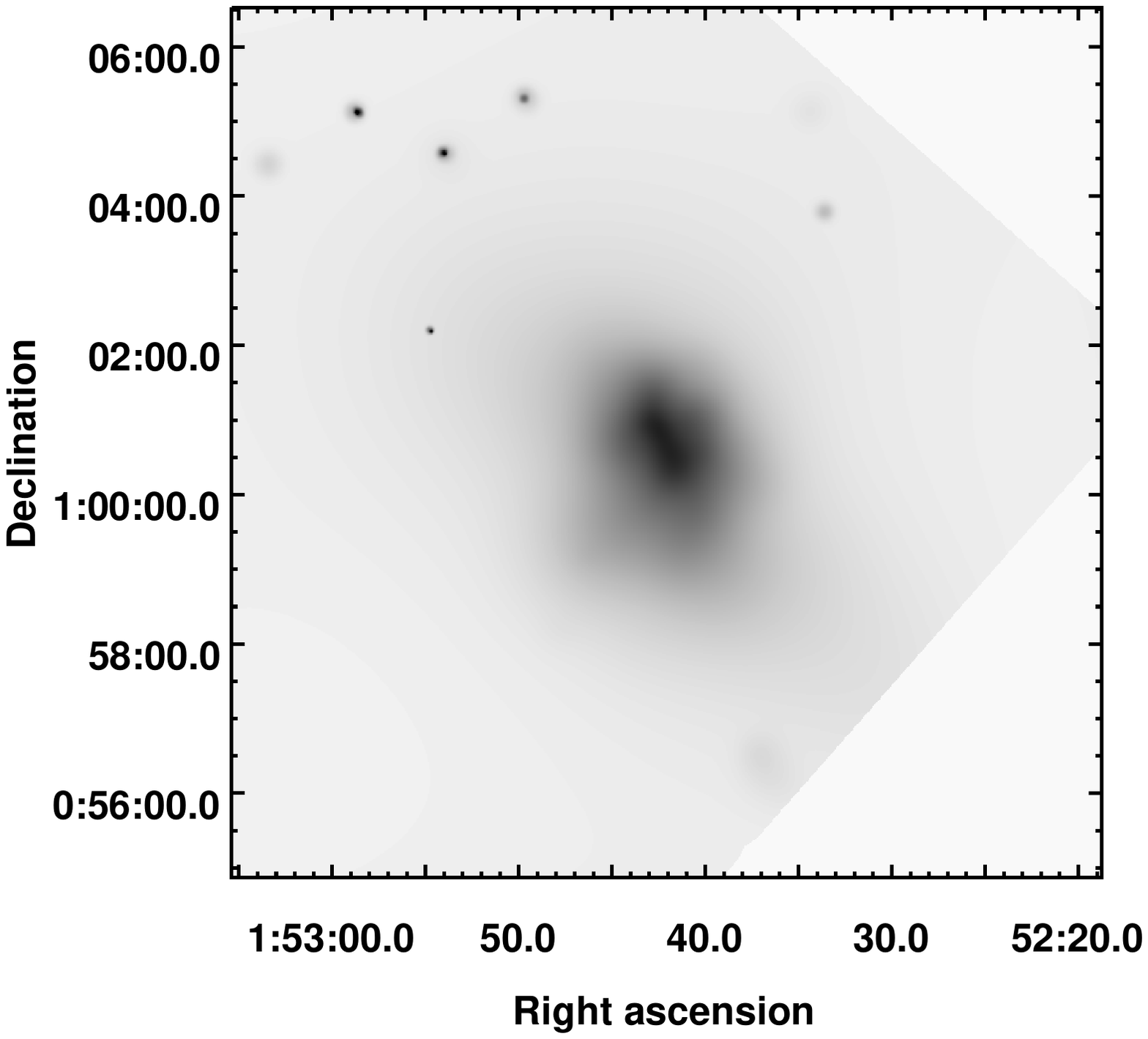}}}  
\put(2.1,-0.21){\scalebox{0.30}{\includegraphics{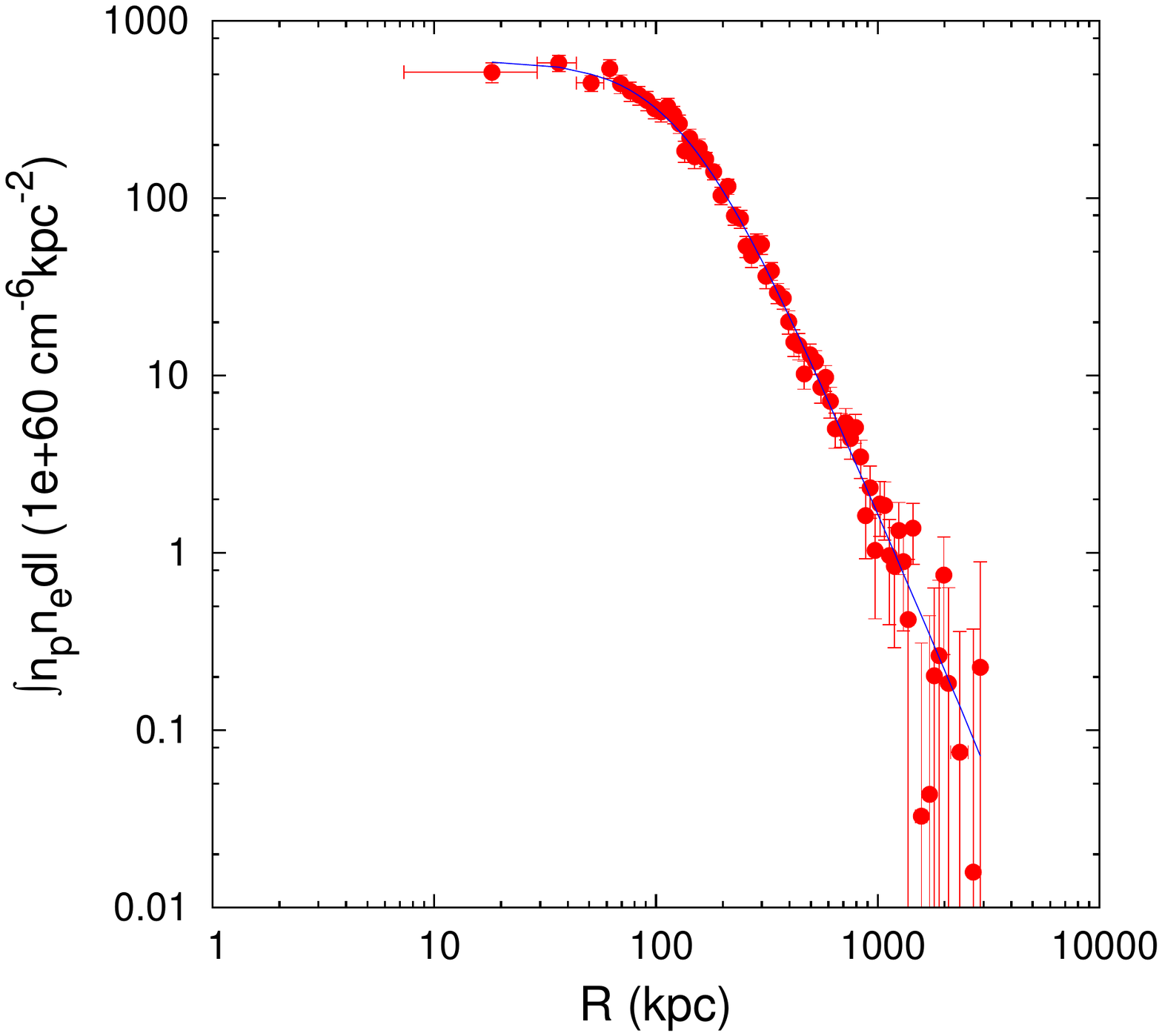}}}
\put(4.2,-0.21){\scalebox{0.30}{\includegraphics{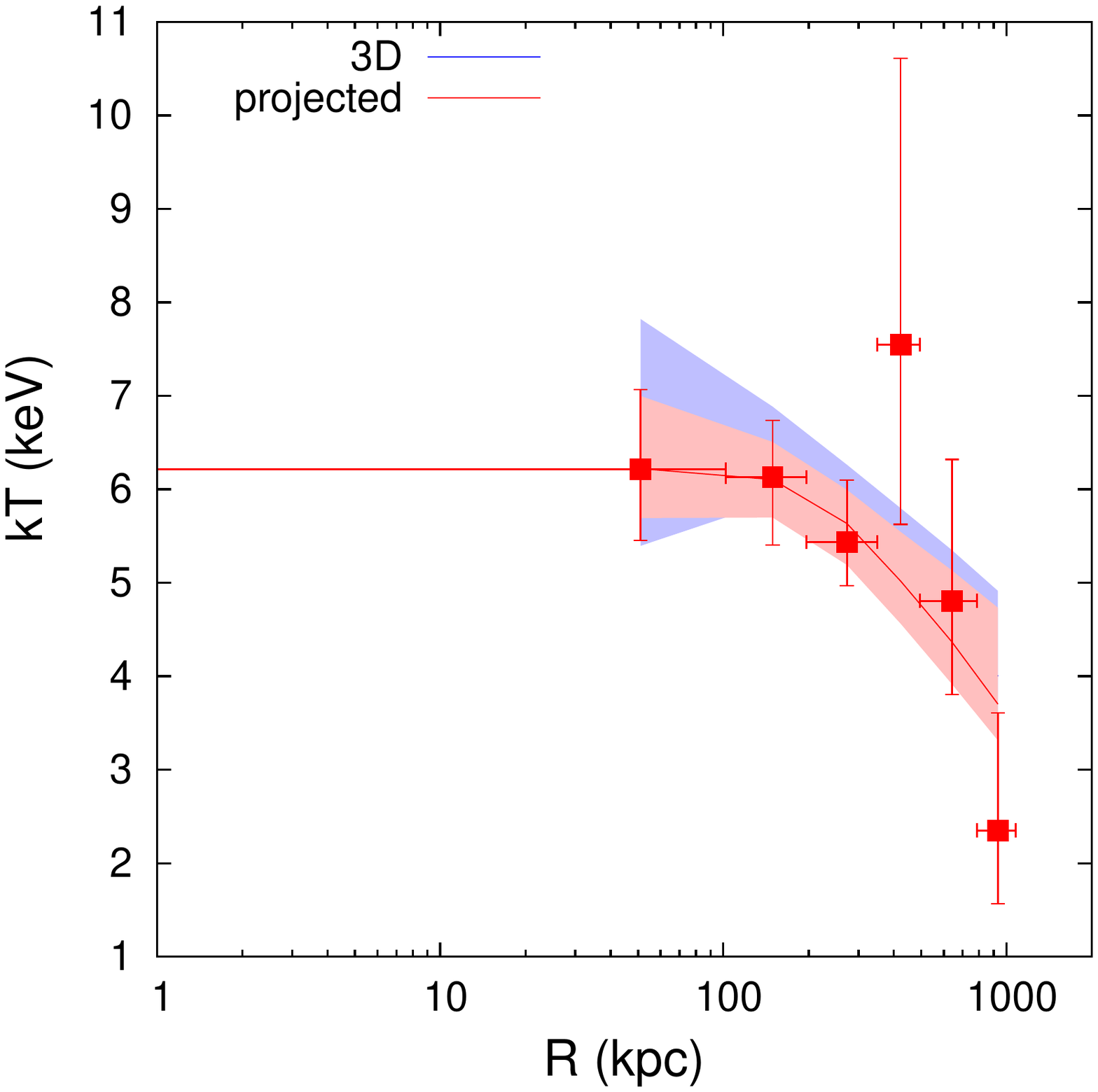}}}
\end{picture}
\end{center}
\caption{\small{Same as Figure~\ref{fig:a2204} but for A267.}\label{fig:a267}}
\end{figure*}

\begin{figure*}
\begin{center}
\setlength{\unitlength}{1in}
\begin{picture}(6.9,2.0)
\put(0.01,-0.8){\scalebox{0.34}{\includegraphics[clip=true]{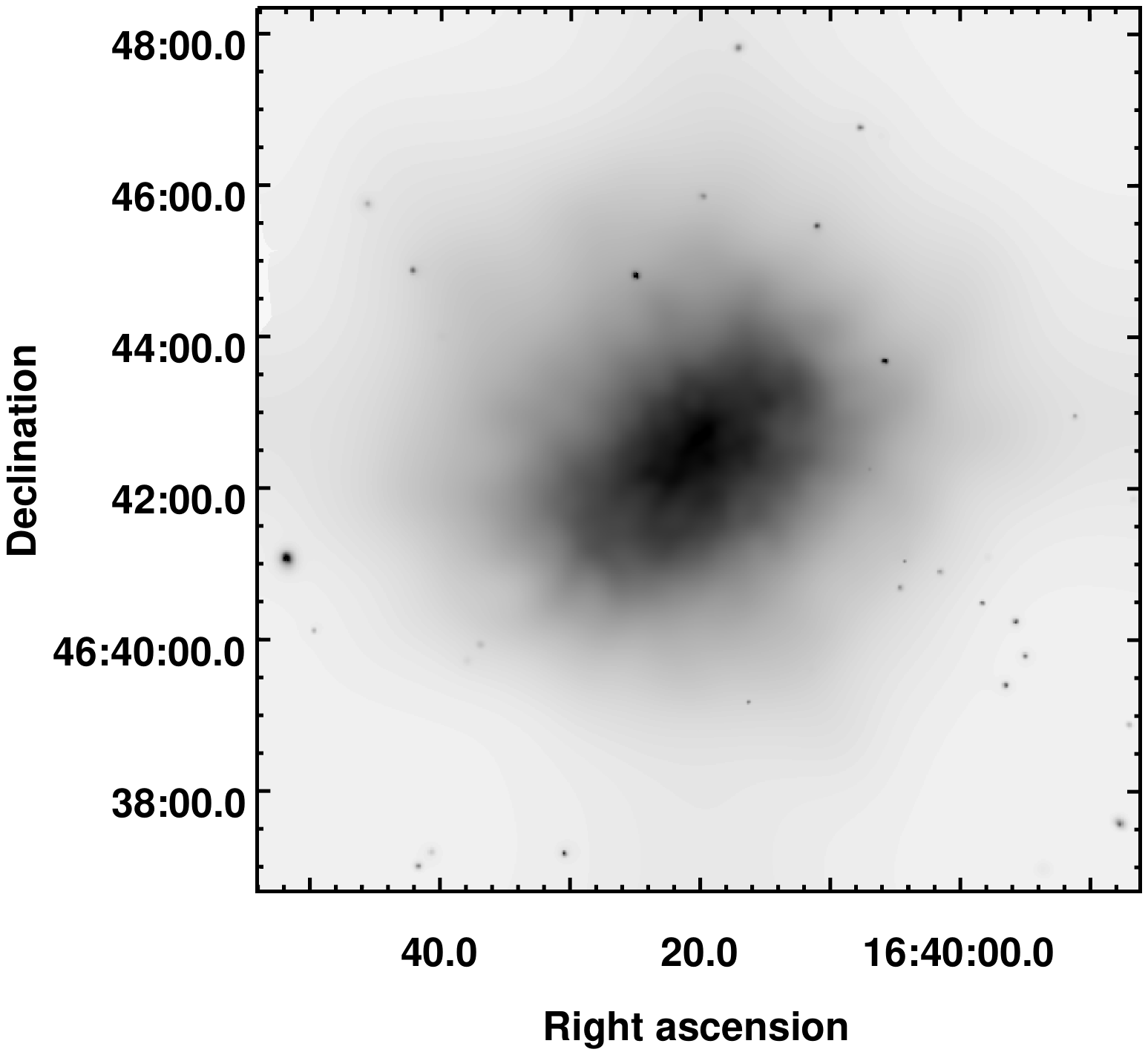}}}  
\put(2.1,-0.21){\scalebox{0.30}{\includegraphics{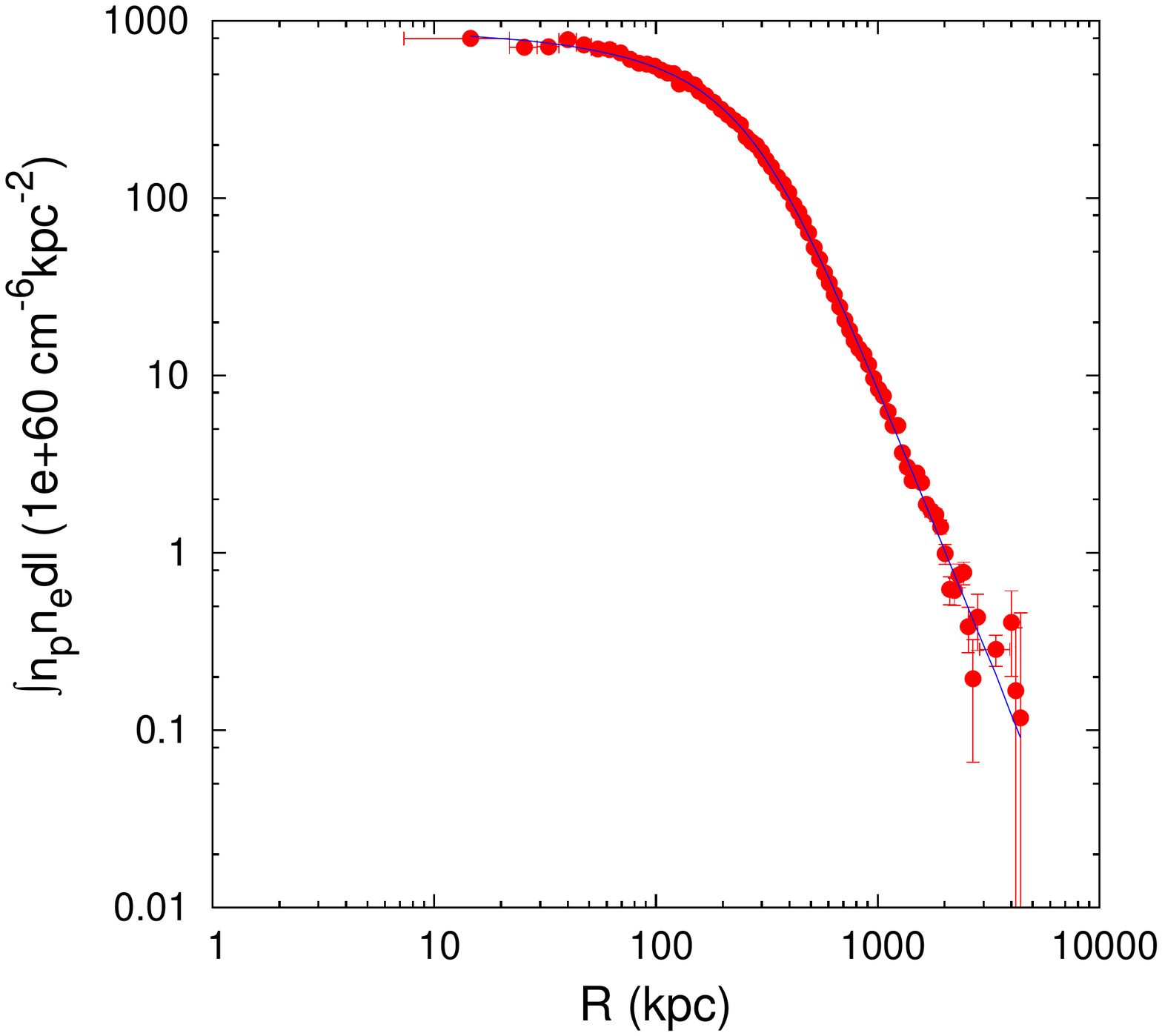}}}
\put(4.2,-0.21){\scalebox{0.30}{\includegraphics{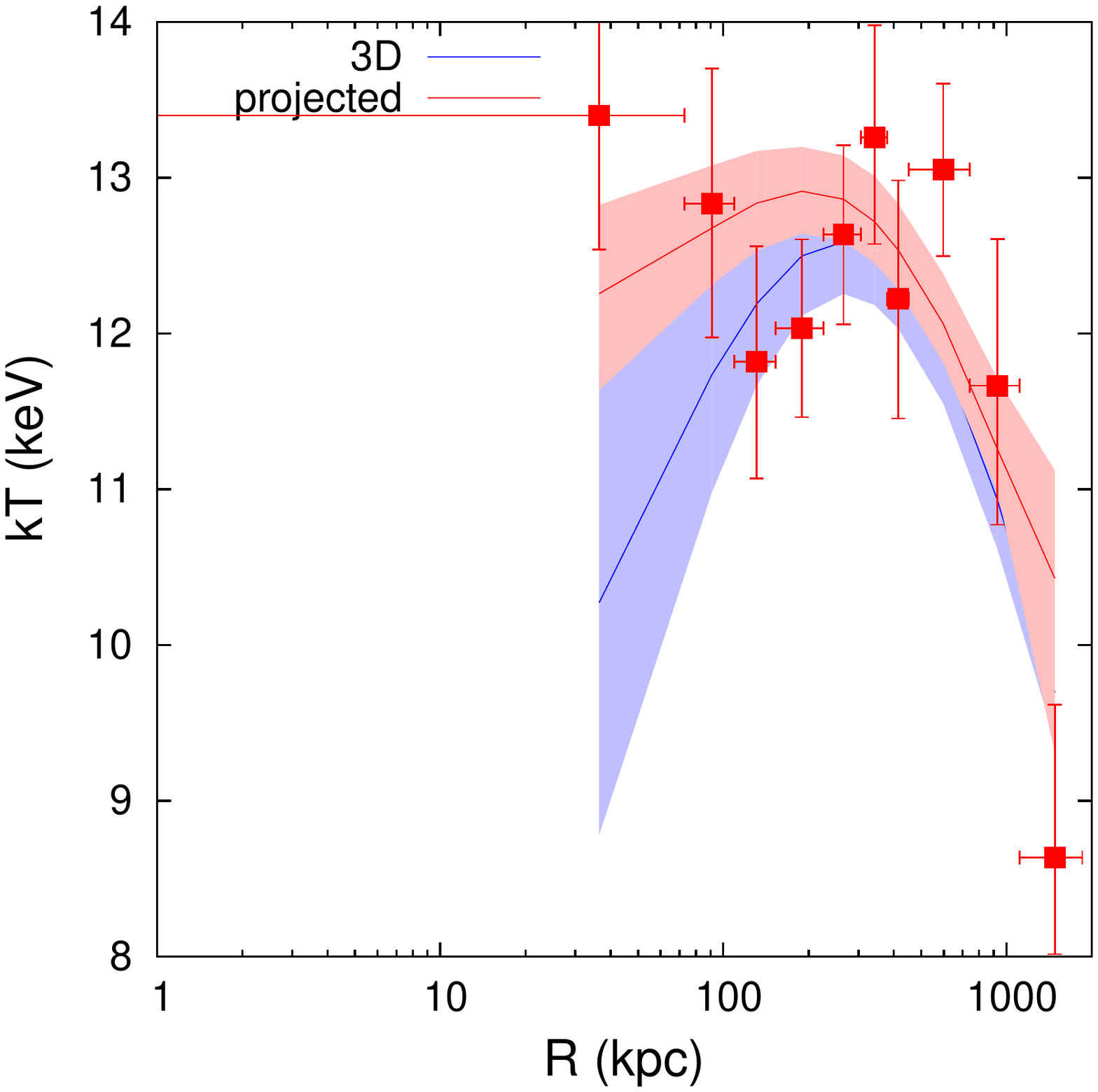}}}
\end{picture}
\end{center}
\caption{\small{Same as Figure~\ref{fig:a2204} but for A2219.}\label{fig:a2219}}
\end{figure*}

\begin{figure*}
\begin{center}
\setlength{\unitlength}{1in}
\begin{picture}(6.9,2.0)
\put(0.01,-0.8){\scalebox{0.34}{\includegraphics[clip=true]{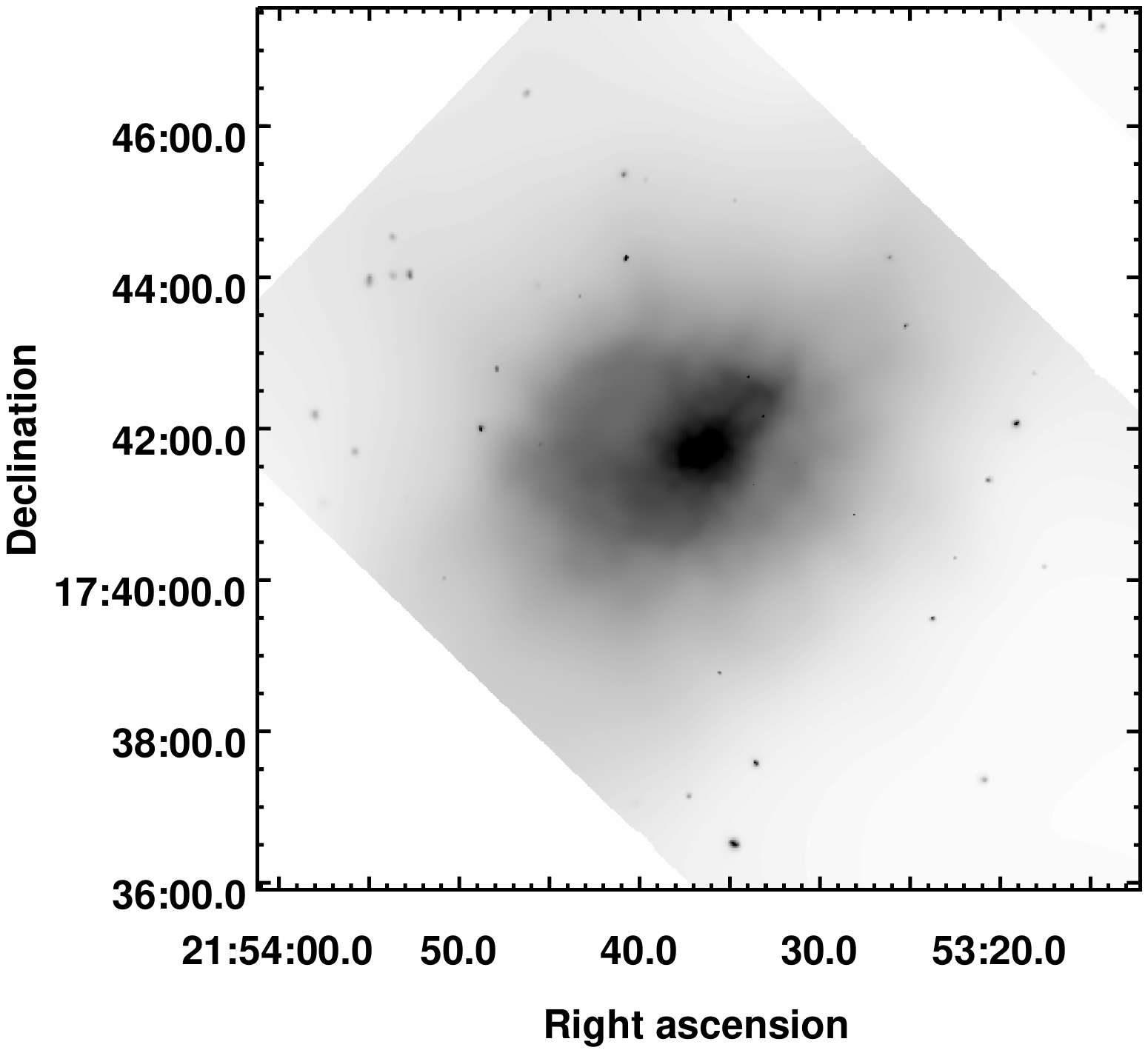}}}  
\put(2.1,-0.21){\scalebox{0.30}{\includegraphics{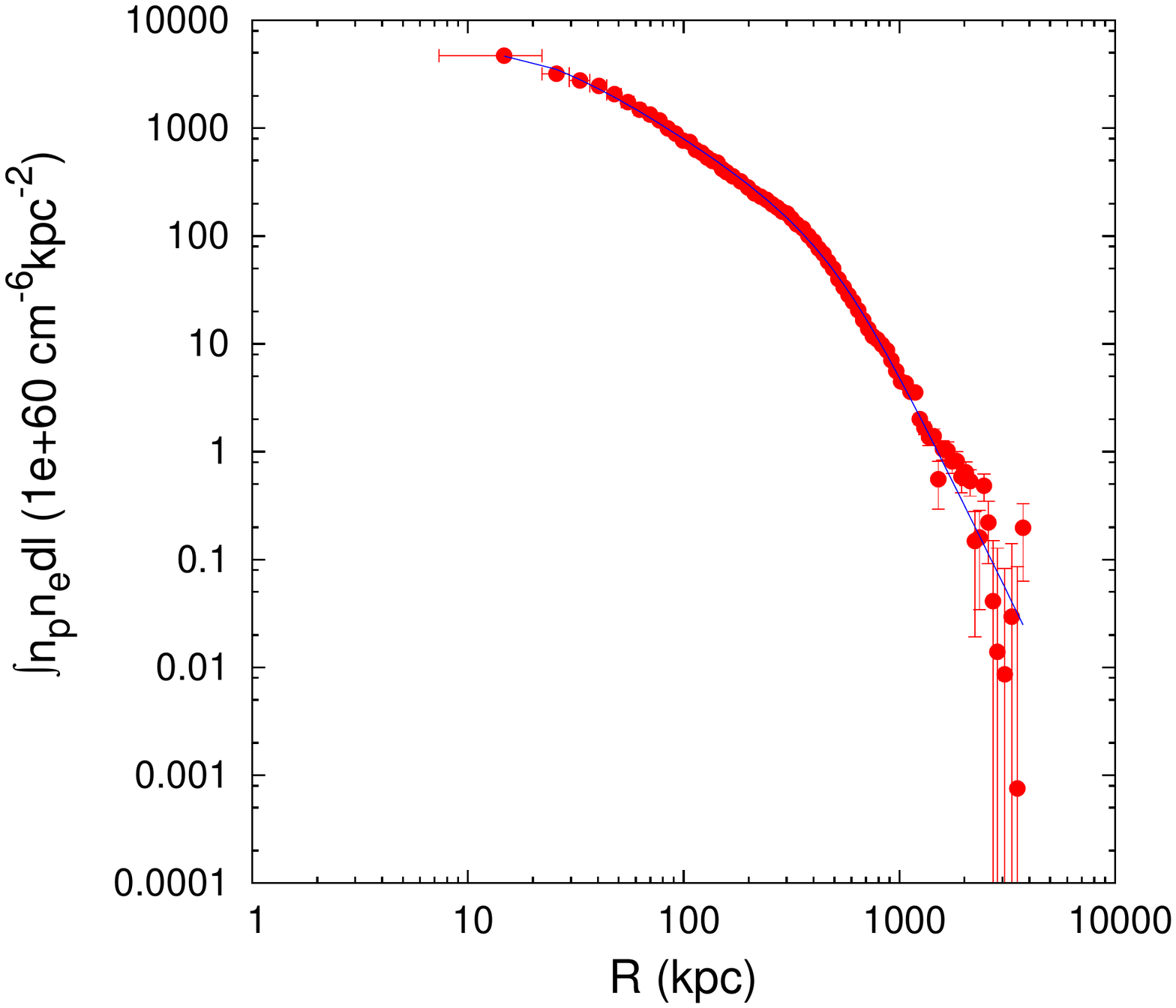}}}
\put(4.2,-0.21){\scalebox{0.30}{\includegraphics{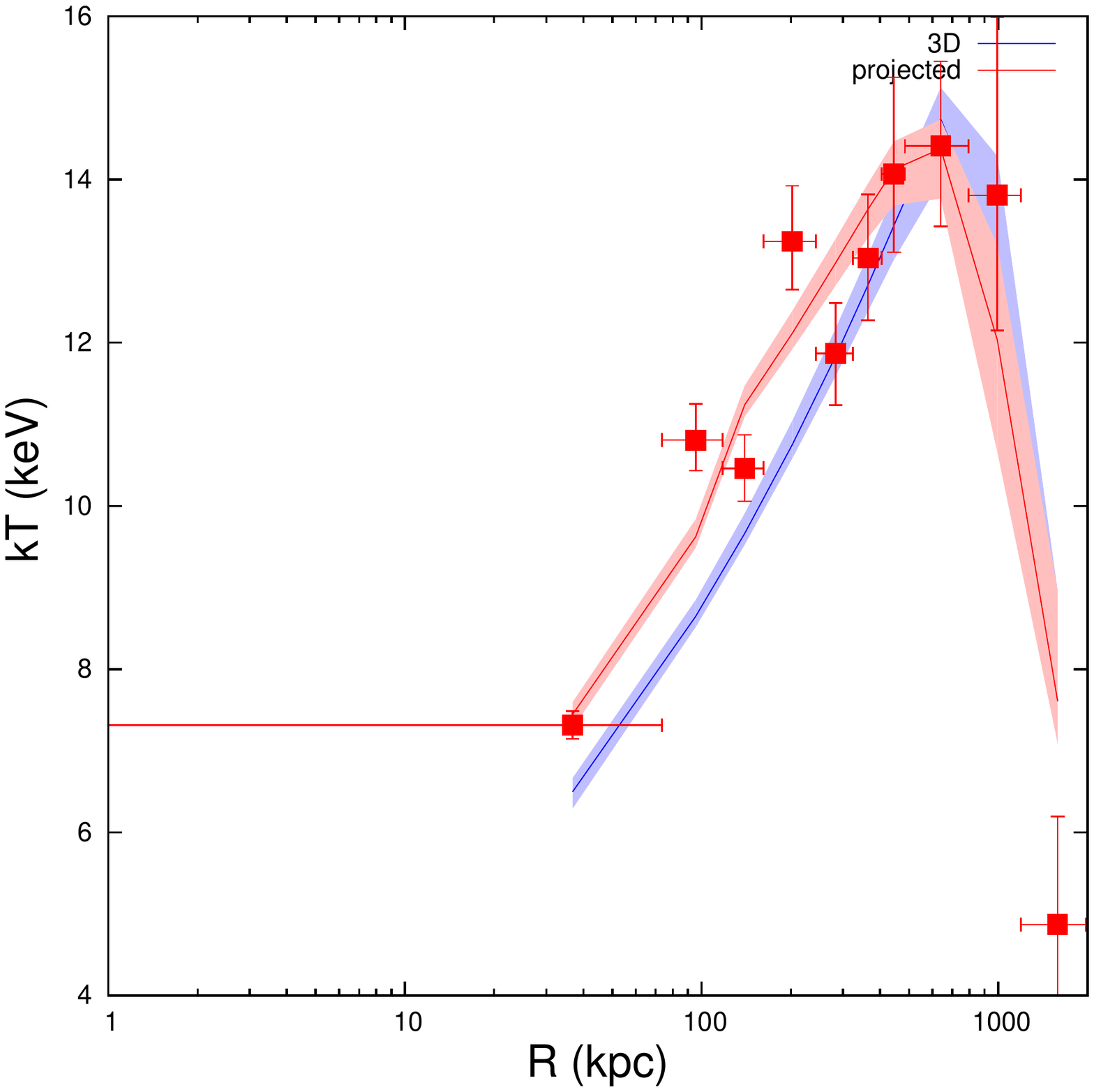}}}
\end{picture}
\end{center}
\caption{\small{Same as Figure~\ref{fig:a2204} but for A2390.}\label{fig:a2390}}
\end{figure*}

\begin{figure*}
\begin{center}
\setlength{\unitlength}{1in}
\begin{picture}(6.9,2.0)
\put(0.01,-0.8){\scalebox{0.34}{\includegraphics[clip=true]{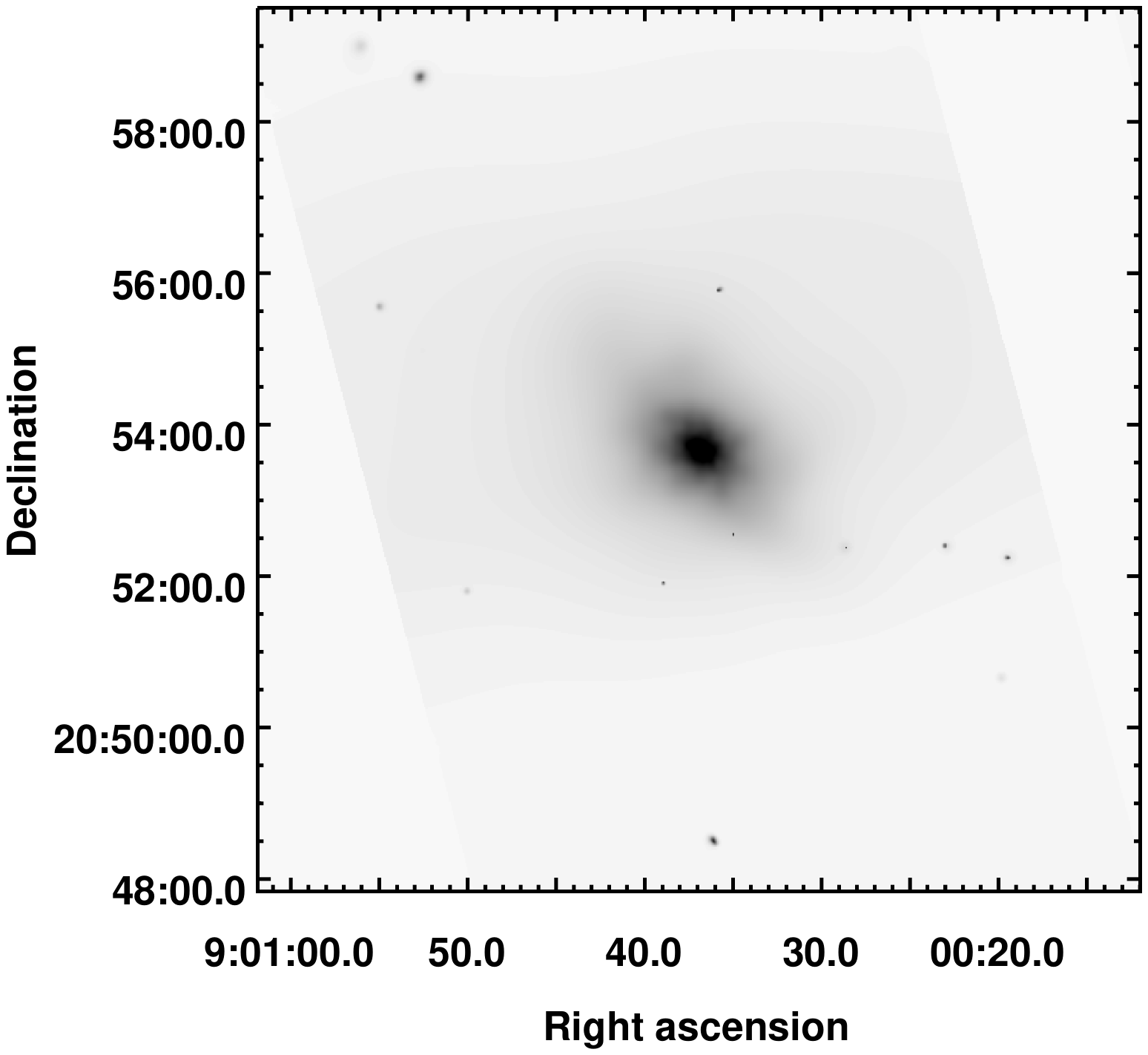}}}  
\put(2.1,-0.21){\scalebox{0.30}{\includegraphics{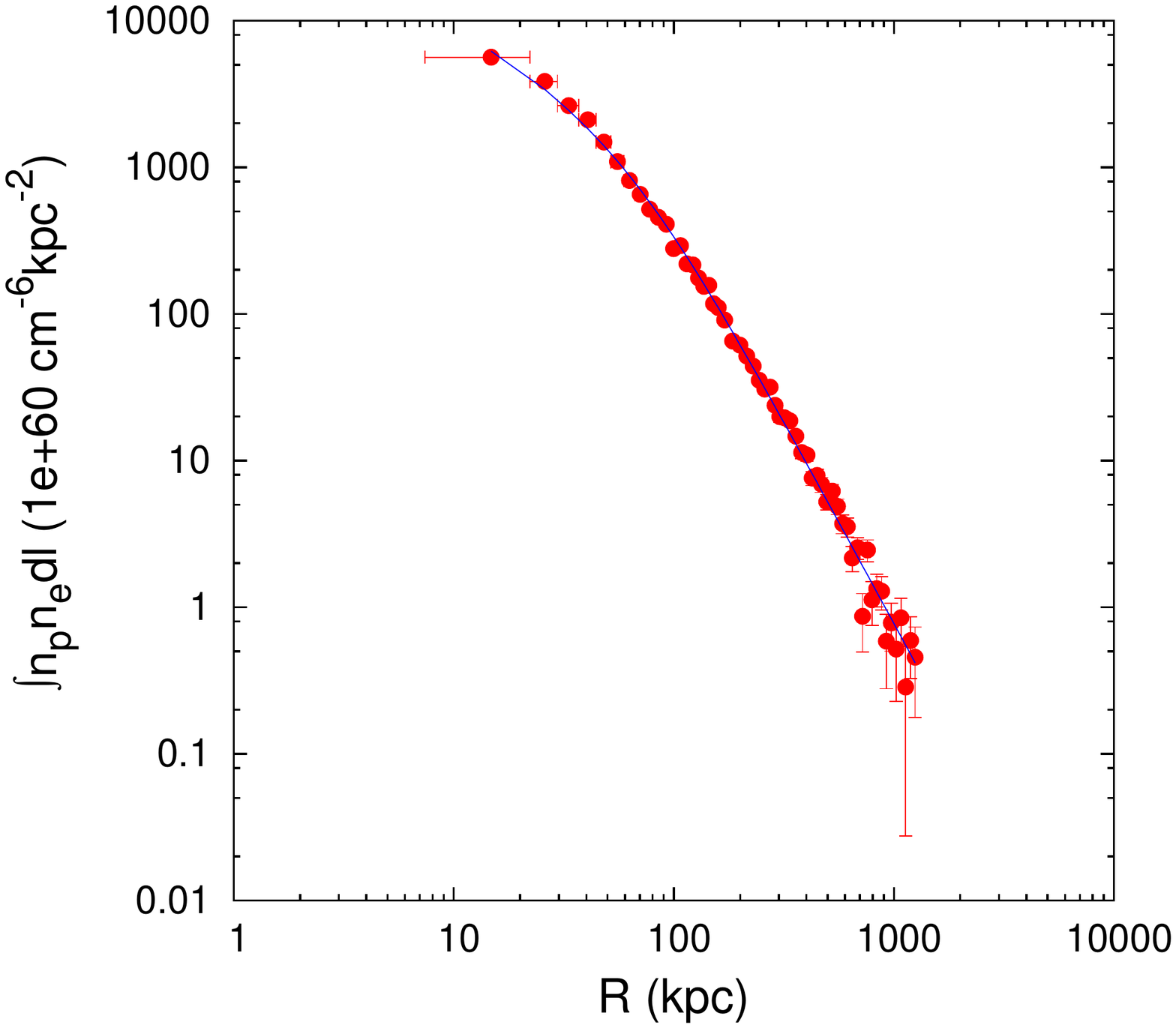}}}
\put(4.2,-0.21){\scalebox{0.30}{\includegraphics{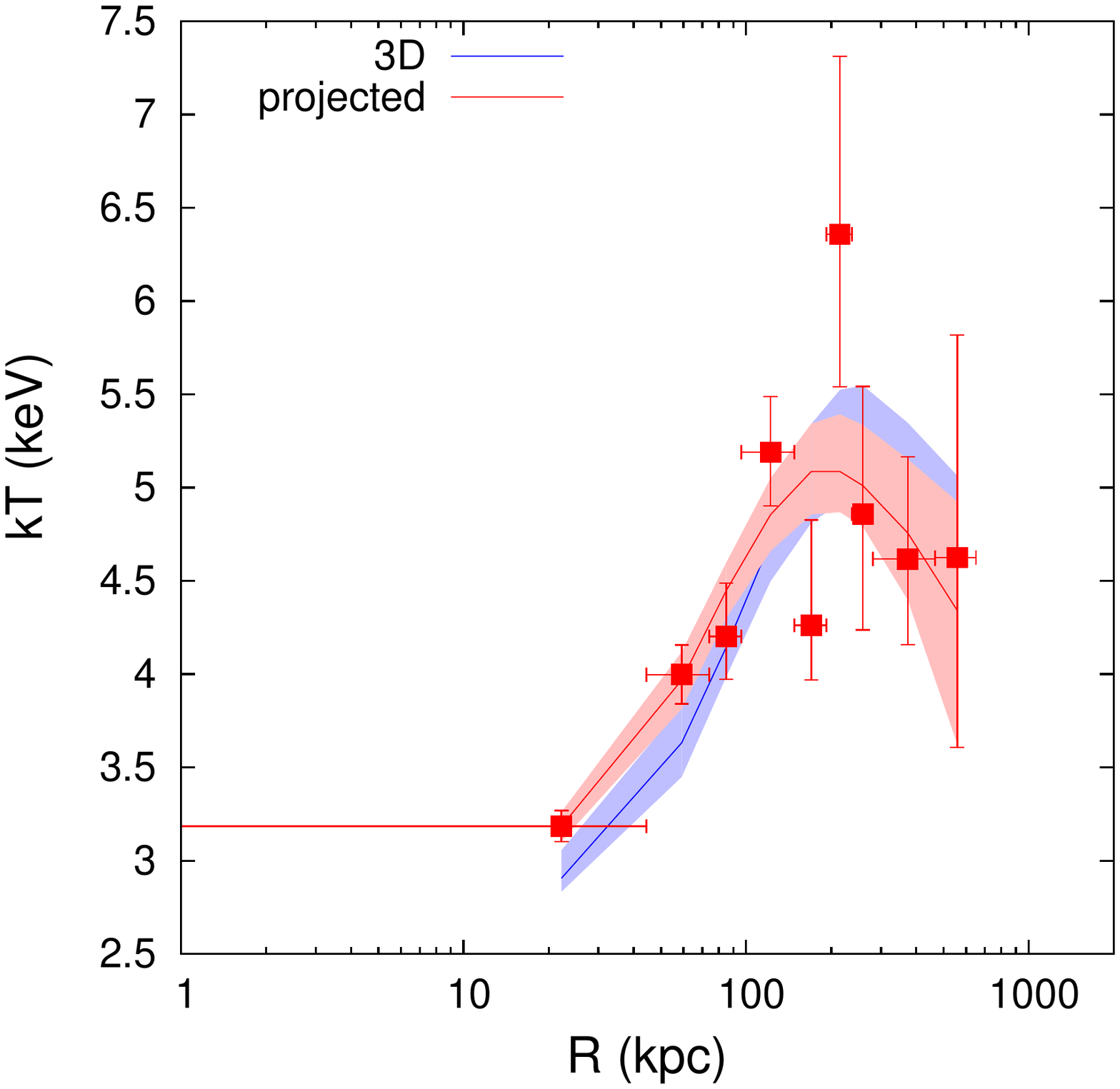}}}
\end{picture}
\end{center}
\caption{\small{Same as Figure~\ref{fig:a2204} but for Z2089.}\label{fig:z2089}}
\end{figure*}

\begin{figure*}
\begin{center}
\setlength{\unitlength}{1in}
\begin{picture}(6.9,2.0)
\put(0.01,-0.8){\scalebox{0.34}{\includegraphics[clip=true]{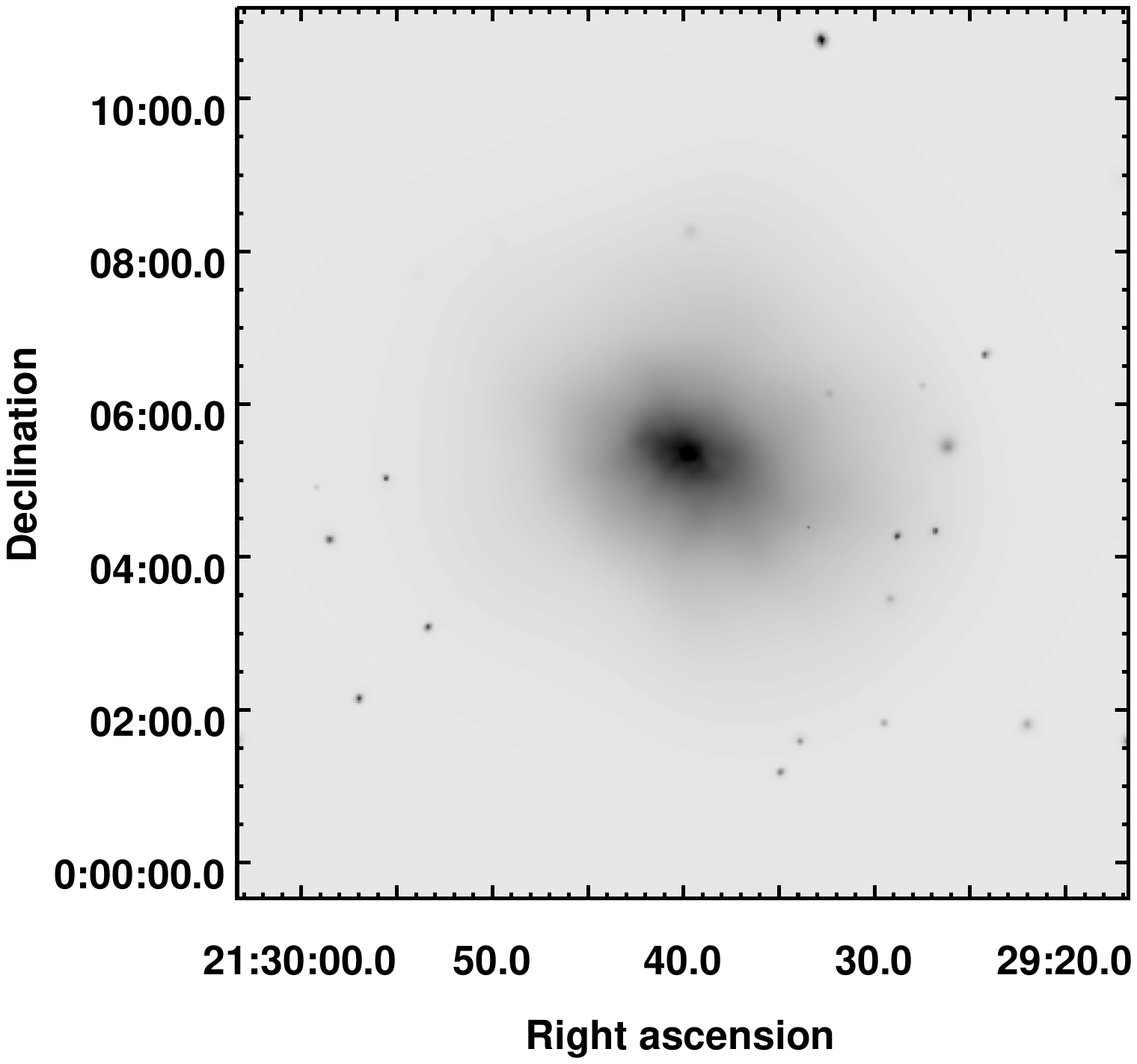}}}  
\put(2.1,-0.21){\scalebox{0.30}{\includegraphics{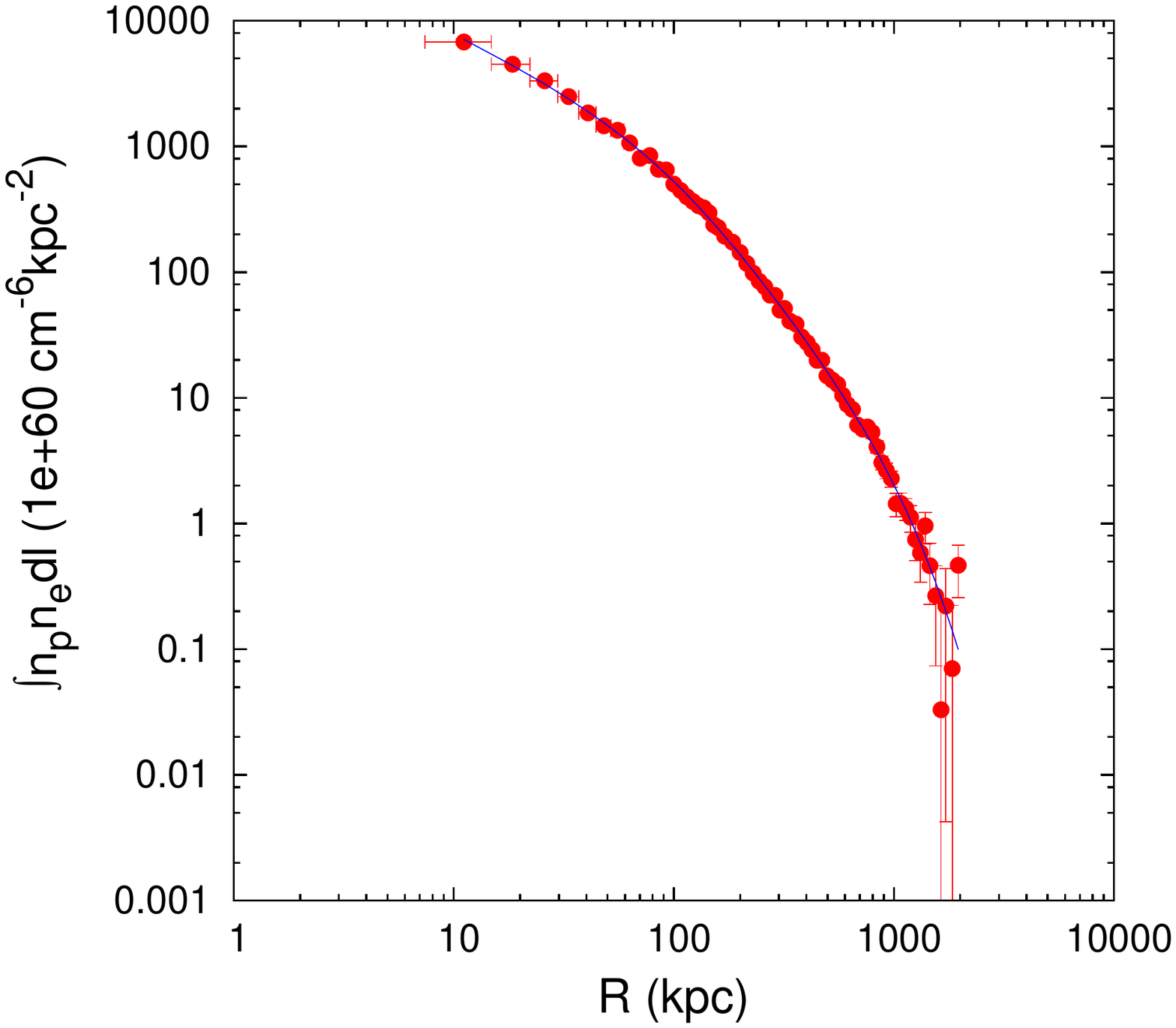}}}
\put(4.2,-0.21){\scalebox{0.30}{\includegraphics{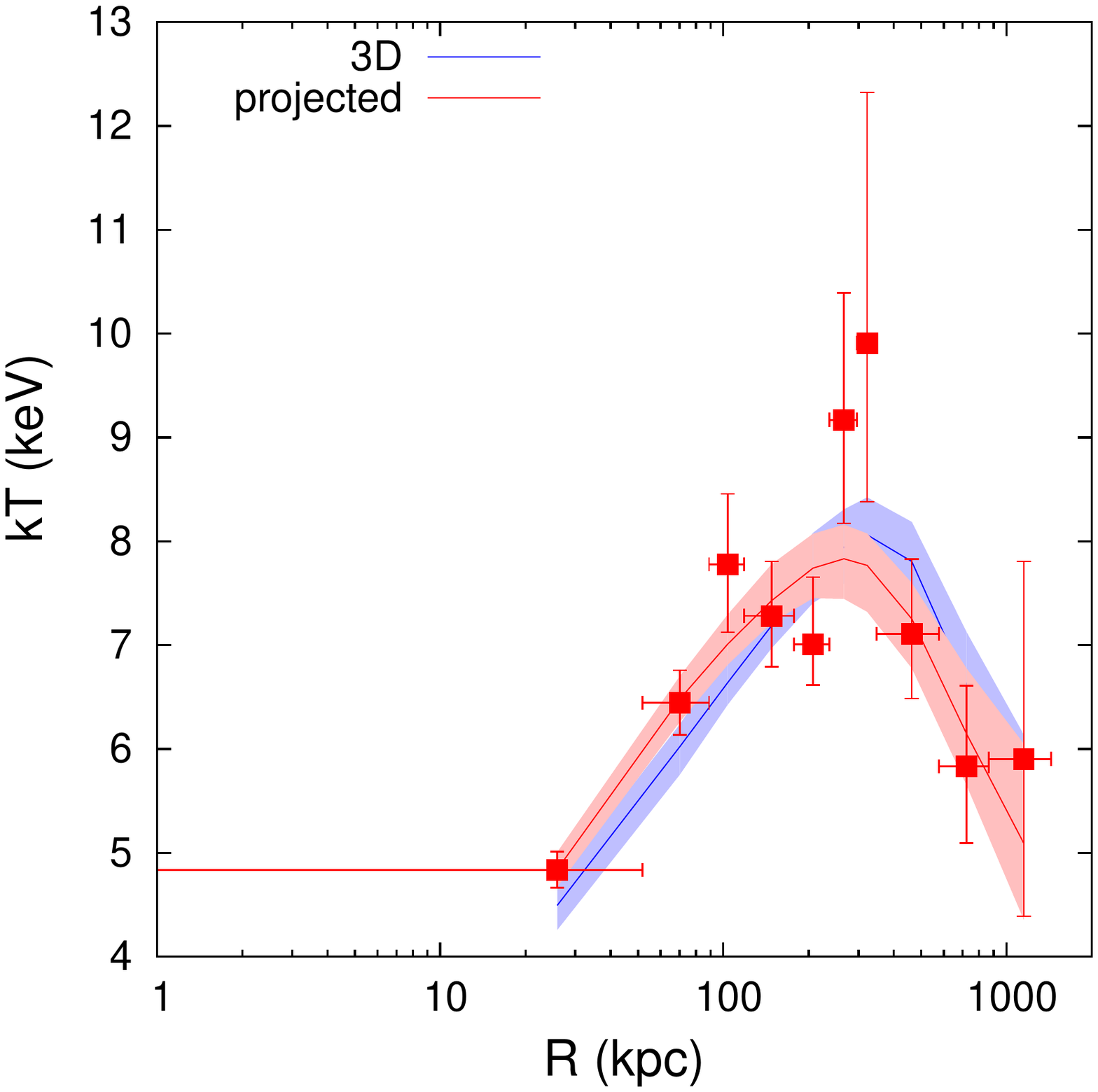}}}
\end{picture}
\end{center}
\caption{\small{Same as Figure~\ref{fig:a2204} but for RXJ2129.6+0005.}\label{fig:rxj2129}}
\end{figure*}

\begin{figure*}
\begin{center}
\setlength{\unitlength}{1in}
\begin{picture}(6.9,2.0)
\put(0.01,-0.8){\scalebox{0.34}{\includegraphics[clip=true]{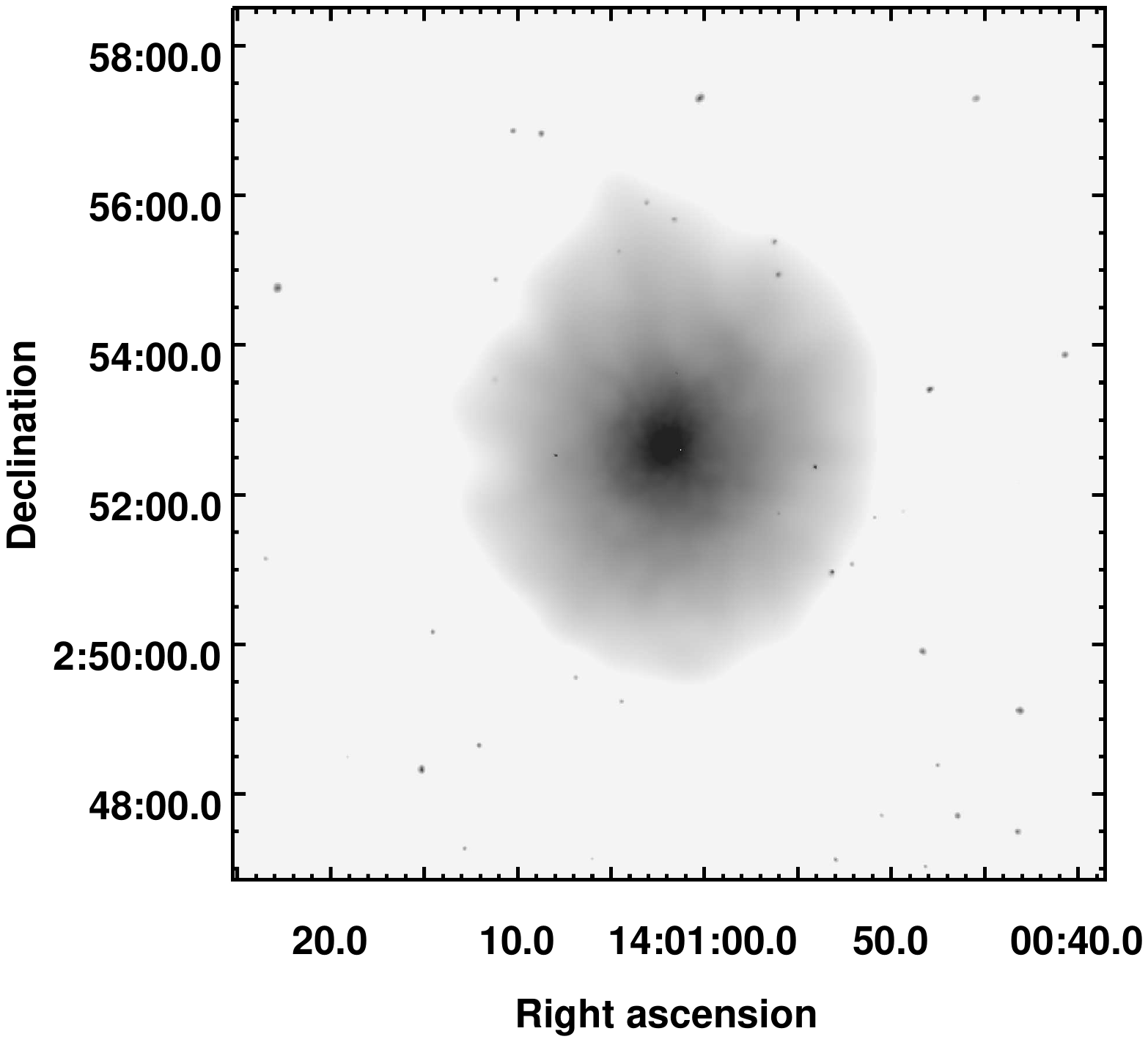}}}  
\put(2.1,-0.21){\scalebox{0.30}{\includegraphics{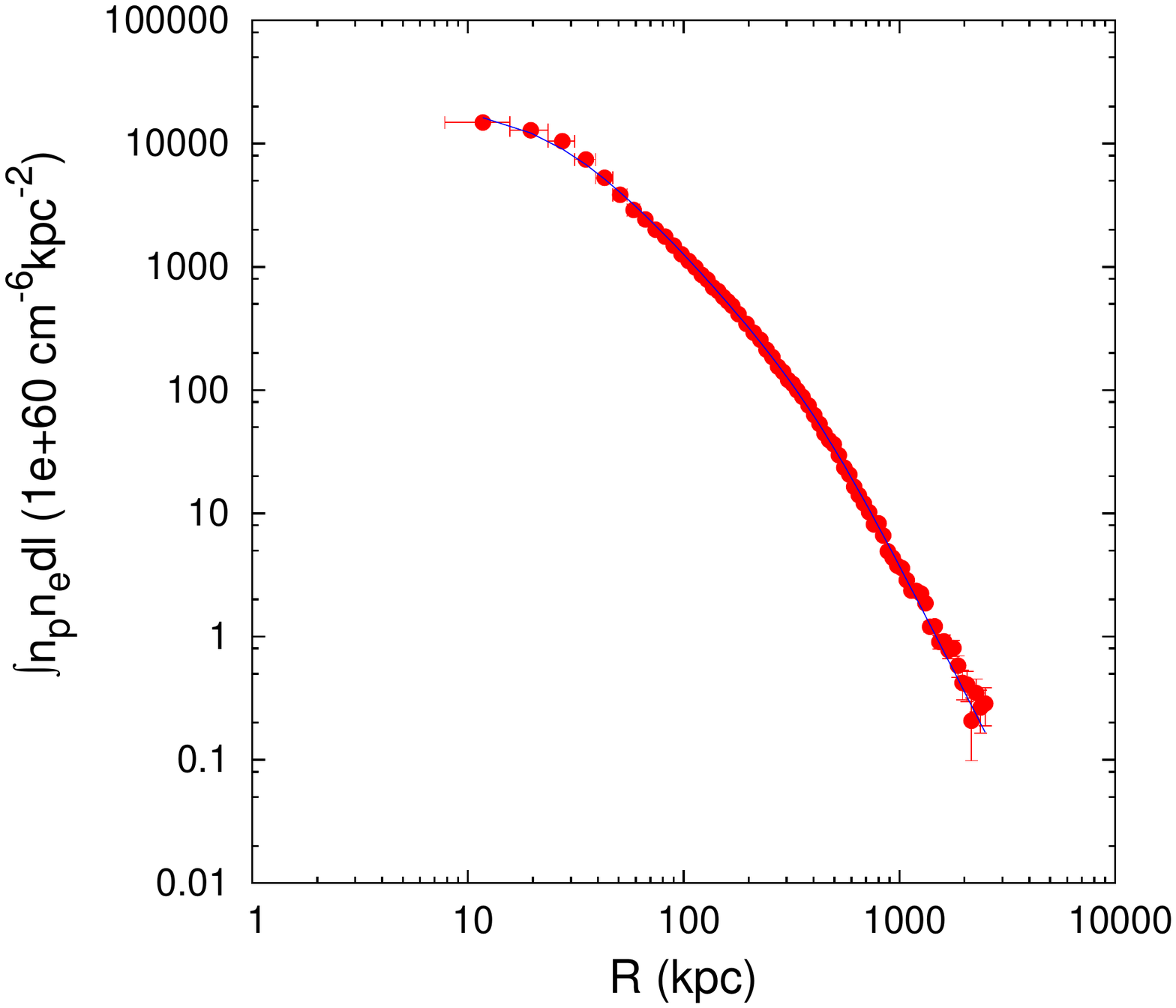}}}
\put(4.2,-0.21){\scalebox{0.30}{\includegraphics{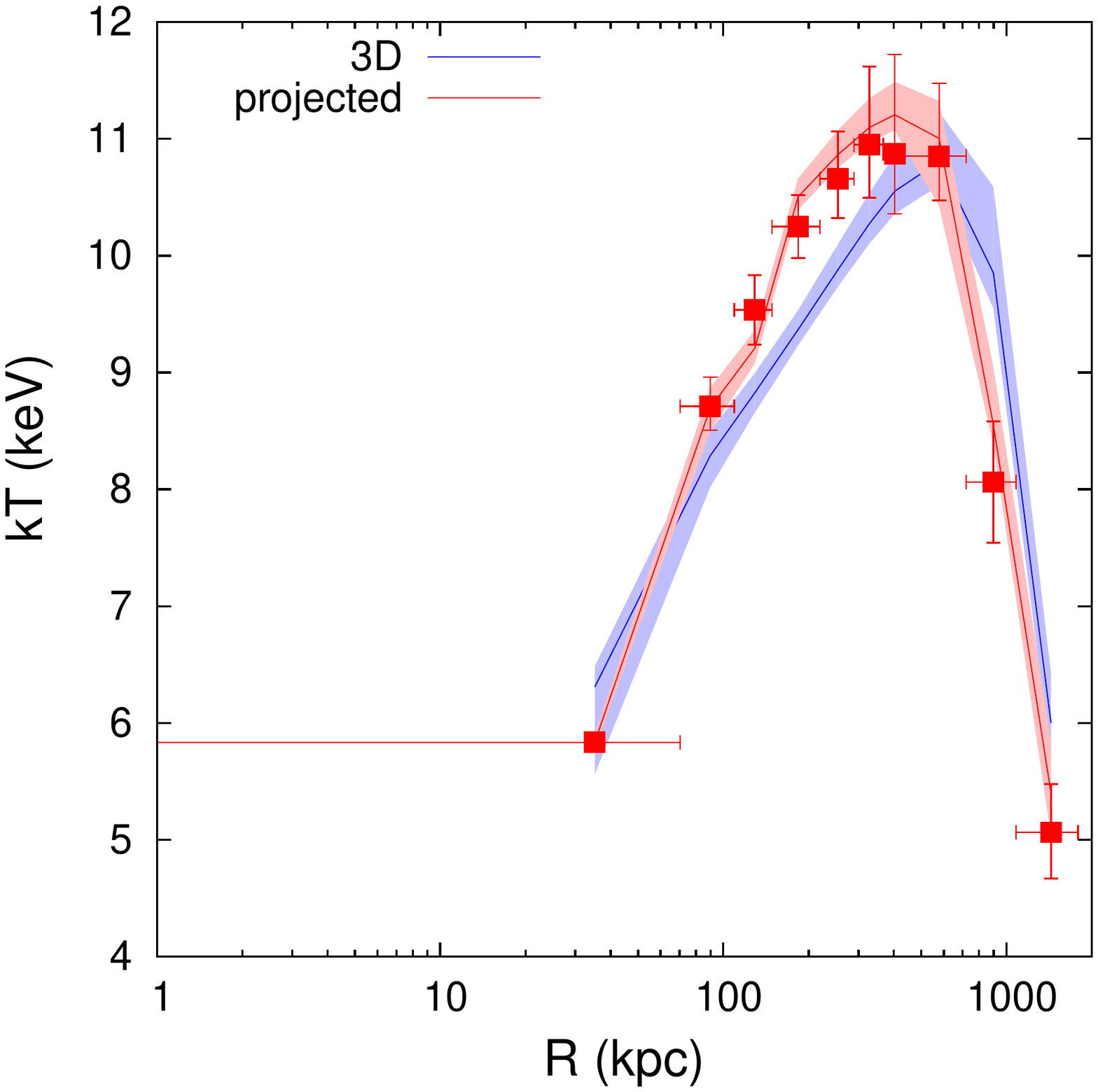}}}
\end{picture}
\end{center}
\caption{\small{Same as Figure~\ref{fig:a2204} but for A1835.}\label{fig:a1835}}
\end{figure*}

\begin{figure*}
\begin{center}
\setlength{\unitlength}{1in}
\begin{picture}(6.9,2.0)
\put(0.01,-0.8){\scalebox{0.34}{\includegraphics[clip=true]{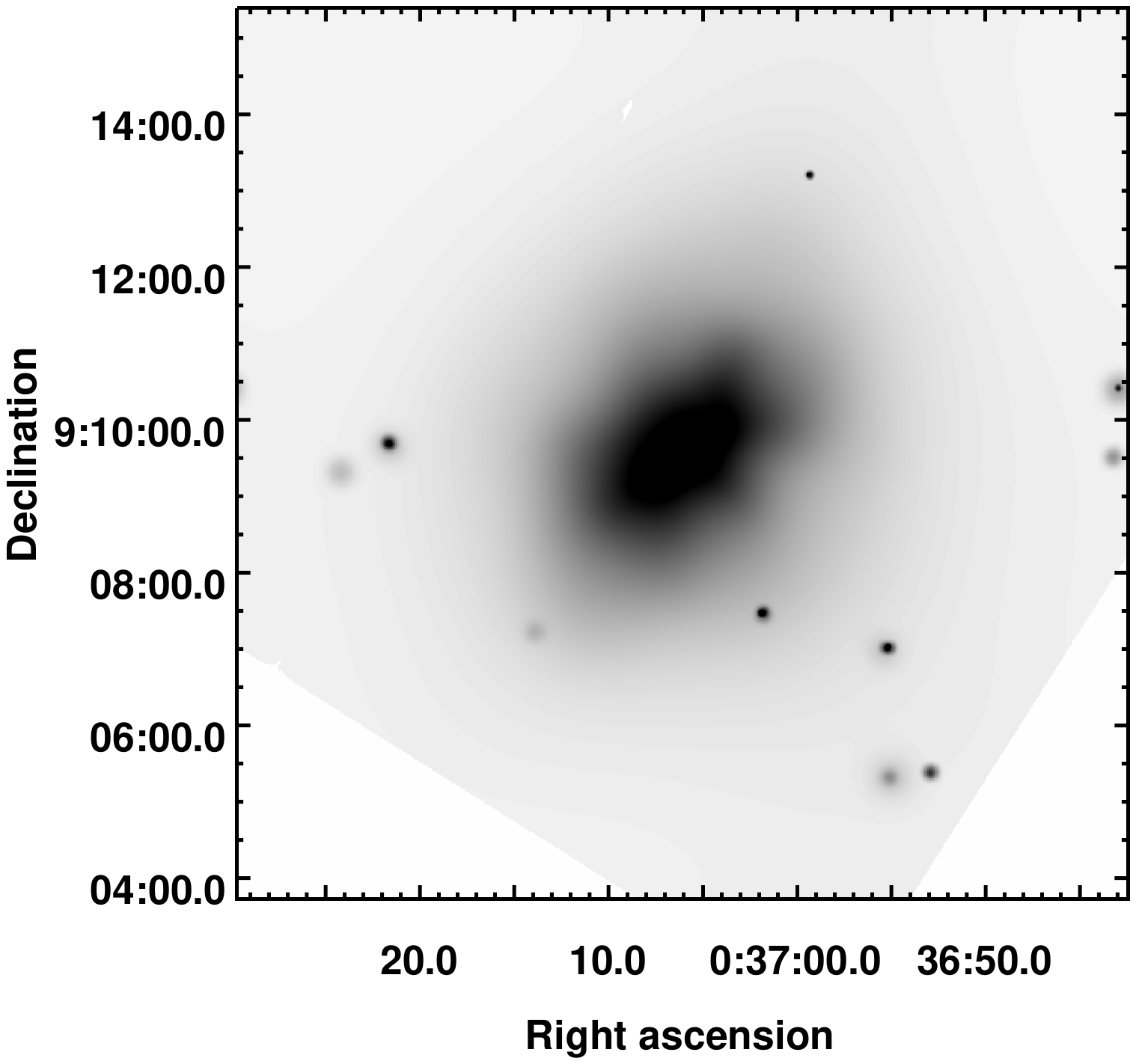}}}  
\put(2.1,-0.21){\scalebox{0.30}{\includegraphics{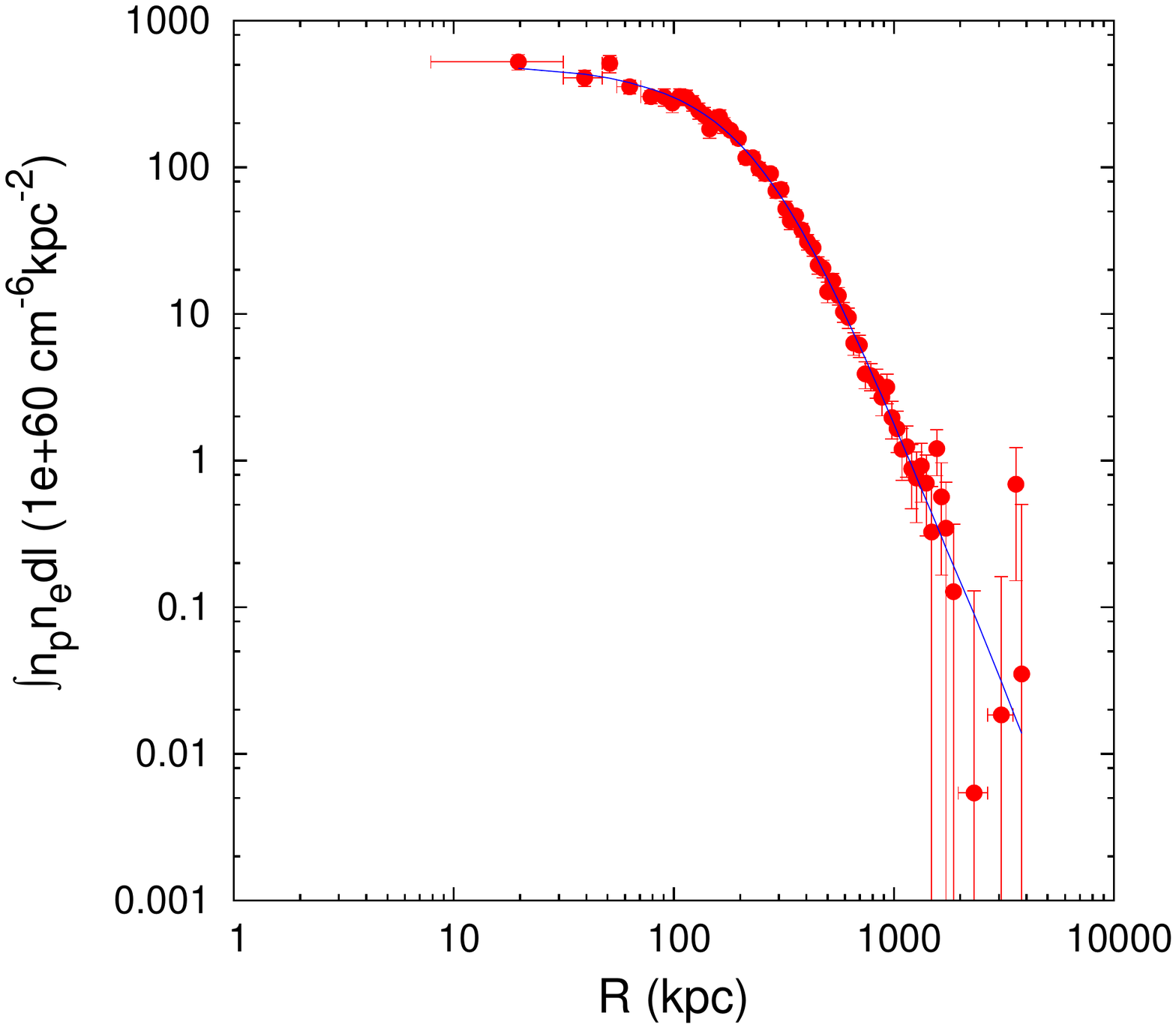}}}
\put(4.2,-0.21){\scalebox{0.30}{\includegraphics{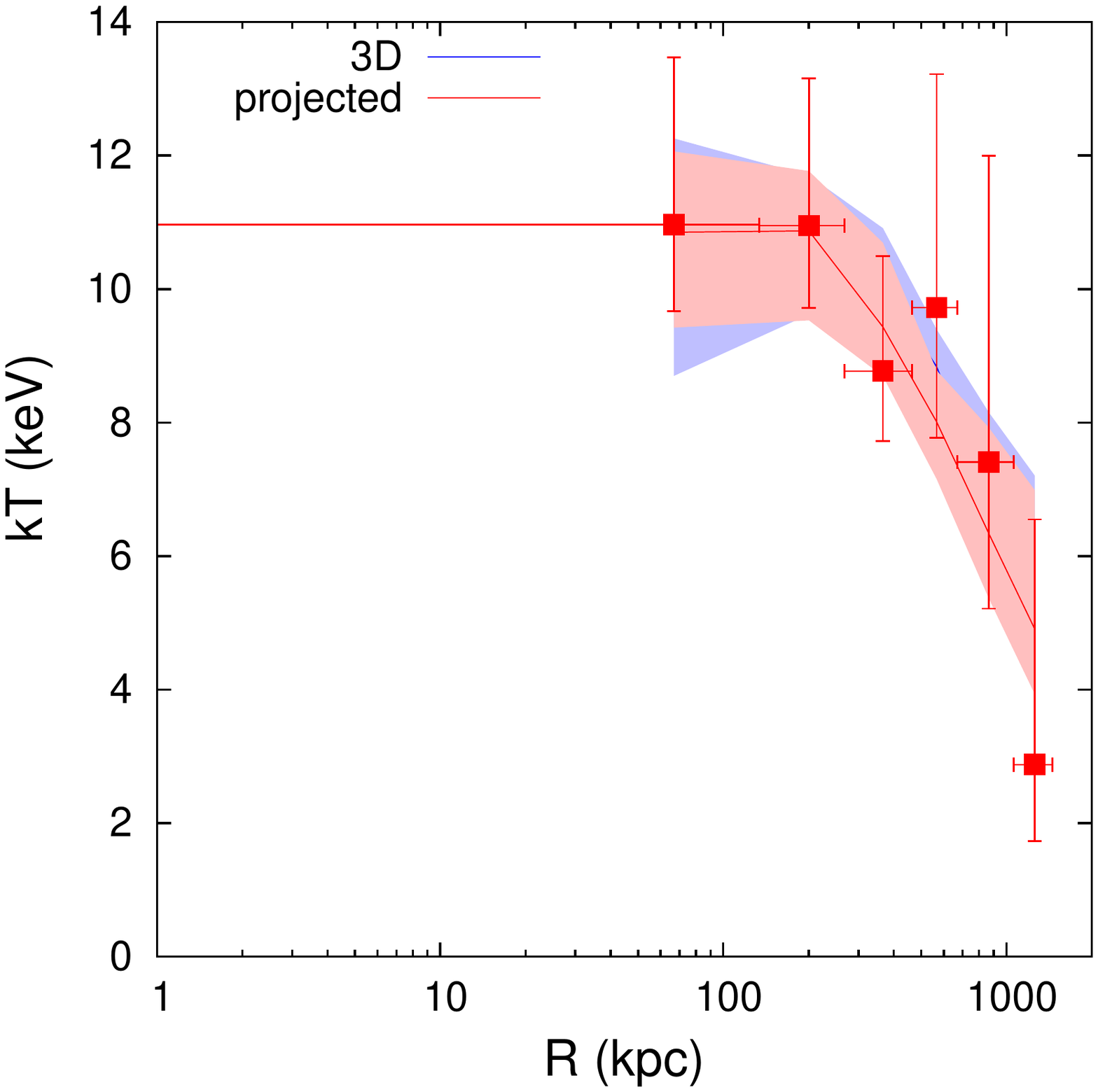}}}
\end{picture}
\end{center}
\caption{\small{Same as Figure~\ref{fig:a2204} but for A68.}\label{fig:a68}}
\end{figure*}

\begin{figure*}
\begin{center}
\setlength{\unitlength}{1in}
\begin{picture}(6.9,2.0)
\put(0.01,-0.8){\scalebox{0.34}{\includegraphics[clip=true]{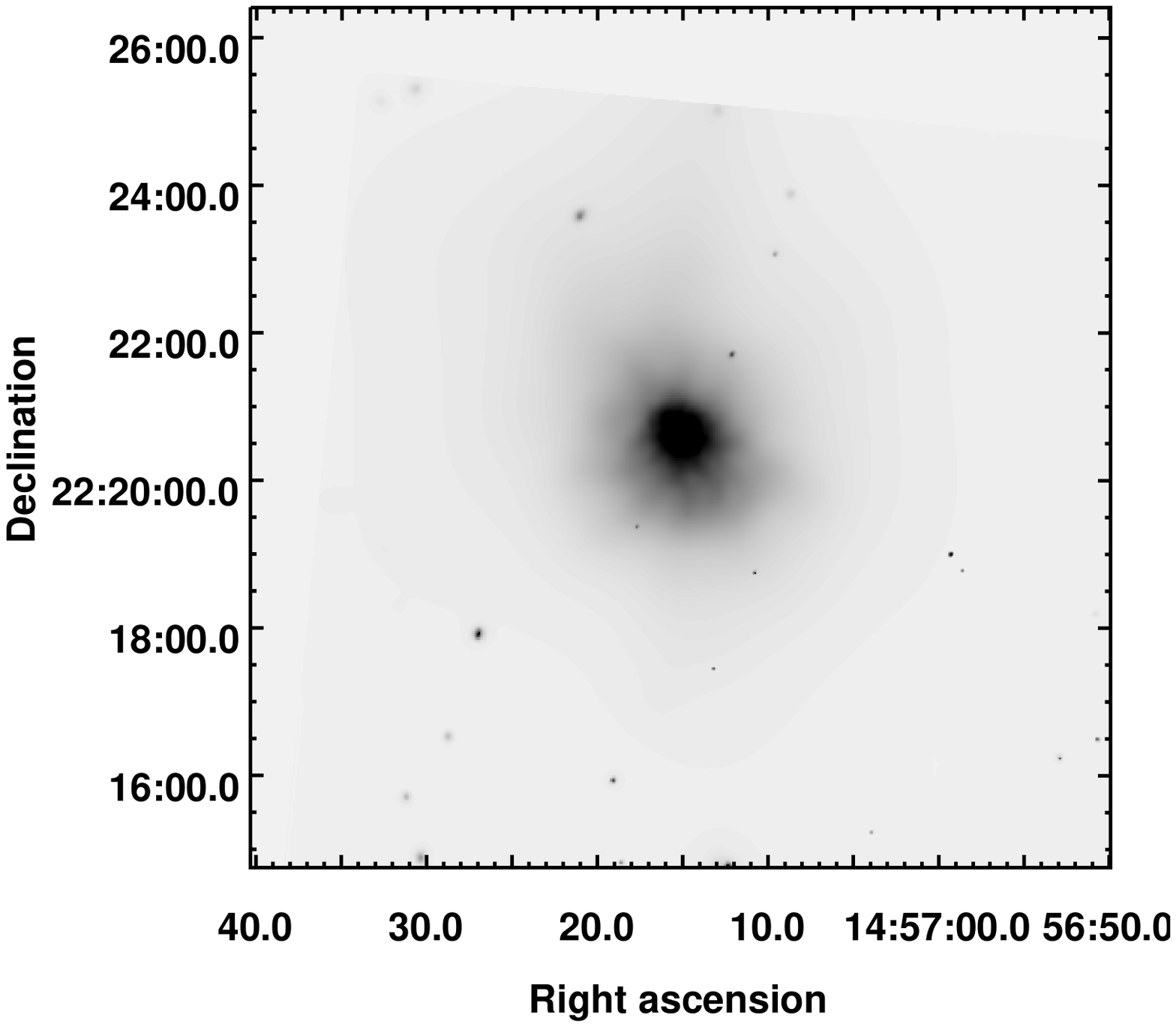}}}  
\put(2.1,-0.21){\scalebox{0.30}{\includegraphics{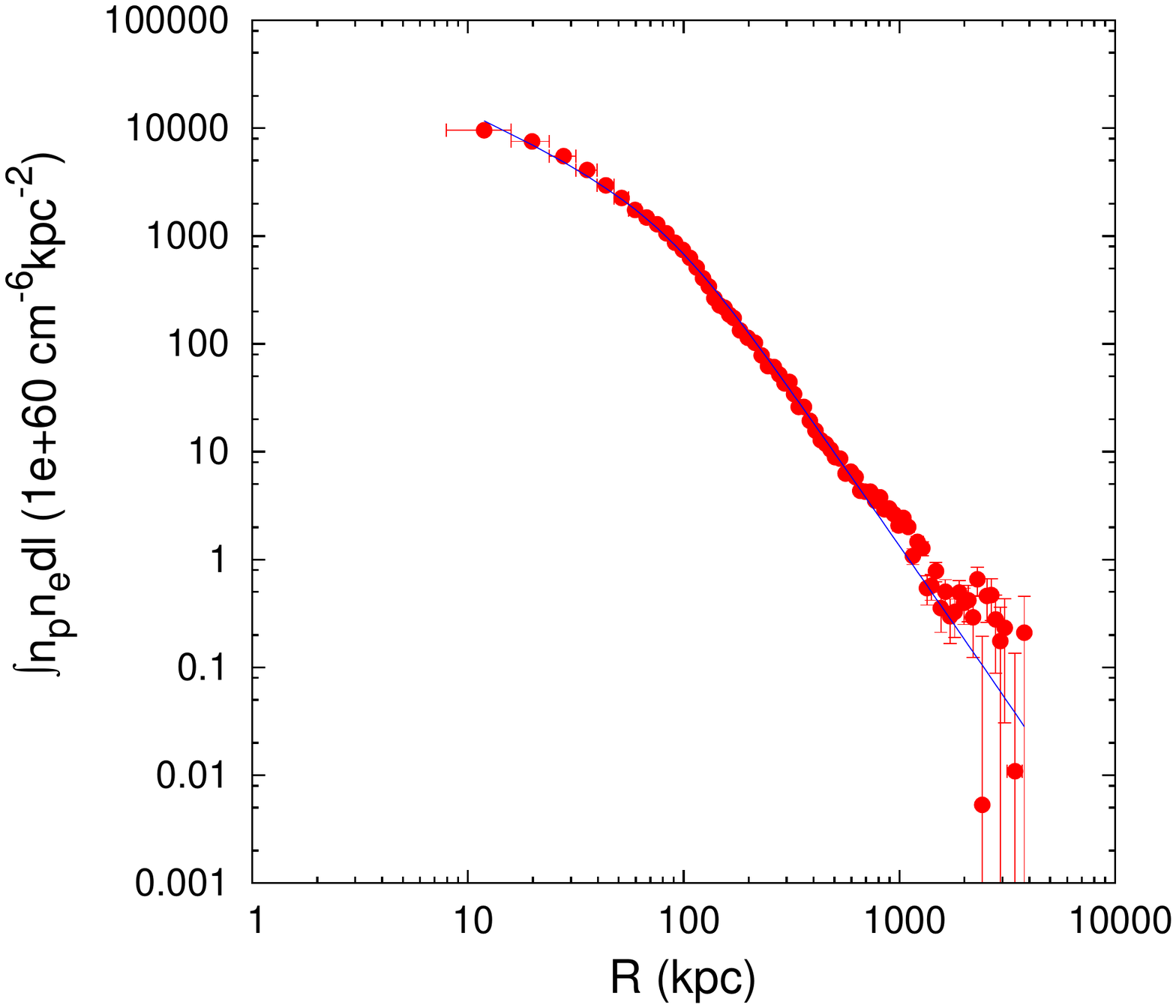}}}
\put(4.2,-0.21){\scalebox{0.30}{\includegraphics{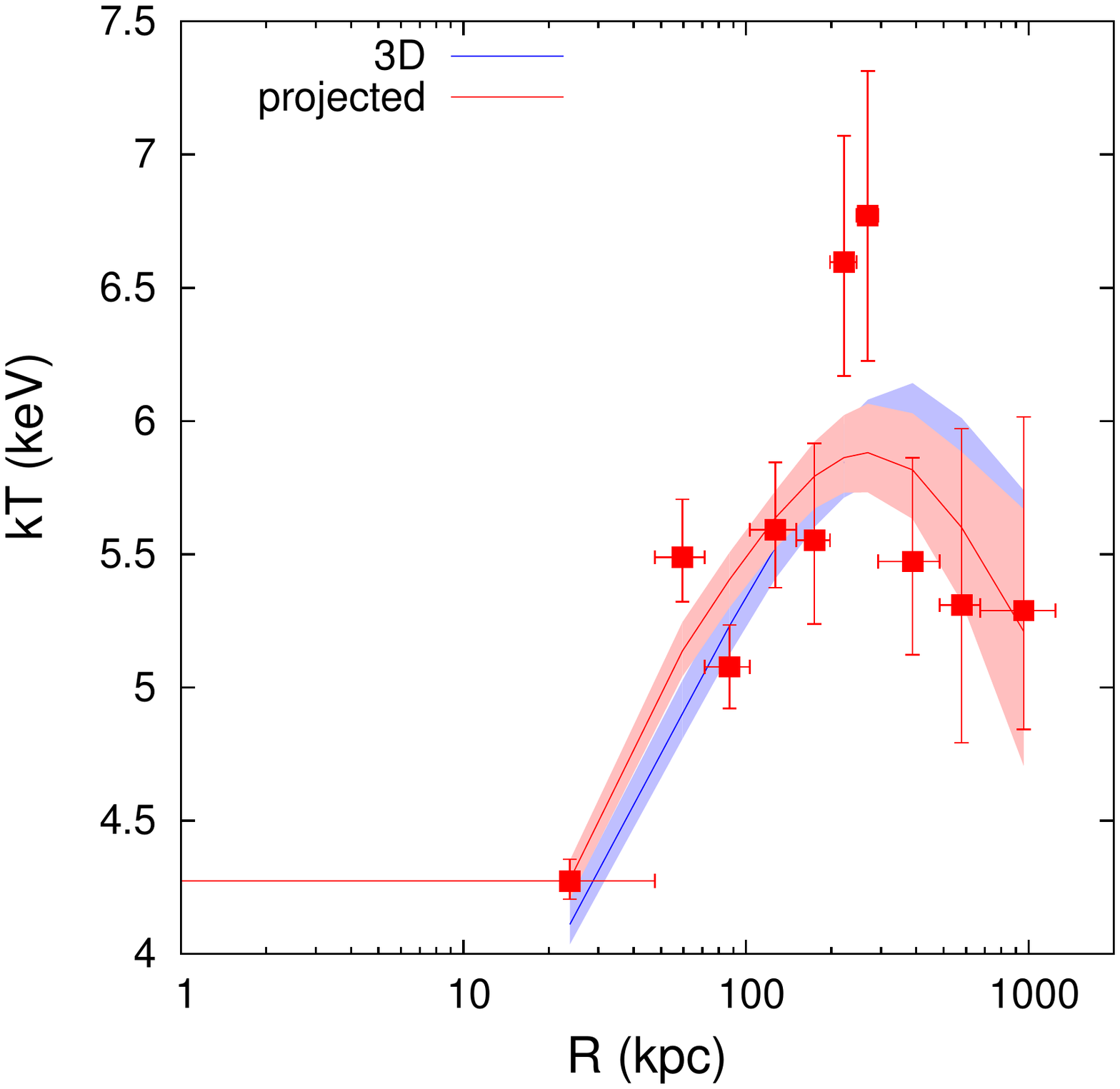}}}
\end{picture}
\end{center}
\caption{\small{Same as Figure~\ref{fig:a2204} but for MS1455.0+2232.}\label{fig:ms1455}}
\end{figure*}

\begin{figure*}
\begin{center}
\setlength{\unitlength}{1in}
\begin{picture}(6.9,2.0)
\put(0.01,-0.8){\scalebox{0.34}{\includegraphics[clip=true]{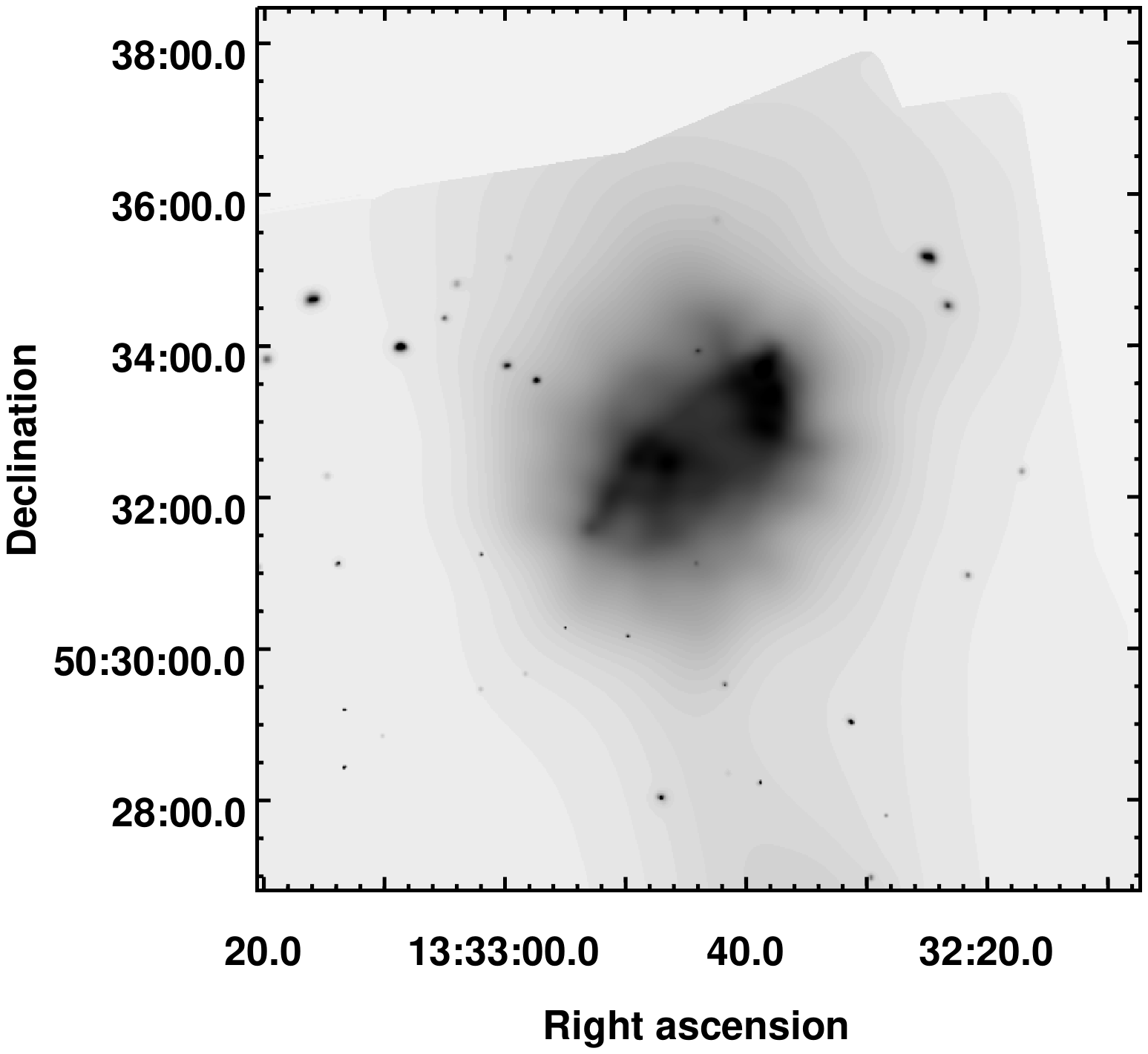}}}  
\put(2.1,-0.21){\scalebox{0.30}{\includegraphics{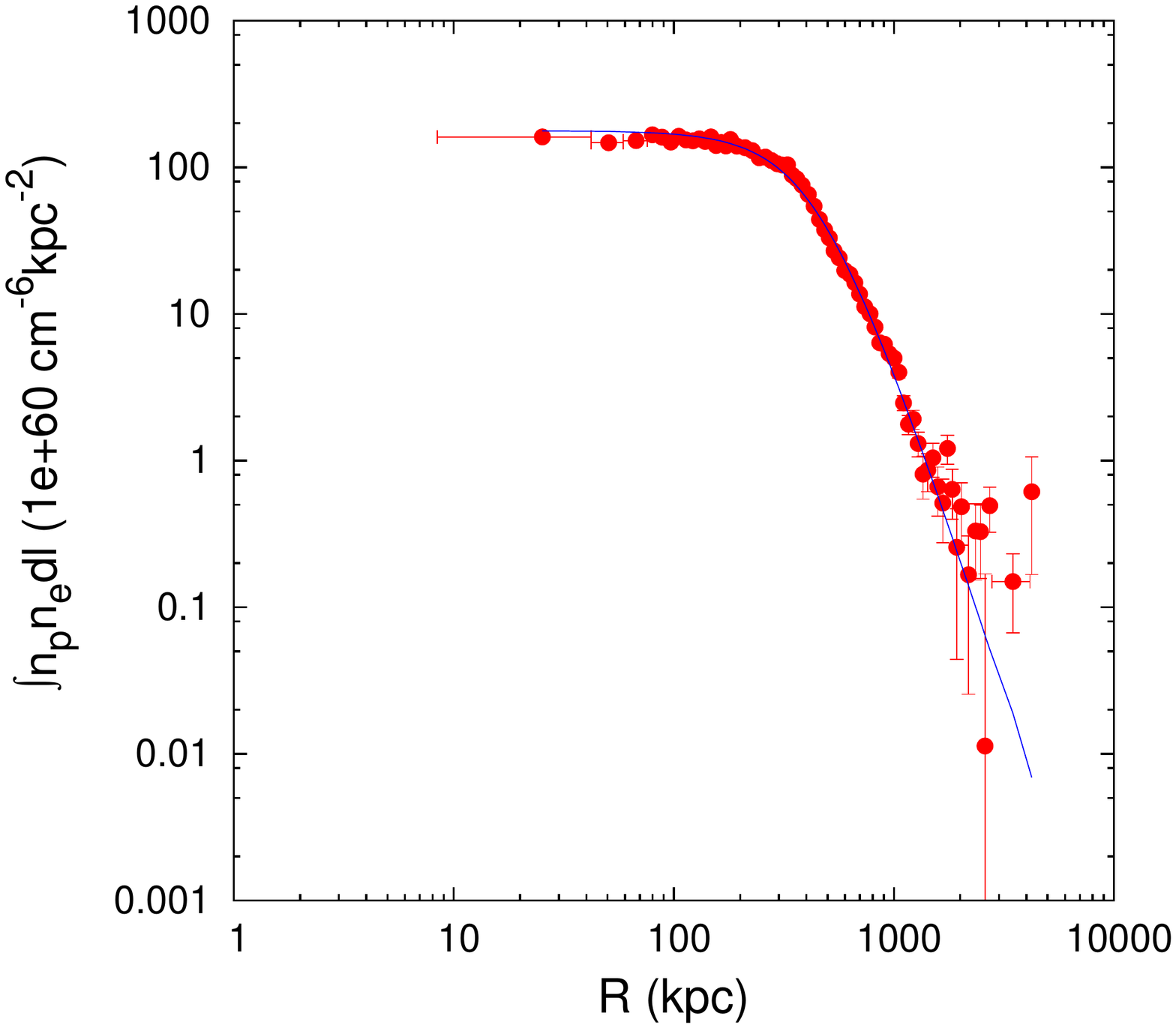}}}
\put(4.2,-0.21){\scalebox{0.30}{\includegraphics{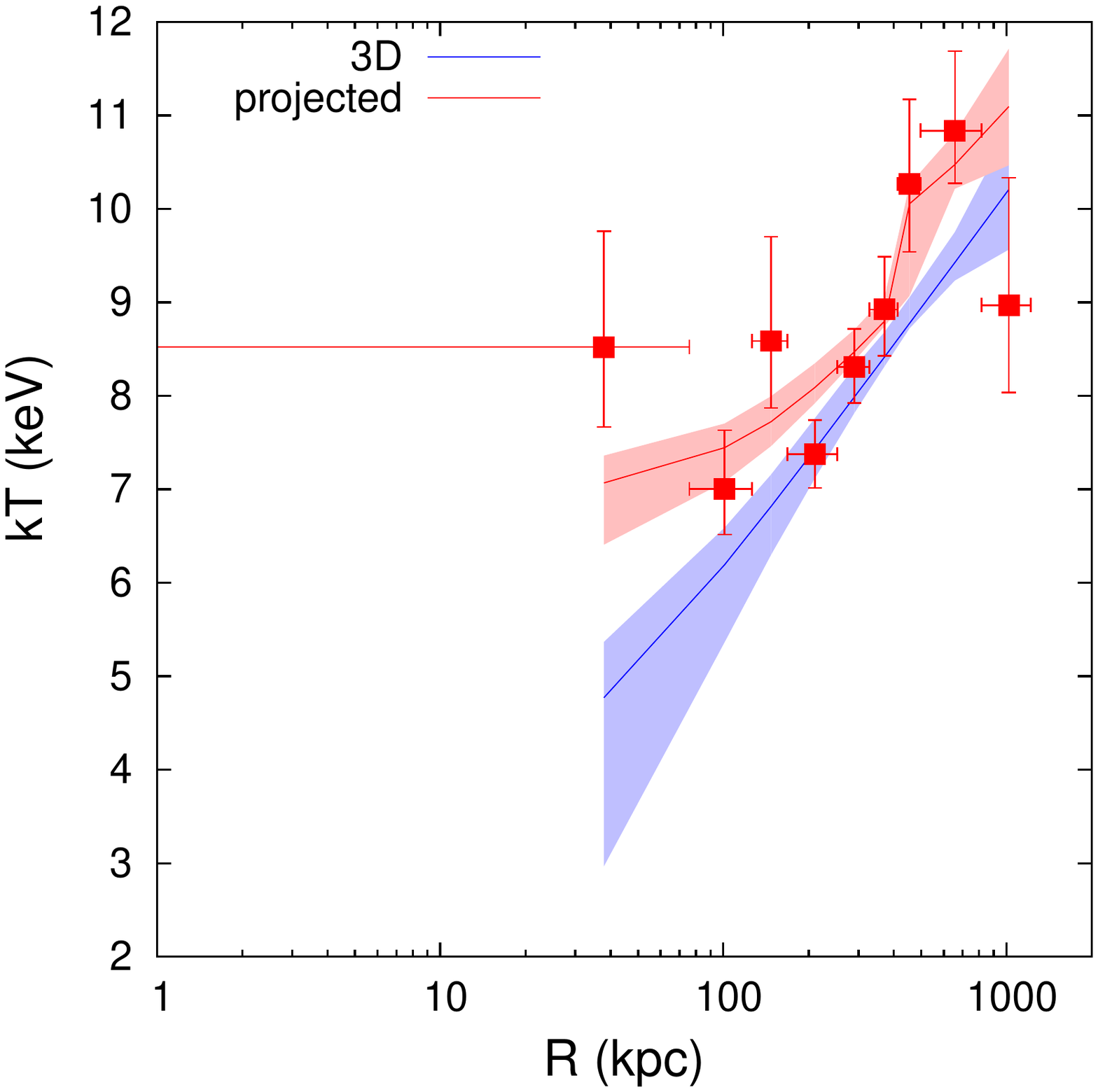}}}
\end{picture}
\end{center}
\caption{\small{Same as Figure~\ref{fig:a2204} but for A1758.}\label{fig:a1758}}
\end{figure*}

\begin{figure*}
\begin{center}
\setlength{\unitlength}{1in}
\begin{picture}(6.9,2.0)
\put(0.01,-0.8){\scalebox{0.34}{\includegraphics[clip=true]{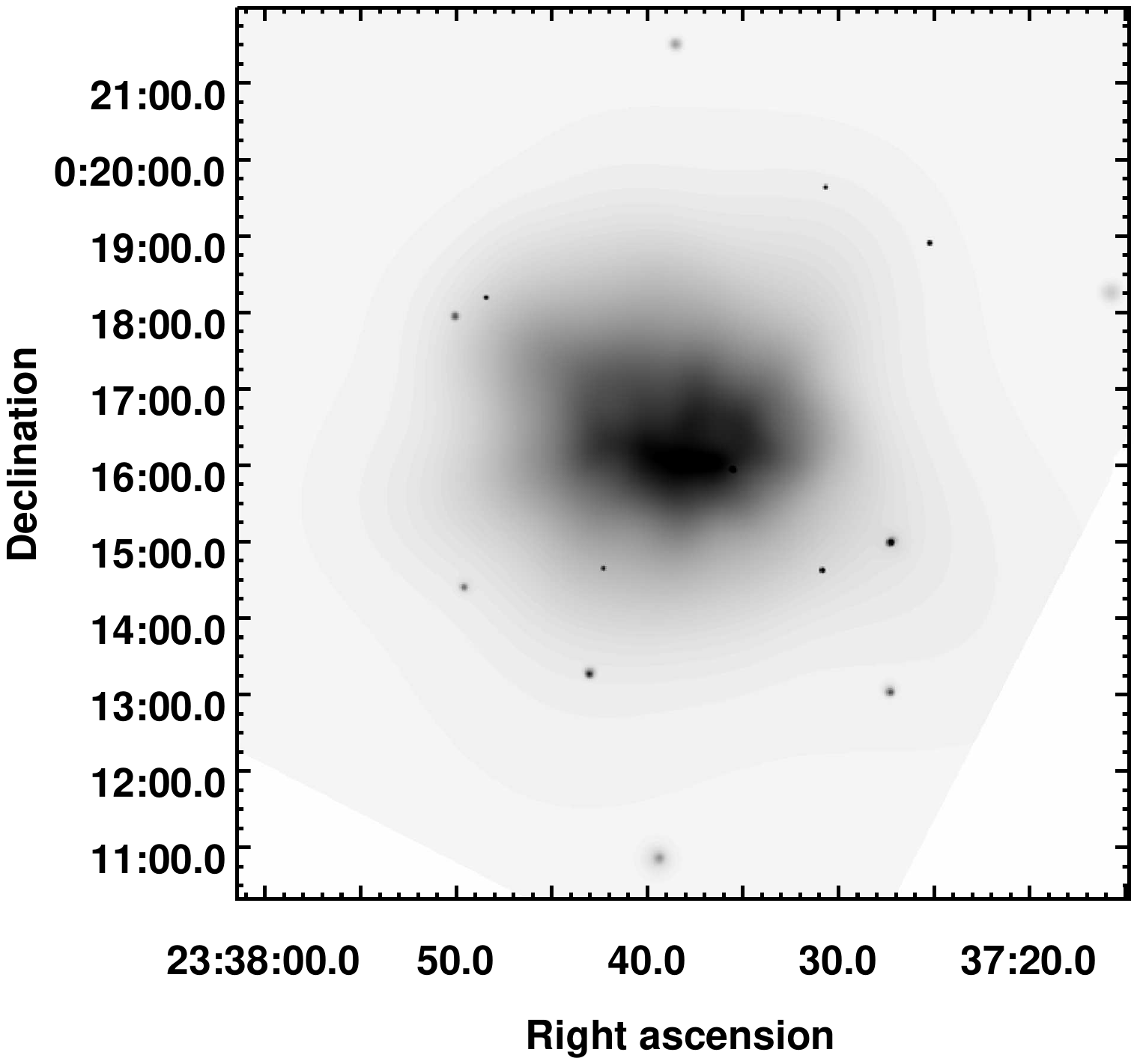}}}  
\put(2.1,-0.21){\scalebox{0.30}{\includegraphics{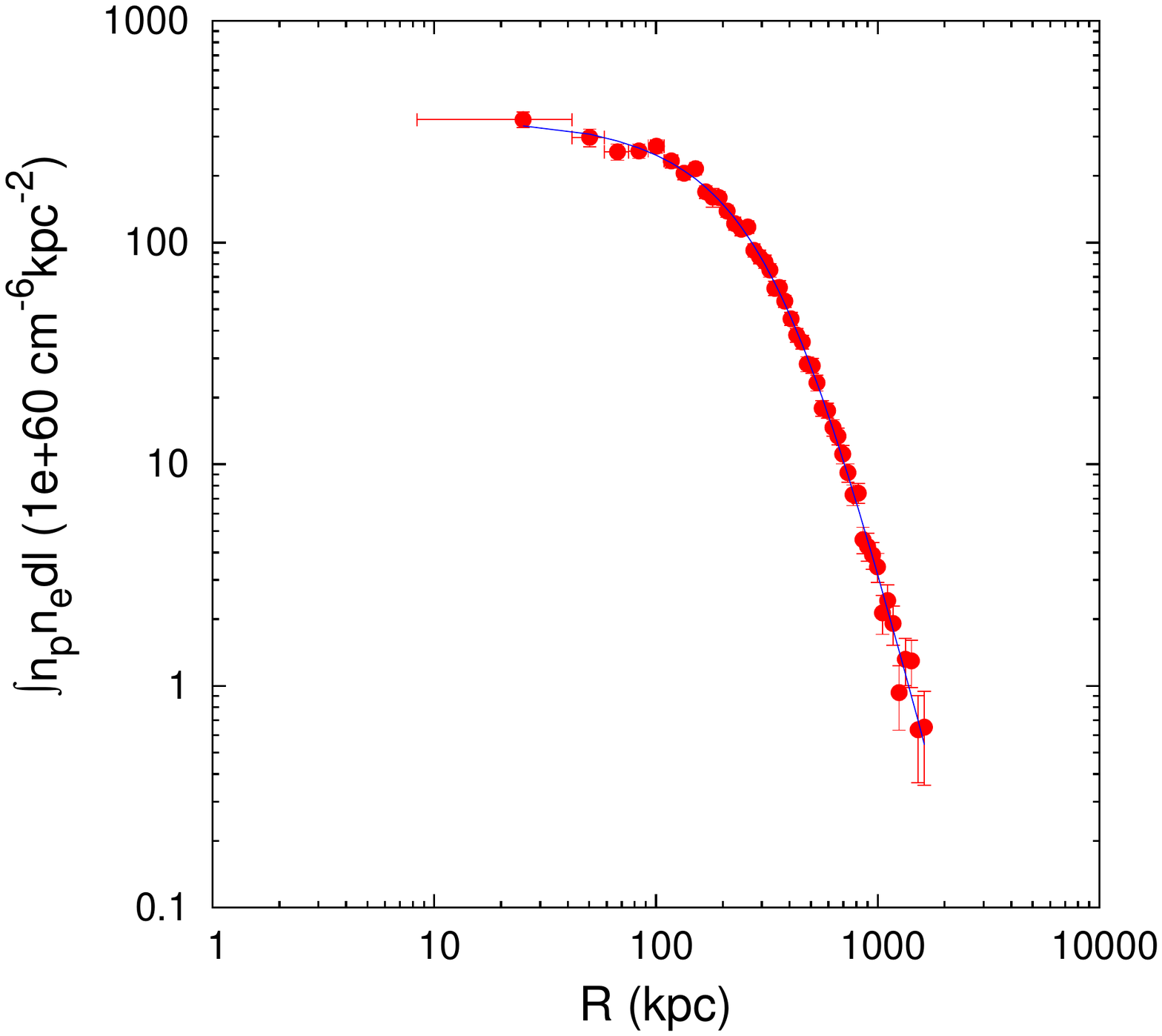}}}
\put(4.2,-0.21){\scalebox{0.30}{\includegraphics{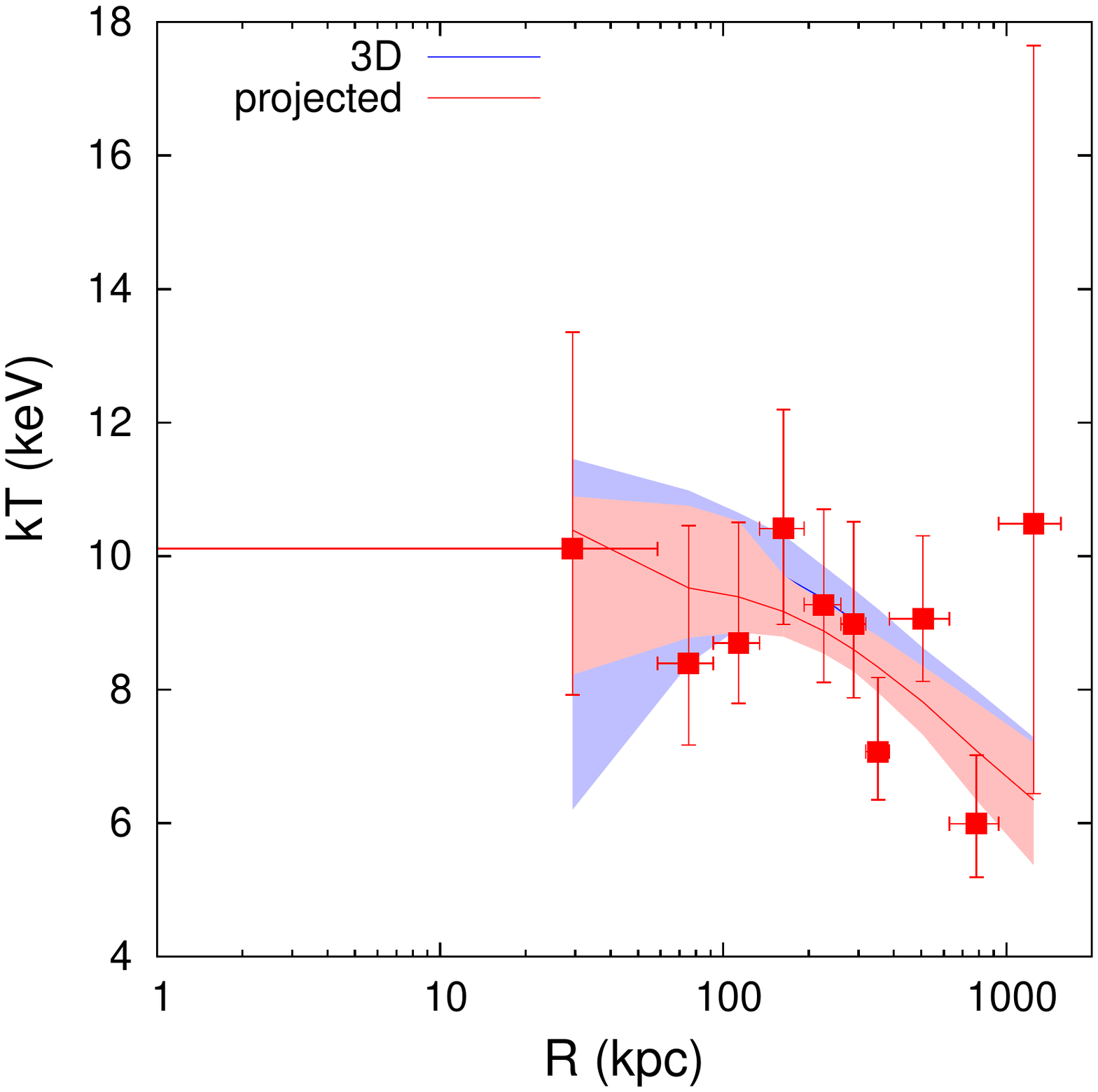}}}
\end{picture}
\end{center}
\caption{\small{Same as Figure~\ref{fig:a2204} but for A2631.}\label{fig:a2631}}
\end{figure*}

\begin{figure*}
\begin{center}
\setlength{\unitlength}{1in}
\begin{picture}(6.9,2.0)
\put(0.01,-0.8){\scalebox{0.34}{\includegraphics[clip=true]{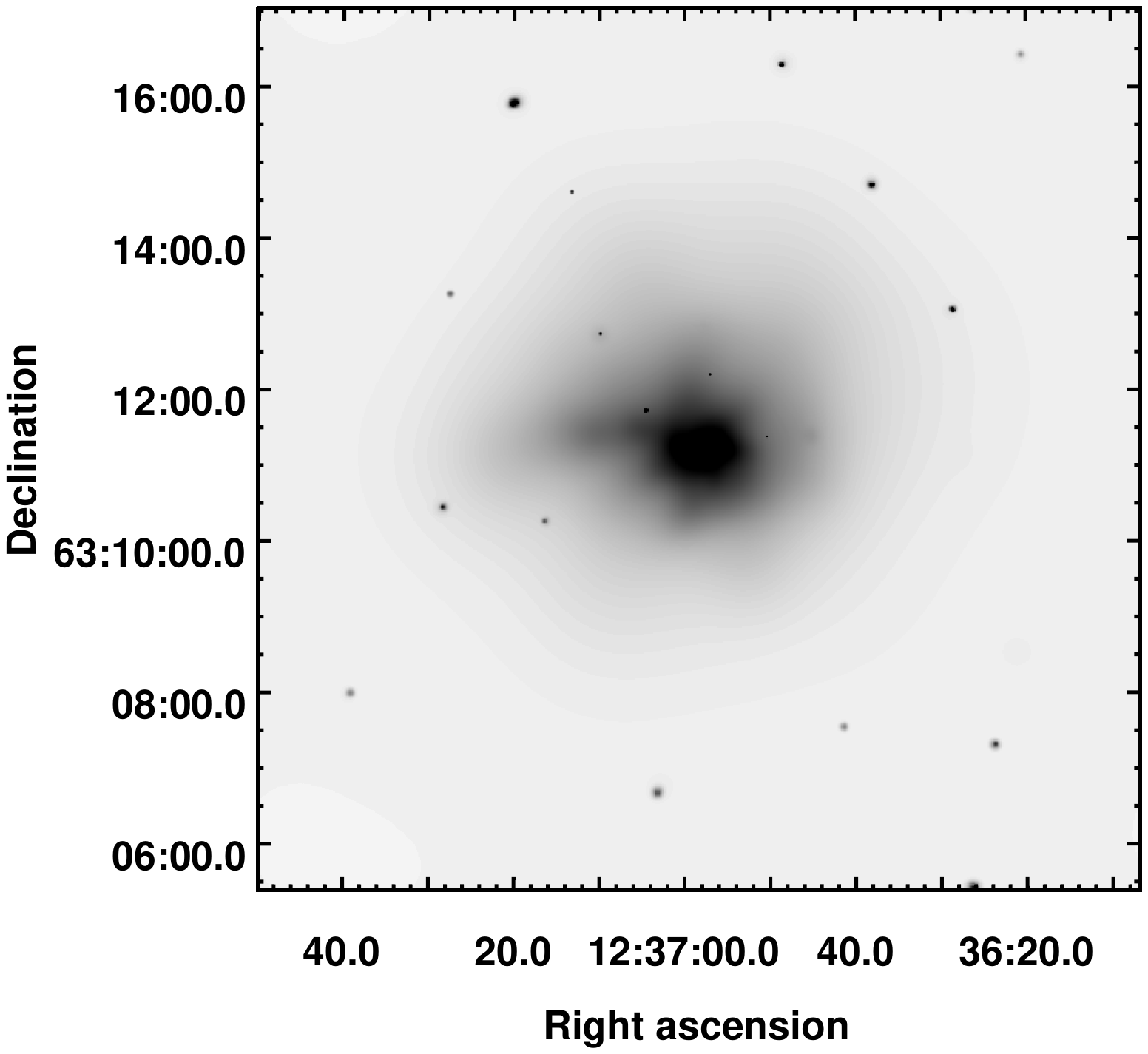}}}  
\put(2.1,-0.21){\scalebox{0.30}{\includegraphics{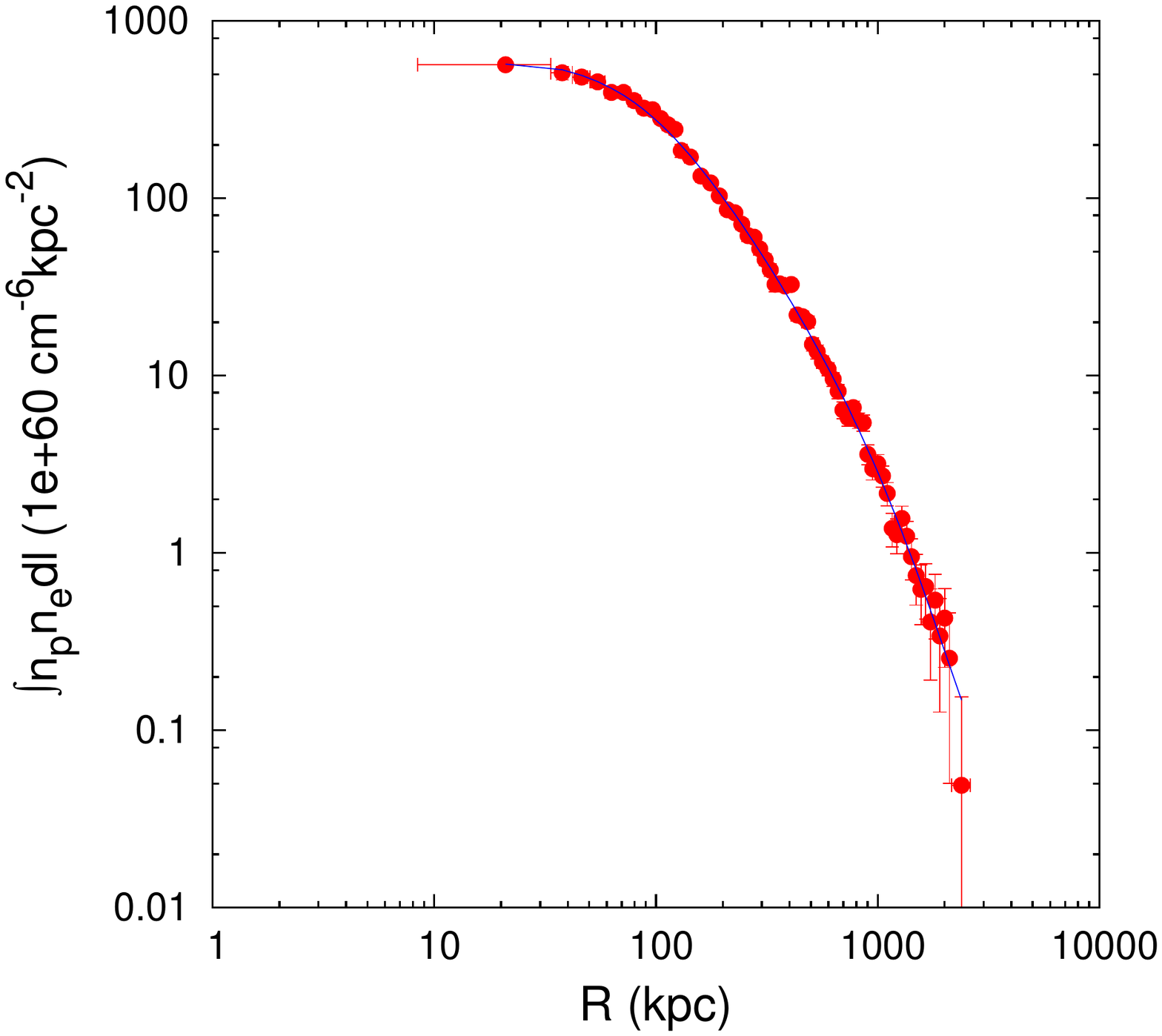}}}
\put(4.2,-0.21){\scalebox{0.30}{\includegraphics{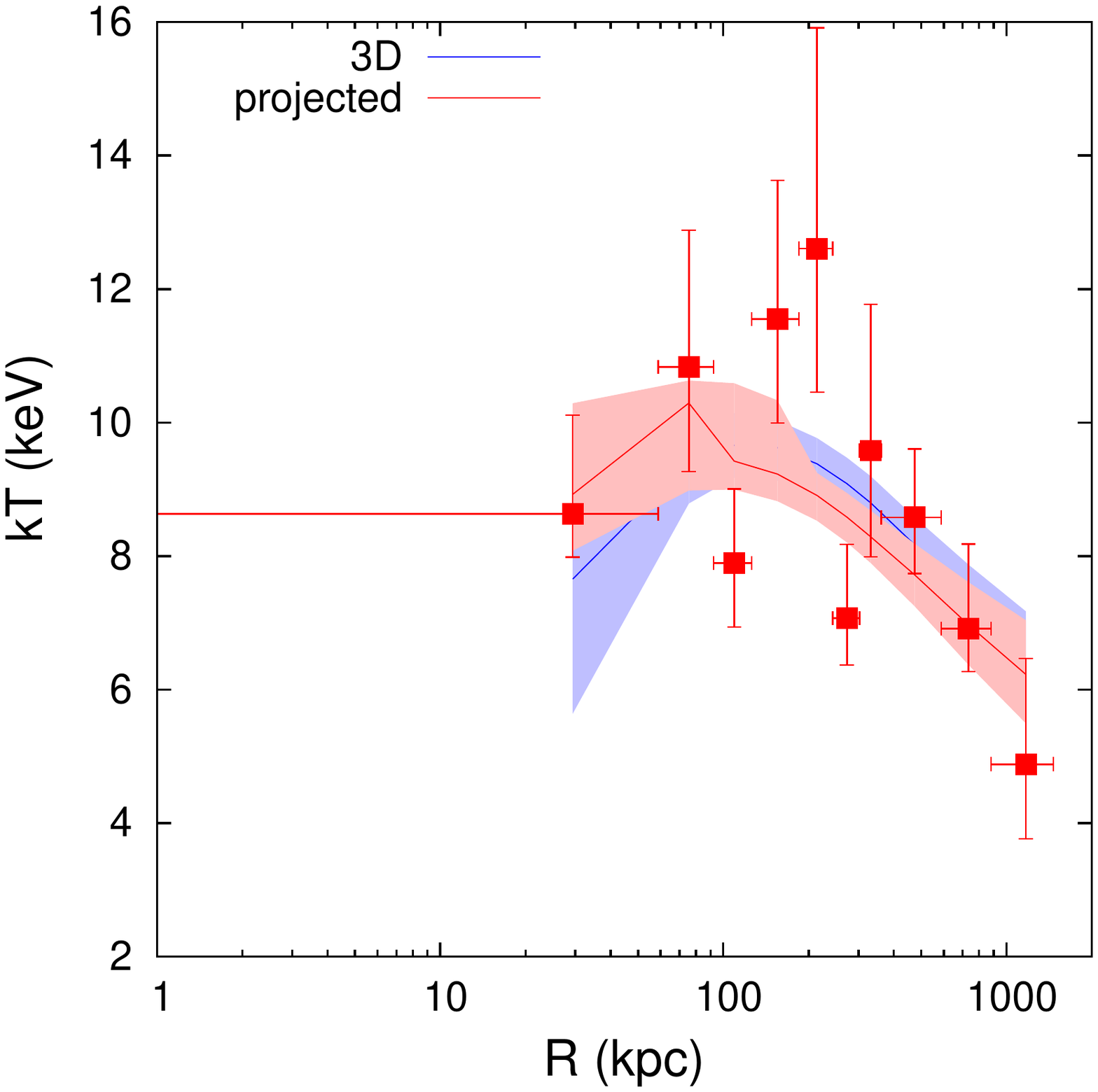}}}
\end{picture}
\end{center}
\caption{\small{Same as Figure~\ref{fig:a2204} but for A1576.}\label{fig:a1576}}
\end{figure*}

\begin{figure*}
\begin{center}
\setlength{\unitlength}{1in}
\begin{picture}(6.9,2.0)
\put(0.01,-0.8){\scalebox{0.34}{\includegraphics[clip=true]{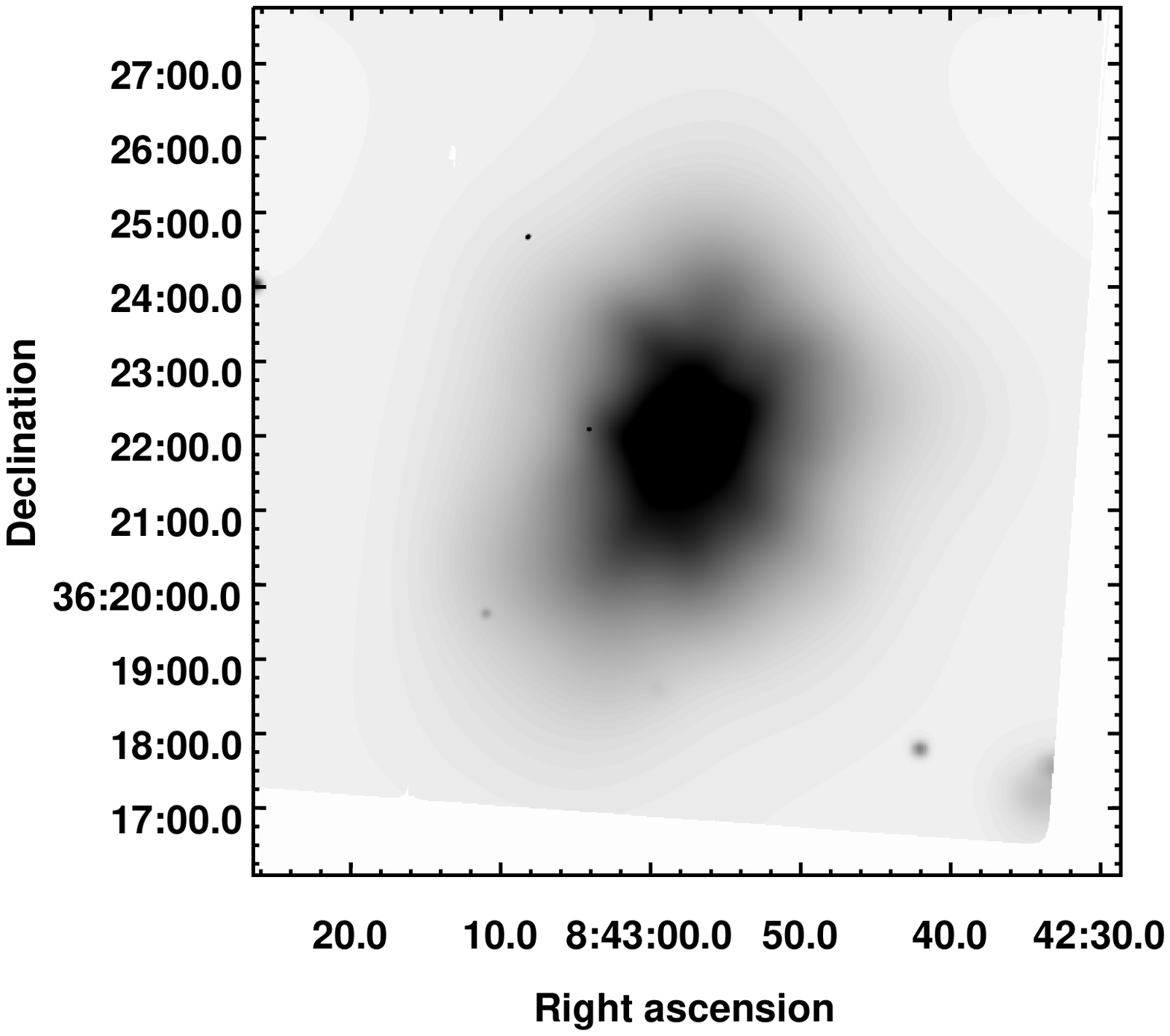}}}  
\put(2.1,-0.21){\scalebox{0.30}{\includegraphics{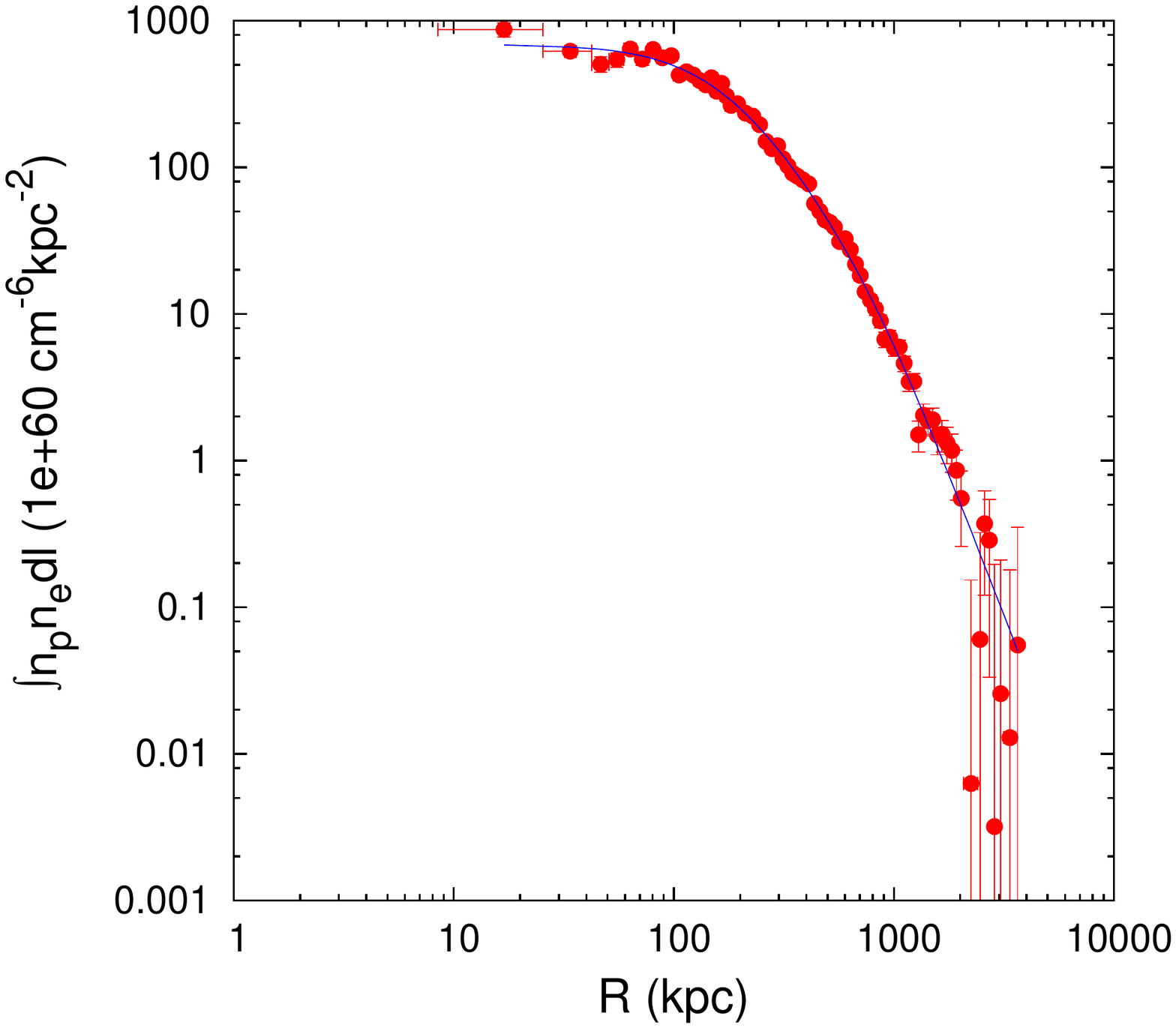}}}
\put(4.2,-0.21){\scalebox{0.30}{\includegraphics{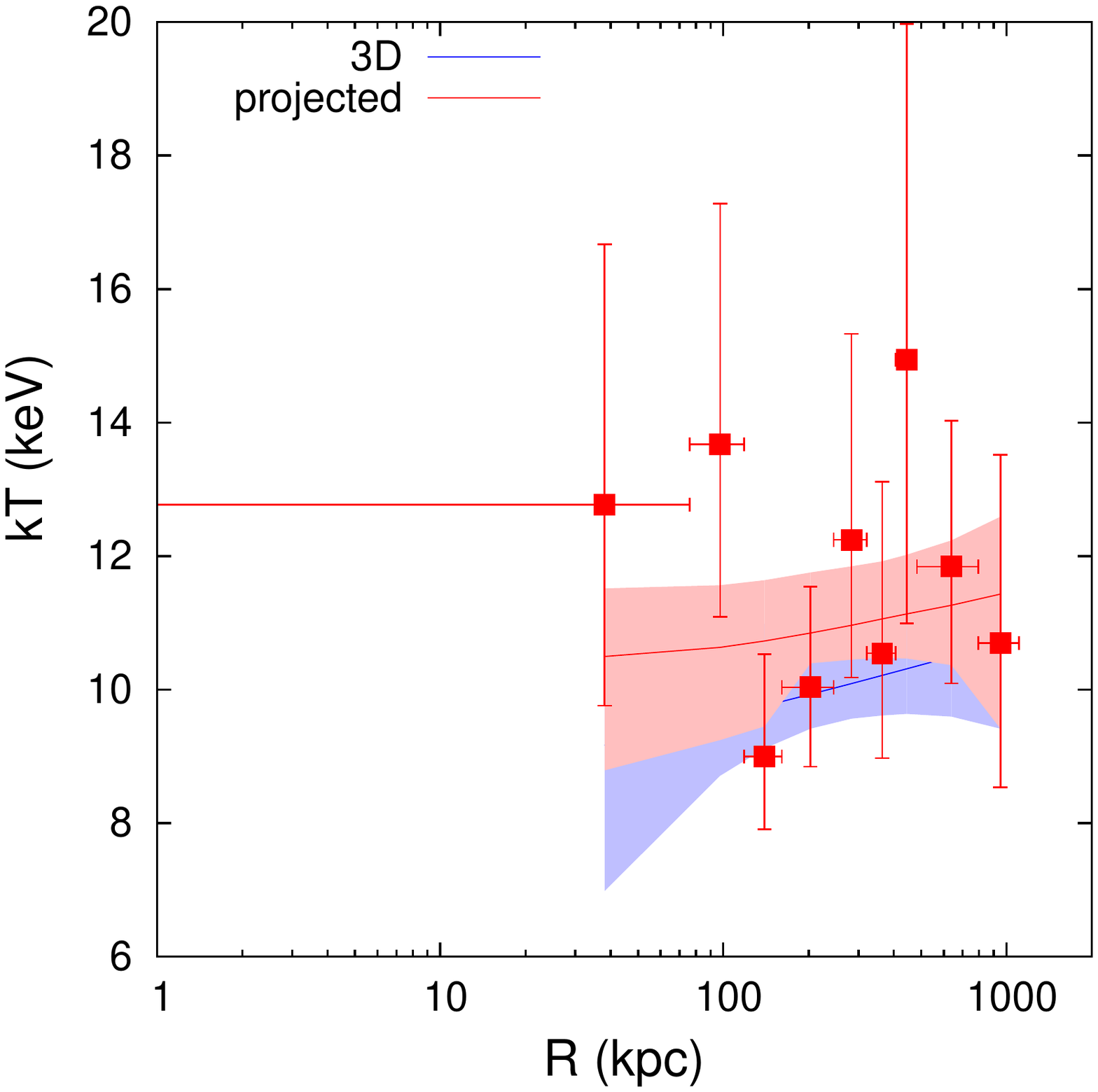}}}
\end{picture}
\end{center}
\caption{\small{Same as Figure~\ref{fig:a2204} but for A697.}\label{fig:a697}}
\end{figure*}

\begin{figure*}
\begin{center}
\setlength{\unitlength}{1in}
\begin{picture}(6.9,2.0)
\put(0.01,-0.8){\scalebox{0.34}{\includegraphics[clip=true]{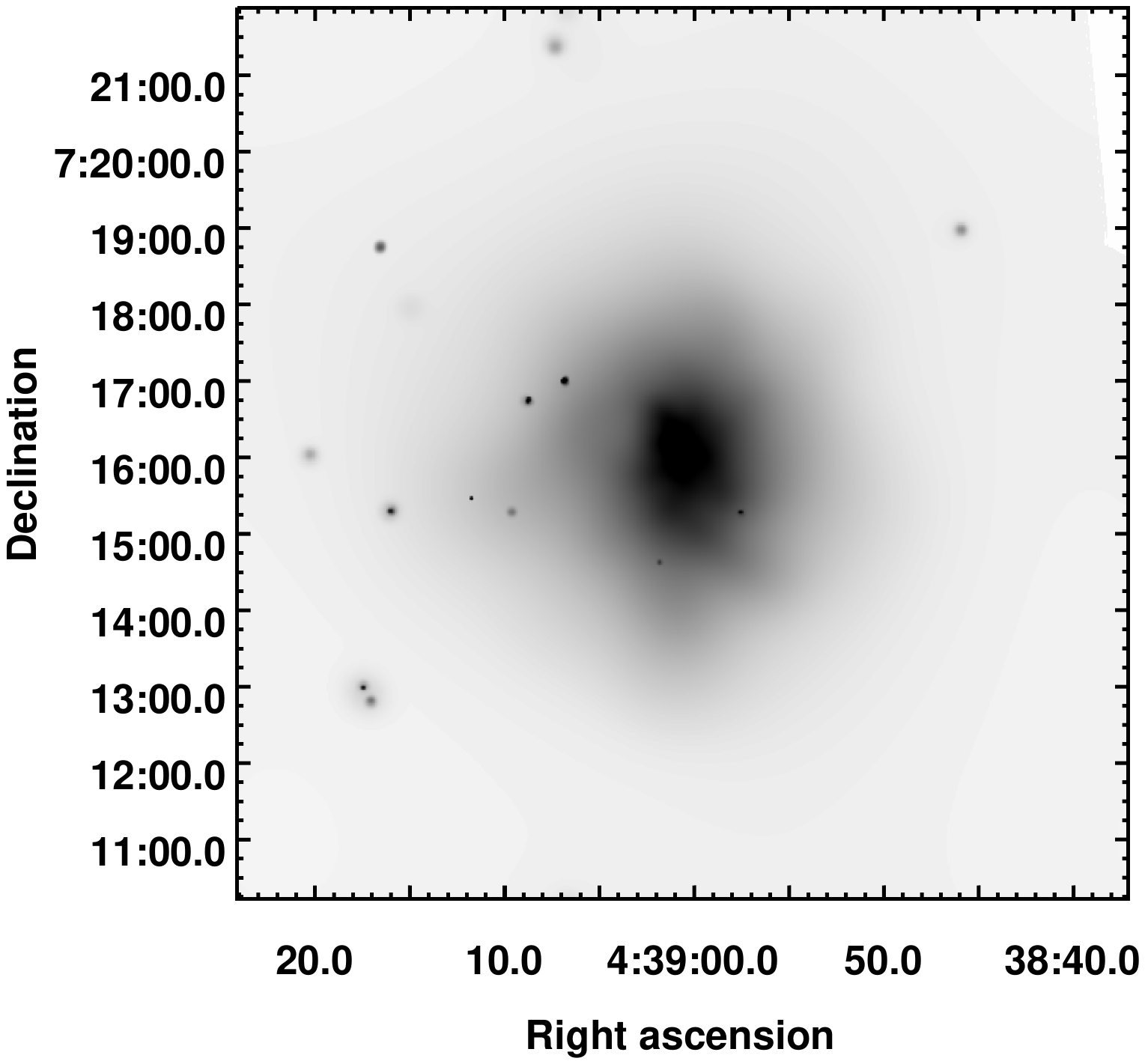}}}  
\put(2.1,-0.21){\scalebox{0.30}{\includegraphics{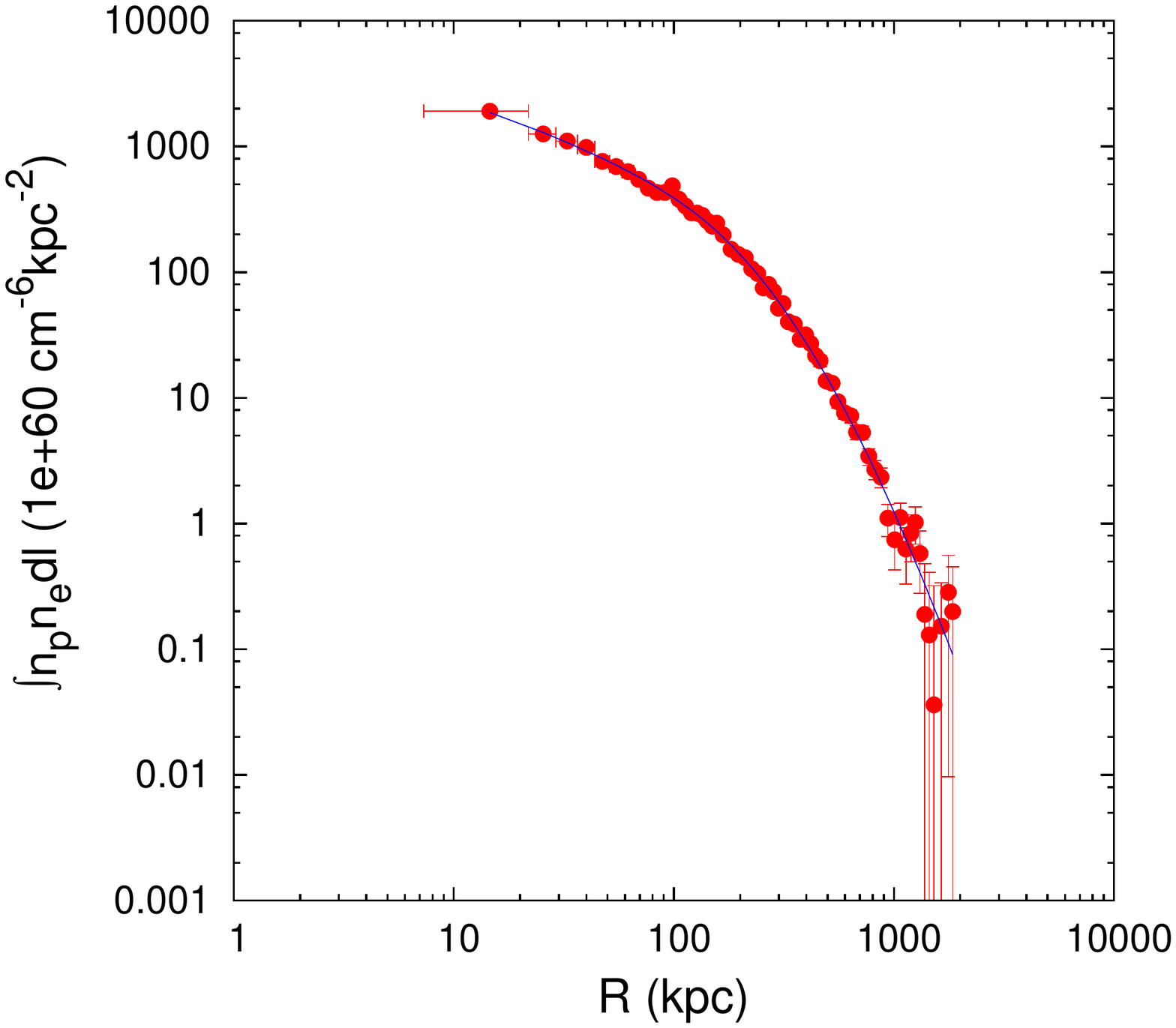}}}
\put(4.2,-0.21){\scalebox{0.30}{\includegraphics{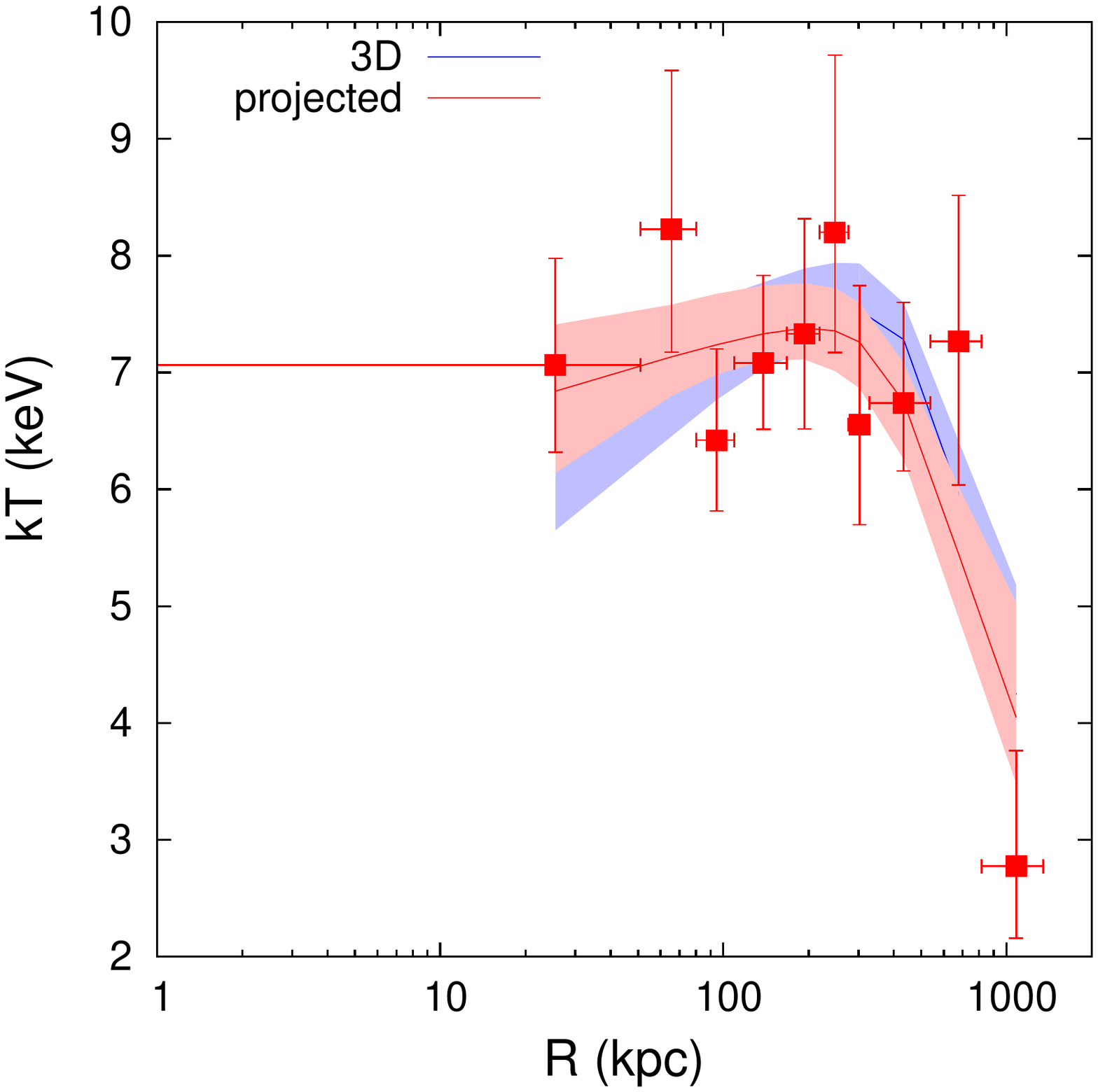}}}
\end{picture}
\end{center}
\caption{\small{Same as Figure~\ref{fig:a2204} but for RXJ0439.0+0715.}\label{fig:rxj0439}}
\end{figure*}

\begin{figure*}
\begin{center}
\setlength{\unitlength}{1in}
\begin{picture}(6.9,2.0)
\put(0.01,-0.8){\scalebox{0.34}{\includegraphics[clip=true]{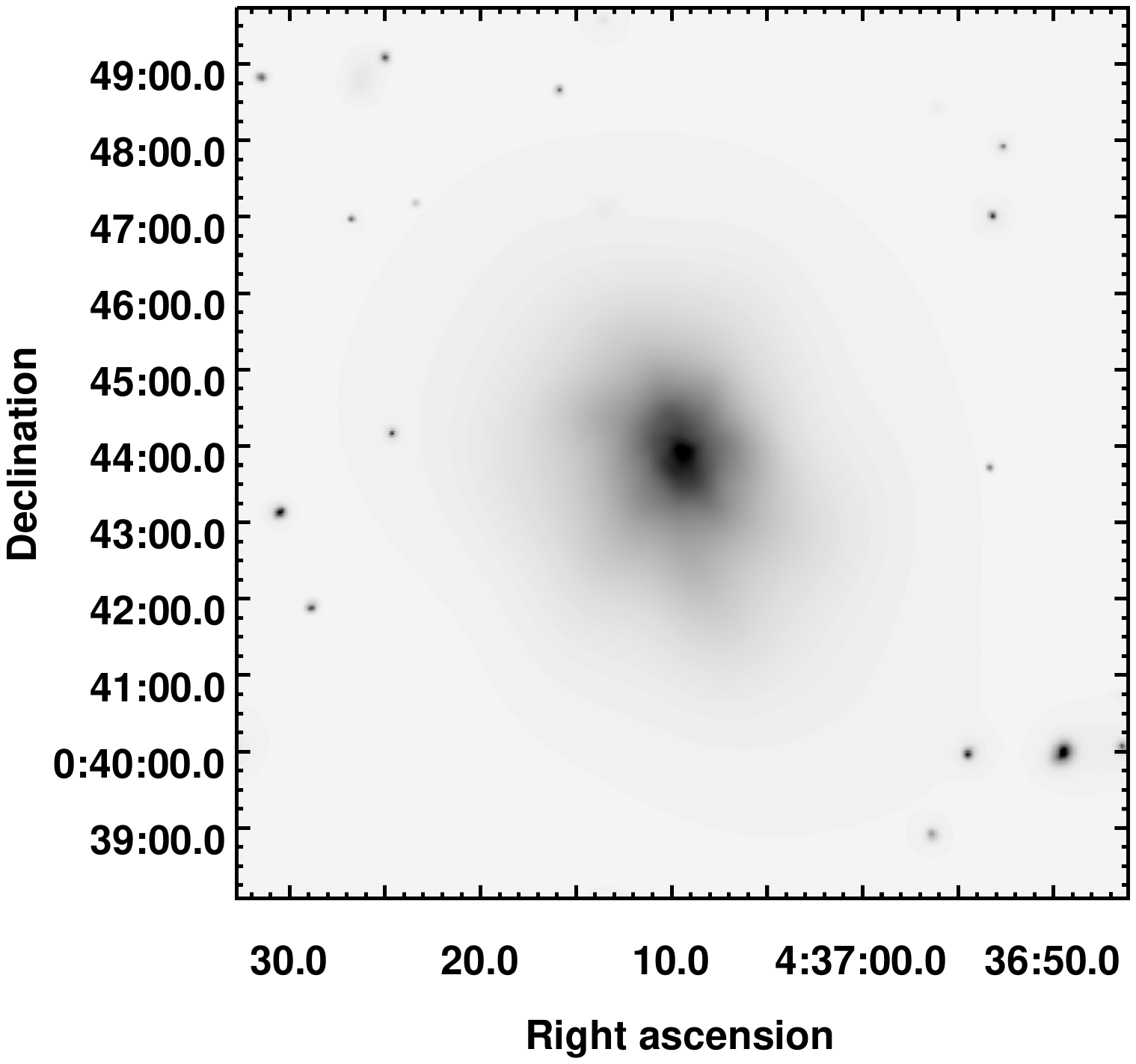}}}  
\put(2.1,-0.21){\scalebox{0.30}{\includegraphics{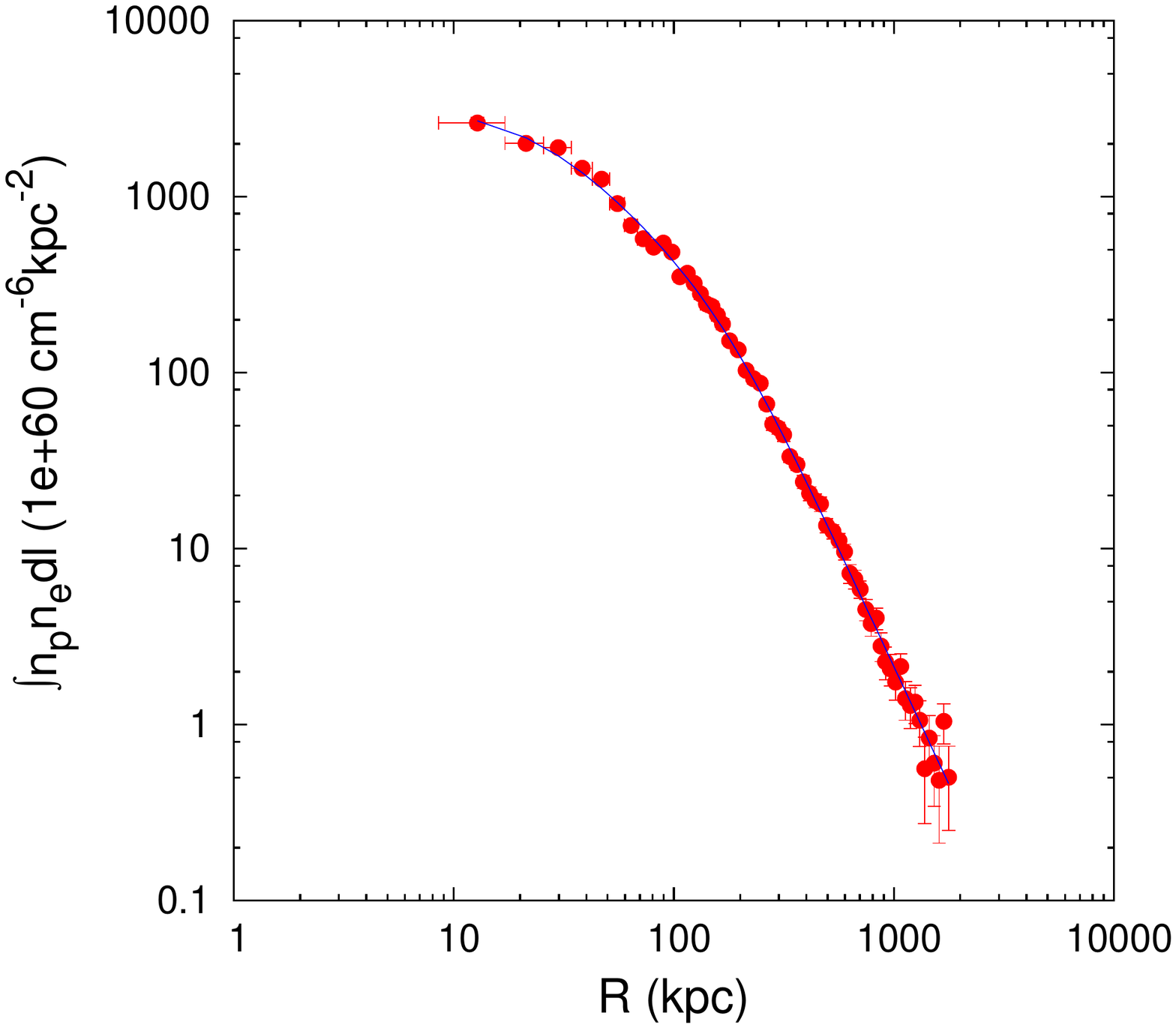}}}
\put(4.2,-0.21){\scalebox{0.30}{\includegraphics{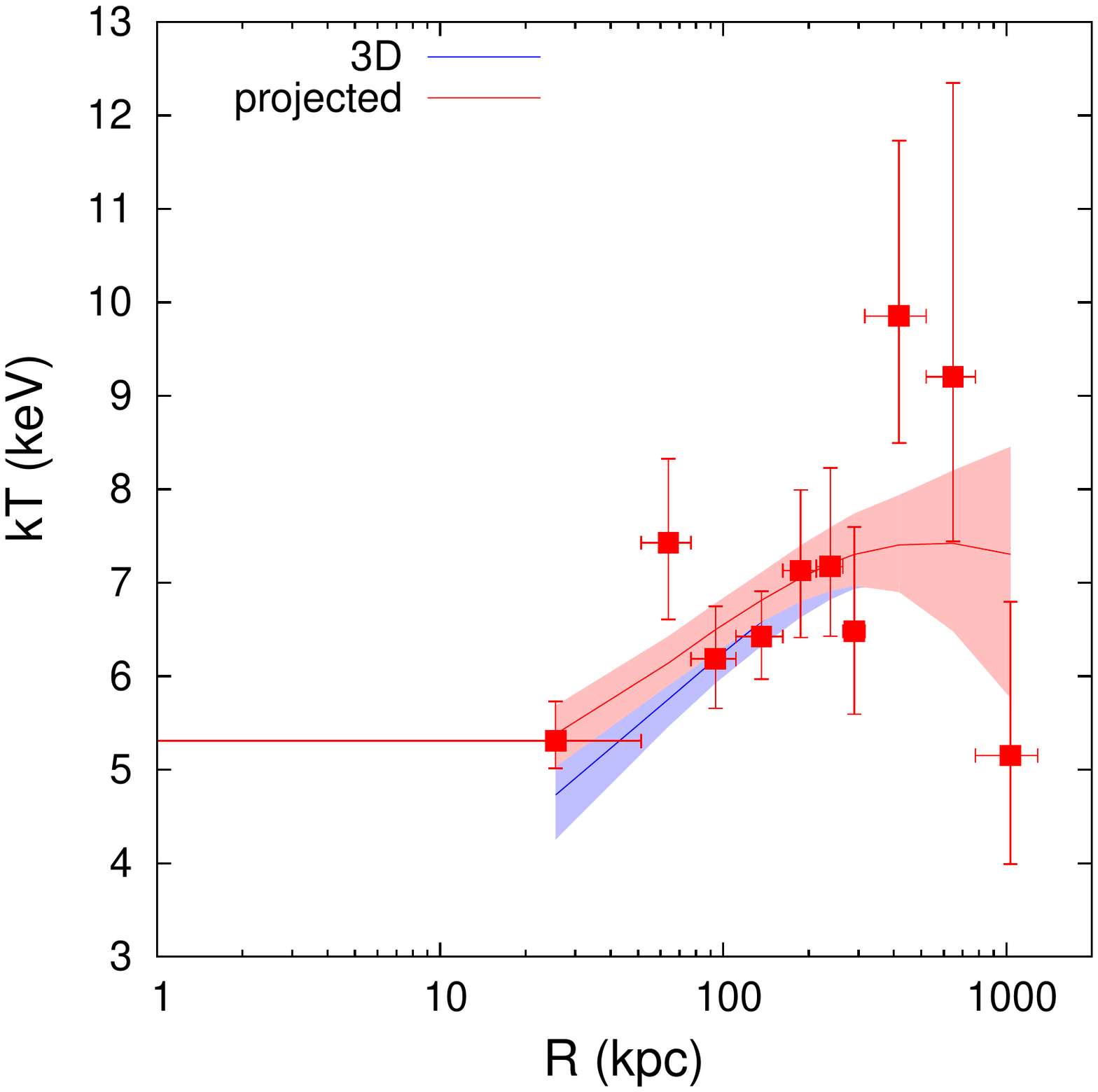}}}
\end{picture}
\end{center}
\caption{\small{Same as Figure~\ref{fig:a2204} but for RXJ0437.1+0043.}\label{fig:rxj0437}}
\end{figure*}

\begin{figure*}
\begin{center}
\setlength{\unitlength}{1in}
\begin{picture}(6.9,2.0)
\put(0.01,-0.8){\scalebox{0.34}{\includegraphics[clip=true]{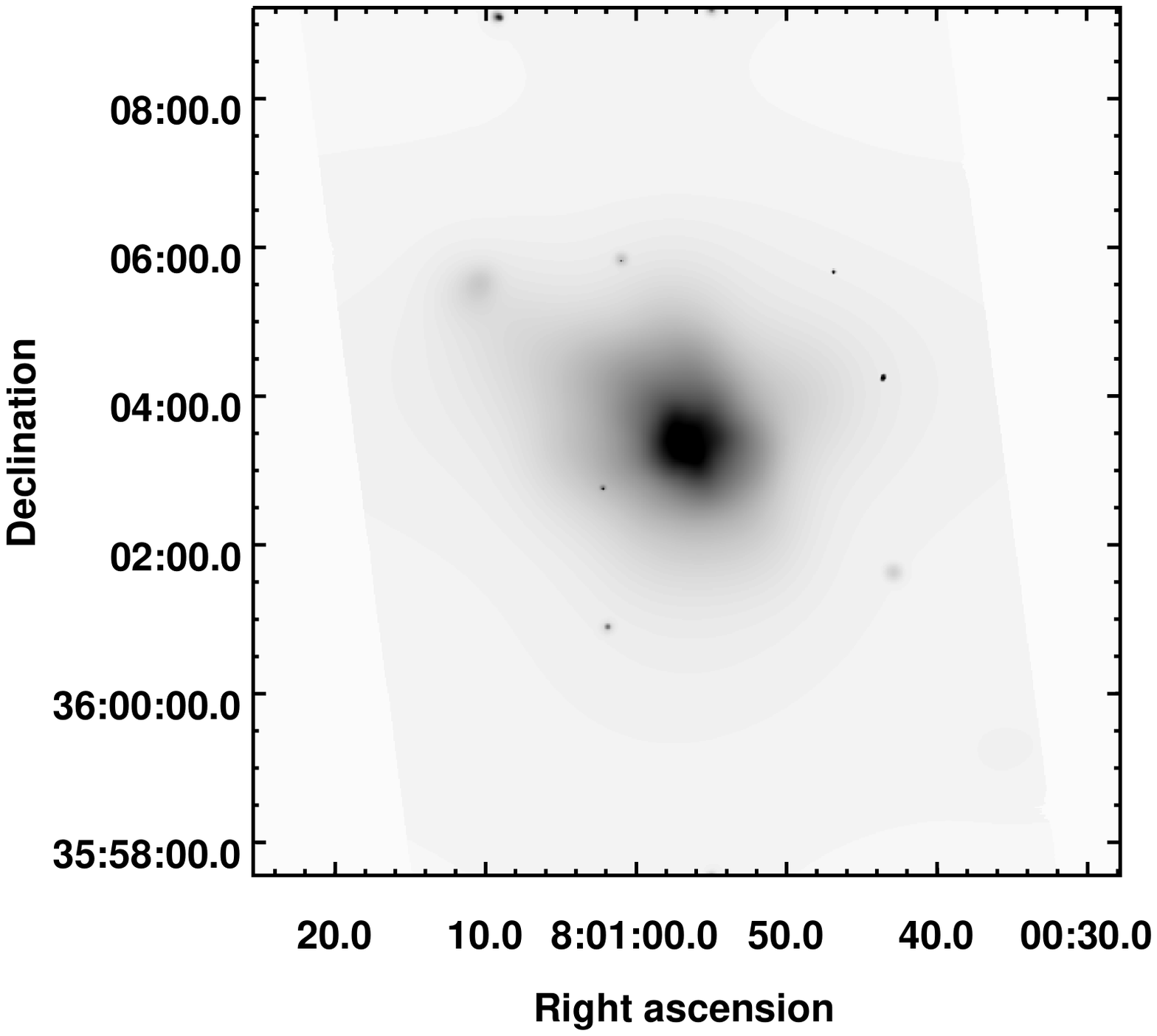}}}  
\put(2.1,-0.21){\scalebox{0.30}{\includegraphics{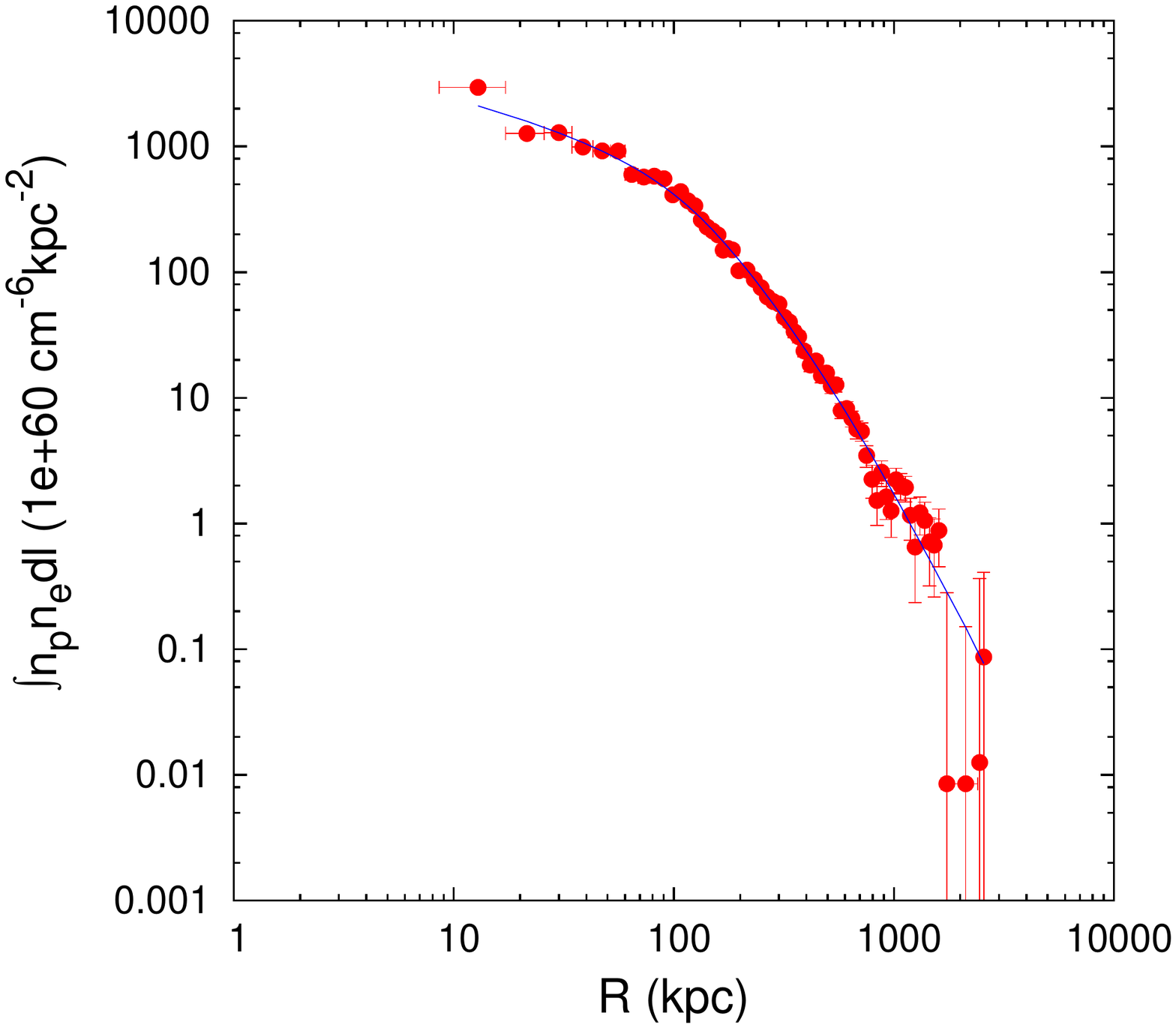}}}
\put(4.2,-0.21){\scalebox{0.30}{\includegraphics{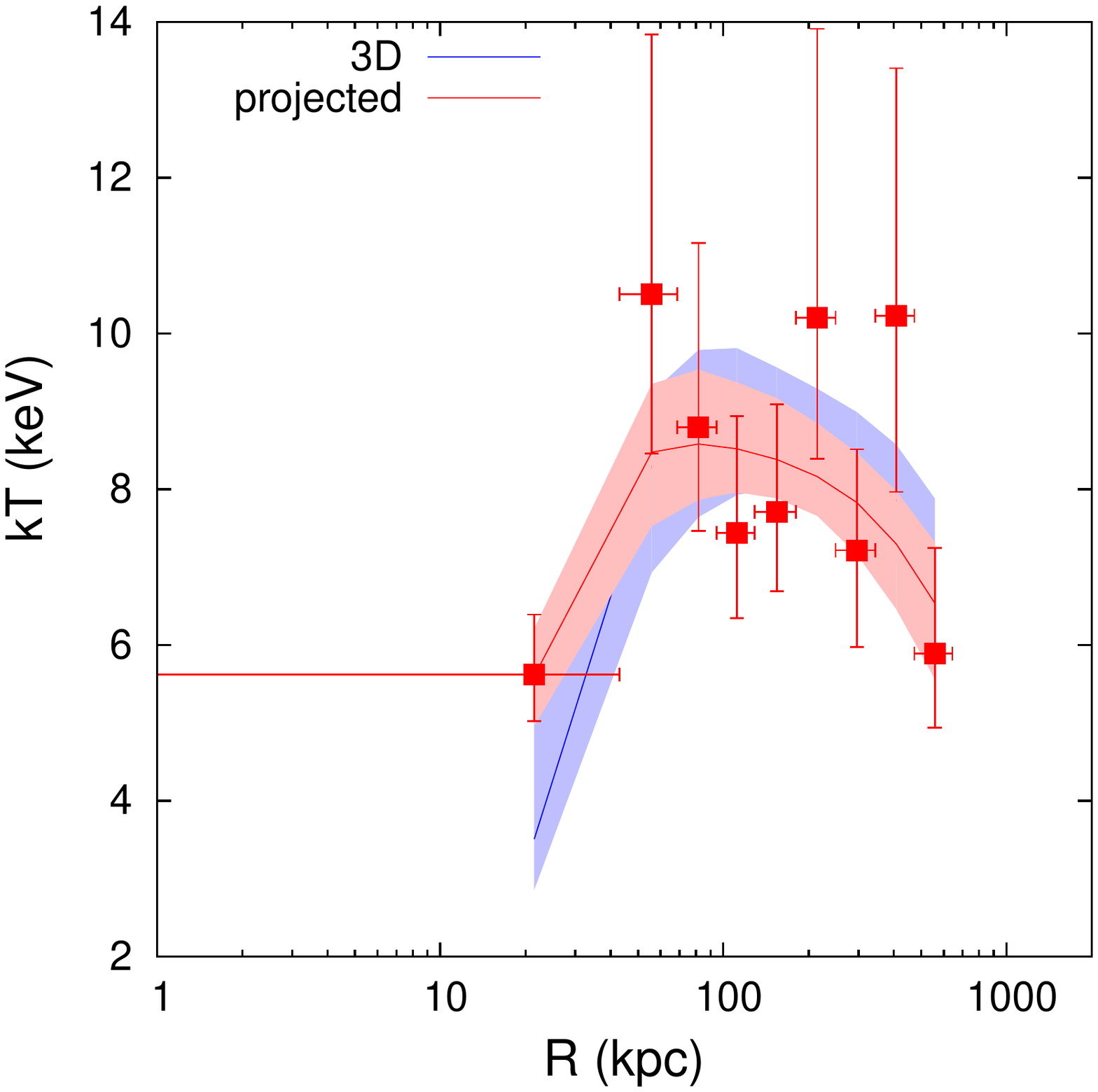}}}
\end{picture}
\end{center}
\caption{\small{Same as Figure~\ref{fig:a2204} but for A611.}\label{fig:a611}}
\end{figure*}

\begin{figure*}
\begin{center}
\setlength{\unitlength}{1in}
\begin{picture}(6.9,2.0)
\put(0.01,-0.8){\scalebox{0.34}{\includegraphics[clip=true]{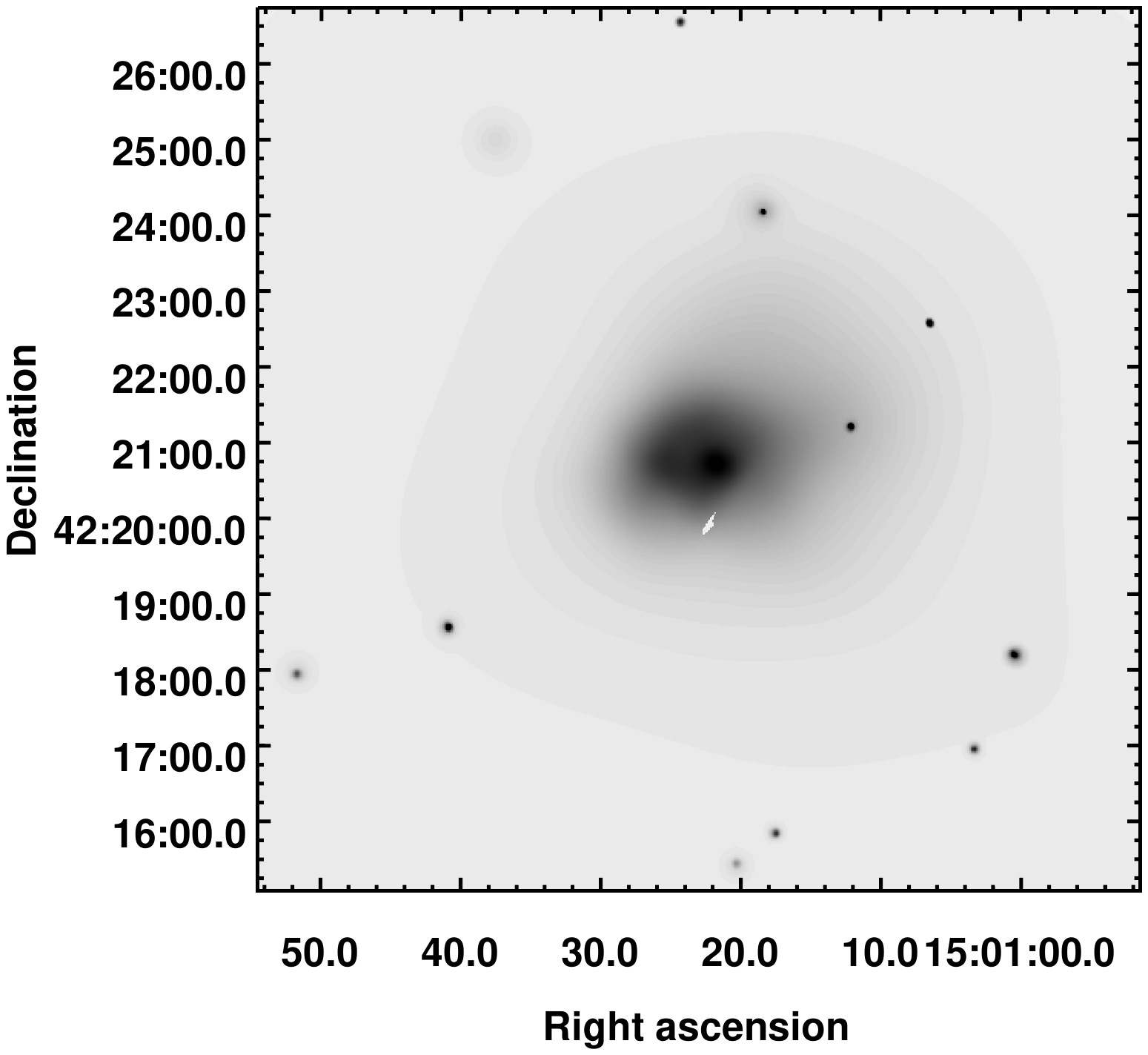}}}  
\put(2.1,-0.21){\scalebox{0.30}{\includegraphics{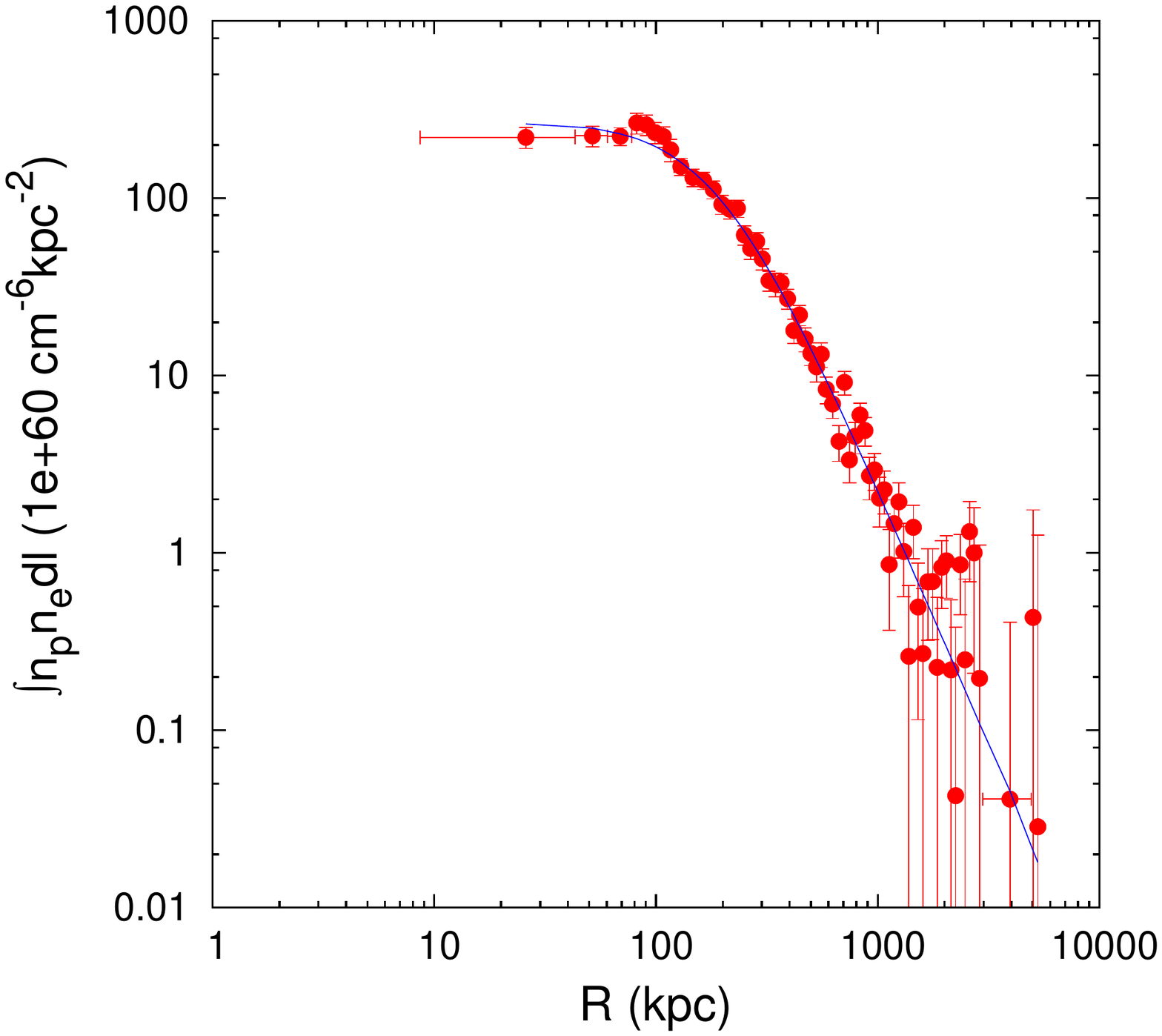}}}
\put(4.2,-0.21){\scalebox{0.30}{\includegraphics{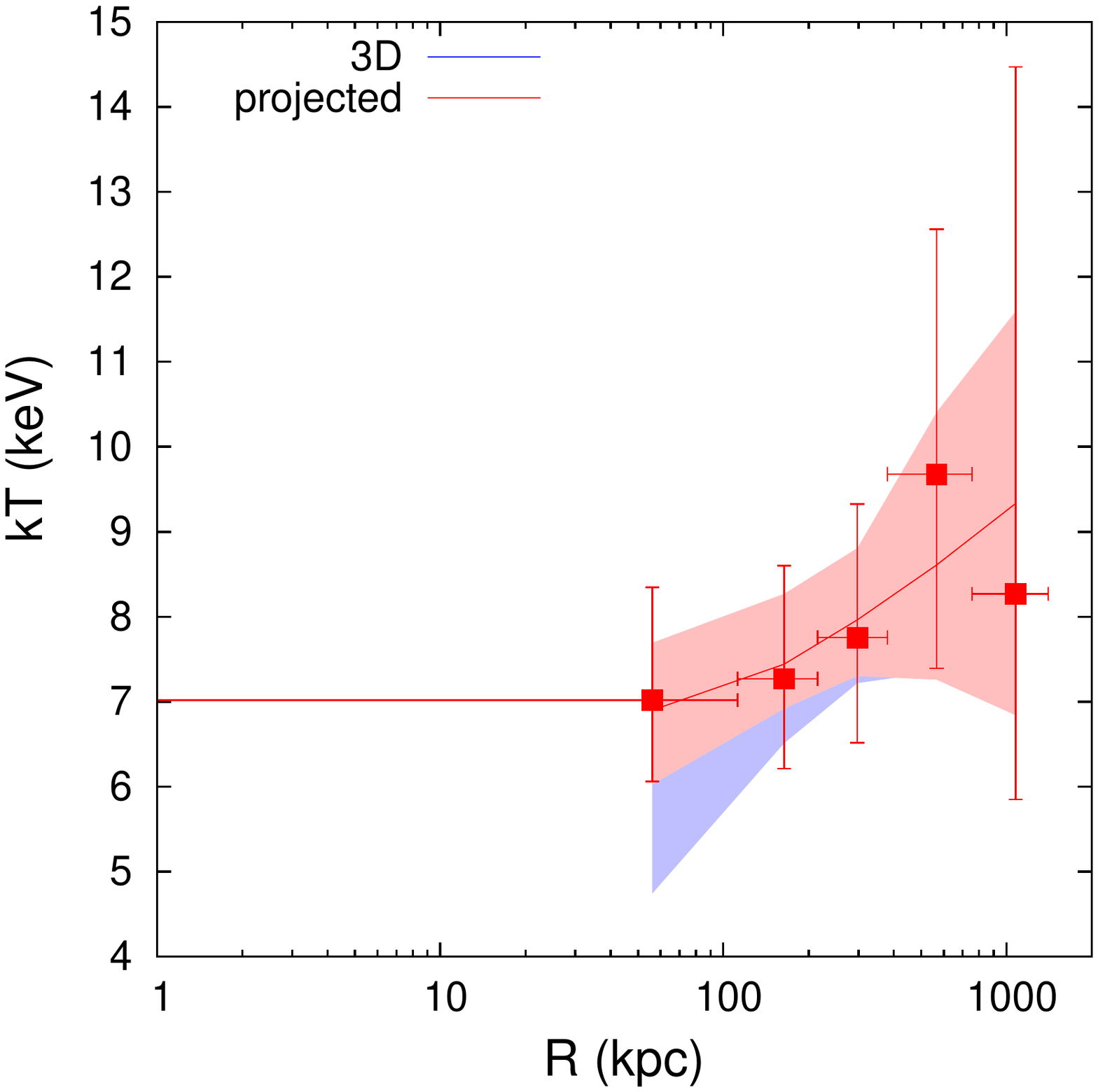}}}
\end{picture}
\end{center}
\caption{\small{Same as Figure~\ref{fig:a2204} but for Z7215.}\label{fig:z7215}}
\end{figure*}

\begin{figure*}
\begin{center}
\setlength{\unitlength}{1in}
\begin{picture}(6.9,2.0)
\put(0.01,-0.8){\scalebox{0.34}{\includegraphics[clip=true]{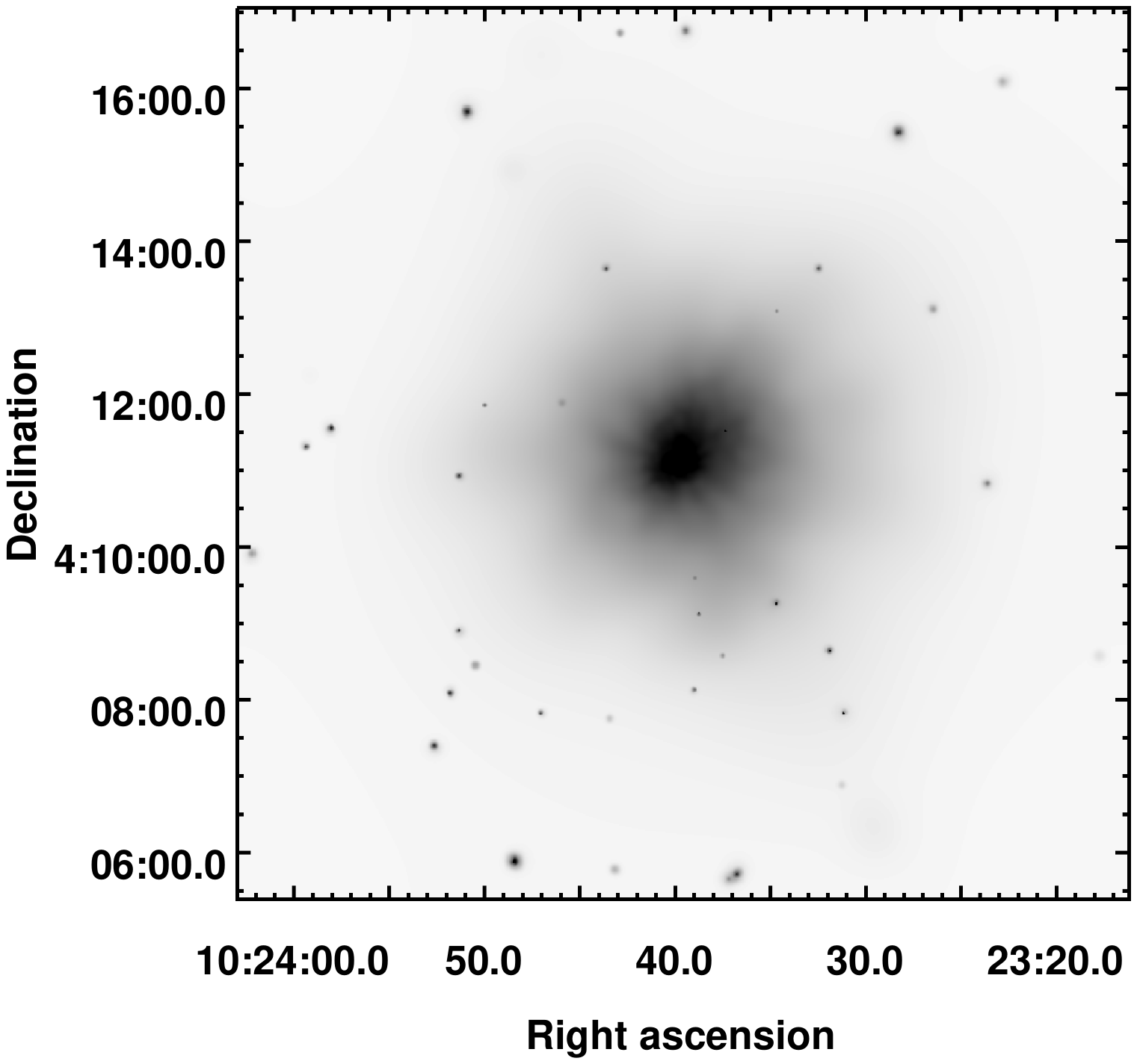}}}  
\put(2.1,-0.21){\scalebox{0.30}{\includegraphics{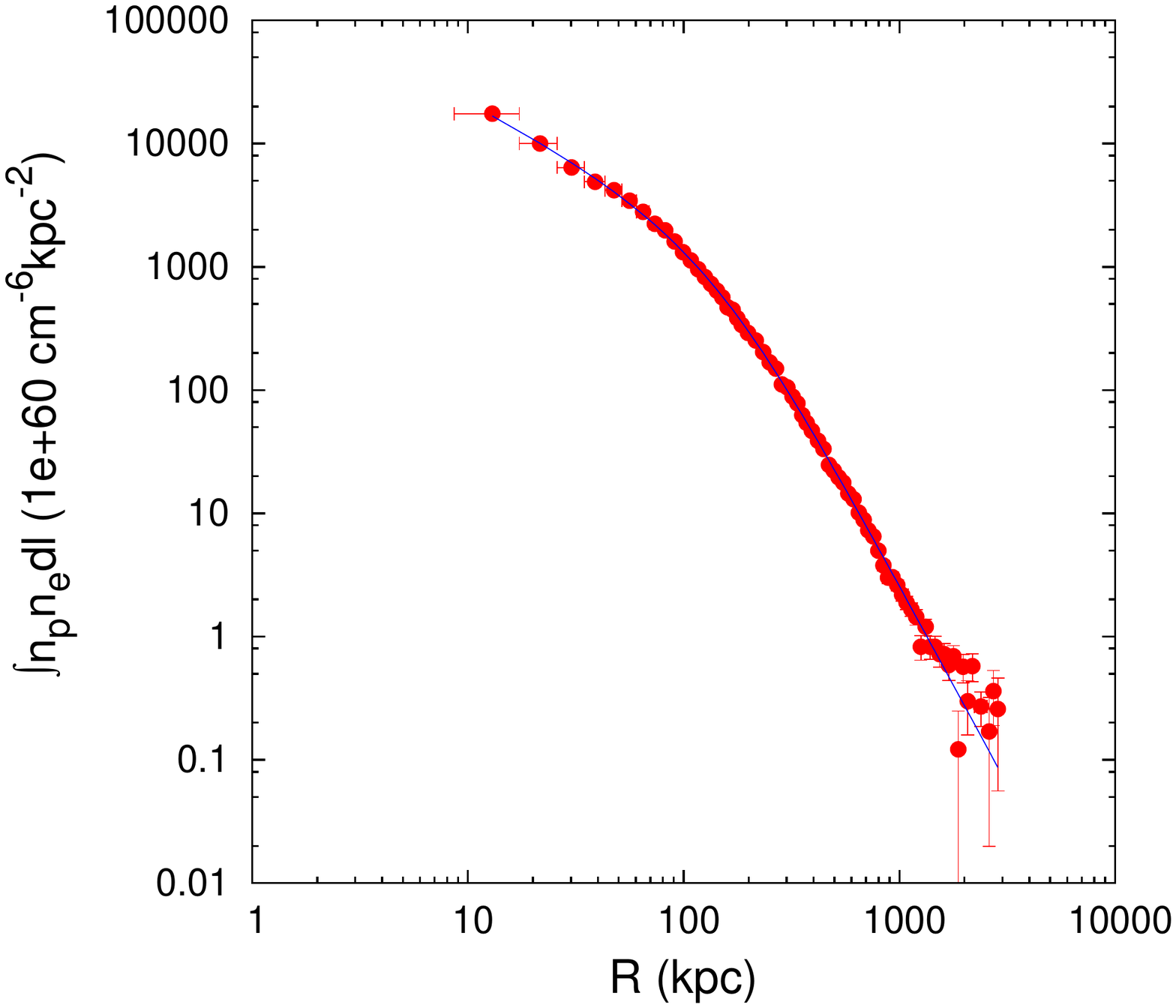}}}
\put(4.2,-0.21){\scalebox{0.30}{\includegraphics{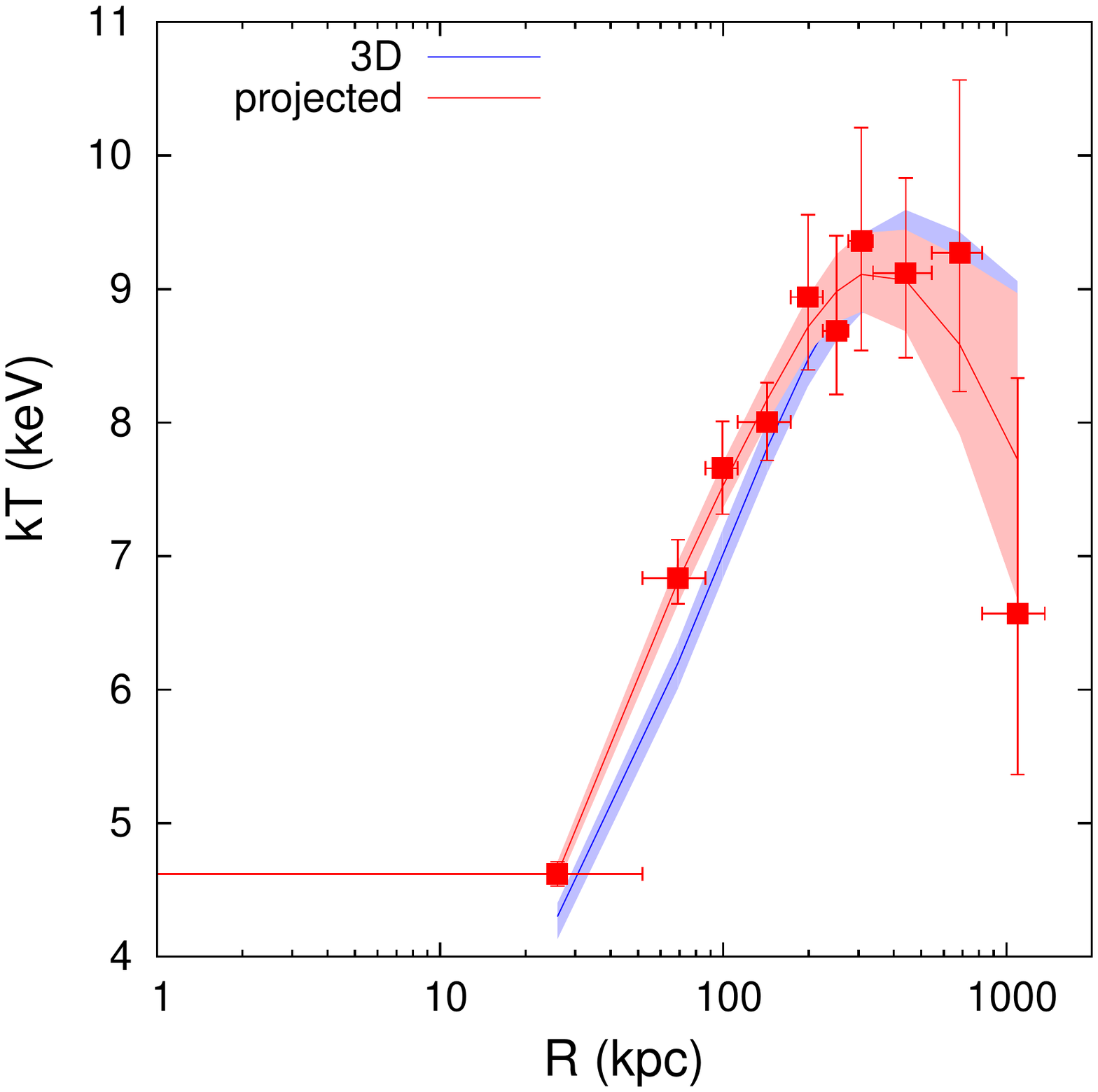}}}
\end{picture}
\end{center}
\caption{\small{Same as Figure~\ref{fig:a2204} but for Z3146.}\label{fig:z3146}}
\end{figure*}

\begin{figure*}
\begin{center}
\setlength{\unitlength}{1in}
\begin{picture}(6.9,2.0)
\put(0.01,-0.8){\scalebox{0.34}{\includegraphics[clip=true]{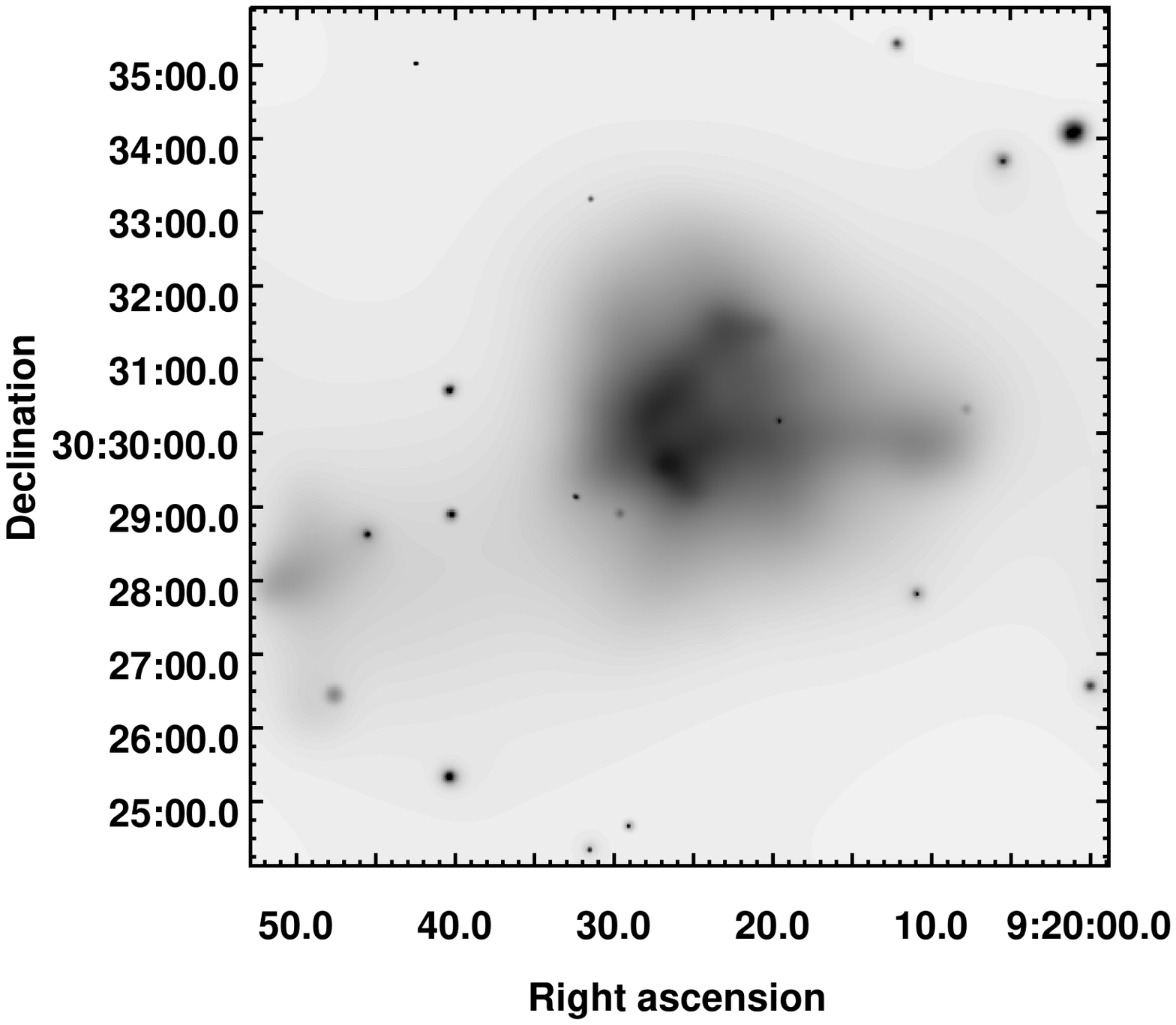}}}  
\put(2.1,-0.21){\scalebox{0.30}{\includegraphics{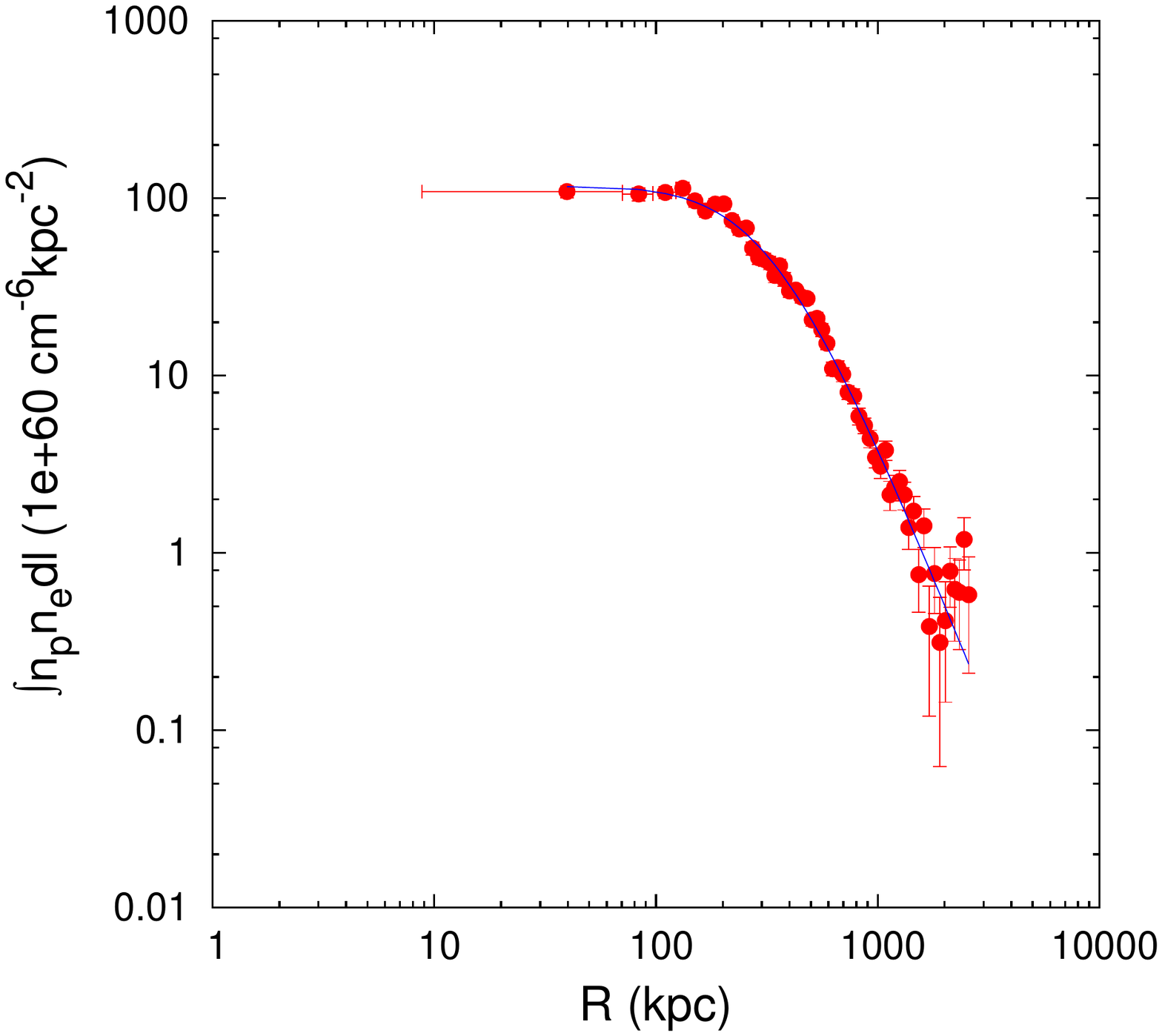}}}
\put(4.2,-0.21){\scalebox{0.30}{\includegraphics{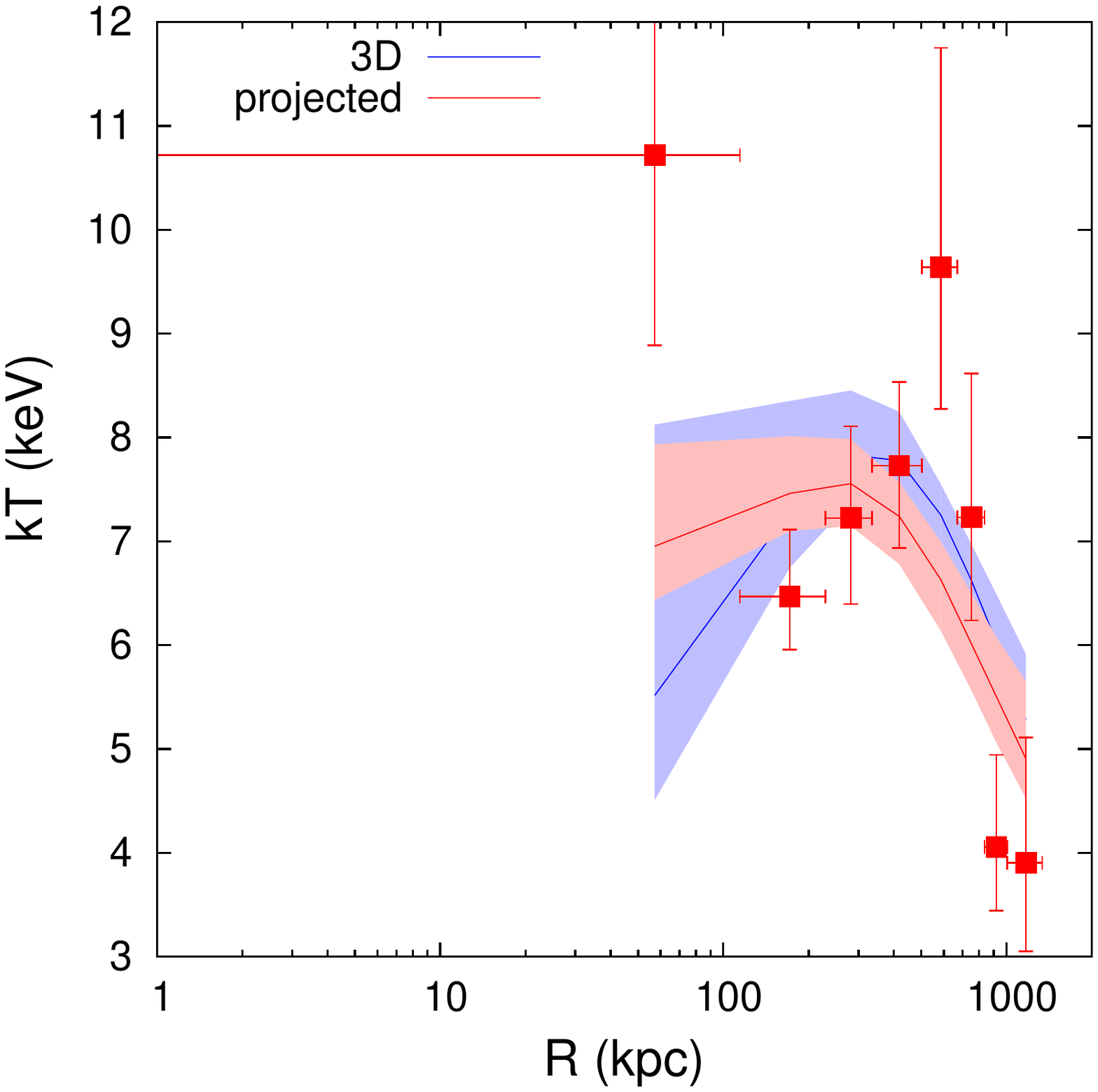}}}
\end{picture}
\end{center}
\caption{\small{Same as Figure~\ref{fig:a2204} but for A781.}\label{fig:a781}}
\end{figure*}

\bsp

\label{lastpage}

\end{document}